\documentclass[10pt,openany,leqno]{book}
\usepackage[dvips]{color}
\usepackage[english]{babel} 
\usepackage{mathrsfs} 
\usepackage{amssymb} 
\usepackage[dvips]{graphicx} 
\usepackage{graphicx} 
\usepackage{latexsym} 
\usepackage{pictex}
\usepackage{amsmath}
\usepackage{indentfirst}
\usepackage{float}
\usepackage{fix-cm}
\usepackage{fancybox}
\usepackage{array}

\newtheorem{theorem}{\bf Theorem} [chapter] 
\newtheorem{definition}{\bf Definition} [chapter] 
\newtheorem{remark}{\bf Remark} [chapter] 
\newtheorem{lemma}{\bf Lemma} [chapter] 
\newcommand{\be}{\begin{equation}}
\newcommand{\ee}{\end{equation}}
\newcommand{\bt}{\begin{theorem} -- }
\newcommand{\et}{\end{theorem}}
\newcommand{\blm}{\begin{lemma} -- }
\newcommand{\elm}{\end{lemma}}
\newcommand{\br}{\begin{remark} -- \rm }
\newcommand{\er}{$\bullet$ \end{remark}}
\newcommand{\erx}{\end{remark}}
\newcommand{\bd}{\begin{definition} -- \rm }
\newcommand{\ed}{\end{definition}}
\newcommand{\ba}{\begin{array}{ll}}
\newcommand{\ea}{\end{array}}
\newcommand{\bc}{\left\{ \begin{array}{l}}
\newcommand{\ec}{\end{array}\right.}
\newcommand{\bq}{\left[ \begin{array}{l}}
\newcommand{\eq}{\end{array}\right.}
\newcommand{\bprf}{\noindent {\small\scshape Proof --} }
\newcommand{\eprf}{$\scc\blacksquare$ \vskip 4pt}
\newcommand{\eprfx}{\quad\scc\blacksquare}
\newcommand{\bs}{\boldsymbol}
\newcommand{\ff}{\mathsf}
\newcommand{\dint}{\displaystyle\int}
\newcommand{\Z}{{\scc 0}}
\newcommand{\uno}{{_1}}
\newcommand{\due}{{_2}}
\newcommand{\tre}{{_3}}
\newcommand{\wt}{\widetilde}

\newcommand{\arccosh}{\hbox{\rm arccosh}}

\def\Dim{\hbox{\rm Dim\,}}
\def\Lim{\displaystyle\lim}
\newcommand{\firma}{$\sc\mathscr{S\! B}$} 
\newcommand{\To}{\Longrightarrow}
\newcommand{\Def}{\overset{\rm def}=}
\newcommand{\bboxed}[1]{\boxed{\vp\;#1\;}}
\def\sc{\scriptstyle}
\def\scc{\scriptscriptstyle}
            
\font\did=cmr10		  
\font\ssmall=cmr7		  
\newcommand{\spesso}{\Huge .}
\newcommand{\spessored}{\hbox{\textcolor{red}{\Huge .}}}
\newcommand{\spessoblue}{\hbox{\textcolor{blue}{\Huge .}}}

\newcommand{\spessino}{\did.}
\newcommand{\spessinoblue}{\hbox{\textcolor{blue}{\huge .}}}
\newcommand{\spessinored}{\hbox{\textcolor{red}{\huge .}}}

\newcommand{\normale}{\setplotsymbol({\fiverm .}) \plotsymbolspacing=.4pt}
\newcommand{\normaleblue}{\hbox{\textcolor{blue}{\large .}}}
\newcommand{\normalered}{\hbox{\textcolor{red}{\large .}}}
\newcommand{\normalegreen}{\textcolor{green}{\large .}}
\newcommand{\normalemag}{\textcolor{magenta}{\large .}}
\newcommand{\vp}{\rule[-1.4mm]{0mm}{5mm}}
\newcommand{\vpx}{\rule[-1.5mm]{0mm}{4.5mm}}
\newcommand{\vpud}{\rule[-4mm]{0mm}{10mm}}

\newcommand{\vspd}{\rule[-2mm]{0mm}{2mm}}
\newcommand{\vspu}{\rule[0mm]{0mm}{4mm}}

\newcommand{\vpt}{\rule[-8pt]{0pt}{20pt}}

\newcommand{\ax}{\\[1mm]}
\newcommand{\ac}{\\[8pt]}
\newcommand{\acc}{\\[12pt]}

\newcommand{\barxi}{\overline{\xi}} 
\newcommand{\wtell}{\widetilde{\ell}} 
\newcommand{\comp}{\kern 1mm \hbox{\ssy \char'016} \kern 1mm}
\graphicspath{%
    {converted_graphics/}
    {/}
}

\begin{document}
\author{Sergio Benenti}
\title{\Huge Mathematical Models in Isotropic Cosmology}
\date{Version May 2016 \\ . \\ 
Abstract. An axiomatic approach to the mathematical models of the isotropic cosmology.
}

\frontmatter

\maketitle
\thispagestyle{empty}

\mainmatter
\pagestyle{myheadings}
\renewcommand{\chaptermark}[1] {\markboth{\did\chaptername \ \thechapter.\ #1}{}} 
\renewcommand{\sectionmark}[1] {\markright{\footnotesize \rm \thesection.\ #1}{}}

\pagenumbering{roman}

\section*{Preface}

1. This work comes from my own need to understand the mathematical basics of modern Cosmology, and it is therefore the result of a research (at the present incomplete) aimed to collect the countless notions of this discipline in a logically ordered system, starting from a number of postulates.

2. The postulates proposed here are very simple and intuitive. They include the {\bf isotropy principle} (the { \it homogeneity principle} comes out as a theorem). As a consequence, the mathematical models that follow cannot describe the complex physical phenomena occurring at the beginning of the universe, as well as other current phenomena  regarding the dark matter and the background radiation. Nevertheless, despite the extreme simplification, one of the surprising facts is that the obtained estimates of the age of the universe as well as of other quantities  are in full agreement with those advanced by astrophysicists.  Anyway, as for any axiomatic theory, our approach is of course open to criticisms, adjustments and extensions. 

3. The philosophy adopted in conducting this work is based on the following rules:

(i) To distinguish the various subjects under investigation between   those having a purely observational or geometrical character (they are incorporated into the chapter of the cosmic kinematics) and those that involve dynamical concepts and laws (as energy-momentum tensors, Einstein field equations, etc.).

(ii) Accompany the mathematical analysis with the geometric vision, and
perform ordinary differential calculus on differential manifolds avoiding the use special kinds of coordinates. It will not be necessary to use, for example,  the typical coordinates of the Friedman-Robertson-Walker metric.

(iii) Pay attention to the logical order in which the definitions and theorems have to be placed.

(iv) Pay a special attention to the dimensionality of the physical or geometrical objects. This is a way for testing the correctness of the calculations and of the numerical evaluations.

(v) Last but not least, be free of preconceptions. For example, do not think {\it ab initio} that the space-time of the relativistic cosmology is a manifold with a Lorentzian metric (as everybody knows) because we have to find out this fact, at the appropriate time, as a theorem. For this reason we cannot start from  the celebrated {\it Weyl principle}, which imposes from the very beginning the existence of a Lorentzian metric for which the world-lines of the galaxies are time-like curves. In my opinion this principle lies at a too advanced position. In  mathematical terms, the Weyl principle ranks in the category of Riemannian (or semi-Riemannian) manifolds,  while it seems more appropriate to begin from a more general category,  as that of the differentiable manifolds. In doing so, we get two advantages: we can locate primary concepts and their relationships in the right order and at the right place and, second, we do not lose secondary but noteworthy concepts, which otherwise would remain hidden.

4. That said, the schedule of this work is the following.

\noindent $\bullet$ {\scshape Chapter 1} (Cosmic kinematics). We compose the basic geometrical structures of the {\bf cosmic space-time}, as the set of all the {\bf cosmic events}. This set is the union of two {\bf singular events} $\alpha$ and $\omega$, representing the beginning (or birth) and the end (or death) of the universe, and a four-dimensional manifold $M$ of {\bf non-singular events}. The manifold will be endowed with two fibrations (Fig.\ \ref{fig:basic}). 
\newcommand{\omino}{
$\beginpicture
\setcoordinatesystem units <1.6mm,1.6mm>
\setplotarea x from -8 to 8, y from -7 to 7
\normalgraphs
\put{\footnotesize $\sc\bullet$} at -1.4 0
\setlinear
\plot -1.4 0 0 0 /
\plot 0 0 .5 .4 1 .5 /
\plot 0 0 1 -.5 /
\plot -.9 0 -.4 .15 -.2 .5 /
\plot -.9 0 -.4 -.4 -.2 -.2 /
\endpicture$}
\begin{figure}[H]
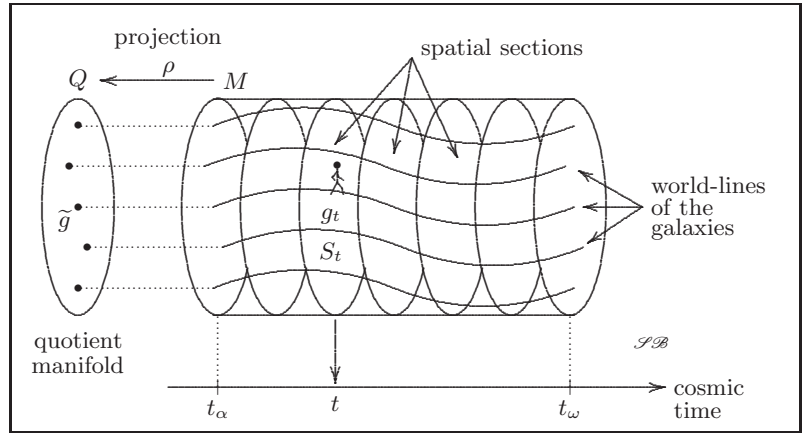

  \centering
$$
\beginpicture
\setcoordinatesystem units <.6cm,.6cm>
\setplotarea x from -8.5 to 9, y from -5 to 4.5
\normalgraphs
\small
\grid 1 1
\startrotation by 0 -1 about 0 0
\setlinear
\plot -2.4 4  -2.4 -4 /
\plot 2.4 4 2.4 -4 /
\ellipticalarc axes ratio 3:1 290 degrees from -1.4 -3.25 center at 0 -3.9
\ellipticalarc axes ratio 3:1 290 degrees from -1.4 -1.95 center at 0 -2.6
\ellipticalarc axes ratio 3:1 290 degrees from -1.4 -.65 center at 0 -1.3
\ellipticalarc axes ratio 3:1 290 degrees from -1.4 .65 center at 0 0
\ellipticalarc axes ratio 3:1 290 degrees from -1.4 1.95 center at 0 1.3
\ellipticalarc axes ratio 3:1 290 degrees from -1.4 3.25 center at 0 2.6
\ellipticalarc axes ratio 3:1 360 degrees from 2.4 3.9 center at 0 3.9
\put{\omino} at -.55 -1.35
\stoprotation
\arrow <6pt> [.2,.6] from -5 -4  to 6 -4  
\put{$S_t$} at -1.4 -1
\put{$g_t$} at -1.4 -.2
\arrow <6pt> [.2,.6] from -1.3 -2.5 to -1.3 -3.9
\put{$t$} [t] at -1.3 -4.2
\setlinear 
\setdots<2pt>
\plot -3.9 -2.45 -3.9 -4 /
\put{$t_\alpha$} [t] at -3.9 -4.3
\put{$\sc |$} at -3.9 -4
\plot 3.9 -2.45 3.9 -4 /
\put{$t_\omega$} [t] at 3.9 -4.3
\put{$\sc |$} at 3.9 -4
\setsolid
\setquadratic
\plot 
-4 0 
-2 .4 
0 0
2 -.4
4 0 /
\put{\beginpicture
\setquadratic
\plot 
-4 0 
-2 .4 
0 0
2 -.4
4 0 /
\endpicture} at -.2 .9
\put{\beginpicture
\setquadratic
\plot 
-4 0 
-2 .4 
0 0
2 -.4
4 0 /
\endpicture} at 0 1.8
\put{\beginpicture
\setquadratic
\plot 
-4 0 
-2 .4 
0 0
2 -.4
4 0 /
\endpicture} at .2 -.9
\put{\beginpicture
\setquadratic
\plot 
-4 0 
-2 .4 
0 0
2 -.4
4 0 /
\endpicture} at 0 -1.8
\ellipticalarc axes ratio 1:3 360 degrees from -7 2.4 center at -7 0
\setlinear 
\setdots<2pt>
\plot -4 -1.8 -7 -1.8 /
\plot -3.8 -.9 -6.8 -.9 /
\plot -4 0 -7 0 /
\plot -4.2 .9 -7.2 .9 /
\plot -4 1.8 -7 1.8 /
\setsolid
\put{$\sc\bullet$} at  -7 -1.8 
\put{$\sc\bullet$} at  -7 1.8 
\put{$\sc\bullet$} at  -6.8 -.9 
\put{$\sc\bullet$} at  -7 0 
\put{$\sc\bullet$} at  -7.2 .9
\put{$M$}  at -3.5 2.8
\put{$Q$}  at -7 2.8
\arrow <6pt> [.2,.6] from -4 2.8 to -6.5 2.8
\put{$\rho$} [b] at -5 2.9
\put{projection} [b] at -5 3.5
\put{world-lines} [l] at 5.7 .5
\put{of the} [l] at 5.7 0
\put{galaxies} [l] at 5.7 -.5
\arrow <5pt> [.2,.6] from 5.5 0  to 4.2 0 
\arrow <5pt> [.2,.6] from 5.5 0  to 4.1 .8
\arrow <5pt> [.2,.6] from 5.5 0  to 4.3 -.8
\put{spatial sections} [l] at .6 3.5
\arrow <5pt> [.2,.6] from .4 3.3  to -1.3 1.4
\arrow <5pt> [.2,.6] from .4 3.3  to 0 1.2
\arrow <5pt> [.2,.6] from .4 3.3  to 1.4 1.1
\put{quotient} at -7 -3
\put{manifold} at -7 -3.5
\put{$\wt g$} at -7.3 -.3
\put{cosmic} [l] at 6.2 -4
\put{time} [l] at 6.2 -4.5
\put{\firma} at 5.7 -3
\endpicture
$$
\vskip -5mm
  \caption{The basic fibration of the cosmic space-time.}
  \label{fig:basic}
\end{figure}

The first fibration is made of three-dimensional manifolds $S_t$, parametrized by a {\bf cosmic time} $t$. They represent the sets of simultaneous events and will be called {\bf spatial sections}. Each $S_t$ will be endowed with a Riemannian structure with metric tensor $g_t$. The {\bf isotropy principle} implies that each metric $g_t$ has a constant curvature, varying with $t$.  The second fibration is made of the {\bf world-lines} of the galaxies, which are transversal to the spatial sections. Since the world-lines do not intersect each other, this theory does not contemplate fragmentations or collisions of galaxies. We postulate that the set $Q$ of the galactic world-lines (which is nothing but the set of all the galaxies) is endowed with the structure of a three-dimensional manifold: it will be called the {\bf quotient manifold} and it will play a basic role in this theory. A theorem establishes that the quotient manifold $Q$ is endowed with  a  metric $\wt g$, called {\bf quotient metric}, in such a way that any spatial section $(S_t,g_t)$ is isomorphic to $(Q,\wt g)$, so that any spatial section can be taken as a representative of the quotient manifold.

In any cosmological theory the so-called {\bf scale factor} plays  a key role. It is a function of the cosmic time $t$, commonly denoted as $a(t)$, that appears as a conformal factor of the spatial metrics. In our theory, however, the notion of scale factor arises as a conformal factor linking two spatial metrics through the formula
$$
\bboxed{g(t_\uno)=a^2(t_\uno,t_\due)\,g(t_\due)}
$$
Therefore it is a two-variable function in the cosmic time: $a(t_\uno,t_\due)$. This fact offers significant opportunities and advantages in the mathematical analysis of cosmological phenomena, which instead escape in using the one-variable version of the scale factor. For example, if we fix a value of the second variable $t_\due$ (which we will call {\bf reference time}) and  leave $t_\uno=t$ free to run as the only independent variable, then a theorem shows that two scale factors obtained in this way, with different values of the reference time, say $a(t,t_2)$ and  $a(t,t'_2)$, differ by a constant factor. As a consequence, we can always impose to the scale factor the {\bf normalization condition} 
$$
a(t_\sharp,t_\sharp)=1
$$
to be satisfied under the free choice of a reference time $t_\sharp$. This allows us to establish an effective test for the physical validity of equations involving the scale factor, which must be invariant under the choice of $t_\sharp$. Once a reference time is fixed, the scale factor $a(t,t_\sharp)$ becomes the {\bf principal cosmic function} from which, in principle, we should derive the evolution of several other observational quantities (like the Hubble parameter, the energy density, the matter density, etc.). For this reason the graph of a scale factor $a(t,t_\sharp)$ will be called {\bf profile of the universe}. 

\noindent $\bullet$ {\scshape Chapter 2} (Cosmic connections). At the end of Chapter 1, through the introduction of the notion of {\bf free particle}, we are led to presume  the existence in space-time of one (or more) linear symmetric connections that, somehow, are `adapted' to the geometric structures introduced so far in $M$. In this chapter we prove the  existence of a family of {\bf cosmic connections} which are the natural prelude to the formulation of a dynamics and  depend on an indeterminate (but not arbitrary) function of the cosmic time. The assignment of such a function through a {\bf bridge-postulate} will mark the passage from kinematics to dynamics. We will examine two possible bridge-postulates. The first one leads towards a generalization of the  Newtonian space-time, where we could build-up a {\bf Newtonian cosmic dynamics}. The second one consists in assuming the existence of {\bf special particles} (read {\bf photons}) wandering in the cosmos with a constant `peculiar velocity'. The surprising result (theorem) is that: 

(i) {\it The cosmic connection is the Levi-Civita connection of a space-time metric}. (ii) {\it This metric has necessarily a Lorentzian signature}. (iii) {\it The galactic world-lines are time-like geodesics and the  world-lines of the photons are null geodesics}. 

This makes us to move towards a {\bf relativistic cosmic dynamics} founded on the Einstein field equations (Chapter 4).

\noindent $\bullet$ {\scshape Chapter 3} is devoted to the preparation of the elements that we need for the formulation of the dynamics
(Ricci tensor, Einstein tensor, etc.). It is pointed out a remarkable fact which greatly simplifies the calculations: due to the isotropy principle any  symmetric two-tensor $T^{\alpha\beta}$ having a geometrical or physical meaning is fully determined by two functions $\phi(t)$ and $\psi(t)$ only, which we call {\bf characteristic functions}. In generic co-moving coordinates the components of such a tensor are
$$
\bboxed{
\bc
T^{\Z\Z}={\phi}(t)=\hbox{\rm a function of $t$ only}
\ac
T^{\Z a}=0
\ac
T^{ab}={\psi}(t)\,\wt g^{ab}(\wt q)=\hbox{\rm a function of $t$ times the quotient metric $\wt g^{ab}$}
\ec}
$$
Then it can be proved that the conservation law and the Einstein field  equations result in ordinary differential equations involving the scale factor and the two characteristic functions of the momentum-energy tensor. Significant  general results are obtained without specifying the form the characteristic functions. 

\noindent $\bullet$ {\scshape Chapter 4}. The choice of the energy-momentum tensor is the first topic of this chapter, devoted to the relativistic cosmic dynamics. We will consider the standard energy-momentum tensor of a {\bf perfect fluid}
$$
\boxed{\vp\;T^{\alpha\beta}=\left(\epsilon+p\right)\,U^\alpha\,U^\beta+p\,g^{\alpha\beta}\;}
$$
where $\epsilon(t)$ is the energy density, $p(t)$ is the intergalactic pressure and $U^\alpha$ is the unitary four-velocity of the galactic fluid  
$$
U^\alpha\Def c^{-1}\,\dfrac{d\gamma^\alpha}{dt}:\, 
\bc
U^\Z=1,
\ax
U^a=0,
\ec
\quad g_{\alpha\beta}\,U^\alpha\,U^\beta=-1.
$$ 
It follows that the conservation law $\nabla_\alpha T^{\alpha\beta}=0$ and the Einstein field equations reduce respectively to two {\bf dynamical equations} only:
$$
\bboxed{\vphantom{\dfrac{\dot a^2}{c^2}} a\,\dot\epsilon+3\,(\epsilon+p)\,\dot a=0}
\;
\bboxed{\dfrac{\dot a^2}{c^2}=\tfrac 13\,a^2\,(\Lambda+\chi\,\epsilon)-\wt K}
$$
where $\wt K$ is the curvature of the quotient manifold. 

\begin{figure} [H]
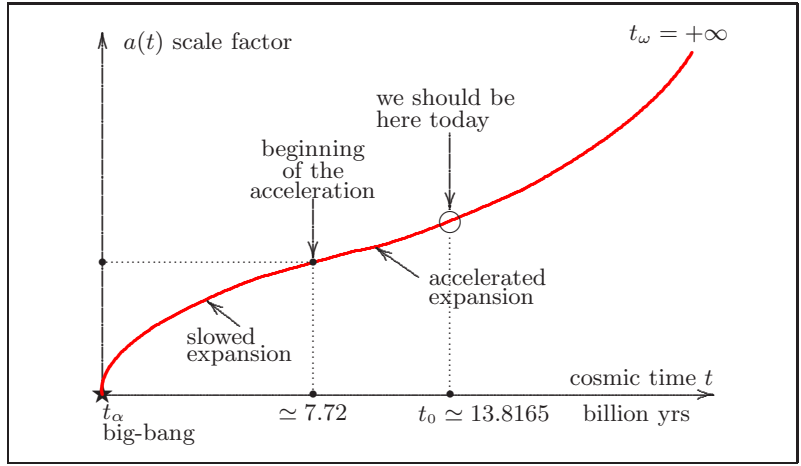

$$
\beginpicture 
\setcoordinatesystem units <1.4cm,1.3cm>
\setplotarea x from -.9 to 6.6, y from -.7 to 4
\normalgraphs
\grid 1 1
\small
\arrow <6pt> [.2,.6] from 0 0 to 5.8 0
\put{$\bigstar$} at 0 0
\put{cosmic time $t$} [rb] at 5.8 .1
\arrow <6pt> [.2,.5] from 0 0 to 0 3.7
\put{$a(t)$ scale factor} [l] at .2 3.6
\put{$t_\omega=+\infty$} [lb] at 5 3.6
\put{$t_\alpha$} [lt] at 0 -.1
\setplotsymbol (\normalered)
\setquadratic
\plot 0 0 
.1 .33
.5 .7
1 .98
1.5 1.2
2 1.35 
2.4 1.47
3 1.65 
4 2.1 
5 2.8 
5.6 3.5 
/
\normale
\setlinear
\setdots<2pt>
\plot 2 0 2 1.35 0 1.35 /
\setsolid
\put{$\sc\bullet$} at 2 1.35
\put{$\sc\bullet$} at 2 0
\put{$\sc\bullet$} at 0 1.35
\put{big-bang} [tl] at 0 -.3
\arrow <6pt> [.2,.6] from 1.3 .7 to 1 .95
\put{slowed} [l] at .8 .6
\put{expansion} [l] at .8 .4
\arrow <6pt> [.2,.6] from 3 1.2 to 2.6 1.5
\put{accelerated} [l] at 3.1 1.2
\put{expansion} [l] at 3.1 .98
\put{$\bigcirc$} at 3.3 1.76
\put{we should be} [l] at 2.6 3.05
\put{here today} [l] at 2.6 2.8
\arrow <6pt> [.2,.6] from 3.3 2.67 to 3.3 1.9
\setdots<2pt>
\plot 3.3 1.6 3.3 0 /
\setsolid
\put{$\sc\bullet$} at 3.3 0
\put{$t_\Z\simeq 13.8165$ \quad billion yrs} [tl] at 3 -.1
\arrow <6pt> [.2,.6] from 2 2 to 2 1.4
\put{beginning} at 2 2.5
\put{of the} at 2 2.3
\put{acceleration} at 2 2.1
\put{$\simeq 7.72$} [t] at 2 -.1
\endpicture
$$
\vskip -3mm
  \caption{\small One of the eligible profiles of the universe.}
  \label{fig:nobel}
\end{figure}

\begin{figure}[H] 
$$
\beginpicture 
\setcoordinatesystem units <.9cm,.9cm>
\setplotarea x from -4.5 to 4, y from -3 to 3
\normalgraphs
\grid 1 1
\put{\includegraphics[width=5.5cm,keepaspectratio]{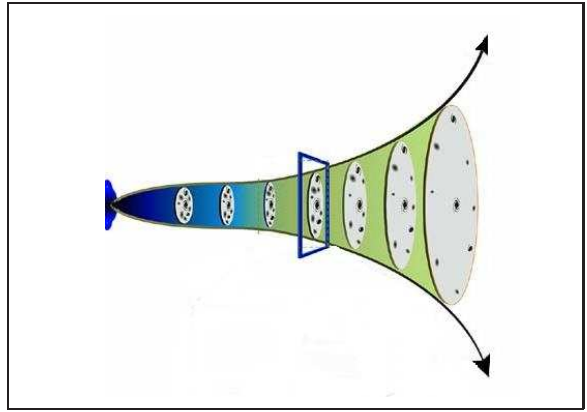}} at 0 0
\endpicture
$$
\vskip -3mm
\caption{\small From Riess Nobel Lecture,}
  \label{fig:riess-2}
\end{figure}

\noindent $\bullet$ {\scshape Chapter 5}. A further crucial step is the choice of an {\bf equation of state} that ties the three unknown functions $\epsilon(t)$, $p(t)$ and $a(t)$. In this chapter we deal with the simplest possible case of a {\bf linear barotropic fluid}, whose equation of state is 
$$
p=w\,\epsilon,
$$
$w$ being a constant parameter. Thus, we are faced with various models whose behaviour depends on the sign of the constant curvature $\wt K$. However, the models with negative spatial curvature are `a priori' excluded from this theory because, since it can be proved that the dynamical equations imply a radial speed expansion permanently greater than the light speed. In turn, also the positive spatial curvature is proved to be inadmissible by a mathematical argument based on a reasonable (although rough) estimate of the present-day matter density.  As a consequence, we confine ourselves to study the models with zero spatial curvature. These models are in fact very simple and do not take into account of all the complex quantum-physical phenomenology occurring in the evolution of the universe, especially in the vicinity of its creation (as said above). Nevertheless they reveal the typical features of the isotropic cosmology and could serve as a starting point for the creation of finer models. For example, in the gallery of the possible profiles of the universe we find that of Fig.\ \ref{fig:nobel}, which is perfect agreement with that appearing in the Nobel Lecture by G.\ Riess  \cite{Riess} (Fig.\ \ref{fig:riess-2}).\footnote{
\kern 2pt The Nobel Prize in Physics 2011 was awarded to Saul Perlmutter, Brian P. Schmidt and Adam G. Riess "for the discovery of the accelerating expansion of the Universe through observations of distant supernovae".} 

\fbox{
\begin{minipage}{.9\linewidth}
\vspu
A crucial and fortunate circumstance is that the dynamical equations of a flat barotropic model are solvable in terms of elementary functions (exponential or hyperbolic functions). This allows us to get the exact expressions of all the  observational variables related to the scale factor. The only approximation is then due to the numerical calculation and to the estimate of the observational  data. \vspd
\end{minipage}}
\vskip 8pt

\noindent $\bullet$ {\scshape Chapter 6}. In order to avoid the common misconceptions on the various notions of horizons, pointed out for example in \cite{Davis-Lineweaver}, we afford the so-called  `horizon problems' on the basis of what we have learned from the previous chapter and with a significant  graphic  method. When applied to the flat dust-matter model, this method gives numerical results in very good  agreement with the current observational data.

S.\ B., May 2016.

\tableofcontents

\chapter{Cosmic kinematics}

\pagenumbering{arabic}

\section{The postulates of the cosmic space-time}

In an axiomatic formulation of cosmology the concept of {\bf event} is a primary (or `primitive') concept  as that of {\bf point} in Euclidean geometry. 

\begin{center}
\fbox{
\begin{minipage}{.95\linewidth} \vspu
{\bf $\bs 1^{st}$ Postulate}. The history of the universe is made of events, whose set will be called {\bf cosmic space-time} and denoted by  $\mathcal E$. There are two {\bf critical events}, denoted by $\alpha$ and $\omega$,  representing the {\bf beginning} and the {\bf end} of the universe, \index{beginning of the universe $\alpha$} \index{end of the universe $\omega$} respectively. All other events are called {\bf regular} and form a four-dimensional manifold $M$:
$$
\bboxed{\mathcal E=\alpha \cup M \cup \omega}
$$
\begin{figure}[H]
  \centering
\vskip -10mm
$$
\beginpicture
\setcoordinatesystem units <.6cm,.6cm>
\setplotarea x from -9.4 to 8, y from -2.5 to 3
\normalgraphs
\small
\ellipticalarc axes ratio 1.1:3 180 degrees from -4 2.4  center at -4 0
\ellipticalarc axes ratio 1.1:3 360 degrees from 4 2.4  center at 4 0
\startrotation by 0 -1 about 0 0
\setlinear
\plot -2.4 4  -2.4 -4 /
\plot 2.4 4 2.4 -4 /
\put{\omino} at -.2 -1.35
\stoprotation
\put{$M$} [b] at -3 1.5
\put{regular events} [b] at 0 1.5
\put{$\alpha$} [b] at -5.5 .3
\put {$\odot$} at -5.5 0
\put {$\sc\bullet$} at -5.5 0
\arrow <6pt> [.2,.6] from -6.3 -.9 to -5.7 -.2
\put {beginning} [r] at -6.5 -1
\put{$\omega$} [b] at 5.5 .3
\put {$\odot$} at 5.5 0
\put {$\sc\bullet$} at 5.5 0
\put {end} [l] at 6.5 -1
\arrow <6pt> [.2,.6] from 6.3 -.9 to 5.7 -.2
\put{\firma} at 6 -2.5
\endpicture
$$
\end{figure}
\end{minipage}}
\end{center}

The second primary concept is that of {\bf cosmic particle}. 

\begin{center}
\fbox{
\begin{minipage}{.95\linewidth} \vspu
{\bf $\bs 2^{nd}$ Postulate}. The life of a {\bf cosmic particle} is a sequence of regular events forming a smooth curve in $M$, called {\bf world-line} of the particle. \vspd
\end{minipage}}
\end{center}

In view of the large-scale approach to cosmology, the cosmic particles are identified with galaxies. The third primary concept is that of {\bf cosmic fluid}.

\begin{center}
\fbox{
\begin{minipage}{.95\linewidth} \vspu
{\bf $\bs 3^{rd}$ Postulate}. (i) The {\bf cosmic fluid} is made of cosmic particles whose world-lines form a {\bf congruence} of curves filling the whole manifold $M$.\footnote{
 By `congruence' we mean that the world-lines do not intersect one another. So, collisions between cosmic particles are excluded.}
(ii) The set $Q$ of all the galactic world-lines has  the structure of a three-dimensional differentiable manifold such that the canonical projection $\rho\colon M \to Q$  is a surjective submersion. We call $Q$ the {\bf quotient manifold}.\footnote{
Note that $Q$ is just the set of all galaxies.}
\vskip -12pt
\begin{figure}[H]
  \centering
$$
\beginpicture
\setcoordinatesystem units <.6cm,.6cm>
\setplotarea x from -9 to 9, y from -3 to 4
\normalgraphs
\small
\startrotation by 0 -1 about 0 0
\setlinear
\plot -2.4 4  -2.4 -4 /
\plot 2.4 4 2.4 -4 /
\ellipticalarc axes ratio 3:1 360 degrees from -1.4 -3.25 center at 0 -3.9
\ellipticalarc axes ratio 3:1 360 degrees from 2.4 3.9 center at 0 3.9
\put{\omino} at -.2 -1.35
\stoprotation
\setquadratic
\plot 
-4 0 
-2 .4 
0 0
2 -.4
4 0 /
\put{\beginpicture
\setquadratic
\plot 
-4 0 
-2 .4 
0 0
2 -.4
4 0 /
\endpicture} at -.2 .9
\put{\beginpicture
\setquadratic
\plot 
-4 0 
-2 .4 
0 0
2 -.4
4 0 /
\endpicture} at 0 1.8
\put{\beginpicture
\setquadratic
\plot 
-4 0 
-2 .4 
0 0
2 -.4
4 0 /
\endpicture} at .2 -.9
\put{\beginpicture
\setquadratic
\plot 
-4 0 
-2 .4 
0 0
2 -.4
4 0 /
\endpicture} at 0 -1.8
\ellipticalarc axes ratio 1:3 360 degrees from -7 2.4 center at -7 0
\setlinear 
\setdots<2pt>
\plot -4 -1.8 -7 -1.8 /
\plot -3.8 -.9 -6.8 -.9 /
\plot -4 0 -7 0 /
\plot -4.2 .9 -7.2 .9 /
\plot -4 1.8 -7 1.8 /
\setsolid
\put{$\sc\bullet$} at  -7 -1.8 
\put{$\sc\bullet$} at  -7 1.8 
\put{$\sc\bullet$} at  -6.8 -.9 
\put{$\sc\bullet$} at  -7 0 
\put{$\sc\bullet$} at  -7.2 .9
\put{$M$}  at -3.5 2.8
\put{$Q$}  at -7 2.8
\arrow <6pt> [.2,.6] from -4 2.8 to -6.5 2.8
\put{$\rho$} [b] at -5 2.9
\put{projection} [b] at -5 3.5
\put{world-lines} [l] at 5.7 .5
\put{of the} [l] at 5.7 0
\put{galaxies} [l] at 5.7 -.5
\arrow <5pt> [.2,.6] from 5.5 0  to 4.2 0 
\arrow <5pt> [.2,.6] from 5.5 0  to 4.1 .8
\arrow <5pt> [.2,.6] from 5.5 0  to 4.3 -.8
\put{quotient} at -7 -3
\put{manifold} at -7 -3.5
\put{\firma} at 6.5 -3
\endpicture
$$
\end{figure}
\end{minipage}}
\end{center}

The third primary concept is that of {\bf cosmic time}. A {\bf cosmic time} is a {\bf regular} mapping from $M$ to an open interval of $\mathbb R$,
$$
t\colon M\to (t_\alpha,t_\omega)\subseteq \mathbb R,
$$
which associates to each regular event $e\in M$ its {\bf date} $t(e)$.
Two cosmic times are said to be {\bf equivalent} if they are related by an affine transformation 
$$
t\mapsto \bar t=\lambda\,t+\mu, \quad \lambda,\mu\in\mathbb R, \quad \boxed{\lambda>0}
$$

The {\it regularity property} of the function $t$ implies that the inverse image of any real number $t\in (t_\alpha,t_\omega))$ is a three-dimensional submanifold $S_t$. Two such submanifolds $S_t$ and $S_{t'}$ do not intersect and form a foliation of the whole manifold $M$.
The submanifolds $S_t$ which we call {\bf spatial sections}.  Two events of an $S_t$ have, by definition, the same date $t$; thus, they are {\bf simultaneous}. Consequently a cosmic time provides a {\bf cosmic chronology}: 
given two events $e_1$ and $e_2$ we say that
$$
\begin{array}{lll}
\hbox{$e_1$ and $e_2$ are {\bf simultaneous}} & {\rm if} & t(e_1)=t(e_2) 
\\
\hbox{$e_1$ occurs {\bf before} $e_2$} & {\rm if} & t(e_1)<t(e_2)  
\\
\hbox{$e_1$ occurs {\bf after} $e_2$} & {\rm if} & t(e_1)>t(e_2)
\ea
$$
It follows that {\it two equivalent cosmic times provides the same   cosmic chronology}. The chronology can be extended to the critical events by setting
$$
t(\alpha)=t_\alpha, \quad t(\omega)=t_\omega.
$$
So, the event $\alpha$ (beginning of the universe) occurs before all other events and $\omega$ (end of the universe) occurs after all other events. 

It is crucial to highlight the following. By means of a reversible smooth transformation $t\mapsto \bar t$ any smooth mapping over an open interval $t\colon M\to (t_\alpha,t_\omega)$ can be transformed into a smooth mapping over the whole real line, $\bar t\colon M\to\mathbb R=(-\infty,+\infty)$, preserving the above chronology. Consequently, if we want to give meaning to a concept like the {\bf duration of the universe} then we must exclude such a general transformation of the cosmic time. So, having assumed as permissible the affine transformations, the {\bf   
life-time-interval} of the universe is of four types: 
\begin{table}[H]
\centering
\begin{tabular}{|c|c|c|} 
 \hline
 $\vpt$ Life-time interval
& Beginning
& End
\\
\hline\hline
$\vpt (-\infty,+\infty)$
& infinite past
& infinite future
\\
\hline
$\vpt (-\infty,t_\omega]$ 
& infinite past
& finite time $t_\omega$
\\
\hline
$\vpt [t_\alpha,t_\omega]$
& finite time $t_\alpha$
& finite time $t_\omega$
\\
\hline
$\vpt [t_\alpha,+\infty)$ 
& finite time $t_\alpha$
& infinite future
\\
\hline 
\end{tabular}
 \caption{Life of the universe.}\label{tab:life}
\end{table}

\begin{center}
\fbox{
\begin{minipage}{.95\linewidth} \vspu
{\bf $\bs 4^{th}$ Postulate}. There exist a cosmic time and a cosmic chronology, as defined  above, such the foliation $S_t$ of the spatial sections is transversal to the world-lines of the cosmic fluid.
\vskip -8pt
\begin{figure}[H]
  \centering
$$
\beginpicture
\setcoordinatesystem units <.6cm,.6cm>
\setplotarea x from -8.5 to 9, y from -4 to 4
\normalgraphs
\small
\startrotation by 0 -1 about 0 0
\setlinear
\plot -2.4 4  -2.4 -4 /
\plot 2.4 4 2.4 -4 /
\ellipticalarc axes ratio 3:1 290 degrees from -1.4 -3.25 center at 0 -3.9
\ellipticalarc axes ratio 3:1 290 degrees from -1.4 -1.95 center at 0 -2.6
\ellipticalarc axes ratio 3:1 290 degrees from -1.4 -.65 center at 0 -1.3
\ellipticalarc axes ratio 3:1 290 degrees from -1.4 .65 center at 0 0
\ellipticalarc axes ratio 3:1 290 degrees from -1.4 1.95 center at 0 1.3
\ellipticalarc axes ratio 3:1 290 degrees from -1.4 3.25 center at 0 2.6
\ellipticalarc axes ratio 3:1 360 degrees from 2.4 3.9 center at 0 3.9
\put{\omino} at -.2 -1.35
\stoprotation
\arrow <6pt> [.2,.6] from -5 -4  to 6 -4  
\put{$S_t$} at -1.4 -1
\arrow <6pt> [.2,.6] from -1.3 -2.5 to -1.3 -3.9
\put{$t$} [t] at -1.3 -4.2
\setlinear 
\setdots<2pt>
\plot -3.9 -2.45 -3.9 -4 /
\put{$t_\alpha$} [t] at -3.9 -4.3
\put{$\sc |$} at -3.9 -4
\plot 3.9 -2.45 3.9 -4 /
\put{$t_\omega$} [t] at 3.9 -4.3
\put{$\sc |$} at 3.9 -4
\setsolid
\setquadratic
\plot 
-4 0 
-2 .4 
0 0
2 -.4
4 0 /
\put{\beginpicture
\setquadratic
\plot 
-4 0 
-2 .4 
0 0
2 -.4
4 0 /
\endpicture} at -.2 .9
\put{\beginpicture
\setquadratic
\plot 
-4 0 
-2 .4 
0 0
2 -.4
4 0 /
\endpicture} at 0 1.8
\put{\beginpicture
\setquadratic
\plot 
-4 0 
-2 .4 
0 0
2 -.4
4 0 /
\endpicture} at .2 -.9
\put{\beginpicture
\setquadratic
\plot 
-4 0 
-2 .4 
0 0
2 -.4
4 0 /
\endpicture} at 0 -1.8
\ellipticalarc axes ratio 1:3 360 degrees from -7 2.4 center at -7 0
\setlinear 
\setdots<2pt>
\plot -4 -1.8 -7 -1.8 /
\plot -3.8 -.9 -6.8 -.9 /
\plot -4 0 -7 0 /
\plot -4.2 .9 -7.2 .9 /
\plot -4 1.8 -7 1.8 /
\setsolid
\put{$\sc\bullet$} at  -7 -1.8 
\put{$\sc\bullet$} at  -7 1.8 
\put{$\sc\bullet$} at  -6.8 -.9 
\put{$\sc\bullet$} at  -7 0 
\put{$\sc\bullet$} at  -7.2 .9
\put{$M$}  at -3.5 2.8
\put{$Q$}  at -7 2.8
\arrow <6pt> [.2,.6] from -4 2.8 to -6.5 2.8
\put{$\rho$} [b] at -5 2.9
\put{projection} [b] at -5 3.5
\put{world-lines} [l] at 5.7 .5
\put{of the} [l] at 5.7 0
\put{galaxies} [l] at 5.7 -.5
\arrow <5pt> [.2,.6] from 5.5 0  to 4.2 0 
\arrow <5pt> [.2,.6] from 5.5 0  to 4.1 .8
\arrow <5pt> [.2,.6] from 5.5 0  to 4.3 -.8
\put{spatial sections} [l] at .6 3.5
\arrow <5pt> [.2,.6] from .4 3.3  to -1.3 1.4
\arrow <5pt> [.2,.6] from .4 3.3  to 0 1.2
\arrow <5pt> [.2,.6] from .4 3.3  to 1.4 1.1
\put{quotient} at -7 -3
\put{manifold} at -7 -3.5
\put{cosmic} [l] at 6.2 -4
\put{time} [l] at 6.2 -4.5
\put{\firma} at 6.5 -3
\endpicture
$$
\end{figure}
\end{minipage}}
\end{center}

The transversality condition means that a world-line is nowhere tangent to a spatial section.  By virtue of well-known arguments of differential geometry, the $3^{rd}$ and the $4^{th}$ postulates imply that the restriction of the projection $\rho$ to any spatial section $S_t$ is a diffeomorphism, and consequently 
\bt\label{t:diff}
The spatial sections and the quotient manifold are diffeomorphic manifolds.
\et

Another  consequence of this postulate is the following.

\bt\label{t:comco}
Any coordinate system $\wt q=(q^a)=(q^1,q^2,q^3)$ on an open domain $U\subseteq Q$ generates a coordinate system $(t,q^a)$ on the open subset of $M$ made of the world-lines determined by $U$ {\rm (Fig.\ \ref{fig:comco}).} 
\et
\begin{figure}[H]
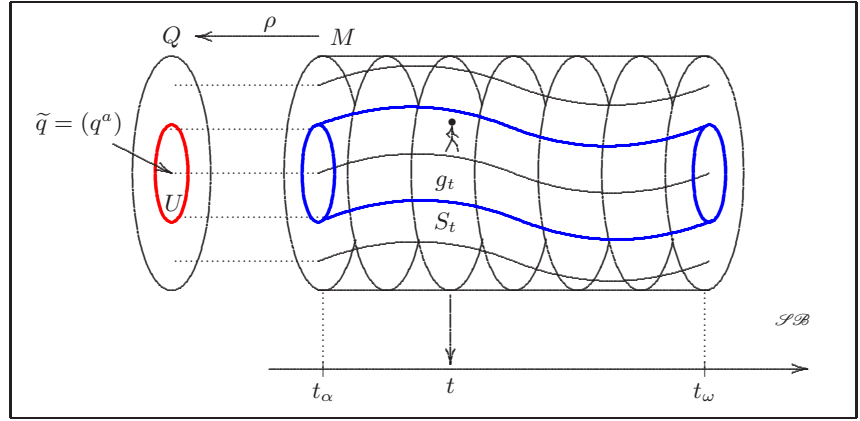

  \centering
$$
\beginpicture
\setcoordinatesystem units <.65cm,.65cm>
\setplotarea x from -10.3 to 7, y from -5 to 3.5
\normalgraphs
\small
\grid 1 1
\startrotation by 0 -1 about 0 0
\setlinear
\plot -2.4 4  -2.4 -4 /
\plot 2.4 4 2.4 -4 /
\ellipticalarc axes ratio 3:1 290 degrees from -1.4 -3.25 center at 0 -3.9
\ellipticalarc axes ratio 3:1 290 degrees from -1.4 -1.95 center at 0 -2.6
\ellipticalarc axes ratio 3:1 290 degrees from -1.4 -.65 center at 0 -1.3
\ellipticalarc axes ratio 3:1 290 degrees from -1.4 .65 center at 0 0
\ellipticalarc axes ratio 3:1 290 degrees from -1.4 1.95 center at 0 1.3
\ellipticalarc axes ratio 3:1 290 degrees from -1.4 3.25 center at 0 2.6
\ellipticalarc axes ratio 3:1 360 degrees from 2.4 3.9 center at 0 3.9
\put{\omino} at -.7 -1.35
\stoprotation
\arrow <6pt> [.2,.6] from -5 -4  to 6 -4  
\put{$S_t$} at -1.4 -1
\put{$g_t$} at -1.4 -.2
\arrow <6pt> [.2,.6] from -1.3 -2.5 to -1.3 -3.9
\put{$t$} [t] at -1.3 -4.2
\setlinear 
\setdots<2pt>
\plot -3.9 -2.45 -3.9 -4 /
\put{$t_\alpha$} [t] at -3.9 -4.3
\put{$\sc |$} at -3.9 -4
\plot 3.9 -2.45 3.9 -4 /
\put{$t_\omega$} [t] at 3.9 -4.3
\put{$\sc |$} at 3.9 -4
\setsolid
\setquadratic
\plot 
-4 0 
-2 .4 
0 0
2 -.4
4 0 /
\put{\beginpicture
\setquadratic
\plot 
-4 0 
-2 .4 
0 0
2 -.4
4 0 /
\endpicture} at 0 1.8
\put{\beginpicture
\setquadratic
\plot 
-4 0 
-2 .4 
0 0
2 -.4
4 0 /
\endpicture} at 0 -1.8
\ellipticalarc axes ratio 1:3 360 degrees from -7 2.4 center at -7 0
\setlinear 
\setdots<2pt>
\plot -4 -1.8 -7 -1.8 /
\plot -3.8 -.9 -6.8 -.9 /
\plot -4 0 -7 0 /
\plot -4.2 .9 -7.2 .9 /
\plot -4 1.8 -7 1.8 /
\setsolid
\put{$M$}  at -3.5 2.8
\put{$Q$}  at -7 2.8
\arrow <6pt> [.2,.6] from -4 2.8 to -6.5 2.8
\put{$\rho$} [b] at -5 2.9
\put{\firma} at 5.7 -3
\setplotsymbol (\normalered)
\ellipticalarc axes ratio 1:3 360 degrees from -7 1 center at -7 0
\setplotsymbol (\normaleblue)
\ellipticalarc axes ratio 1:3 360 degrees from -4 1 center at -4 0
\ellipticalarc axes ratio 1:3 360 degrees from 4 1 center at 4 0
\put{\beginpicture
\setquadratic
\plot 
-4 0 
-2 .4 
0 0
2 -.4
4 0 /
\endpicture} at .1 -.95
\put{\beginpicture
\setquadratic
\plot 
-4 0 
-2 .4 
0 0
2 -.4
4 0 /
\endpicture} at -.1 .96
\put{$\wt q=(q^a)$} [r] at -8 1
\put{$U$} at -6.95 -.6
\normale
\arrow <6pt> [.2,.6] from -8.2 .6 to -7 0
\endpicture
$$
\vskip -5mm
  \caption{Co-moving coordinates and spatial metrics.}
  \label{fig:comco}
\end{figure}
Coordinates on $M$ of this type are called {\bf co-moving coordinates}.\footnote{
\kern 2pt  Since $t$ is constant on each spatial section, co-moving coordinates are also called {\bf synchronous coordinates}.} The coordinates $\wt q=(q^a)$ are Lagrangian coordinates of the galactic fluid: they have a constant value on each world-line of $U$.  In this way the coordinates $\wt q$ can be interpreted as coordinates on each spatial section $S_t$. So, they will be called {\bf spatial coordinates}.\footnote{
\kern 2pt Greek indices $\alpha,\beta,...$ will run from 0 to 3. Latin indices $a,b,...$ will run from 1 to 3.} 

At this point of our route the cosmic space-time is equipped with two {\bf transversal trivial fibrations}: 

(i) A fibration over the open real interval $(t_\alpha,t_\beta)$ (beginning and and of the universe) with fibers diffeomorphic to the quotient manifold $Q$. 

(ii) A fibration over the quotient manifold $Q$ with fibers  the galactic world-lines which are diffeomorphic to the interval $(t_\alpha,t_\beta)$.

\begin{center}
\fbox{
\begin{minipage}{.95\linewidth} \vspu
 {\bf $\bs 5^{th}$ Postulate}. Each spatial section $S_t$  is endowed with a positive-definite metric tensor $g_t$ smoothly depending on $t$. \vspd
\end{minipage}}
\end{center}
In other words, we think of each $S_t$ as a three-dimensional Riemannian manifold where the metric tensor components $g_{ab}(t,\wt q)$ in co-moving coordinates are smooth functions of $t$.

The {\bf Copernican principle} assumes that neither the Sun nor the Earth are in a central, specially favored position in the universe. This principle is extended to cosmology with the following {\bf isotropy principle}. 
\begin{center}
\fbox{
\begin{minipage}{.95\linewidth} \vspu
{\bf $\bs 6^{th}$ Postulate}. On each spatial section $S_t$ there is no distinguished vector field having any   epistemological  meaning. \vspd
\end{minipage}}
\end{center}
\bt\label{t:scalar}
 Any scalar field on $M$ having an epistemological  meaning is a function of the cosmic time $t$ only i.e., it is constant on each $S_t$.
\et 
\bprf 
By means of the metric $g_t$ we can define the gradient of such a scalar field which is then a distinguished vector field on $S_t$. This is in contrast with the isotropy principle. \eprf

\bt \label{t:spcurv}
Each spatial section $(S_t,g_t)$ is a manifold with constant curvature $K(t)$.
\et
\bprf 
The Ricci tensor $R_t$ of the metric $g_t$ must be proportional to the metric tensor itself, $R_t=\lambda_t\,g_t$, otherwise the existence of distinguished Ricci directions would be in contrast with the isotropy principle. In turn, the factor $\lambda_t$ must be constant on $S_t$ because of Theorem \ref{t:scalar}. Thus  $(S_t,g_t)$ is an Einstein manifold. It is known that an Einstein manifold of dimension 3 has constant curvature.  \eprf

\section{Manifolds with constant curvature}\label{s:ccm}
\index{constant curvature} \index{Riemann!tensor} \index{linear connection}
\index{connection!linear}

In this section we recall the main features of the manifolds with constant curvature.\footnote{
\kern 2pt  A high level reference to this topic is \cite{Besse}. A classical reference for a simpler approach, sufficient for our needs, is   \cite{Wolf}.} 
 For the Riemann curvature tensor and the Ricci tensor of a linear \underbar{s}y\underbar{mmetric} connection $\Gamma$ we will refer to the following 
definitions:\footnote{
\kern 2pt Our definition of the Riemann tensor is that of \cite{MTW} and \cite{Eisenhart}. On the contrary, our Ricci tensor is the opposite of that of  \cite{Eisenhart}.} 
\be\label{e:Riem1}
\bboxed{R^\nu_{\;\,\alpha\mu\beta}\Def\partial_\mu\Gamma^\nu_{\alpha\beta}-\partial_\beta \Gamma^\nu_{\mu\alpha}
+\Gamma^\ell_{\alpha\beta}\;\Gamma^\nu_{\mu\ell}
-\Gamma^\ell_{\mu\alpha}\;\Gamma^\nu_{\beta\ell}}
\ee
\be\label{e:Ricci-def}
\bboxed{R_{\alpha\beta}\Def R^\mu_{\;\,\alpha\mu\beta}
=\partial_\mu\Gamma^\mu_{\alpha\beta}
-\partial_\beta\Gamma^\mu_{\alpha\mu}
+\Gamma^\sigma_{\alpha\beta}\,\Gamma^\mu_{\sigma\mu}
-\Gamma^\sigma_{\alpha\mu}\,\Gamma^\mu_{\sigma\beta}}
\ee
where $\Gamma_{\alpha\beta}^\gamma=\Gamma_{\beta\alpha}^\gamma$ are the symbols of $\Gamma$ in any coordinate system. 

A Riemannian manifold with metric tensor $g_{\alpha\beta}$ is said to be of {\bf constant curvature} $K$ when the totally covariant Riemann
tensor
 \be\label{e:Riem-cov}
\bboxed{R_{\lambda\alpha\mu\beta}\Def g_{\lambda\nu}\,R^\nu_{\;\,\alpha\mu\beta}}
\ee
of the Levi-Civita connection satisfies the equation\footnote{
\kern 2pt \cite{Eisenhart}, Section 26.} 
\be\label{e:K2}
\bboxed{R_{\alpha\beta\gamma\delta}=K\,(g_{\alpha\gamma}\,g_{\beta\delta}-g_{\alpha\delta}\,g_{\beta\gamma})}
\ee
equivalent to
\be\label{e:ccm1}
R^\alpha_{\;\beta\gamma\delta}=K\,(\delta^\alpha_\gamma\,g_{\beta\delta}-\delta^\alpha_\delta\,g_{\beta\gamma}),
\ee
It follows that
\be\label{e:ccm2}
\boxed{\;\vp R_{\alpha\beta}=(n-1)\,K\,g_{\alpha\beta}\;}\;\;
\boxed{\;\vp R=n\,(n-1)\,K\;}
\ee
where $n$ is the dimension of the manifold and $R$ is the {\bf Ricci scalar} (or {\bf scalar curvature})
\be\label{e:Ricci-sc-def}
\bboxed{R\Def g^{\alpha\beta}\,R_{\alpha\beta}}
\ee
For a conformal transformation $\bar g_{\alpha\beta}=\alpha\,g_{\alpha\beta}$ with a {\it constant} factor $\alpha$ the Riemann tensor components are invariant: $\bar R^\alpha_{\;\beta\gamma\delta}=R^\alpha_{\;\beta\gamma\delta}$. Hence, from \eqref{e:ccm1},
$$
\bar K\,(\delta^\alpha_\gamma\,\bar g_{\beta\delta}-\delta^\alpha_\delta\,\bar g_{\beta\gamma})=K\,(\delta^\alpha_\gamma\,g_{\beta\delta}-\delta^\alpha_\delta\,g_{\beta\gamma}),
$$
and consequently $\alpha\,\bar K=K$. This shows that
\be\label{e:ccm3}
\bboxed{\bar g_{\alpha\beta}=\alpha\,g_{\alpha\beta}\;\;\hbox{($\alpha=\rm constant$)}\;\To \bar K=\dfrac{K}\alpha}
\ee
With a similar argument one can show that: {\it two metric tensors $\bar g_{\alpha\beta}$ and $g_{\alpha\beta}$ on a same manifold, with the same signature and with constant curvatures of the same sign are conformal $\bar g_{\alpha\beta}=\alpha\,g_{\alpha\beta}$ with a p\underline{ositive} constant  conformal factor $\alpha$}.

\section{Dimensional analysis}\label{s:Dim}

To test the correctness of the formulas that we will write, it is important to consider the physical dimension of the involved objects. We will denote the physical dimension of an object $X$ 
by the symbol $\Dim(X)$.\footnote{
\kern 2pt The symbol $[X]$ is more commonly used.} 
 The basic  physical dimensions are 
$$
\bc
\Dim(\hbox{time})=T
\\[1mm]
\Dim(\hbox{length})=L
\ec
\;\;
\bc
\Dim(\hbox{mass})= M
\\[1mm]
\Dim(\hbox{dimensionless quantity})= 1
\ec
$$
Then the dimension of any object $X$ will be expressed by the product of positive or negative integer powers of these simbols,
$$
\Dim(X)=T^a\,L^b\,M^c, \quad a,b,c\in\mathbb Z.
$$
For instance:
\vskip -5mm
\begin{table}[H]
\centering
\begin{tabular}{|c|c|} 
 \hline
 $\vpt$ Object
& $\Dim$
\\
\hline\hline
$\vpt$ area
& $L^2$
\\
\hline
$\vpt$ volume
& $L^3$
\\
\hline
$\vpt$ velocity
& $L\,T^{-1}$
\\
\hline
$\vpt$ acceleration
& $L\,T^{-2}$
\\
\hline
$\vpt$ angle
& 1
\\
\hline
$\vpt$ angular velocity
& $T^{-1}$
\\
\hline 
\end{tabular}
 \caption{Dimension of the basic geometric and kinematic quantities.}\label{tab:Dim-1}
\end{table}

\begin{table}[H]
\centering
\begin{tabular}{|c|c|c|c|}
 \hline
 $\vpt$ Object
& Symbol
& $\Dim$
& Note
\\
\hline\hline
$\vpt$ force
& $F$
& $M\,L\,T^{-2}$
& mass $\times$ acceleration
\\
\hline
$\vpt$ pressure
& $P$
& $M\,L^{-1}\,T^{-2}$
& force/area
\\
\hline
$\vpt$ energy (work)
& $E$
& $M\,L^2\,T^{-2}$
& work $=$ force $\times$ length
\\
\hline
$\vpt$ energy density
& $\epsilon$
&$M\,L^{-1}\,T^{-2}$
& energy/volume
\\
\hline
$\vpt$ mass density
& $\rho$
& $M\,L^{-3}$
& mass/volume
\\
\hline 
\end{tabular}
 \caption{Dimension of the basic physical quantities.}\label{tab:Dim-2}
\end{table}

The coordinates of a manifold can be dimensionless (e.g.\ angles) or with a physical dimension (time, length,...). About the co-moving coordinates introduced by Theorem \ref{t:comco} we assume that:
$$
\ba
\hbox{(i) The cosmic time $t$ is time-dimensional: $\Dim(t)=T$}.
\\[1mm]
\hbox{(ii) The spatial coordinates $q^a$ are length-dimensional: $\Dim(q^a)=L$}.
\ea
$$
Then it will be convenient to replace the time coordinate $t$ with a length-dimensional coordinate $q^\Z$ via the simple relationship 
\be\label{e:q0t}
\boxed{\vp\;q^\Z=\kappa\,t\;}
\ee
where $\kappa$ is an arbitrary constant with the dimension of a velocity: $\Dim(\kappa)=L\,T^{-1}$. It is immaterial the numerical value of this constant.\footnote
{\kern 2pt Later, when dealing with cosmic dynamics, we shall be led to consider $\kappa=c$, the light speed.} 
 In this way we get {\bf length-dimensional co-moving coordinates}.

\begin{center}
\fbox{
\begin{minipage}{.95\linewidth} \vspu
In the following we will always refer to length-dimensional co-moving coordinates of this kind. We will call them {\bf homogeneous coordinates}.\vspd
\end{minipage}}
\end{center}
Consequently:
\be\label{e:Dim-R} 
\bboxed{\ba
\Dim(g_{\alpha\beta})=\Dim(g^{\alpha\beta})=1.
\ac
\Dim(\Gamma_{\alpha\beta,\gamma})=\Dim(\Gamma_{\alpha\beta}^\gamma)=L^{-1}.
\ac
\Dim(R^\nu_{\;\,\alpha\mu\beta})=L^{-2}.
\ac
\Dim(R_{\alpha\beta})=\Dim(R)=L^{-2}.
\ea}
\ee

\br
Do not pay attention to the dimension of the coordinates is unfortunately a widespread harmful habit, which produces confusion and mistakes in the writing and interpreting of  the tensorial-type formulas. For example, if you use angular coordinates (for instance on a sphere or hyper-sphere with fixed radius) which are dimensionless, then $\Dim(g_{\alpha\beta})=L^2$ and consequently, 
$$
\bc
\Dim(g^{\alpha\beta})=L^{-2},
\ac
\Dim(\Gamma_{\alpha\beta,\gamma})=L^2,
\ac
\Dim(\Gamma_{\alpha\beta}^\gamma)=1,
\ec
\quad
\bc
\Dim(R^\nu_{\;\,\alpha\mu\beta})=1,
\ac
\Dim(R_{\alpha\beta})=1,
\ac
\Dim(R)=L^{-2}.
\ec
$$
Note that the Ricci scalar still maintains its dimension $L^{-2}$, as should be. But if you use coordinates of mixed type (dimensionless, length-dimensional, time-dimensional,...) then the dimension of the objects listed above depends on the indices. This may create a big confusion. \er


\section{Scale factor and quotient metric}\label{s:qma}

\bt\label{t:fact-1}
There exists a two-variable function $a(t_\uno,t_\due)$ such that
\be\label{e:scf-1}
\bboxed{g(t_\uno)=a^2(t_\uno,t_\due)\,g(t_\due)}
\ee
where $g(t_\uno)$ and $g(t_\due)$ are the metric tensors on the spatial sections $S_{t_\uno}$ and $S_{t_\due}$.
\et

\bprf
Each spatial section $S_t$ has a constant curvature $K(t)$. Since two spatial sections $S_{t_\uno}$ and $S_{t_\due}$ are diffeomorphic, the constant curvatures $K(t_\uno)$ and $K(t_\due)$ have the same sign (or are both equal to $0$).  In accordance with what has been said at the end of Section \ref{s:ccm}, the metrics $g(t_\uno)$ and  $g(t_\due)$ are conformal with a  positive constant conformal factor $a^2$: $g(t_\uno)=a^2\,g(t_\due)$. \eprf 
 
Actually this constant is a dimensionless positive function $a^2(t_\uno,t_\due)$ of the two dates $t_\uno$ and $t_\due$. We assume, without loss of generality, that this function is p\underbar{ositive}. 
From the definition \eqref{e:scf-1} it follows that it obeys the following rules:\footnote{
\kern 2pt Note that it is not commutative in $(t_\uno,t_\due)$. The use of the scale factor must be attentive to the location of the two variables $(t_\uno,t_\due)$.
}
\be\label{e:scf-2} 
\bboxed{\ba 
a(t,t)=1,
\ac
a(t_\uno,t_\due)\;a(t_\due,t_\tre)=a(t_\uno,t_\tre) \quad
\hbox{(composition rule)},
\ax
a(t_\uno,t_\due)=\dfrac 1{a(t_\due,t_\uno)}. 
\ea} 
\ee

If we fix a {\bf reference time} $t_\sharp$ then by virtue of Theorem \ref{t:diff} (page\ \pageref{t:diff}) we can take the spatial section $S_{t_\sharp}$ as a representative of the quotient manifold $Q$. Then $Q$ is endowed with the {\bf quotient metric} 
\be\label{e:defqm} 
\bboxed{\wt g\Def g(t_\sharp)}
\ee
and \eqref{e:scf-1} provides the equation
\be\label{e:fact} 
\bboxed{g(t)=a^2(t,t_\sharp)\,\wt g}
\ee

\begin{center}
\fbox{\begin{minipage}{.95\linewidth} \vspu
{\bf Notation}. The geometrical objects associated with the quotient metric -- or which belong to the geometry of the quotient manifold -- will be marked with a  tilde $\;\wt{}\;$. An exception will be the  arc element of the quotient metric $\wt g$, denoted by $ds$ -- see for instance equation \eqref{e:dstds} below. \vspd
\end{minipage}}
\end{center}

This {\bf factorization equation} will be widely applied in the following. An equivalent form is 
\be\label{e:dstds}
\bboxed{ds_t=a(t,t_\sharp)\,ds}
\ee
where $ds_t$ is the arc element of $g(t)$.

\begin{figure} [H]
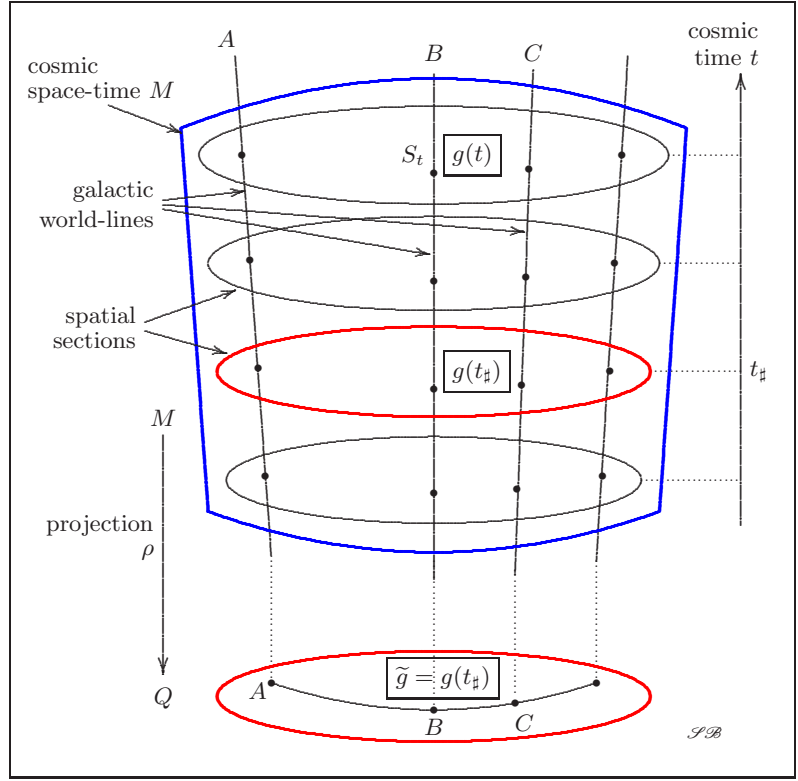

  \centering
\kern -3mm 
$
\beginpicture
\setcoordinatesystem units <1.2cm,1.2cm>
\setplotarea x from -4.7 to 4, y from -3.8 to 4.8
\normalgraphs
\small
\grid 1 1
\setlinear
\plot 0 -1.6 0 4 /
\plot .9 -1.55 1.1 4.05 /
\plot 1.8 -1.4 2.15 4.2 /
\plot -1.8 -1.35 -2.2 4.2 /
\setdots<2pt>
\setlinear
\plot 0 -1.4 0 -3 /
\plot .9 -1.5 .9 -2.9 /
\plot 1.8 -1.4 1.8 -2.7 /
\plot -1.8 -1.4 -1.8 -2.7 /
\plot 2.3 -.5 3.4 -.5 /
\plot 2.4 .7 3.4 .7 /
\plot 2.5 1.9 3.4 1.9 /
\plot 2.6 3.1 3.4 3.1 /
\setsolid
\arrow <6pt> [.2,.6] from 3.4 -1 to 3.4 4
\put{cosmic} [br] at 3.6 4.4
\put{time $t$} [br] at 3.6 4.1
\setplotsymbol (\normaleblue)
\setquadratic
\plot -2.8 3.4 
0 3.95
2.8 3.4 /
\plot -2.5 -.85 
0 -1.3
2.5 -.85 /
\setlinear
\plot -2.8 3.4 -2.5 -.85 /
\plot 2.8 3.4 2.5 -.85 /
\normale
\ellipticalarc axes ratio 4.8:1 360 degrees from 2.3 -.5 center at 0 -.5
\ellipticalarc axes ratio 4.8:1 360 degrees from 2.5 1.9 center at 0 1.9
\ellipticalarc axes ratio 4.8:1 360 degrees from 2.6 3.1 center at 0 3.1
\put{$Q$} [t] at -3 -2.8
\setplotsymbol (\normalered)
\ellipticalarc axes ratio 4.8:1 360 degrees from 2.4 -2.9 center at 0 -2.9
\ellipticalarc axes ratio 4.8:1 360 degrees from 2.4 .7 center at 0 .7
\normale
\put{$\sc\bullet$} at -1.8 -2.75
\put{$\sc\bullet$} at 0 -3.05
\put{$\sc\bullet$} at .9 -2.97
\put{$\sc\bullet$} at 1.8 -2.75
\setquadratic
\plot -1.8 -2.75 0 -3.05 1.8 -2.75 /
\put{$M$} [b] at -3 .1
\arrow <6pt> [.2,.6] from -3 0 to -3 -2.65
\put{projection} [r] at -3.1 -1
\put{$\rho$} [r] at -3.1 -1.3
\put{galactic} [r] at -3.1 2.7
\put{world-lines} [r] at -3.1 2.4
\arrow <5pt> [.2,.6] from -3 2.6 to -2.1 2.7
\arrow <5pt> [.2,.6] from -3 2.55 to 1 2.25
\arrow <5pt> [.2,.6] from -3 2.5 to -.05 2
\put{spatial} [r] at -3.3 1.3
\put{sections} [r] at -3.3 1.05
\arrow <5pt> [.2,.6] from -3.2 1.25 to -2.2 1.6
\arrow <5pt> [.2,.6] from -3.2 1.15 to -2.3 .9
\put{$A$} [b] at -2.3 4.3
\put{$A$} [rt] at -1.85 -2.75
\put{$B$} [b] at 0 4.15
\put{$B$} [t] at 0 -3.15
\put{$C$} [b] at 1.1 4.15
\put{$C$} [t] at 1 -3.1
\put{$\sc\bullet$} at -2.13 3.1
\put{$\sc\bullet$} at 0 2.9
\put{$\sc\bullet$} at 1.05 2.95
\put{$\sc\bullet$} at 2.08 3.1
\put{$\sc\bullet$} at -2.04 1.94
\put{$\sc\bullet$} at 0 1.7
\put{$\sc\bullet$} at 1.02 1.75
\put{$\sc\bullet$} at 2 1.9
\put{$\sc\bullet$} at -1.95 .74
\put{$\sc\bullet$} at 0 .5 
\put{$\sc\bullet$} at .97 .55
\put{$\sc\bullet$} at 1.95 .7
\put{$\sc\bullet$} at -1.87 -.46
\put{$\sc\bullet$} at 0 -.65
\put{$\sc\bullet$} at .92 -.6
\put{$\sc\bullet$} at 1.87 -.46
\normale
\put{$S_t$} [r] at -.1 3.1
\put{$\boxed{g(t)}$} [l] at .1 3.1
\put{$\boxed{g(t_\sharp)}$} [l] at .1 .7
\put{$t_\sharp$} [l] at 3.5 .7
\put{$\boxed{\wt g=g(t_\sharp)}$}  at .1 -2.7
\put{cosmic} [l] at -4.5 4.1
\put{space-time $M$} [l] at -4.5 3.8
\arrow <5pt> [.2,.6] from -3.6 3.65 to -2.8 3.35
\put{\firma} at 3 -3.3
\endpicture
$
  \caption{The quotient manifold and the quotient metric.}
  \label{fig:Qman}
\end{figure}

We call {\bf scale factor} the positive function $a(t,t_\sharp)$.\footnote{
\kern 2pt It is also called {\bf scale parameter} or {\bf expansion factor}. It is also denoted by the symbol $R(t)$, here used for the radius of the universe.} 
 It has to be regarded as a function of $t$, but determined by the preset value of $t_\sharp$. When there is no danger of confusion, we can simply denote it by $a(t)$,  
\be\label{e:def-at}
\bboxed{a(t)\Def a(t,t_\sharp)}
\ee
Note that $a(t_\sharp)=a(t_\sharp,t_\sharp) =1$. For this reason $t_\sharp$ will be also called {\bf normalization time}.

\br\label{r:a=1}
The composition rule in \eqref{e:scf-2} implies that if $a(t_\uno,t_\due)$ is constant then this constant is necessarily equal to 1. For $a(t)={\rm constant}=1$ we have the so-called {\bf static universe}: $g(t)=g^\sharp$ for all $t$. 
\er

\br\label{r:a=0}
The scale factor plays a central role in cosmology because, as a function of the cosmic time, it contains all the information concerning the {\bf evolution of the universe}. It is determined by dynamical equations established in the next chapter.   \er

\begin{center}
\fbox{\begin{minipage}{.95\linewidth}
\br\label{r:test}\vspu
 In the current literature it is not sufficiently emphasized the fact that the scale factor $a(t)$ is defined up to the choice of a reference time. This oversight precludes the recognition of some important facts. It should be borne in mind that, in order to have a physical meaning, any formula involving $a(t)$ and its derivatives must be invariant under the change of a reference time.\vspd\quad \er
\end{minipage}}
\end{center}

\br
Let $a(t,t_\sharp)$ and $a(t,t_\flat)$ be the scale parameters associated with two  different reference times $t_\sharp$ and $t_\flat$. By applying 
the composition rule \eqref{e:scf-2} we get
\be\label{e:at0t*}
\bboxed{a(t,t_\sharp)=a(t,t_\flat)\,a(t_\flat,t_\sharp)}
\ee
This shows that the scale factors associated with two different reference dates differ by a constant factor (depending on the reference dates). 
\er

\br
The metric tensor $g(t)=a^2(t,t_\sharp)\,g(t_\sharp)$ does not depend, by definition, on the choice of the reference time $t_\sharp$, so that
\be\label{e:agt0t*}
a^2(t,t_\sharp)\,g(t_\sharp)=a^2(t,t_\flat)\,g(t_\flat)
\ee
For $t=t_\sharp$ we get the relation
\be\label{e:qma-3}
\bboxed{g(t_\sharp)=a^2(t_\sharp,t_\flat)\,g(t_\flat)}
\ee
which is in agreement with \eqref{e:fact}. \er

\br
We can write the factorization equation \eqref{e:fact} in terms of the coordinate $q^\Z=\kappa\,t$ by setting
\begin{gather}
\label{e:Aq0}
\bboxed{a(t)=A(q^\Z)}
\\
\label{e:factq0}
\bboxed{g_{ab}(q^\Z,\wt q)=A^2(q^\Z)\;\wt g_{ab}(\wt q)}
\end{gather}
As a rule, we will use a small letter to denote a scalar function of $t$ and the corresponding capital letter to denote the same scalar as a function of $q^\Z$: $f(t)=F(q^\Z)$. 
\er

\br\label{r:test-G}
The quotient manifold is endowed with the Levi-Civita connection  coming from the quotient metric, henceforth denoted by $\wt\Gamma$, whose Christoffel symbols are
\be\label{e:wtGamma}
\wt\Gamma_{ab}^c\Def \tfrac 12\,\wt g^{cd}\,\big(\partial_a\wt g_{bd}+\partial_b\wt g_{da}-\partial_d \wt g_{ab}\big),
\ee
where $\wt g^{ab}$ are the contravariant components of the quotient metric. 
It is easy to check that these symbols are invariant under the change of the reference date. It follows that the $\wt\Gamma$-geodesics, as unparametrized curves, are not affected by the change of the reference time.
\er

\br\label{r:Dim-Gamma}
Since the co-moving coordinates $q^a$ are $L$-dimensional (Section \ref{s:Dim}) the metric tensor components $\wt g_{ab}$ are dimensionless: 
$$
\bc
\Dim(\wt g^{ab})=1.
\ax
\Dim(\wt\Gamma_{ab}^c)=L^{-1}. \quad \bullet
\ec
$$
\erx

\br
On the quotient manifold the {\bf arc-element} $ds$ \index{arc-element $ds$} of any curve $q^a=\gamma^a(\xi)$  (with generic parameter $\xi$) is defined by 
\be\label{e:def-ds}
ds\Def\sqrt{\wt g_{ab}\,dq^a\,dq^b}=\sqrt{\wt g_{ab}\,\dfrac{d\gamma^a}{d\xi}\,\dfrac{d\gamma^b}{d\xi}
\;}\; d\xi.
\ee
In turn, the {\bf arc-length} \index{arc-length $s$} $s$ is defined by
\be\label{e:def-s}
s(\xi_\uno)-s(\xi_\Z)\Def\dint_{\xi_\Z}^{\xi_\uno}\!ds
=\dint_{\xi_\Z}^{\xi_\uno}\!\sqrt{\wt g_{ab}\,\dfrac{d\gamma^a}{d\xi}\,\dfrac{d\gamma^b}{d\xi}
\;}\; d\xi.
\ee
It is a privileged parameter for any curve on $Q$, since 
\be\label{e:ds}
\bboxed{\wt g_{ab}\,\dfrac{d\gamma^a}{ds}\,\dfrac{d\gamma^b}{ds}=1}
\ee
This follows from \eqref{e:def-ds}
$$
\dfrac{ds}{d\xi}=\sqrt{\wt g_{ab}\,\dfrac{d\gamma^a}{d\xi}\,\dfrac{d\gamma^b}{d\xi}
\;}
$$
and
$$
\wt g_{ab}\,\dfrac{d\gamma^a}{ds}\,\dfrac{d\gamma^b}{ds}
=\wt g_{ab}\,\dfrac{d\gamma^a}{d\xi}\,\dfrac{d\gamma^b}{d\xi}
\left(\dfrac{d\xi}{ds}
\right)^{\!2}=1.
$$
However, the arc-element and the arc-length are not invariant under the change of the reference time $t_\sharp$. If we denote by $ds_\sharp$ and $ds_*$ the arc-elements defined in the quotient metrics $g(t_\sharp)$ and $g(t_\flat)$, then equation \eqref{e:qma-3} is equivalent to
\be\label{e:dst0t*}
\quad\bboxed{ds_\sharp=\dfrac{1}{a(t_\sharp,t_\flat)}\,ds_*}\quad\bullet
\ee
\erx

\section{The Hubble law}\label{s:Hl}

Any curve in the quotient manifold, of length $\ell_\sharp$, is carried along the galactic world-lines and generates curves of length $\ell(t)$ on each spatial section $S_t$. By virtue of \eqref{e:dstds}, this length is given by
\be \label{e:ell-t}
\bboxed{\ell(t)=a(t,t_\sharp)\,\ell_\sharp}
\ee
It follows that
\be\label{e:elldot} 
\dot\ell(t)=\dot a(t,t_\sharp)\,\ell_\sharp.
\ee

\bt \label{t:Hnotsharp} 
The derivative $\dot\ell(t)\,$\footnote{
\kern 2pt A dot over a symbol will denote the derivative with 
respect to $t$.}   
 obeys the rule
\be \label{e:H-ell-2}
\bboxed{\dot\ell(t)=h(t)\,\ell(t)}
\ee
where 
\be\label{e:def-ht}
\bboxed{h(t)\Def\dfrac{\dot a(t,t_\sharp)}{a(t,t_\sharp)}}
\ee
does not depend on the reference time $t_\sharp$.
\et

\bprf
From \eqref{e:elldot} and \eqref{e:ell-t} we get
\be \label{e:H-ell-1}
\dot\ell(t)=\dfrac{\dot a(t,t_\sharp)}{a(t,t_\sharp)}\,\ell(t).
\ee
The scale factors relative to different normalization times differ by a multiplicative constant. So, the ratio  \eqref{e:def-ht} remains unchanged. \eprf

\begin{figure} [H]
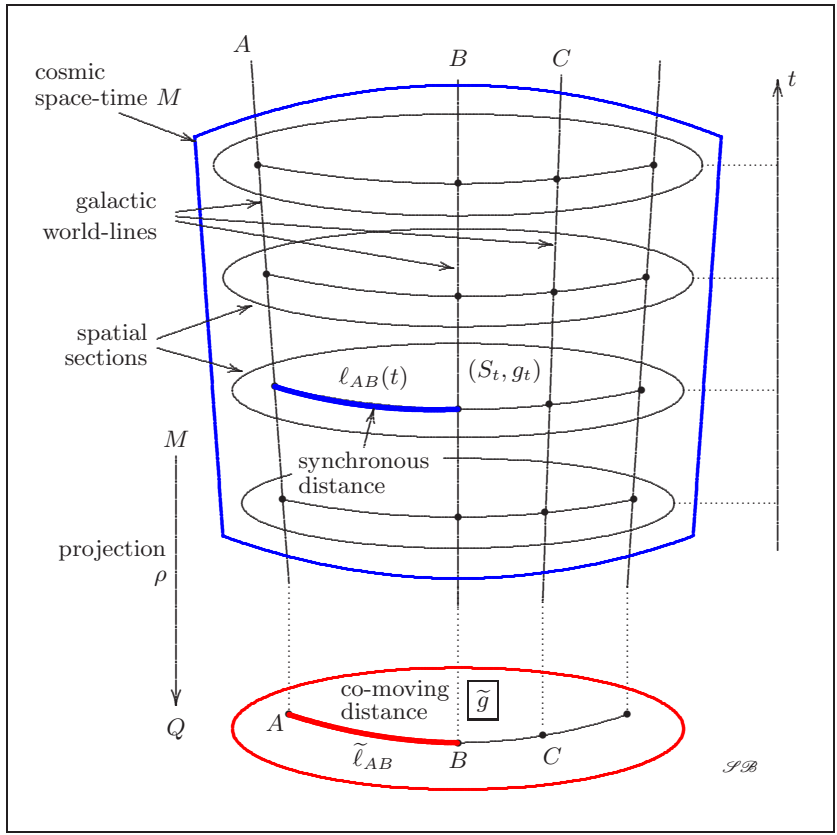

  \centering
\kern -3mm 
$
\beginpicture
\setcoordinatesystem units <1.25cm,1.25cm>
\setplotarea x from -4.8 to 4, y from -4 to 4.8
\normalgraphs
\small
\grid 1 1
\setlinear
\plot 0 -1.6 0 4 /
\plot .9 -1.55 1.1 4.05 /
\plot 1.8 -1.4 2.15 4.2 /
\plot -1.8 -1.35 -2.2 4.2 /
\setdots<2pt>
\setlinear
\plot 0 -1.4 0 -3 /
\plot .9 -1.5 .9 -2.9 /
\plot 1.8 -1.4 1.8 -2.7 /
\plot -1.8 -1.4 -1.8 -2.7 /
\plot 2.3 -.5 3.4 -.5 /
\plot 2.4 .7 3.4 .7 /
\plot 2.5 1.9 3.4 1.9 /
\plot 2.6 3.1 3.4 3.1 /
\setsolid
\arrow <6pt> [.2,.6] from 3.4 -1 to 3.4 4
\put{$t$} [tr] at 3.6 4.1
\setplotsymbol (\normaleblue)
\setquadratic
\plot -2.8 3.4 
0 3.95
2.8 3.4 /
\plot -2.5 -.85 
0 -1.3
2.5 -.85 /
\setlinear
\plot -2.8 3.4 -2.5 -.85 /
\plot 2.8 3.4 2.5 -.85 /
\normale
\ellipticalarc axes ratio 4.8:1 94 degrees from 2.3 -.5 center at 0 -.5
\ellipticalarc axes ratio 4.8:1 -220 degrees from 2.3 -.5 center at 0 -.5
\ellipticalarc axes ratio 4.8:1 360 degrees from 2.4 .7 center at 0 .7
\ellipticalarc axes ratio 4.8:1 360 degrees from 2.5 1.9 center at 0 1.9
\ellipticalarc axes ratio 4.8:1 360 degrees from 2.6 3.1 center at 0 3.1
\put{$Q$} [t] at -3 -2.8
\setplotsymbol (\normalered)
\ellipticalarc axes ratio 3.7:1 360 degrees from 2.4 -2.9 center at 0 -2.9
\normale
\put{$\sc\bullet$} at -1.8 -2.75
\put{$\sc\bullet$} at 0 -3.05
\put{$\sc\bullet$} at .9 -2.97
\put{$\sc\bullet$} at 1.8 -2.75
\setquadratic
\plot -1.8 -2.75 0 -3.05 1.8 -2.75 /
\put{$M$} [b] at -3 .1
\arrow <6pt> [.2,.6] from -3 0 to -3 -2.65
\put{projection} [r] at -3.1 -1
\put{$\rho$} [r] at -3.1 -1.3
\put{galactic} [r] at -3.2 2.7
\put{world-lines} [r] at -3.2 2.4
\arrow <5pt> [.2,.6] from -3 2.6 to -2.1 2.7
\arrow <5pt> [.2,.6] from -3 2.55 to 1 2.25
\arrow <5pt> [.2,.6] from -3 2.5 to -.05 2
\put{$A$} [b] at -2.3 4.3
\put{$A$} [rt] at -1.85 -2.75
\put{$B$} [b] at 0 4.15
\put{$B$} [t] at 0 -3.15
\put{$C$} [b] at 1.1 4.15
\put{$C$} [t] at 1 -3.1
\put{spatial} [r] at -3.3 1.3
\put{sections} [r] at -3.3 1.05
\arrow <5pt> [.2,.6] from -3.2 1.25 to -2.2 1.6
\arrow <5pt> [.2,.6] from -3.2 1.15 to -2.3 .9
\put{$\sc\bullet$} at -2.13 3.1
\put{$\sc\bullet$} at 0 2.9
\put{$\sc\bullet$} at 1.05 2.95
\put{$\sc\bullet$} at 2.08 3.1
\setquadratic
\plot -2.13 3.1 0 2.9 2.08 3.1 /
\put{$\sc\bullet$} at -2.04 1.94
\put{$\sc\bullet$} at 0 1.7
\put{$\sc\bullet$} at 1.02 1.75
\put{$\sc\bullet$} at 2 1.9
\plot -2.04 1.94 0 1.7 2 1.9 /
\put{$\sc\bullet$} at -1.95 .74
\put{$\sc\bullet$} at 0 .5 
\put{$\sc\bullet$} at .97 .55
\put{$\sc\bullet$} at 1.95 .7
\plot -1.95 .74 0 .5 1.95 .7 /
\put{$\sc\bullet$} at -1.87 -.46
\put{$\sc\bullet$} at 0 -.65
\put{$\sc\bullet$} at .92 -.6
\put{$\sc\bullet$} at 1.87 -.46
\plot -1.87 -.46 0 -.65 1.87 -.46 /
\setplotsymbol (\spessoblue)
\setquadratic
\plot -1.95 .74 -1 .54 0 .5  /
\setplotsymbol (\spessored)
\plot -1.8 -2.75 -.9 -2.97  0 -3.05 /
\normale
\put{\small synchronous} [lt] at -1.7 .05
\put{\small distance} [l] at -1.7 -.3
\arrow <5pt> [.2,.6] from -1 .1 to -.9 .5
\put{$\ell_{AB}(t)$} [b] at -.9 .7
\put{$(S_t,g_t)$} [l] at .1 .9
\put{\small co-moving} [lb] at -1.25 -2.6 
\put{\small distance} [lb] at -1.25  -2.8 
\put{$\wtell_{AB}$} [t] at -.9 -3.05
\put{$\boxed{\wt g}$} [l] at .1 -2.6
\put{cosmic} [l] at -4.5 4.1
\put{space-time $M$} [l] at -4.5 3.8
\arrow <5pt> [.2,.6] from -3.6 3.65 to -2.8 3.35
\put{\firma} at 3 -3.3
\endpicture
$
\vskip -2mm
  \caption{
Synchronous and co-moving distance.
}
  \label{fig:Q}
\end{figure}

The above argument can be applied to the case of geodesic curves, where the `lengths' are `distances'. In particular, we can consider two distances between two galaxies $A$ and $B$:

(i) The {\bf co}--{\bf moving distance} $\wt \ell_{AB}$: this is the length of the geodesic joining $A$ to $B$ in the quotient metric $\wt g$. This distance is a constant depending on $A$ and $B$ only (Fig.\ \ref{fig:Q}).

(ii) The {\bf synchronous distance} \index{synchronous!distance} $\ell_{AB}(t)$ at the time $t$:  this is the length of the shortest geodesic joining $A$ to $B$ in the spatial section $S_t$ with metric $g(t)$. The derivative
  $\dot \ell_{AB}(t)$ with respect to $t$ of the synchronous distance $\ell_{AB}(t)$ is called the 
\index{recession!velocity}
{\bf recession velocity} of the galaxies $A$ and $B$.

By applying equations \eqref{e:ell-t} and \eqref{e:H-ell-2} to these distances we find the two relationships
\begin{gather}
\label{e:Hl0}
\bboxed{\ell_{AB}(t)=a(t_\sharp)\;\wtell_{AB}}
\\
\label{e:Hl}
\bboxed{\dot \ell_{AB}(t)=h(t)\;\ell_{AB}(t)}
\end{gather}
In the second of these we recognize the celebrated {\bf Hubble law}, and in $h(t)$ the {\bf Hubble parameter}.

\begin{center}
\fbox{\begin{minipage}{.95\linewidth}
\br  \vspu The Hubble parameter $h(t)$ is not affected by the choice of the reference time $t_\sharp$.  In the present theory the Hubble law is a matter of pure Kinematics: it is a consequence of the  postulates so far stated for the structure of the cosmic space-time. In other words, {\it it does not depend on whatsoever dynamical setting}.  \vspd \quad\er
\end{minipage}}
\end{center}
  
\section{Isotropic vectors and tensors}\label{s:ivt}

\bd
A tensor on space-time is {\bf isotropic} \index{isotropic!tensor} if it meets the isotropy principle: it does not generates  distinguished vector fields tangent to the spatial sections. 
\ed

Since the isotropy principle is one of the main postulates, only isotropic tensors are {\bf admissible} \index{admissible tensor} in the present cosmological theory. The {\bf isotropic scalar} \index{isotropic!scalar} have been considered in Theorem \ref{t:scalar}. Now we characterize the isotropic vectors and two-tensors.

\bt\label{t:V}
A vector field $V^\alpha$ is isotropic if and only if 
its components in a co-moving coordinate system $(q^\Z,q^a)$ are of the following type:
$$
\bc
V^\Z=\hbox{\rm function of $q^\Z$ only},
\ax
V^a=0.
\ec
$$
\et 

\bprf
If $V^\Z$ is also a function of the coordinates $\wt q$ then its gradient defines a distinguished direction on each spatial section. If $V^a\neq 0$ then a distinguished direction field is defined on each spatial section. \eprf

\bt\label{t:Sq0}
A contravariant symmetric two-tensor $T^{\alpha\beta}$ is isotropic if and only if its components  are of the following type:
\be\label{e:Tq0}
\bboxed{\bc
T^{\Z\Z}={\Phi}(q^\Z)=\hbox{\rm a function of $q^\Z$ only},
\ax
T^{\Z a}=0,
\ax
T^{ab}={\Psi}(q^\Z)\,\wt g^{ab}(\wt q)=\hbox{\rm a function of $q^\Z$ times $\wt g^{ab}$},
\ec}
\ee
where $\wt g^{ab}$ are the contravariant components of the quotient metric $\wt g_{ab}$. 
A similar result holds for an admissible covariant symmetric tensor:
\be\label{e:Tcovq0}
\bboxed{\bc
T_{\Z\Z}={\Phi}(q^\Z)=\hbox{\rm a function of $q^\Z$ only},
\ax
T_{\Z a}=0,
\ax
T_{ab}={\Psi}(q^\Z)\,\wt g_{ab}(\wt q)=\hbox{\rm a function of $q^\Z$ only times $\wt g_{ab}$} 
\ec}
\ee
\et

\bprf
$
T^{\alpha\beta}
=J_{\alpha'}^\alpha\,J_{\beta'}^\beta\,T^{\alpha'\beta'},
\quad J^{\alpha'}_\alpha\Def\dfrac{\partial q^{\alpha'}}{\partial q^\alpha},
\quad J_{\alpha'}^\alpha\Def\dfrac{\partial q^\alpha}{\partial q^{\alpha'}}.
$

For a transformation of co-moving coordinates leaving $q^\Z$ invariant we have $J^{\Z'}_\Z=1 ,\quad J^{\Z'}_a=0,\quad  J^{a'}_\Z=0, \quad J^\Z_{c'}=0, \quad J^c_{\Z'}=0$. Thus
$$
\bq
T^{\Z\Z}=J_{\alpha'}^\Z\,J_{\beta'}^\Z\,T^{\alpha'\beta'}
=(J_{\Z'}^\Z)^2\,T^{\Z'\Z'}=T^{\Z'\Z'}.
\ac
T^{\Z b}=J_{\alpha'}^\Z\,J_{\beta'}^b\,T^{\alpha'\beta'}
=J_{\Z'}^\Z\,J_{b'}^b\,T^{\Z'b'}=J_{b'}^b\,T^{\Z'b'}.
\ac
T^{ab}=J_{\alpha'}^a\,J_{\beta'}^b\,T^{\alpha'\beta'}
=J_{a'}^a\,J_{b'}^b\,T^{a'b'}.
\eq
$$
This shows that: $T^{\Z\Z}$ is a scalar, so it must be a function of $q^\Z$ only; $T^{\Z b}$ is a vector, so it must vanish; $T^{ab}$ is a symmetric tensor on each spatial section generating eigenfields, so it must be proportional to the spatial metric $g_{ab}(q^\Z,\wt q)=A^2(q^\Z)\,\wt g_{ab}(\wt q)$. \eprf

\br\label{r:PhiPsi}
According to this theorem any admissible symmetric two-tensor $T^{\alpha\beta}$ is fully determined by two functions $\Phi$ and $\Psi$ of $q^\Z$ only.
\er

\bt\label{t:noskew}
No skew-symmetric isotropic two-tensor is admissible in the cosmic space-time.\footnote{
\kern 2pt In other words: {\it only symmetric two-tensors are admissible in the isotropic models of the universe}.
}
\et 

\bprf
A skew-symmetric two-tensor $A^{\alpha\beta}$ gives rise to a spatial vector field $A^{\Z a}$ and to a spatial skew-symmetric tensor field $A^{ab}$. The isotropy principle implies $A^{\Z a}=0$. Any antisymmetric tensor field $A^{ab}$ on a three-dimensional Riemannian space has a real eigenvector. This is in contrast with the  isotropy principle. Thus $A^{ab}=0$. \eprf

\br
Theorem \ref{t:noskew} shows that the existence in space-time of a single or a finite number of electro-magnetic fields is incompatible with the isotropic cosmology. However, a continuous distribution of electro-magnetic fields such as those emitted from galaxies, may not generate any particular direction. Such fields are then admissible. \er

\br\label{r:symmconn}
Since the torsion $\Gamma_{\alpha\beta}^\gamma-\Gamma_{\beta\alpha}^\gamma$ of a linear connection is a skew-symmetric tensor in the lower indices, only symmetric connections are allowed in space-time. \er

\section{The cosmic radius and the angular distance}\label{s:crag}

For cosmological models with non-flat spatial sections -- $K(t)\neq 0$ -- we can define the {\bf cosmic radius} 
\index{cosmic!radius} $R(t)>0$  through the equation
\be\label{e:KR}
\boxed{\vp\;K(t)=\dfrac \varepsilon{R^2(t)}, \quad \varepsilon=\pm 1\;}
\ee
with $\varepsilon=\pm 1$ according to the sign of the curvature. If we set 
\be\label{e:Rtilde}
\wt K\Def K(t_\sharp),\quad \wt R\Def R(t_\sharp),
\ee
then we have\footnote{
\kern 2pt Note that the product $\varepsilon\,\wt K$ is always positive.} 
\be\label{e:Radius}
\bboxed{\varepsilon\,\wt K=\dfrac 1{\wt R^2}}
\ee
and, due to the general formula \eqref{e:ell-t}, we have
\begin{gather}
\label{e:aRR}
\bboxed{R(t)=a(t,t_\sharp)\,\wt R}
\\
\label{e:wtK}
\bboxed{K(t)=\dfrac{\wt K}{a^2(t,t_\sharp)},\;\;\forall\;t}
\end{gather}
By putting in particular $t=t_\flat$ in these equations we get
\begin{gather}
\label{e:Rt*Rt0}
\bboxed{R(t_\flat)=a(t_\flat,t_\sharp)\,\wt R}
\\
\label{e:Kt*Kt0}
\bboxed{K(t_\flat)=\dfrac{\wt K}{a^2(t_\flat,t_\sharp)}}
\end{gather}
Furthermore, from $h=\dot a/a$ we get
\be\label{e:h2}
\bboxed{h(t)=\dfrac{\dot R(t)}{R(t)}=\dfrac{d\log R(t)}{dt}}
\ee
This formula gives the Hubble parameter in terms of the cosmic radius. We call $\dot R$ the {\bf radial velocity} of the universe. 
Finally, from \eqref{e:Hl0} and  \eqref{e:aRR} it follows that for any pair  of galaxies 
\be\label{e:defpsi}
\bboxed{\psi_{AB}\Def\dfrac{\wtell_{AB}}{\wt R}=\dfrac {\ell_{AB}(t)}{R(t)}=\rm constant}
\ee
We call {\bf angular distance}  of two galaxies the dimensionless constant $\psi_{AB}$. Its geometrical meaning is explained in the next section. As a consequence, the Hubble law can be written as
\be\label{e:Hl-psi}
\bboxed{\dot \ell_{AB}(t)=\psi_{AB}\,\dot R(t)}
\ee
Note that the angular distance $\psi_{AB}$ does not depend on the choice of the reference date $t_\sharp$. 

\section{Topological types of the universe}

The isotropic cosmological models that can be constructed on the basis of the postulates so far stated  have two characteristic elements: (i) the scale parameter $a(t)$, a function of the cosmic time $t$ defined on a real (bounded or unbounded) interval $(t_\alpha,t_\omega)$ (beginning and end of the universe) which contains all the information about the evolution of the cosmos  (expansion, contraction, etc.) and (ii) the sign of the constant curvature $\wt K$ of the quotient manifold. However, it must be observed that this description is incomplete because it lacks a third characteristic element: the topology of the quotient manifold (that is the topology of the spatial sections).

There are several topological types of three-dimensional manifolds with constant curvature. The types that are commonly considered in cosmology are the following three ones: 

(i)  $Q\simeq \mathbb S_\tre$, the three-dimensional sphere, positive curvature,

(ii) $Q\simeq \mathbb H_\tre$, the three-dimensional pseudo-sphere , negative curvature,

(iii) $Q\simeq \mathbb E_\tre$, the three-dimensional Euclidean space, zero curvature. 

The corresponding models are called (i) {\bf closed universe}, (ii) {\bf open universe} and (iii) {\bf flat universe}.\footnote{
\kern 2pt Some other topological structures have been object of speculations. Among these, the closest to our understanding is the three-dimensional torus $\mathbb T_\tre$. But, although $\mathbb T_\tre$ is a good representative of a manifold with zero curvature, such a topological model is contrary to the isotropy principle: there are just three distinguished and independent vector fields whose integral curves are circles.} 
 In this section we take a look at these three types of universes, but 
 we have to point out right now  that the non-flat universes are incompatible with the barotropic dynamics studied in Chapter \ref{c:bar}.

\subsection{Closed universe}

\begin{figure} [H]
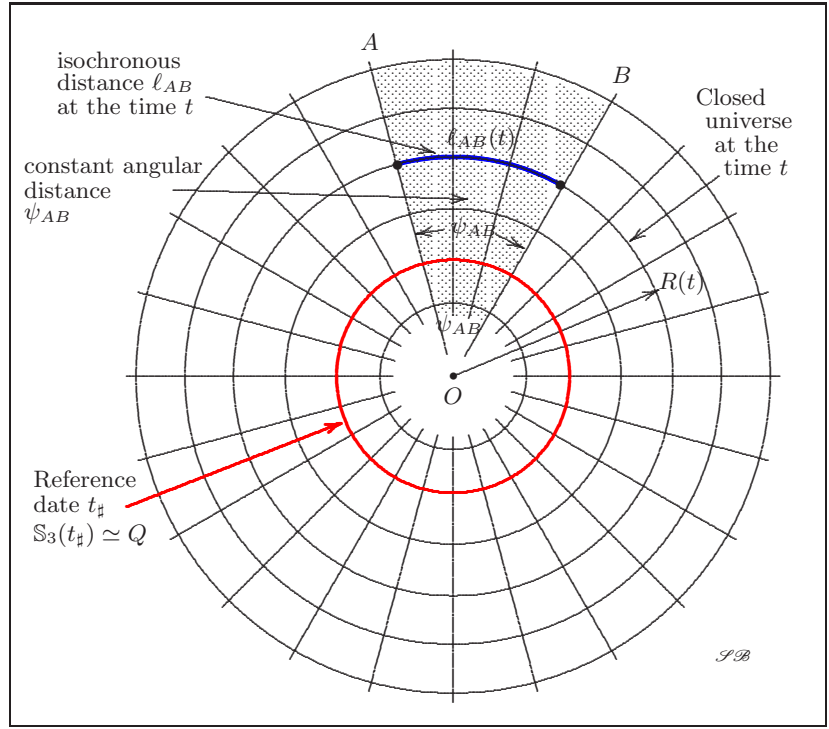
 
$$
\beginpicture
\setcoordinatesystem units <.62cm, .62cm>
\setplotarea x from -9 to 9, y from -7 to 7
\normalgraphs
\putrectangle corners at -9.5 8 and 8 -7.5
\small
\put{$O$} [t] at 0 -.25
\put{$\sc\bullet$} at 0 0
\setlinear 
\circulararc 360 degrees from 0 1.57 center at 0 0
\circulararc 360 degrees from 0 3.6 center at 0 0
\circulararc 360 degrees from 0 5.75 center at 0 0
\circulararc 360 degrees from 0 6.8 center at 0 0
\setplotsymbol (\spessinoblue)
\circulararc 15 degrees from 0 4.71 center at 0 0
\circulararc -30 degrees from 0 4.71 center at 0 0
\put{$\bullet$} at -1.2 4.53
\put{$\bullet$} at 2.3 4.1
\put{$\ell_{AB}(t)$} at .6 5.1
\put{$\psi_{AB}$} at .45 3.2
\put{$\psi_{AB}$} at .13 1.1
\normale
\circulararc -360 degrees from 0 4.71 center at 0 0
\setsolid
\setplotsymbol ({\fiverm .})
\arrow <6pt> [.2,.6] from -.2 3.15  to -.8 3.05
\arrow <6pt> [.2,.6] from .92 3.05 to 1.5 2.75
\put{\firma} at 6 -6
\plot 0 1.3 0 7 /  
\startrotation by .965 -.258 about 0 0 
\plot 0 1.3 0 7 /
\stoprotation
\startrotation by .866 -.5 about 0 0 
\plot 0 .5 0 7 /
\stoprotation
\startrotation by .707 -.707 about 0 0 
\plot 0 1.3 0 7 /
\stoprotation
\startrotation by .5 -.866 about 0 0 
\plot 0 1.3 0 7 /
\stoprotation
\startrotation by .965 .258 about 0 0 
\plot 1.3 0 7 0 /
\stoprotation
\plot 1.3 0 7 0 /  
\startrotation by .965 -.258 about 0 0 
\plot 1.3 0 7 0 /
\stoprotation
\startrotation by .866 -.5 about 0 0 
\plot 1.3 0 7 0 /
\stoprotation
\startrotation by .707 -.707 about 0 0 
\plot 1.3 0 7 0 /
\stoprotation
\startrotation by .5 -.866 about 0 0 
\plot 1.3 0 7 0 /
\stoprotation
\startrotation by .965 .258 about 0 0 
\plot 0 -1.3 0 -7 /
\stoprotation
\plot 0 -1.3 0 -7 /   
\plot -1.3 0 -7 0 /
\startrotation by .965 -.258 about 0 0 
\plot -1.3 0 -7 0 /
\stoprotation
\startrotation by .866 -.5 about 0 0 
\plot -1.3 0 -7 0 /
\stoprotation
\startrotation by .707 -.707 about 0 0 
\plot -1.3 0 -7 0 /
\stoprotation
\startrotation by .5 -.866 about 0 0 
\plot -1.3 0 -7 0 /
\stoprotation
\startrotation by .965 .258 about 0 0 
\plot 0 .5 0 7 /
\stoprotation
\startrotation by .965 .258 about 0 0 
\plot -1.3 0 -7 0 /
\stoprotation
\startrotation by .866 .5 about 0 0 
\plot -1.3 0 -7 0 /
\stoprotation
\startrotation by .707 .707 about 0 0 
\plot -1.3 0 -7 0 /
\stoprotation
\startrotation by .5 .866 about 0 0 
\plot -1.3 0 -7 0 /
\stoprotation
\startrotation by .965 -.258 about 0 0 
\plot 0 -1.3 0 -7 /
\stoprotation
\setshadegrid span <1.35pt>
\vshade 
-1.7 6.5 6.55
-.365 1.3 6.75
-.4 1.3 6.7
0 1.3 6.8
.75 1.2 6.7
2.1 3.8 6.4
3.3 5.8 5.8
 /
\put{constant angular} [l] at -9.2 4.5
\put{distance} [l] at -9.2 4
\put{$\psi_{AB}$} [l] at -9.2 3.5
\arrow <6pt> [.2,.6] from -6 4 to .3 3.8
\put{isochronous} [l] at -8.5 6.8
\put{distance $\ell_{AB}$} [l] at -8.5 6.3
\put{at the time $t$} [l] at -8.5 5.8
\arrow <6pt> [.2,.6] from -5.3 6 to -.4 4.8
\put{$A$} [br] at -1.6 7
\put{$B$} [bl] at 3.4 6.3
\put{$R(t)$} [l] at 4.4 2
\arrow <6pt> [.2,.6] from 0 0  to 4.4 1.85
\put{Closed} [l] at 5.2 6
\put{universe} [l] at 5.4 5.5
\put{at the} [l] at 5.6 5
\put{time $t$} [l] at 5.8 4.5
\arrow <6pt> [.2,.6] from 5.7 4.3  to 3.8 2.9
\setplotsymbol (\normalered)
\circulararc 360 degrees from 0 2.5 center at 0 0
\put{Reference} [l] at -9 -2.2
\put{date $t_\sharp$} [l] at -9 -2.8
\setplotsymbol (\normalered)
\arrow <6pt> [.2,.6] from -7 -2.8  to -2.4 -1
\put{$\mathbb S_3(t_\sharp)\simeq Q$} [l] at -9 -3.4
\endpicture
$$
\vskip -4mm 
  \caption{Radial diagram of the closed universe.}
  \label{fig:cu}
\end{figure}
In a closed universe the spatial sections $S_t$ are three-dimensional sphere of radius $R(t)$  immersed in the Euclidean affine space  $\mathbb R^4=(w,x,y,z)$ and centered at the origin $O$ of the coordinates; they represented by the equation
$$
w^2+x^2+y^2+z^2=R^2(t).
$$
Their curvature is $K(t)=R^{-2}(t)>0$. This geometrical vision differs from that of the space-time where the submanifolds $S_t$ form a foliation. Here they contract and expand in time according to the function $R(t)$. What we get is a sort of movie which we call \index{radial!diagram} {\bf radial diagram} (Fig.\ \ref{fig:cu}). One of the spheres, corresponding to a reference date $t_\sharp$, can be identified with the quotient manifold $Q$. Any galaxy is represented by a point moving along a straight line crossing the origin $O$. At any date $t$ two galaxies $A$ and $B$ stay on the sphere of radius $R(t)$ and are separated by a circular arc of maximal radius (i.e.\ by a geodesic arc) of length $\ell_{AB}(t)$.  Then the straight lines joining $A$ and $B$ to the center $O$ form, in agreement with \eqref{e:defpsi}, an angle $\psi_{AB}$ such that $\ell_{AB}(t)=\psi_{AB}\,R(t)$, which remains constant in time. The maximal distance between two galaxies is $\pi\,R$, the half of the length $2\,\pi\,R$ of a maximal (geodesic) circle. Then the maximal angular distance is $\psi_{\rm max}=\pi$. 
\subsection{Open universe}
In this model the spatial sections $S_t$ are three-dimensional hyperboloid $\mathbb H_3$ of radius $R_t$
$$
w^2-x^2-y^2-z^2=R^2(t)
$$
immersed in the Minkowski affine space $M_4=(w,x,y,z)$ with signature $(-+++)$. The hyperboloid $\mathbb H_3$ is the set of points $P$ such that the vector $OP$ is time-like  and with positive component with respect the time-like coordinate $w$. It results to be 
 a space-like three-dimensional surface whose curvature at the date $t$ is $K(t)=-R^{-2}(t)<0$ (Fig.\  \ref{fig:ou}). Remarks similar to those concerning the closed universe hold for the open universe. The only difference is that in the open universe the angular distance is not bounded.

\begin{figure}[H]
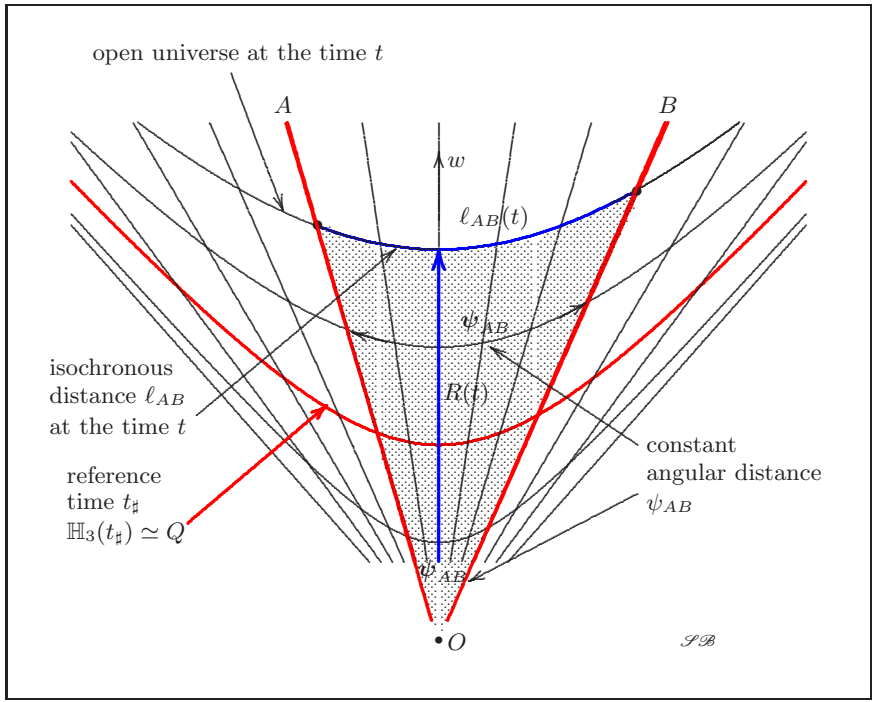
 
$$\kern -1mm
\beginpicture
\setcoordinatesystem units <1.15cm, 1.3cm>
\setplotarea x from -4.25 to 4.25, y from .8 to 5.3
\normalgraphs
\small
\putrectangle corners at -5 6.5 and 5 -.6
\inboundscheckon
\setlinear
\plot 
0 0 5 5 /
\plot
0 0 -5 5 /
\setquadratic
\plot
0	1
0.2	1.02
0.4	1.077
0.6	1.166
0.8	1.281
1	1.414
1.2	1.562
1.4	1.72
1.6	1.887
1.8	2.059
2	2.236
2.2	2.417
2.4	2.6
2.6	2.786
2.8	2.973
3	3.162
3.2	3.353
3.4	3.544
3.6	3.736
3.8	3.929
4	4.123
4.2	4.317
4.4	4.512
/

\plot
0	1
-0.2	1.02
-0.4	1.077
-0.6	1.166
-0.8	1.281
-1	1.414
-1.2	1.562
-1.4	1.72
-1.6	1.887
-1.8	2.059
-2	2.236
-2.2	2.417
-2.4	2.6
-2.6	2.786
-2.8	2.973
-3	3.162
-3.2	3.353
-3.4	3.544
-3.6	3.736
-3.8	3.929
-4	4.123
-4.2	4.317
-4.4	4.512
/

\setplotsymbol(\normalered)
\plot
0	2
0.2	2.01
0.4	2.04
0.6	2.088
0.8	2.154
1	2.236
1.2	2.332
1.4	2.441
1.6	2.561
1.8	2.691
2	2.828
2.2	2.973
2.4	3.124
2.6	3.28
2.8	3.441
3	3.606
3.2	3.774
3.4	3.945
3.6	4.118
3.8	4.294
4	4.472
4.2	4.652
4.4	4.833
/
\plot
0	2
-0.2	2.01
-0.4	2.04
-0.6	2.088
-0.8	2.154
-1	2.236
-1.2	2.332
-1.4	2.441
-1.6	2.561
-1.8	2.691
-2	2.828
-2.2	2.973
-2.4	3.124
-2.6	3.28
-2.8	3.441
-3	3.606
-3.2	3.774
-3.4	3.945
-3.6	4.118
-3.8	4.294
-4	4.472
-4.2	4.652
-4.4	4.833
/
\normale
\setsolid
\plot
0	3
0.2	3.007
0.4	3.027
0.6	3.059
0.8	3.105
1	3.162
1.2	3.231
1.4	3.311
1.6	3.4
1.8	3.499
2	3.606
2.2	3.72
2.4	3.842
2.6	3.97
2.8	4.104
3	4.243
3.2	4.386
3.4	4.534
3.6	4.686
3.8	4.841
4	5
4.2	5.161
4.4	5.325
/
\plot
0	3
-0.2	3.007
-0.4	3.027
-0.6	3.059
-0.8	3.105
-1	3.162
-1.2	3.231
-1.4	3.311
-1.6	3.4
-1.8	3.499
-2	3.606
-2.2	3.72
-2.4	3.842
-2.6	3.97
-2.8	4.104
-3	4.243
-3.2	4.386
-3.4	4.534
-3.6	4.686
-3.8	4.841
-4	5
-4.2	5.161
-4.4	5.325
/
\plot
0	4
0.2	4.005
0.4	4.02
0.6	4.045
0.8	4.079
1	4.123
1.2	4.176
1.4	4.238
1.6	4.308
1.8	4.386
2	4.472
2.2	4.565
2.4	4.665
2.6	4.771
2.8	4.883
3	5
3.2	5.122
3.4	5.25
3.6	5.381
3.8	5.517
4	5.657
4.2	5.8
4.4	5.946
/
\setlinear
\plot 0 3 0 6 /
\plot 0 0 1 6 /
\plot 0 0 2 6 /
\plot 0 0 3 6 /
\plot 0 0 4 6 /
\plot 0 0 5 6 /
\plot 0 0 -1 6 /
\plot 0 0 -2 6 /
\plot 0 0 -3 6 /
\plot 0 0 -4 6 /
\plot 0 0 -5 6 /
\setplotsymbol(\normalered)
\plot 0 0 -2 6 /
\plot 0 0 3 6 /
\setplotsymbol(\normaleblue)
\setlinear
\arrow <8pt> [.2,.6] from 0 0  to 0 4
\plot
0	4
0.2	4.005
0.4	4.02
0.6	4.045
0.8	4.079
1	4.123
1.2	4.176
1.4	4.238
1.6	4.308
1.8	4.386
2	4.472
2.2	4.562
2.3	4.59
/
\put{$\bullet$} at 2.29 4.59
\plot
0	4
-0.2	4.005
-0.4	4.02
-0.6	4.045
-0.8	4.079
-1	4.123
-1.2	4.176
-1.4	4.238
-1.4	4.238
/
\put{$\bullet$} at -1.4	4.238
\normale
\setquadratic
\plot
0	4
-0.2	4.005
-0.4	4.02
-0.6	4.045
-0.8	4.079
-1	4.123
-1.2	4.176
-1.4	4.238
-1.6	4.308
-1.8	4.386
-2	4.472
-2.2	4.565
-2.4	4.665
-2.6	4.771
-2.8	4.883
-3	5
-3.2	5.122
-3.4	5.25
-3.6	5.381
-3.8	5.517
-4	5.657
-4.2	5.8
-4.4	5.946
/
\plot
2	4.472
2.2	4.565
2.4	4.665
2.6	4.771
2.8	4.883
3	5
3.2	5.122
3.4	5.25
3.6	5.381
3.8	5.517
4	5.657
4.2	5.8
4.4	5.946
/
\setsolid
\inboundscheckoff
\setplotsymbol(\normalered)
\setlinear
\plot -.083 .2 -1.75 5.3 /
\plot .1 .2 2.62 5.3 /
\put{$\bs\psi_{AB}$} [b] at .05 .6
\normale
\arrow <6pt> [.2,.6] from -.6 3.05  to -1 3.15
\arrow <6pt> [.2,.6] from 1.4 3.3 to 1.7 3.45
\put{$O$} [lb] at .1 -.1
\put{$\sc\bullet$} at 0 0
\put{$A$} [b] at -1.8 5.4
\put{$B$} [b] at 2.65 5.4
\put{$\bs\psi_{AB}$} [lb] at 0.25 3.15
\put{$R(t)$} [lb] at .05 2.4
\put{$\ell_{AB}(t)$} [lb] at .24 4.2
\arrow <6pt> [.2,.6] from 0 4.3 to 0 5
\put{$w$} [l] at .1 4.9
\put{open universe at the time  $t$} [l] at -4 6
\arrow <6pt> [.2,.6] from -2.4 5.8  to -1.8 4.4
\put{constant} [l] at 2.4 2
\put{angular distance} [l] at 2.4 1.7
\put{$\psi_{AB}$} [l] at 2.4 1.4
\arrow <6pt> [.2,.6] from 2.3 2 to .6 3
\arrow <6pt> [.2,.6] from 2.3 1.5 to .35 .6
\put{isochronous} [l] at -4.5 2.8
\put{distance $\ell_{AB}$} [l] at -4.5 2.5
\put{at the time $t$} [l] at -4.5 2.2
\arrow <6pt> [.2,.6] from -2.8 2.3  to -.5 4
\put{\firma} at 3 0
\setplotsymbol (\normalered)
\arrow <6pt> [.2,.6] from -2.9 1.2 to -1.3 2.4
\put{reference} [l] at -4.3 1.7
\put{time $t_\sharp$} [l] at -4.3 1.4
\put{$\mathbb H_3(t_\sharp)\simeq Q$} [l] at -4.3 1.1
\normale
\setshadegrid span <1.25pt>
\vshade 
-1.4 4.3 4.3
-.7 2 4
0 0 4
1.1 2.3 4.1
2.3 4.5 4.6
 /
\endpicture
$$
\vskip -4mm  
  \caption{Radial diagram of the open universe.}
  \label{fig:ou}
\end{figure}

\subsection{Flat universe}

In this model the concept of radius of the universe does not make sense. So we must refer to the scale parameter $a(t)$ only. The spatial sections are the three-dimensional planes $w=\rm constant$ immersed in the Euclidean affine space $\mathbb R^4=(w,x,y,z)$.

\section{Co-moving volumes and conserved densities}\label{s:cVsd}

Let $U$ be a domain in the quotient manifold $Q$ with a finite volume 
$$
\wt{\mathcal V}(U)=\dint_U \sqrt{\det[\wt g_{ab}]\;}
\;dq^\uno\land dq^\due\land dq^\tre.
$$
Carried along the world-lines, $U$ generates domains $U_t\subset S_t$ with finite volumes
$$
\mathcal V(U,t)=\dint_{U_t} \sqrt{\det[g_{ab}(t,\wt q)]\;}\;dq^\uno\land dq^\due\land dq^\tre.
$$ 

\bt\label{t:Va3}
The ratio $\mathcal V(U,t)/a^3(t)$ does not depend on $t$:
\be\label{e:Va3}
\bboxed{\dfrac{\mathcal V(U,t)}{a^3(t)}=\wt{\mathcal V}(U)={\rm constant} >0}
\ee
\et

\bprf
The factorization $g_{ab}(t,\wt q)=a^2(t)\;\wt g_{ab}(\wt q)$ implies 
$$
\ba
\mathcal V(U,t)=\dint_{U_t}\! \sqrt{\det[g_{ab}(t,\wt q)]\;}\;dq^\uno\land dq^\due\land dq^\tre
\ac
=a^3(t)\dint_{U_t} \!\sqrt{\det[\wt g_{ab}(\wt q)]\;}\;dq^\uno\land dq^\due\land dq^\tre
=a^3(t)\,\wt{\mathcal V}(U). \eprfx
\ea
$$ 

\bt\label{t:muu} 
For any scalar function  of the cosmic time $\mu(t)$ the following equations are equivalent,
\be\label{e:muu}
\begin{array}{ll}
\bboxed{\mu(t)\,\mathcal V^u(U,t)={\rm const.} \;\; \forall\; U}
\iff
\bboxed{\mu(t)\;a^{3u}(t)=\rm const.\;}
\ac
\iff 
\bboxed{a\,\dot\mu+3\,u\,\mu\,\dot a=0}
\iff 
\bboxed{h=-\tfrac 1{3u}\,\dfrac{\dot\mu}\mu} 
\ea
\ee
\et

When these equations are satisfied we say that $\mu(t)$ is a {\bf conserved density of order $u$}. 

\bprf 
Write equation \eqref{e:Va3} as $\mathcal V(U,t)=\wt{\mathcal V}(U)\,a^3(t)$. Then the condition $\mu(t)\,\mathcal V^u(U,t)=\rm const.$ is equivalent to  $\mu(t)\,\wt{\mathcal V}^u(U)\,a^{3u}(t)=\rm const.$, hence to $\mu(t)\,a^{3u}(t)=\rm const$. By differentiation we get the second equivalence. The last equivalence is due to the definition of the Hubble factor: $h=\dot a/a$. \eprf

The typical case is the mass  (or matter) density for which $u=1$. But there are other density of physical interest for which $u\neq 1$ (see Chapter \ref{c:bar}).

\section{Cosmic monitor and free particles}\label{s:cosmmon}

The quotient manifold will play a crucial role in the sequel. In order to make such an abstract concept more accessible we can think of it as a \index{cosmic!monitor} {\bf cosmic monitor} whose {\it pixels}, which are bright fixed points, represent the galaxies. Of course we need an effort of imagination because such a monitor is neither flat nor two-dimensional: it is a (possibly curved) three-dimensional screen. On the cosmic monitor there is also  a clock showing the cosmic time $t$.

Imagine a very special person, the \index{cosmic!watcher} {\bf cosmic watcher}, sitting in front of  (or better, sitting inside) the monitor. Since the cosmic monitor is endowed with a metric (the quotient metric $\wt g$) the cosmic watcher is able to recognize distinguished curves, called {\bf geodesics}, connecting any pair of galaxies with a minimal (or stationary) distance. Then the cosmic watcher is able to measure the co-moving distance $\wtell_{AB}$ of two galaxies (Fig.\ \ref{fig:cosmcon-0}).

\begin{figure} [H]
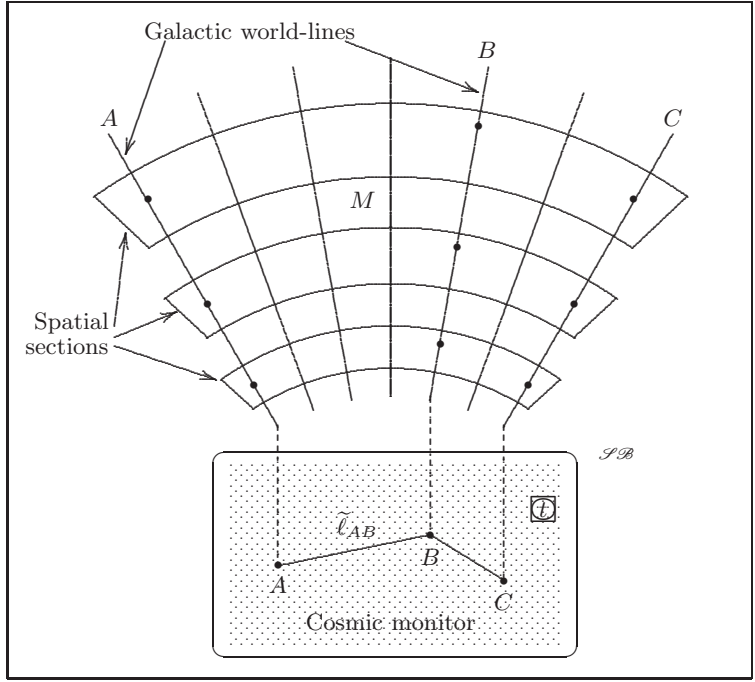
 
  \centering
$
\beginpicture
\setcoordinatesystem units <.75cm,.75cm>
\setplotarea x from -6.8 to 6.4, y from -4 to 8
\normalgraphs
\small
\grid 1 1
\setlinear
\plot 0 1 0 7 /
\startrotation by .984 -.173 about 0 -3 
\plot 0 1 0 7 /
\stoprotation
\put{$\sc\bullet$} at 1.55 5.8
\put{$\sc\bullet$} at 1.18 3.65
\put{$\sc\bullet$} at .88 1.93
\put{$\sc\bullet$} at -4.3 4.5
\put{$\sc\bullet$} at -3.25 2.64
\put{$\sc\bullet$} at -2.43 1.2
\put{$\sc\bullet$} at 4.3 4.5
\put{$\sc\bullet$} at 3.25 2.64
\put{$\sc\bullet$} at 2.43 1.2
\startrotation by .939 -.342 about 0 -3 
\plot 0 1 0 7 /
\stoprotation
\startrotation by .866 -.5 about 0 -3 
\plot 0 1 0 7 /
\stoprotation
\setlinear
\plot 0 1 0 7 /
\startrotation by .984 .173 about 0 -3 
\plot 0 1 0 7 /
\stoprotation
\startrotation by .939 .342 about 0 -3 
\plot 0 1 0 7 /
\stoprotation
\startrotation by .866 .5 about 0 -3 
\plot 0 1 0 7 /
\stoprotation
\circulararc 33 degrees from 0 1.5 center at 0 -3
\circulararc -33 degrees from 0 1.5 center at 0 -3
\circulararc 35 degrees from 0 2.25 center at 0 -3
\circulararc -35 degrees from 0 2.25 center at 0 -3
\setlinear
\plot -2.45 .8 -3 1.3 /
\plot 2.45 .8 3 1.3 /
\circulararc 33 degrees from 0 3 center at 0 -3
\circulararc -33 degrees from 0 3 center at 0 -3
\circulararc 35 degrees from 0 4 center at 0 -3
\circulararc -35 degrees from 0 4 center at 0 -3
\plot -3.25 2.05 -4 2.75 /
\plot 3.25 2.05 4 2.75 /
\circulararc 33 degrees from 0 4.9 center at 0 -3
\circulararc -33 degrees from 0 4.9 center at 0 -3
\circulararc 35 degrees from 0 6.2 center at 0 -3
\circulararc -35 degrees from 0 6.2 center at 0 -3
\plot -4.3 3.65 -5.25 4.55 /
\plot 4.3 3.65 5.25 4.55 /
\put{$M$} at -.5 4.5
\put{$A$} [b] at -5 5.8
\put{$B$} [b] at 1.7 7
\put{$C$} [b] at 5 5.8
\put{Galactic world-lines} at -2.5 7.5
\arrow <6pt> [.2,.6] from -4 7.3 to -4.7 5.3
\arrow <6pt> [.2,.6] from -.8 7.3 to 1.55 6.4
\put{Spatial} [r] at -5 2.3
\put{sections} [r] at -5 1.9
\arrow <6pt> [.2,.6] from -4.9 2.25 to -4.7 3.9
\arrow <6pt> [.2,.6] from -4.9 2.05 to -3.8 2.5
\arrow <6pt> [.2,.6] from -4.9 1.9 to -3.1 1.3
\put{
\cornersize{.1}
\ovalbox{
\begin{minipage}{4.5cm}
\vskip -2mm
\begin{minipage}{4cm}
\rule[-2.7cm]{0mm}{0cm}
\end{minipage}
\end{minipage}}
} at 0 -1.8 
\put{
\setshadegrid span <2pt>
\setlinear
\vshade -3 -1.5 1.8 3 -1.5 1.8 /
} at 0 -1.8
\put{Cosmic monitor} at 0 -3
\put{$\sc\bullet$} [t] at -2 -1.92
\put{$A$} [t] at -2 -2.2
\put{$\sc\bullet$} [t] at 2 -2.2
\put{$C$} [t] at 2 -2.5
\put{$\sc\bullet$} [t] at .7 -1.4
\put{$B$} [t] at .7 -1.7
\setlinear
\setdashes <2pt>
\plot -2 .5 -2 -2 /
\plot .7 1 .7 -1.4 /
\plot 2 .5 2 -2.2 /
\setsolid
\plot -2 -2 .7 -1.45 2 -2.25 /
\put{$\wtell_{AB}$} [b] at -.6 -1.5
\put{\Large$\square$} at 2.7 -1
\put{$\bs\bigcirc$} at 2.7 -1
\put{$t$} at 2.7 -1
\put{\firma} at 4 0
\endpicture
$
 \caption{The cosmic space-time and the cosmic monitor.}
  \label{fig:cosmcon-0}
\end{figure}

\begin{center}
\fbox{\begin{minipage}{.95\linewidth} \vspu
{\bf $\bs 6^{th}$ Postulate}. \label{p:particle} There are bodies,  other than galaxies, running in the universe. We call them \index{particles} {\bf particles}. The life of a particle  is represented by a world-line in space-time transversal to the spatial sections. \vspd
\end{minipage}}
\end{center}

The cosmic watcher cannot see the  world-line $\gamma(t)$ of a particle. What he can see on the monitor is only the projection $\wt\gamma(t)$ of $\gamma(t)$. He looks at  $\wt\gamma(t)$ as the {\bf motion of a point}.  Then he can measure the \index{monitor speed of a particle} {\bf monitor speed} $v(t)$ 
\be\label{e:mon-speed}
\bboxed{v(t)=\dfrac{ds}{dt}}
\ee
where $s$ is the arc-length along $\wt\gamma(t)$.

Among the various curves $\wt\gamma(t)$ that the cosmic watcher can see there are also geodesics. He argues that they represent distinguished particles whose world lines in space-time have some special features. But he cannot claim that these  world-lines are geodesics because he does not know if in  space-time there is any special metric or a connection.\label{p:cosmcon}  Anyway, he proposes the following formal definition:

\bd\label{d:free-part}
A {\bf free particle} (or {\bf free-falling particle}) is a particle whose monitor motion $\wt\gamma(t)$ is a geodesic. No matter if the cosmic time $t$ is an affine parameter or not.\footnote{
\kern 2pt In fact this distinction will be matter of a postulate (page \pageref{s:1st}).} 
\ed

\begin{figure} [H]
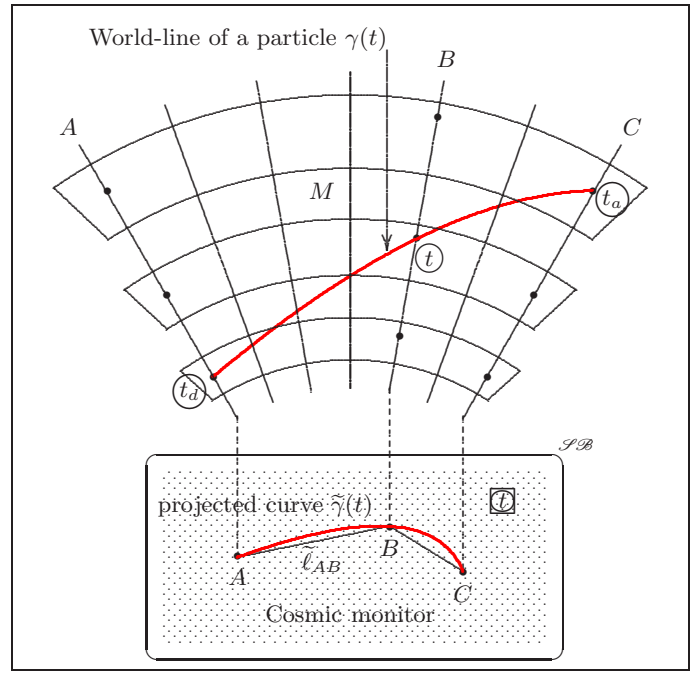
 
$$
\beginpicture
\setcoordinatesystem units <.75cm,.75cm>
\setplotarea x from -6.2 to 6.2, y from -4 to 7.8
\normalgraphs
\small
\putrectangle corners at -6 7.8 and 6 -4 
\setlinear
\plot 0 1 0 6.6 /
\startrotation by .984 -.173 about 0 -3 
\plot 0 1 0 6.6 /
\stoprotation
\put{$\sc\bullet$} at 1.55 5.8
\put{$\sc\bullet$} at 1.18 3.65
\put{$\sc\bullet$} at .88 1.93
\put{$\sc\bullet$} at -4.3 4.5
\put{$\sc\bullet$} at -3.25 2.64
\put{$\sc\bullet$} at -2.43 1.2
\put{$\sc\bullet$} at 4.3 4.5
\put{$\sc\bullet$} at 3.25 2.64
\put{$\sc\bullet$} at 2.43 1.2
\setplotsymbol (\normalered)
\setquadratic
\plot 
-2.43 1.2 
1.18 3.65
4.3 4.5 /
\normale
\setlinear
\startrotation by .939 -.342 about 0 -3 
\plot 0 1 0 6.6 /
\stoprotation
\startrotation by .866 -.5 about 0 -3 
\plot 0 1 0 6.6 /
\stoprotation
\setlinear
\plot 0 1 0 6.6 /
\startrotation by .984 .173 about 0 -3 
\plot 0 1 0 6.6 /
\stoprotation
\startrotation by .939 .342 about 0 -3 
\plot 0 1 0 6.6 /
\stoprotation
\startrotation by .866 .5 about 0 -3 
\plot 0 1 0 6.6 /
\stoprotation
\circulararc 33 degrees from 0 1.5 center at 0 -3
\circulararc -33 degrees from 0 1.5 center at 0 -3
\circulararc 35 degrees from 0 2.25 center at 0 -3
\circulararc -35 degrees from 0 2.25 center at 0 -3
\setlinear
\plot -2.45 .8 -3 1.3 /
\plot 2.45 .8 3 1.3 /
\circulararc 33 degrees from 0 3 center at 0 -3
\circulararc -33 degrees from 0 3 center at 0 -3
\circulararc 35 degrees from 0 4 center at 0 -3
\circulararc -35 degrees from 0 4 center at 0 -3
\plot -3.25 2.05 -4 2.75 /
\plot 3.25 2.05 4 2.75 /
\circulararc 33 degrees from 0 4.9 center at 0 -3
\circulararc -33 degrees from 0 4.9 center at 0 -3
\circulararc 35 degrees from 0 6.2 center at 0 -3
\circulararc -35 degrees from 0 6.2 center at 0 -3
\plot -4.3 3.65 -5.25 4.55 /
\plot 4.3 3.65 5.25 4.55 /
\put{$M$} at -.5 4.5
\put{$A$} [b] at -5 5.5
\put{$B$} [b] at 1.7 6.7
\put{$C$} [b] at 5 5.5
\put{
\cornersize{.1}
\ovalbox{
\begin{minipage}{5.2cm}
\begin{minipage}{4.7cm}
\rule[-2.5cm]{0mm}{0cm}
\end{minipage}
\end{minipage}}
} at 0 -2
\put{
\setshadegrid span <2pt>
\setlinear
\vshade -3.5 -1.5 1.6 3.5 -1.5 1.6 /
} at 0 -2
\put{Cosmic monitor} at 0 -3
\put{$\sc\bullet$} [t] at -2 -1.92
\put{$A$} [t] at -2 -2.2
\put{$\sc\bullet$} [t] at 2 -2.2
\put{$C$} [t] at 2 -2.5
\put{$\sc\bullet$} [t] at .7 -1.4
\put{$B$} [t] at .7 -1.7
\setlinear
\setdashes <2pt>
\plot -2 .5 -2 -2 /
\plot .7 1 .7 -1.4 /
\plot 2 .5 2 -2.2 /
\setsolid
\plot -2 -2 
.7 -1.45 
2 -2.25 /
\put{$\wtell_{AB}$} [t] at -.5 -1.8
\setplotsymbol (\normalered)
\setquadratic
\plot -2 -2 
.7 -1.45 
2 -2.25 /
\normale
\put{\Large $\square$} at 2.7 -1
\put{$\bs\bigcirc$} at 2.7 -1
\put{$t$} at 2.7 -1
\put{World-line of a particle $\gamma(t)$} [r] at .65 7.2
\arrow <6pt> [.2,.6] from .65 7 to .65 3.45
\put{projected curve $\wt\gamma(t)$} [rb] at .4 -1.3
\put{\Large $\bigcirc$} at -2.85 .95
\put{$t_d$} at -2.85 .95
\put{\large $\bigcirc$} at 1.4 3.3
\put{$t$} at 1.4 3.3
\put{\Large $\bigcirc$} at 4.65 4.35
\put{$t_a$} at 4.65 4.35
\put{\firma} at 4 0
\endpicture
$$
\vskip -3mm
 \caption{Particle observed on the cosmic monitor.}
  \label{fig:cosmcon-1}
\end{figure}

Thus, we are encouraged to investigate on the existence  of linear connections on space-time that are, in a sense, {\it adapted} to the geometric structures so far determined by the postulates -- the spatial foliation, the spatial metrics, the congruence of the galactic world-lines, the quotient manifold and the quotient metric. This will be done in Chapter \ref{c:cosmcon}. At the end of our analysis we will be faced with two types of connections (Sections \ref{s:1st} and \ref{s:2nd}) which open two distinct ways for developing the cosmic dynamics: the Newtonian way and the relativistic way.

$$
\beginpicture
\setcoordinatesystem units <.8cm,.8cm>
\setplotarea x from -5 to 5, y from -7 to 6
\normalgraphs
\footnotesize
\put{\includegraphics[width=8cm,keepaspectratio]{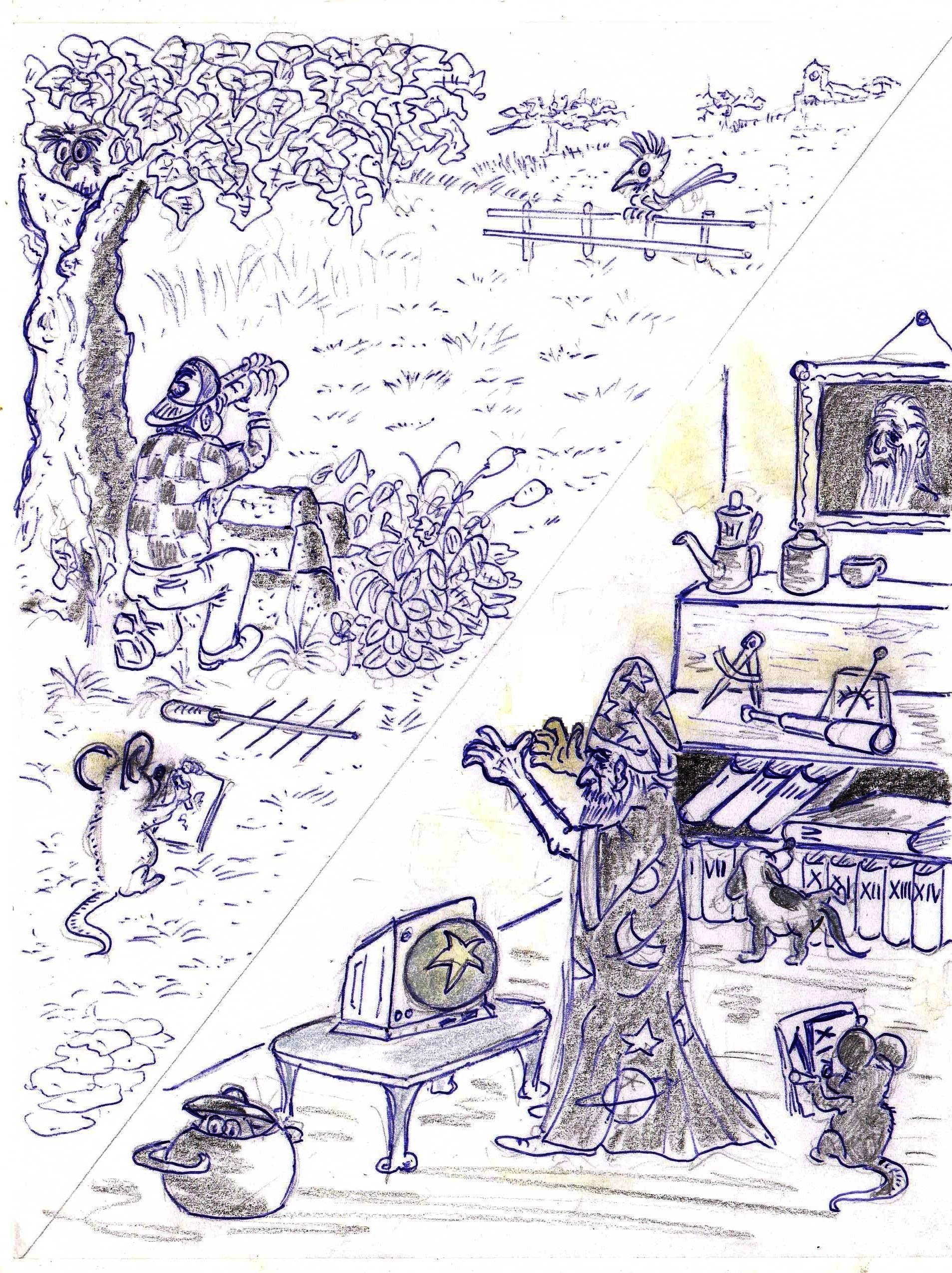}} at 0 0
\setlinear
\plot
-5 6.4 5 6.4 5 -6.4 -5 -6.4 -5 6.4 /
\put{The cosmic monitor and the cosmic watcher} at 0 -6.9
\put{(interpreted by my schoolmate Tony Magala)} at 0 -7.4
\put{a bird watcher} [lb] at -1 2.9
\arrow <6pt> [.2,.6] from -1.2 2.9 to -1.8 2.8
\put{the cosmic} [l] at -1.8 -1.5
\put{watcher} [l] at -1.3 -1.8
\put{the cosmic} [l] at -2.5 -2.5
\put{monitor} [l] at -2 -2.8
\endpicture
$$

\section{Local reference frames and peculiar velocity}

In order to locate an event in space-time we need to know \underbar{where} it occurs and \underbar{when}. To do this we need to assign a \index{reference frame} {\bf reference frame} made of the congruence of world-lines of particles of an ideal `body' and of a transversal foliation $S_t$ of simultaneous events parametrized by a `time' $t$. Then we can say that a certain event occurs in a point of a certain `body' and at a certain `date' $t$.  

In cosmology we have a \index{privileged reference frame} {\bf privileged reference frame}: the galactic world-lines and the cosmic time $t$. Note that this is similar to what happens in Newtonian space-time (which is an affine space): there is an absolute time $t$, a foliation $S_t$ made of three-dimensional affine subspaces, and a class of equivalence of reference frames, whose world-lines are parallel straight lines, called {\bf inertial frames}.  Instead, in Special Relativity i.e., in the Minkowski space-time (which is still an affine space) there is not a privileged time but a class of equivalence of reference frames, whose world-lines are parallel straight lines, still called \index{inertial!frame} {\bf inertial frames}. Each one of them determines an orthogonal foliation of three-dimensional affine subspaces, hence a time $t$ which is `relative' to the frame. 

Consider now the intergalactic journey of a particle $P$ from a galaxy $A$ to a galaxy $C$. Let  $t_d$ and $t_a$ be the dates of departure and arrival (see Fig.\ \ref{fig:cosmcon-1}, page \pageref{fig:cosmcon-1}). At a date $t\in (t_d,t_a)$ the particle $P$ crosses a galaxy $B_t$. Then we are faced with the following data:
$$
\left\{\begin{array}{lll}
\ell_{AB_t}(t) & = & \hbox{the isochronous distance from $A$ to $B_t$ measured} \\  & & \hbox{at the actual time $t$}.
\\
\wtell_{AB_t}(t)  & = & \hbox{the co-moving distance from $A$ to $B_t$ measured} \\ & & \hbox{by the cosmic watcher}.
\\
a(t)& = & \hbox{the scale parameter, unknown to the cosmic watcher.}
\ea\right.
$$
They are linked by equation \eqref{e:Hl0}  
$$
\ell_{AB_t}(t)=a(t)\,\wtell_{AB_t}(t).
$$
It follows that $\dot\ell_{AB_t}(t)=\dot a(t)\,\wtell_{AB_t}(t)+a(t)\,
\dot{\wtell}_{AB_t}(t)\;\To$
\be\label{e:rf-3}
\dot \ell_{AB_t}(t)=h(t)\,\ell_{AB_t}(t)+a(t)\,\dot{\wtell}_{AB_t}(t), \quad
h(t)\Def\dfrac{\dot a(t)}{a(t)}.
\ee
These formulas are quite similar to the  composition law of velocities in classical mechanics. The galaxy $B_t$ can be interpreted as a local reference frame that moves with respect to the main (fixed) reference frame $A$. Then the first term $\dot \ell_{AB_t}$ is the \index{absolute!velocity} {\bf absolute velocity} of the particle $P$ (the velocity with respect to the main frame $A$). The second term $\dot a\,\wtell_{AB_t}=h\,\ell_{AB_t}$ plays the role of \index{dragging velocity} {\bf dragging velocity}. The third term $a\,\dot{\wtell}_{AB_t}$ is the \index{relative!velocity} {\bf relative velocity}, that is the velocity of the particle with respect to the moving frame $B_t$. Hence equation \eqref{e:rf-3} can be read as
$$
\bc
\hbox{absolute velocity of $P$ w.\ r.\ to the frame $A$} 
\\[1mm]
= \hbox{dragging velocity of the galaxy $B_t$} 
\\[1mm]
+\; \hbox{relative velocity of $P$ w.\ r.\ to the frame $B_t$.}
\ec
\iff \bc
\dot \ell_{AB_t}
\\[1mm]
=\dot a\,\wtell_{AB_t}
\\[1mm]
+\;a\,\dot{\wtell}_{AB_t}.
\ec
$$
In cosmology these three velocities are called \index{total velocity} {\bf total velocity}, \index{recession velocity} {\bf recession velocity} and \index{peculiar velocity} {\bf peculiar velocity}, respectively:
\be\label{e:rf-5}
\bboxed{
\bc
\hbox{total velocity $v_{\rm tot}(P/A)$ } 
\\[1mm]
= \hbox{recession velocity $v_{\rm rec}(B_t/A)$} 
\\[1mm]
+\; \hbox{peculiar velocity $v_{\rm pec}(P/B_t)$}
\ec
\iff
\bc
\dot \ell_{AB_t}
\\[1mm]
=\dot a\,\wtell_{AB_t}
\\[1mm]
+\;a\,\dot{\wtell}_{AB_t}
\ec
}
\ee
Here the slash symbol $/$ stands for {\it with respect to}. The total velocity pertains the galaxy $A$ and the particle $P$ when it crosses the galaxy $B_t$ at the date $t$. The recession velocity pertains the two galaxies $A$ and $B_t$ only. The peculiar velocity  pertains the particle $P$ and the galaxy $B_t$ only. The cosmic watcher is able to measure $\wtell_{AB_t}=s(t)-s(t_d)$ and $\dot{\wtell}_{AB_t}=\dot{s}(t)$ only. 
\be\label{e:rec}
\bboxed{v_{\rm rec}(B_t/A)=h(t)\,\ell_{AB_t}=\dot a(t)\,\big[s(t)-s(t_d)\big]}
\ee
\vskip -9mm
$$
\kern -5mm 
\beginpicture
\setcoordinatesystem units <1cm,.5cm>
\setplotarea x from -5 to 5, y from -1 to 2
\normalgraphs
\small
\arrow <6pt> [.2,.6] from 1 1 to 1 2
\setlinear
\plot 0 1 1 1 /
\put{$\dot a(t)$ unknown to the cosmic watcher} [r] at -.2 1
\arrow <6pt> [.2,.6] from 2 0 to 2 2
\plot 0 0 2 0 /
\put{$s(t)$ known to the cosmic watcher} [r] at -.2 0
\endpicture
$$
\vskip -7mm
\be\label{e:vpec}
\bboxed{v_{\rm pec}(P/B_t)=a(t)\,\dot{\bar\ell}_{AB_t}=a(t)\,\dfrac{ds}{dt}}
\ee
\vskip -8mm
$$
\kern -5mm
\beginpicture
\setcoordinatesystem units <1cm,.5cm>
\setplotarea x from -5 to 5, y from -1 to 2
\normalgraphs
\small
\arrow <6pt> [.2,.6] from 2 1 to 2 2
\setlinear
\plot 0 1 2 1 /
\put{$a(t)$ unknown to the cosmic watcher} [r] at -.1 1
\arrow <6pt> [.2,.6] from 2.6 0 to 2.6 2
\plot 0 0 2.6 0 /
\put{$\dot{s}(t)$ known to the cosmic watcher} [r] at -.1 0
\endpicture
$$
This last formula provides the definition of {\bf peculiar velocity} of a particle.


\chapter{Cosmic connections}\label{c:cosmcon}

\section{Preamble}

A {\bf connection} $\Gamma$ on a manifold $M$ is a mathematical device which determines a {\bf transport}\footnote{
\kern 2pt Sometimes called {\bf parallel transport}.} 
 of vectors along curves. A connection is {\bf linear} if the transport commute with linear combinations of vectors (with constant coefficients). In the domain of any given coordinate system a linear connection is determined by {\bf symbols} or {\bf coefficients} $\Gamma_{\alpha\beta}^\gamma$ which   are functions of the coordinates. Then the {\bf transport equations} of a vector $v^\alpha(\xi)$ along any curve $q^\alpha=\gamma^\alpha(\xi)$ are
\be\label{e:transp-eq}
\bboxed{\dfrac{dv^\gamma}{d\xi} +\Gamma_{\alpha\beta}^\gamma\,v^\alpha\,\dfrac{d\gamma^\beta}{d\xi}=0}
\ee
By means of a linear connection we can also define the {\bf acceleration vector} along any curve $q^\alpha=\gamma^\alpha(\xi)$: 
\be\label{e:acc-xi}
\bboxed{a^\gamma(\xi)\Def\dfrac{d^2\gamma^\gamma}{d\xi^2}+\Gamma_{\alpha\beta}^\gamma\,\dfrac{d\gamma^\alpha}{d\xi}\,\dfrac{d\gamma^\beta}{d\xi}}
\ee
When the acceleration is parallel to the velocity, namely when
\be\label{e:geod-xi-0}
\bboxed{a^\gamma(\xi)=\lambda(\xi)\,\dfrac{d\gamma^\gamma}{d\xi}}
\ee
then the curve is said to be a {\bf geodesic of the connection} $\Gamma$ or, briefly, a $\Gamma$-{\bf geodesic}. Thus the geodesics are characterized by the {\bf geodesic equations}
\be\label{e:geod-xi-1}
\bboxed{\dfrac{d^2\gamma^\gamma}{d\xi^2}+\Gamma_{\alpha\beta}^\gamma\,\dfrac{d\gamma^\alpha}{d\xi}\,\dfrac{d\gamma^\beta}{d\xi}=\lambda(\xi)\,\dfrac{d\gamma^\gamma}{d\xi}}
\ee
The acceleration \eqref{e:acc-xi} of a curve (as well as the velocity)  strictly depends on the choice of the parameter $\xi$ of the curve. As shown by the following theorem, it is always possible to determine a parameter $\barxi$ for which the acceleration vanishes, so that the geodesic equations read
\be\label{e:geod-aff}
\bboxed{\dfrac{d^2\gamma^\gamma}{d\barxi{}^2}+\Gamma_{\alpha\beta}^\gamma\,\dfrac{d\gamma^\alpha}{d\barxi}\,\dfrac{d\gamma^\beta}{d\barxi}=0}
\ee
Such a parameter is said to be an {\bf affine parameter}, since it is determined up to an affine transformation (with constant coefficients). 
The comparison with the transport equations \eqref{e:transp-eq} shows that  a $\Gamma$-geodesic is a curve whose velocity vector, with respect to an affine parameter, is invariant under the parallel transport. 

\bt
The transformations $\barxi=\barxi(\xi)$ satisfying the differential equation
\be\label{e:aff-eq}
\dfrac{d\barxi}{d\xi}=\exp\dint\!\!\lambda(\xi)\,d\xi.
\ee
determine all the affine parameters preserving the orientation of the geodesic.
\et

\bprf
Under a change of parameter equations \eqref{e:geod-xi-1} transform to
$$
\dfrac{d\barxi}{d\xi}\dfrac{d\,}{d\barxi}\left(\dfrac{d\barxi}{d\xi}\dfrac{d\gamma^\gamma}{d\barxi}\right)
+\Gamma_{\alpha\beta}^\gamma\,\dfrac{d\gamma^\alpha}{d\barxi}\,\dfrac{d\gamma^\beta}{d\barxi}
\left(\dfrac{d\barxi}{d\xi}\right)^2
=\lambda(\xi)\,\dfrac{d\barxi}{d\xi}\dfrac{d\gamma^\gamma}{d\barxi}
$$
$$
\To\;\left(\dfrac{d\barxi}{d\xi}\right)^2\dfrac{d\,}{d\barxi}\left(\dfrac{d\gamma^\gamma}{d\barxi}\right)
+
\dfrac{d\barxi}{d\xi}\dfrac{d\,}{d\barxi}\left(\dfrac{d\barxi}{d\xi}\right)\dfrac{d\gamma^\gamma}{d\barxi}
+\Gamma_{\alpha\beta}^\gamma\,\dfrac{d\gamma^\alpha}{d\barxi}\,\dfrac{d\gamma^\beta}{d\barxi}
\left(\dfrac{d\barxi}{d\xi}\right)^2
=\lambda(\xi)\,\dfrac{d\barxi}{d\xi}\dfrac{d\gamma^\gamma}{d\barxi}
$$
$$
\To\;\left[\dfrac{d\,}{d\barxi}\left(\dfrac{d\gamma^\gamma}{d\barxi}\right)
+\Gamma_{\alpha\beta}^\gamma\,\dfrac{d\gamma^\alpha}{d\barxi}\,\dfrac{d\gamma^\beta}{d\barxi}\right]\left(\dfrac{d\barxi}{d\xi}\right)^2
=\left[\lambda(\xi)\,\dfrac{d\barxi}{d\xi}
-\dfrac{d\barxi}{d\xi}\dfrac{d\,}{d\barxi}\left(\dfrac{d\barxi}{d\xi}\right)\right]\dfrac{d\gamma^\gamma}{d\barxi}.
$$
The parameter $\barxi$ is affine if and only if
$$
\lambda(\xi)\,\dfrac{d\barxi}{d\xi}
-\dfrac{d\barxi}{d\xi}\dfrac{d\,}{d\barxi}\left(\dfrac{d\barxi}{d\xi}\right)=0
\;\iff\;\lambda(\xi)=\dfrac{d\,}{d\barxi}\left(\dfrac{d\barxi}{d\xi}\right)
\;\iff\;\lambda(\xi)=\dfrac{d\,}{d\barxi}\left(\dfrac{d\barxi}{d\xi}\right)
$$
$$
\iff\;\lambda(\xi)=\dfrac{d\xi}{d\barxi}\dfrac{d\,}{d\xi}\left(\dfrac{d\barxi}{d\xi}\right)
\;\iff\;\lambda(\xi)=\dfrac{d\,}{d\xi}\left(\log\dfrac{d\barxi}{d\xi}\right)
$$
under the assumption that $d\barxi/d\xi>0$. \eprf

\section{Connections in the cosmic space-time}
Due to the isotropy principle only symmetric connections, for which  $\Gamma_{\alpha\beta}^\gamma=\Gamma_{\beta\alpha}^\gamma$, are admissible.\footnote{
\kern 2pt See Remark \ref{r:symmconn}, page \pageref{r:symmconn}.} 
 In homogeneous co-moving coordinates $(q^\Z,q^a)$ we classify the symbols according to the number of the $\sc 0$--indices:
\be\label{e:symbols-0}
\bc
\Gamma_{\Z\Z}^\Z, 
\\[1mm]
\Gamma_{\Z\Z}^c, \quad \Gamma_{a\Z}^\Z=\Gamma_{\Z a}^\Z, 
\\[1mm]
\Gamma_{a\Z}^c=\Gamma_{\Z a}^c, \quad \Gamma_{ab}^\Z=\Gamma_{ba}^\Z,
\\[1mm] 
\Gamma_{ab}^c=\Gamma_{ba}^c.
\ec
\ee
Then the transport equations \eqref{e:transp-eq} and the geodesic equations \eqref{e:geod-xi-1} read, respectively,
\begin{gather}
\label{e:transp}
\bc
\dfrac{dv^\Z}{d\xi} 
+\Gamma_{\Z\Z}^\Z\,v^\Z\,\dfrac{d\gamma^\Z}{d\xi}
+\Gamma_{a\Z}^\Z\left(\!v^a\,\dfrac{d\gamma^\Z}{d\xi}+v^\Z\,\dfrac{d\gamma^a}{d\xi}\!
\right)
+\Gamma_{ab}^\Z\,v^a\,\dfrac{d\gamma^b}{d\xi}
=0,
\ac
\dfrac{dv^c}{d\xi} 
+\Gamma_{\Z\Z}^c\,v^\Z\,\dfrac{d\gamma^\Z}{d\xi}
+\Gamma_{a\Z}^c\left(\!v^a\,\dfrac{d\gamma^\Z}{d\xi}+v^\Z\,\dfrac{d\gamma^a}{d\xi}\right)
+\Gamma_{ab}^c\,v^a\,\dfrac{d\gamma^b}{d\xi}
=0.
\ec
\\[1mm]
\label{e:geod-xi-2}
\bc
\dfrac{d^2\gamma^\Z}{d\xi^2}
+\Gamma_{ab}^\Z\,\dfrac{d\gamma^a}{d\xi}\,\dfrac{d\gamma^b}{d\xi}
+2\,\Gamma_{a\Z}^\Z\,\dfrac{d\gamma^a}{d\xi}\,\dfrac{d\gamma^\Z}{d\xi}
+\Gamma_{\Z\Z}^\Z\,\left(\dfrac{d\gamma^\Z}{d\xi}\right)^2
=\lambda\,\dfrac{d\gamma^\Z}{d\xi},
\ac
\dfrac{d^2\gamma^c}{d\xi^2}
+\Gamma_{ab}^c\,\dfrac{d\gamma^a}{d\xi}\,\dfrac{d\gamma^b}{d\xi}
+2\,\Gamma_{a\Z}^c\,\dfrac{d\gamma^a}{d\xi}\,\dfrac{d\gamma^\Z}{d\xi}
+\Gamma_{\Z\Z}^c\,\left(\dfrac{d\gamma^\Z}{d\xi}\right)^2
=\lambda\,\dfrac{d\gamma^c}{d\xi}.
\ec
\end{gather}

\br
If a curve is transversal to the spatial sections, then the coordinate $q^\Z$ can be taken as a parameter. So, $\gamma^\Z(q^\Z)=q^\Z$ and consequently $d\gamma^\Z/dq^\Z=1$. Then the transport equations \eqref{e:transp} and the geodesic equations \eqref{e:geod-xi-2} reduce respectively to
\begin{gather}
\label{e:transp-q0}
\bc
\dfrac{dv^\Z}{dq^\Z} 
+\Gamma_{\Z\Z}^\Z\,v^\Z
+\Gamma_{a\Z}^\Z\left(v^a+v^\Z\,\dfrac{d\gamma^a}{dq^\Z}
\right)+\Gamma_{ab}^\Z\,v^a\,\dfrac{d\gamma^b}{dq^\Z}
=0,
\ac
\dfrac{dv^c}{dq^\Z} 
+\Gamma_{\Z\Z}^c\,v^\Z
+\Gamma_{a\Z}^c\left(v^a+v^\Z\,\dfrac{d\gamma^a}{dq^\Z}
\right)
+\Gamma_{ab}^c\,v^a\,\dfrac{d\gamma^b}{dq^\Z}
=0.
\ec
\\
\label{e:geod-q0-1}
\kern 3mm \bc
\Gamma_{ab}^\Z\,\dfrac{d\gamma^a}{dq^\Z}\,\dfrac{d\gamma^b}{dq^\Z}
+2\,\Gamma_{a\Z}^\Z\,\dfrac{d\gamma^a}{dq^\Z}
+\Gamma_{\Z\Z}^\Z
=\lambda,
\ac
\dfrac{d\,}{dq^\Z}\dfrac{d\gamma^c}{dq^\Z}
+\Gamma_{ab}^c\,\dfrac{d\gamma^a}{dq^\Z}\,\dfrac{d\gamma^b}{dq^\Z}
+2\,\Gamma_{a\Z}^c\,\dfrac{d\gamma^a}{dq^\Z}
+\Gamma_{\Z\Z}^c
=\lambda\,\dfrac{d\gamma^c}{dq^\Z}.\quad\bullet
\ec
\end{gather}
\erx

\section{The basic requirements}

We look for a linear connection $\Gamma=(\Gamma_{\alpha\beta}^\gamma)$ that meets the postulates of the cosmic kinematics or, in other words, that is adapted to the geometrical structures introduced on the space-time: the congruence of the galactic world-lines and the spatial foliation. We will translate these features into precise requirements.  

\begin{center}
\fbox{\begin{minipage}{.95\linewidth} \vspu
{\bf $\bs 1^{st}$ Requirement}. \label{req:1} The galactic world-lines are geodesics of $\Gamma$ with affine parameter $q^\Z$. \vspd
\end{minipage}}
\end{center}

\begin{center}
\fbox{\begin{minipage}{.95\linewidth} \vspu
{\bf $\bs 2^{nd}$ Requirement}. The property of a vector to be tangent to the spatial foliation is preserved by the $\Gamma$-transport along the galactic world-lines. \vspd
\end{minipage}}
\end{center}

Henceforth we will call {\bf spatial} \index{spatial!vector} every vector tangent to a spatial section. In co-moving coordinates a spatial vector is characterized by the condition $v^\Z=0$.

\bt
The $1^{st}$ requirement is satisfied if and only if $\Gamma_{\Z\Z}^\Z=0$ and $\Gamma_{\Z\Z}^c=0$.
\et

\bprf
A galactic world-line is parametrized by $q^\Z$ and characterized by the condition $q^a=\rm constant$. In this case the geodesic equations \eqref{e:geod-q0-1} give
$$
\Gamma_{\Z\Z}^\Z=\lambda,\quad
\Gamma_{\Z\Z}^c=0.
$$
The parameter $q^\Z$ is affine if and only if $\lambda=0$. \eprf

\bt
The $2^{nd}$ requirement is satisfied if and only if {\rm (i)} $\Gamma_{a\Z}^\Z=0$ and {\rm (ii)} the transport of a spatial vector along a galactic world-line is represented by the equations
\be\label{e:transp*}
\dfrac{dv^c}{dq^\Z} +\Gamma_{a\Z}^c\,v^a=0.
\ee
\et

\bprf
For a spatial vector $v^\Z=0$ and the transport equations 
\eqref{e:transp-q0} with $\gamma^a=\rm constant$ reduce to
$$
\bc
\Gamma_{a\Z}^\Z\,v^a=0,
\ac
\dfrac{dv^c}{dq^\Z} +\Gamma_{a\Z}^c\,v^a=0. \eprfx
\ec
$$
By virtue of the $3^{rd}$ postulate each spatial section $S_t$ is endowed with a positive-definite metric tensor $g_t$. Hence, a further `natural' requirement is the following.

\begin{center}
\fbox{\begin{minipage}{.95\linewidth} \vspu
{\bf $\bs 3^{rd}$ Requirement}. The norm of the spatial vectors is preserved by the $\Gamma$-transport along the galactic world-lines. \vspd
\end{minipage}}
\end{center}
  
\begin{figure} [H]
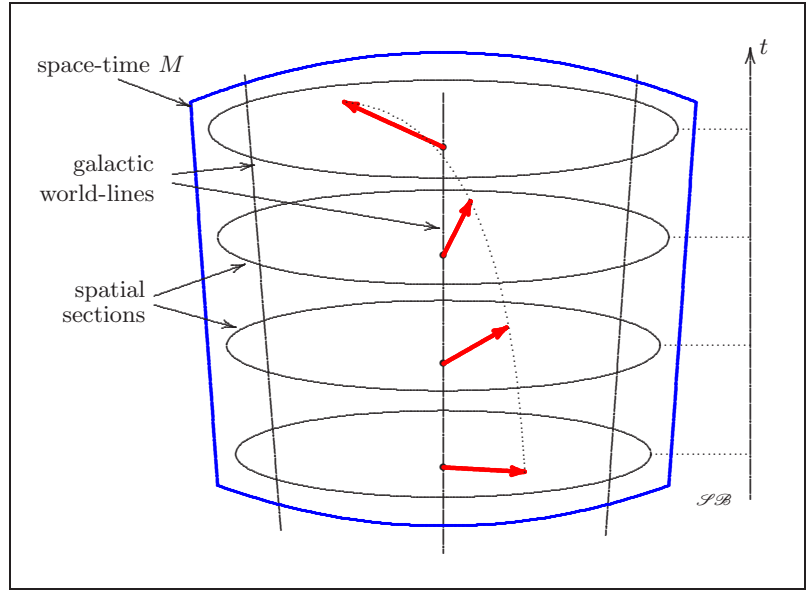

  \centering
\kern -3mm 
$
\beginpicture
\setcoordinatesystem units <1.2cm,1.2cm>
\setplotarea x from -4.8 to 4, y from -2 to 4.5
\normalgraphs
\small
\grid 1 1
\setlinear
\plot 0 -1.6 0 3.5 /
\plot 1.8 -1.4 2.15 3.7 /
\plot -1.8 -1.35 -2.2 3.7 /
\setdots<2pt>
\plot 2.3 -.5 3.4 -.5 /
\plot 2.4 .7 3.4 .7 /
\plot 2.5 1.9 3.4 1.9 /
\plot 2.6 3.1 3.4 3.1 /
\setsolid
\arrow <6pt> [.2,.6] from 3.4 -1 to 3.4 4
\put{$t$} [tr] at 3.6 4.1
\setplotsymbol (\normaleblue)
\setquadratic
\plot -2.8 3.4 
0 3.95
2.8 3.4 /
\plot -2.5 -.85 
0 -1.3
2.5 -.85 /
\setlinear
\plot -2.8 3.4 -2.5 -.85 /
\plot 2.8 3.4 2.5 -.85 /
\normale
\ellipticalarc axes ratio 4.8:1 360 degrees from 2.3 -.5 center at 0 -.5
\ellipticalarc axes ratio 4.8:1 360 degrees from 2.4 .7 center at 0 .7
\ellipticalarc axes ratio 4.8:1 360 degrees from 2.5 1.9 center at 0 1.9
\ellipticalarc axes ratio 4.8:1 360 degrees from 2.6 3.1 center at 0 3.1
\put{galactic} [r] at -3.2 2.7
\put{world-lines} [r] at -3.2 2.4
\arrow <5pt> [.2,.6] from -3 2.6 to -2.1 2.7
\arrow <5pt> [.2,.6] from -3 2.5 to -.05 2
\put{spatial} [r] at -3.3 1.3
\put{sections} [r] at -3.3 1.05
\arrow <5pt> [.2,.6] from -3.2 1.25 to -2.2 1.6
\arrow <5pt> [.2,.6] from -3.2 1.15 to -2.3 .9
\setplotsymbol (\spessinored)
\put{$\sc\bullet$} at 0 2.9
\arrow <5pt> [.2,.6] from 0 2.9 to -1.1 3.4
\put{$\sc\bullet$} at 0 1.7
\arrow <5pt> [.2,.6] from 0 1.7 to .3 2.3
\put{$\sc\bullet$} at 0 .5
\arrow <5pt> [.2,.6] from 0 .5 to .7 .9
\put{$\sc\bullet$} at 0 -.65
\arrow <5pt> [.2,.6] from 0 -.65 to .9 -.7
\normale
\setdots<2pt>
\setquadratic
\plot .9 -.7 .3 2.3 -1.1 3.4 /
\setsolid
\put{space-time $M$} [l] at -4.5 3.8
\arrow <5pt> [.2,.6] from -3.6 3.65 to -2.8 3.35
\put{\firma} at 3 -1

\endpicture
$
\vskip -1mm
  \caption{The parallel transport requirement.}
  \label{fig:pt}
\end{figure}

The \index{norm of a spatial vector} {\bf norm} of a spatial vector $v^a(\xi)$ along a curve $q^\alpha(\xi)$ is defined by\footnote{
\kern 2pt Recall \eqref{e:Aq0}, $a(t)=A(q^\Z)$.}
\be\label{e:norm}
\|v(\xi)\|
\Def g_{ab}(q^\Z,\wt q)\,v^a(\xi)\,v^b(\xi)=A^2(q^\Z)\,\wt g_{ab}(\wt q)\,v^a(\xi)\,v^b(\xi).
\ee

\bt
If the $2^{nd}$ requirement is satisfied then the connection meets the $3^{rd}$ requirement if and only if $\;\Gamma_{a\Z}^b=H\,\delta_a^b$, where $H(q^\Z)$ is the Hubble parameter.
\et

\bprf
A galactic world-line can be parametrized by $q^\Z$. From the definition \eqref{e:norm} it follows that ($'=d/dq^\Z$)
$$
\bq
\dfrac{d\|v\|}{dq^\Z}=
2\,A\,A'\,\wt g_{ab}\,v^a\,v^b
+2\,A^2\,\wt g_{ab}\,v^a\,\dfrac{dv^b}{dq^\Z}
\ac
\hbox{use the transport equations \eqref{e:transp*}  $\dfrac{dv^b}{dq^\Z} 
=-\Gamma_{c\Z}^b\,v^c$,}
\acc
=2\,A\,A'\,\wt g_{ab}\,v^a\,v^b
-2\,A^2\,\wt g_{ab}\,v^a\,\Gamma_{c\Z}^b\,v^c
=2\,A\,A\,\wt g_{ab}\,v^a\left(A'\,v^b
-H^{-1}\Gamma_{c\Z}^b\,v^c\right).
\eq
$$
The $3^{rd}$ requirement is equivalent to 
$$
\dfrac{d\|v\|}{dq^\Z}=0
\iff
\wt g_{ab}\,v^a\,(v^b-H^{-1}\Gamma_{c\Z}^b\,v^c)=0.
$$
According to this last equations the spatial vectors $v^a$ and $v^a+x^a$, with $x^a\Def -H^{-1}\Gamma_{c\Z}^a\,v^c $, must be orthogonal. This is absurd unless $v^a+x^a=0$ i.e., $v^b=H^{-1}\Gamma_{c\Z}^b\,v^c$ $\iff$
$\Gamma_{a\Z}^b=H\,\delta_a^b$. \eprf

At this point, the table of the symbols of the connection $\Gamma$ is the following:
\be\label{e:tab-1}
\boxed{\;\ba
\Gamma_{\Z\Z}^\Z=0,\quad \Gamma_{a\Z}^\Z=0, \quad \Gamma_{\Z\Z}^c=0
\ac
\Gamma_{a\Z}^b=H(q^\Z)\,\delta_a^b
\ac
\hbox{$\Gamma^\Z_{ab}$ and $\Gamma^c_{ab}$ to be determined}
\ea\;}
\ee
The geodesic equations \eqref{e:geod-xi-2} in a generic parameter $\xi$ reduce to
\be\label{e:geod-xi-3}
\bc
\dfrac{d^2\gamma^\Z}{d\xi^2}
+\Gamma_{ab}^\Z\,\dfrac{d\gamma^a}{d\xi}\,\dfrac{d\gamma^b}{d\xi}
=\lambda\,\dfrac{d\gamma^\Z}{d\xi},
\ac
\dfrac{d^2\gamma^c}{d\xi^2}
+\Gamma_{ab}^c\,\dfrac{d\gamma^a}{d\xi}\,\dfrac{d\gamma^b}{d\xi}
+2\,H\,\dfrac{d\gamma^c}{d\xi}\,\dfrac{d\gamma^\Z}{d\xi}
=\lambda\,\dfrac{d\gamma^c}{d\xi}.
\ec
\ee
For $\xi=q^\Z$,
\be\label{e:geod-q0-L}
\bc
\Gamma_{ab}^\Z\,\dfrac{d\gamma^a}{dq^\Z}\,\dfrac{d\gamma^b}{dq^\Z}
=\lambda,
\ac
\dfrac{d\,}{dq^\Z}\dfrac{d\gamma^c}{dq^\Z}
+\Gamma_{ab}^c\,\dfrac{d\gamma^a}{dq^\Z}\,\dfrac{d\gamma^b}{dq^\Z}
+2\,H\,\dfrac{d\gamma^c}{dq^\Z}
=\lambda\,\dfrac{d\gamma^c}{dq^\Z}.
\ec
\ee
Assuming that the above three requirements are met, now we ask the connection $\Gamma$ to meet the $4^{th}$ postulate: the isotropy principle.

\begin{center}
\fbox{\begin{minipage}{.95\linewidth} \vspu
{\bf $\bs 4^{th}$ Requirement}. The connection $\Gamma$ is {\bf isotropic} \index{isotropic!connection} 
\index{connection!isotropic} in the sense that it does not give rise to distinguished vector fields tangent to the spatial sections. \vspd
\end{minipage}}
\end{center}

\bt
If the connection is isotropic then 
\be\label{e:Gamma-q0}
\Gamma_{ab}^\Z=F(q^\Z)\,\wt g_{ab}(\wt q)
\ee
where $F(q^\Z)$ is a function of $q^\Z$ only.
\et

\bprf
Going back to the definition \eqref{e:geod-xi-0} of geodesic, $a^\alpha=\lambda\,d\gamma^\alpha/d\xi$, we observe that the multiplier $\lambda$ is a scalar since $a^\alpha$ and $d\gamma^\alpha/d\xi$ are vectors. Then the left hand side of the first equation \eqref{e:geod-q0-L} is a scalar. It follows that $\Gamma_{ab}^\Z(q^\Z,\wt q)$ are the components of a spatial covariant symmetric tensor for each fixed $q^\Z$. It generates distinguished eigenvectors unless it is of the form \eqref{e:Gamma-q0} (see also Section \ref{s:ivt} below). \eprf

Now the table of the $\Gamma$-symbols is 
\be\label{e:tab-2}
\boxed{\;\ba
\Gamma_{\Z\Z}^\Z=0,\quad \Gamma_{a\Z}^\Z=0, \quad \Gamma_{\Z\Z}^c=0
\ac
\Gamma_{a\Z}^b=H(q^\Z)\,\delta_a^b, \quad \Gamma^\Z_{ab}=F(q^\Z)\,\wt g_{ab}
\ac
\hbox{$F(q^\Z)$ and $\Gamma^c_{ab}$ to be determined}
\ea\;}
\ee
The $\Gamma$-geodesic equations \eqref{e:geod-xi-3} in a generic parameter $\xi$ reduce to
\be\label{e:geod-xi-4}
\bc
\dfrac{d^2\gamma^\Z}{d\xi^2}
+F\,\wt g_{ab}\,\dfrac{d\gamma^a}{d\xi}\,\dfrac{d\gamma^b}{d\xi}
=\lambda\,\dfrac{d\gamma^\Z}{d\xi},
\ac
\dfrac{d^2\gamma^c}{d\xi^2}
+\Gamma_{ab}^c\,\dfrac{d\gamma^a}{d\xi}\,\dfrac{d\gamma^b}{d\xi}
+2\,H\,\dfrac{d\gamma^c}{d\xi}\,\dfrac{d\gamma^\Z}{d\xi}
=\lambda\,\dfrac{d\gamma^c}{d\xi}.
\ec
\ee
and for $\xi=q^\Z$ to
\be\label{e:geod-q0-4}
\bc
F\,\wt g_{ab}\,\dfrac{d\gamma^a}{dq^\Z}\,\dfrac{d\gamma^b}{dq^\Z}
=\lambda,
\ac
\dfrac{d\,}{dq^\Z}\dfrac{d\gamma^c}{dq^\Z}
+\Gamma_{ab}^c\,\dfrac{d\gamma^a}{dq^\Z}\,\dfrac{d\gamma^b}{dq^\Z}
+2\,H\,\dfrac{d\gamma^c}{dq^\Z}
=\lambda\,\dfrac{d\gamma^c}{dq^\Z}.
\ec
\ee

\section{Projection of geodesics}

Let us return to the end of Section  \ref{s:cosmmon} where the cosmic watcher noted the presence of geodesics on his monitor. Arguing that these $\wt\Gamma$-geodesics may result from the projection of geodesics of a connection in space-time we are led to the following 
\begin{center}
\fbox{\begin{minipage}{.75\linewidth} \vspu
{\bf $\bs 5^{th}$ Requirement}. $\Gamma$-geodesics project onto $\wt\Gamma$-geodesics. \vspd
\end{minipage}}
\end{center}

This means that if $\gamma(\xi)$ is a geodesic for the connection $\Gamma$ then the projected curve $\wt\gamma(\xi)$ must be a geodesic for the Levi-Civita connection $\wt\Gamma$ of the quotient metric $\wt g$. Note that this requirement is already satisfied by the galactic world-lines. In fact they are $\Gamma$-geodesics which projects onto single points of the quotient manifold  which may be interpreted as \index{singular geodesic} {\bf singular geodesics}.\footnote{
\kern 2pt A parametrized curve is interpreted as a motion. So, a point represents the motion of a point at rest.} 
  
Impose this $5^{th}$ requirement to those geodesics $\gamma$ which can be parametrized by $q^\Z$ and by the arc-length $s$ of their projected curves $\wt\gamma$ (see Fig.\ \ref{fig:geored}). We will call them \index{regular!$\Gamma$-geodesics} {\bf regular $\Gamma$-geodesics}. The arc-length $s$ is a distinguished parameter for any curve on $Q$ since
\be\label{e:ds}
\wt g_{ab}\,\dfrac{d\gamma^a}{ds}\,\dfrac{d\gamma^b}{ds}=1
\ee
and is an affine parameter for all $\wt\Gamma$-geodesics,
\be\label{e:Q-geo-s}
\hbox{$q^a=\gamma^a(s)$ is a $\wt\Gamma$-geodesic} \iff \dfrac{d^2\gamma^c}{ds^2}+\wt\Gamma_{ab}^c\,\dfrac{d\gamma^a}{ds}\,\dfrac{d\gamma^b}{ds}=0.
\ee

\begin{figure} [H]
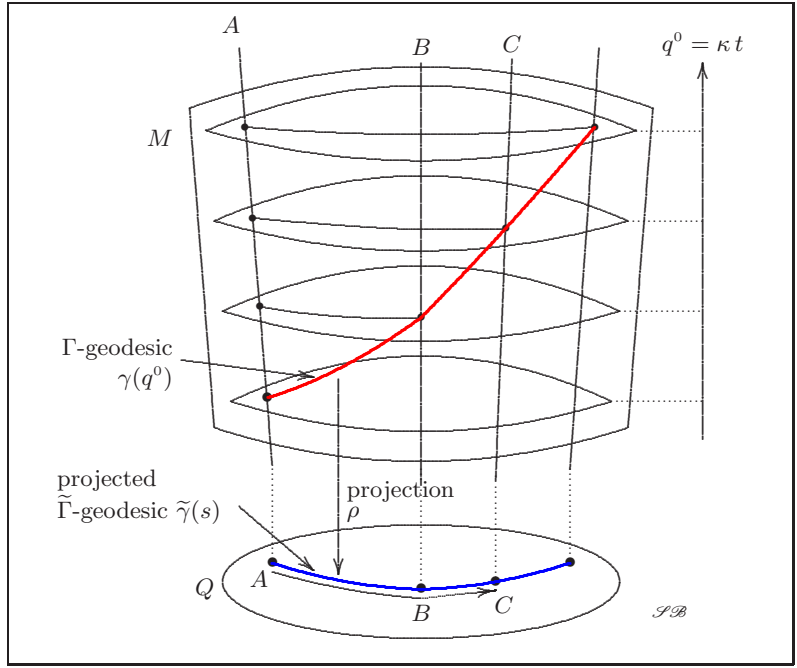

  \centering
\kern 3mm
$
\beginpicture
\setcoordinatesystem units <1.1cm,1cm>
\setplotarea x from -5 to 4.5, y from -4 to 4.8
\normalgraphs
\small
\grid 1 1
\setquadratic
\plot -2.3 -.5 0 .1 2.3 -.5 / 
\plot -2.3 -.5 0 -.9 2.3 -.5 /
\put{\beginpicture
\setquadratic
\plot -2.4 -.5 0 .1 2.4 -.5 /
\plot -2.4 -.5 0 -.9 2.4 -.5 /
\endpicture} at 0 1.2
\put{\beginpicture
\setquadratic
\plot -2.5 -.5 0 .1 2.5 -.5 /
\plot -2.5 -.5 0 -.9 2.5 -.5 /
\endpicture} at 0 2.4
\put{\beginpicture
\setquadratic
\plot -2.6 -.5 0 .1 2.6 -.5 /
\plot -2.6 -.5 0 -.9 2.6 -.5 /
\endpicture} at 0 3.6
\setlinear
\plot 0 -1.6 0 4 /
\plot .9 -1.55 1.1 4.05 /
\plot 1.8 -1.4 2.15 4.2 /
\plot -1.8 -1.35 -2.2 4.2 /
\setdots<2pt>
\setlinear
\plot 0 -1.4 0 -3 /
\plot .9 -1.5 .9 -2.9 /
\plot 1.8 -1.4 1.8 -2.7 /
\plot -1.8 -1.4 -1.8 -2.7 /
\plot 2.3 -.5 3.4 -.5 /
\plot 2.4 .7 3.4 .7 /
\plot 2.5 1.9 3.4 1.9 /
\plot 2.6 3.1 3.4 3.1 /
\setsolid
\arrow <6pt> [.2,.6] from 3.4 -1 to 3.4 4
\put{$q^\Z=\kappa\,t$} [b] at 3.4 4.1
\put{$M$} [r] at -3 3
\setquadratic
\plot -2.8 3.4 
0 3.95
2.8 3.4 /
\plot -2.5 -.85 
0 -1.3
2.5 -.85 /
\setlinear
\plot -2.8 3.4 -2.5 -.85 /
\plot 2.8 3.4 2.5 -.85 /
\setquadratic 
\put{$\bullet$} at -1.86 -.45
\plot -1.95 .76 -1 .63 0 .615 /
\put{$\sc\bullet$} at -1.95 .76
\put{$\sc\bullet$} at 0 .615 
\plot -2.04 1.94 -.3 1.8     1.02 1.8 /
\put{$\sc\bullet$} at -2.04 1.94
\put{$\sc\bullet$} at 1.02 1.8
\plot -2.13 3.15 0 3.05 2.1 3.15 /
\put{$\sc\bullet$} at -2.13 3.15
\put{$\sc\bullet$} at 2.1 3.15
\normale
\put{$\Gamma$-geodesic} [r] at -3 .2
\put{$\gamma(q^\Z)$} [r] at -3 -.2
\arrow <6pt> [.2,.6] from -2.9 .05 to -1.3 -.18
\setplotsymbol (\normalered)
\setquadratic
\plot 
-1.85 -.45 
-.95 -.04
0 .615 
1.02 1.8
2.1 3.15 /
\put{$Q$} [r] at -2.5 -3
\normale
\ellipticalarc axes ratio 3.5:1 360 degrees from 2.4 -2.9 center at 0 -2.9
\normale
\put{$\bullet$} at -1.8 -2.65
\put{$\bullet$} at 0 -3
\put{$\bullet$} at .9 -2.9
\put{$\bullet$} at 1.8 -2.65
\put{projected} [l] at -4.4 -1.55
\put{$\wt\Gamma$-geodesic $\wt\gamma(s)$} [lt] at -4.4 -1.75
\arrow <6pt> [.2,.6] from -2.2 -1.9 to -1.3 -2.75
\arrow <6pt> [.2,.6] from -1 -.2 to -1 -2.8
\put{projection} [l] at -.9 -1.7
\put{$\rho$} [l] at -.9 -2
\setplotsymbol (\normaleblue)
\setquadratic
\plot
-1.8 -2.65
-.9 -2.9
0 -3 
.9 -2.9
1.8 -2.65 /
\normale
\plot
-1.8 -2.77
-.9 -3
0 -3.12 
.75 -3.035
.87 -3.02 /
\arrow <6pt> [.2,.6] from  .7 -3.04  to .87 -3.02
\put{$A$} [rt] at -1.85 -2.75
\put{$A$} [b] at -2.3 4.4
\put{$B$} [b] at 0 4.1
\put{$B$} [t] at 0 -3.2
\put{$C$} [b] at 1.1 4.15
\put{$C$} [t] at 1 -3.1
\put{\firma} at 3 -3.3
\endpicture
$
  \caption{Regular $\Gamma$-geodesics project onto $\wt\Gamma$-geodesics.}
  \label{fig:geored}
\end{figure}

The reversible relationship between the parameters $q^\Z$ and $s$ is represented by a function\footnote{
\kern 2pt Actually, the reversibility condition is $\dfrac{dq^\Z}{ds}\neq 0$. The condition $\dfrac{dq^\Z}{ds}>0$ is not restrictive: it simply means that the two parameters $q^\Z$ and $s$ are assumed to be with the same orientation.}
\be\label{e:defVq0}
\bboxed{{V}(q^\Z)\Def\dfrac{ds}{dq^\Z}>0}
\ee
\bt\label{t:redgeo}
The $5^{th}$ requirement implies that the symbols $\Gamma_{ab}^c$  coincide with the Christoffel symbols $\wt\Gamma_{ab}^c$ of the quotient metric
\be\label{e:G=tG}
\boxed{\vp\;\Gamma_{ab}^c(q^\Z,\wt q)=\wt\Gamma_{ab}^c(\wt q)\;}
\ee
and that  the function $V(q^\Z)>0$ \eqref{e:defVq0} satisfies the equation
\be\label{e:VGH} 
\bboxed{\dfrac{d\log V}{dq^\Z}-F\,V^2+2\,H=0}
\ee
\et 

\bprf
Taking into account equation \eqref{e:ds} the geodesic equations \eqref{e:geod-q0-4} are equivalent to 
\be\label{e:geod-q0-4**}
\bc
V^2\,F=\lambda,
\ax
V^2\left(\dfrac{d\,}{ds}\dfrac{d\gamma^c}{ds}
+\Gamma_{ab}^c\,\dfrac{d\gamma^a}{ds}\,\dfrac{d\gamma^b}{ds}\right)
=\left(V^3\,F-V'-2\,H\,V\right)\dfrac{d\gamma^c}{ds}.
\ec
\ee
By replacing the expression of $\lambda$ given by the first equation in the second set we get the {\bf characteristic equations of the regular} $\Gamma$-{\bf geodesics} 
\index{characteristic!equations of \\ regular $\Gamma$-geodesics}
in the parameter $s$: 
\be\label{e:geod-rs}
\dfrac{d\,}{ds}\dfrac{d\gamma^c}{ds}
+\Gamma_{ab}^c\,\dfrac{d\gamma^a}{ds}\,\dfrac{d\gamma^b}{ds}
=V^{-1}\left(\vpx V^2\,F-(\log V)'-2\,H\right)\dfrac{d\gamma^c}{ds}.
\ee
These equations involve only the parametric equations $q^a=\gamma^a(q^\Z)$ and their first and second derivatives. Consequently, they are satisfied by the projected curve $\wt\gamma(s)$. But, in acoordance with the $5^{th}$ requirement, this projected curve must be a $\wt\Gamma$-geodesic. As a consequence, by the comparison with the $\wt\Gamma$-geodesic equations \eqref{e:Q-geo-s} we get 
$$
\left(\vpx\Gamma_{ab}^c-\wt\Gamma_{ab}^c\right)\dfrac{d\gamma^a}{ds}\,\dfrac{d\gamma^b}{ds}
=V^{-1}\left(\vpx V^2\,F-(\log V)'-2\,H\right)\dfrac{d\gamma^c}{ds}.
$$
These equations must hold for all geodesics. Since the left sides are quadratic in the $s$-velocities and the right sides are linear, both sides must vanish. \eprf

The following  is in a sense the inverse of the previous theorem.

\bt\label{t:cored}
Let $\gamma$ be a \index{regular!curve} {\bf regular curve} on space-time, that is a curve which can be parametrized by $q^\Z$ as well as by the arc-length $s$ of the projected curve $\wt\gamma$. Assume that {\rm (i)} equations \eqref{e:G=tG} and \eqref{e:VGH} are satisfied  and that {\rm (ii)} the projected curve $\wt\gamma$ is a $\wt\Gamma$-geodesic. Then $\gamma$ is a $\Gamma$-geodesic.
\et

\bprf
Under the assumptions (i) and (ii)
$$
\bc
\Gamma_{ab}^c=\wt\Gamma_{ab}^c
\ax
\dfrac{d\,}{ds}\dfrac{d\gamma^c}{ds}
+\wt\Gamma_{ab}^c\,\dfrac{d\gamma^a}{ds}\,\dfrac{d\gamma^b}{ds}=0
\ec
$$
and equations \eqref{e:geod-q0-4**} reduce to
$$
\bc
V^2\,F=\lambda,
\ax
V^2\,F-(\log V)'-2\,H\,V=0
\ec
$$
The second equation is satisfied because of the assumption (i). The first equation gives the expression of $\lambda$ and in this context is irrelevant.
Hence, equations \eqref{e:geod-q0-4**} are satisfied. On the other hand, ss we have seen in the previous proof, equations \eqref{e:geod-q0-4**} are equivalent to equations \eqref{e:geod-q0-4} which in turn are the $\Gamma$-geodesic equations in the parameter $q^\Z$. \eprf

\section{Cosmic connections}\label{s:CoCo}

The arguments of this chapter can be summarized in the following definition and theorem.

\bd\label{d:CoCo}
A {\bf cosmic connection} \index{cosmic!connection} is a linear symmetric connection on the cosmic space-time satisfying the five requirements listed above, which are compatible with the postulates of the cosmic kinematics stated in the first chapter.
\ed

\bt
In any co-moving coordinate system the symbols of a cosmic connection are 
\be\label{e:tab-3}
\boxed{\;\ba
\Gamma_{\Z\Z}^\Z=0,\quad \Gamma_{a\Z}^\Z=0, \quad \Gamma_{\Z\Z}^c=0
\ac
\Gamma_{a\Z}^b=H(q^\Z)\,\delta_a^b, \quad \Gamma^\Z_{ab}=F(q^\Z)\,\wt g_{ab}
\ac
\Gamma^c_{ab}=\wt\Gamma^c_{ab}
\ea\;}
\ee
where $H(q^\Z)$ is the Hubble parameter, $\wt\Gamma^c_{ab}$ are the Christoffel symbols of the quotient metric $\wt g$, and $F(q^\Z)$ is a function satisfying the equation 
\begin{gather}
\label{e:VGH*} 
\bboxed{\dfrac{d\log V}{dq^\Z}+2\,H=F\,V^2}
\\
\label{e:defVq0*}
\bboxed{{V}(q^\Z)\Def\dfrac{ds}{dq^\Z}}
\end{gather}
along any regular $\Gamma$-geodesic. 
\et
 
Despite the complexity of the above discussion, the result \eqref{e:tab-3} is very simple. 

\br
Having in mind Remark \ref{r:test} we observe that the five requirements of a cosmic connection are expressed in a geometrical way which is manifestly invariant with respect to the choice of a reference date (take also into account Remark \ref{r:test-G}). Nevertheless, we can observe this invariance directly from the expressions \eqref{e:tab-3} of the symbols. (i) The symbols $\Gamma_{a\Z}^b=H(q^\Z)\,\delta_a^b$ are invariant since the Hubble factor is invariant (Theorem \ref{t:Hnotsharp}). (ii) The coefficients $\Gamma^c_{ab}=\wt\Gamma^c_{ab}$ are invariant (Remark \ref{r:test-G}). (iii) About the undetermined function $F$ entering the symbols $\Gamma^\Z_{ab}=F(q^\Z)\,\wt g_{ab}$, we observe that due to equation \eqref{e:dst0t*}
$$
ds(t_\Z)=\dfrac{1}{a(t_\Z,t_*)}\,ds(t_*)
$$
from the definition \eqref{e:defVq0*} it follows that
\be\label{e:Vt0t*}
V(t_\Z,q^\Z)=\dfrac{ds(t_\Z)}{dq^\Z}=\dfrac 1{a(t_\Z,t_*)}\dfrac{ds(t_*)}{dq^\Z}=\dfrac 1{a(t_\Z,t_*)}\,V(t_*,q^\Z),
\ee
thus
$$
d\log V(t_\Z,q^\Z)=d\log V(t_*,q^\Z).
$$
Hence, the left-hand side of equation \eqref{e:VGH*} is invariant, so that $F\,V^2$ must be invariant:
$$
F(t_\Z,q^\Z)\,V^2(t_\Z,q^\Z)=F(t_*,q^\Z)\,V^2(t_*,q^\Z).
$$ 
Due to \eqref{e:Vt0t*}, this equation is equivalent to
\be\label{e:Ft0t*}
\bboxed{F(t_\Z,q^\Z)=a^2(t_\Z,t_*)\,F(t_*,q^\Z)}
\ee
In turn, due to \eqref{e:qma-3}, this last equation is equivalent to the invariance of $\Gamma^\Z_{ab}=F(q^\Z)\,\wt g_{ab}$. \er

\br 
The role of the undetermined function $F(q^\Z)$ raises a subtle argument. In fact  equation \eqref{e:VGH*} involve not only the functions $F$ and $H$, which participate in the definition of the connection, but also the function $V$ which instead, by definition, is linked to the structure of the regular geodesics. This is a paradox that may cast doubt on the correctness of this equation which in a sense is `hybrid'. This paradox will be clarified in the following. \er
$$
\beginpicture
\setcoordinatesystem units <1.05cm,1.4cm>
\setplotarea x from -5.05 to 5.05, y from -.3 to 7
\normalgraphs
\grid 1 1
\small
\setquadratic
\plot  
.6 4.2 
1.05 4.7
1.6 4.8 /
\setlinear
\plot .23 4 .6 4.2 /
\plot 1.6 4.8 2 5 /
\setquadratic 
\plot  
.3 4.35 
.9 4.85
1.4 4.88 /
\setlinear
\plot 0 4.2 .3 4.35 /
\plot 1.4 4.88 1.77 5.03 /
\setquadratic
\plot  
-.6 4.2 
-1.05 4.7
-1.6 4.8 /
\setlinear
\plot -.23 4 -.6 4.2 /
\plot -1.6 4.8 -2 5 /
\setquadratic 
\plot  
-.3 4.35 
-.9 4.85
-1.4 4.88 /
\setlinear
\plot 0 4.2 -.3 4.35 /
\plot -1.4 4.88 -1.77 5.03 /
\setlinear
\plot -.55 .8 -.23 4 /
\plot .55 .8 .23 4 /
\put{\beginpicture
\put{$\boxed{\hbox{\footnotesize 1.st Bridge postulate}}$} [rt] at -.5 2.1
\plot -.5 1.5 -.5 2.1 /
\endpicture} at -2.55 4.53
\put{$\sc\bullet$} at -1.13 4.13
\put{\beginpicture
\put{$\boxed{\hbox{Newtonian connection}}$} [rt] at -.5 2.1
\plot -.5 1.5 -.5 2.1 /
\endpicture} at -3.2 5.25
\put{$\sc\bullet$} at -1.6 4.8
\put{\includegraphics[width=1.5cm,keepaspectratio]{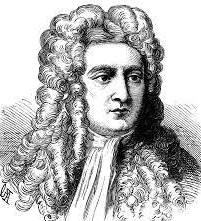}} [rb] at -1.5 5.5
\put{\footnotesize\scshape Realm of} [l] at -4.7 6.5
\put{\footnotesize\scshape Newtonian} [l] at -4.7 6.2
\put{\footnotesize\scshape Cosmology} [l] at -4.7 5.9
\put{\beginpicture
\put{$\boxed{\hbox{\footnotesize 2.nd Bridge connection}}$} [lt] at .37 3.6
\plot .37 3 .37 3.6 /
\endpicture} at 2.65 4.53
\put{$\sc\bullet$} at 1.1 4.13
\put{\beginpicture
\put{$\boxed{\hbox{Relativistic connection}}$} [lt] at .37 3.6
\plot .37 3 .37 3.6 /
\endpicture} at 3.25 5.25
\put{$\sc\bullet$} at 1.6 4.8
\put{\includegraphics[width=1.5cm,keepaspectratio]{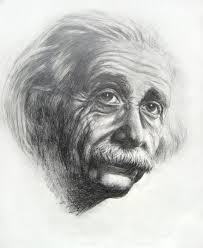}} [lb] at 1.5 5.5
\put{\footnotesize\scshape Realm of} [r] at 4.7 6.5
\put{\footnotesize\scshape Relativistic} [r] at 4.7 6.2
\put{\footnotesize\scshape Cosmology} [r] at 4.7 5.9
\put{\huge ?} at 0 6
\setplotsymbol (\spessino)
\plot -.6 4.2 -4.5 3.7  / 
\plot .6 4.2 4.5 3.7  / 
\plot .05 4.3 1.5 7 /
\plot -.05 4.3 -1.5 7 /
\put{$\boxed{\hbox{\textcolor{red}{\bf We are here}}}$} [rt] at -.6 3.9
\put{\textcolor{red}{$\bullet$}} at 0 3.7
\setplotsymbol (\spessinored)
\plot 0 3.7 0 3.9  /
\plot -.6 3.7 0 3.7 /
\put{\scshape Realm of Kinematics} at -3 1 
\setplotsymbol (\spesso)
\put{$\boxed{\hbox{Quotient manifold}}$} [lt] at .55 1.6
\plot .55 1 .55 1.6 /
\put{$\sc\bullet$} [b] at .55 .95
\put{$\boxed{\hbox{Scale parameter}}$} [rt] at -.5 2.1
\plot -.5 1.5 -.5 2.1 /
\put{$\sc\bullet$} [b] at -.5 1.45
\put{$\boxed{\hbox{Quotient metric}}$} [lt] at .45 2.6
\plot .45 2 .45 2.6 /
\put{$\sc\bullet$} [b] at .44 1.95
\put{$\boxed{\hbox{Hubble law}}$} [rt] at -.4 3.1
\plot -.4 2.5 -.4 3.1 /
\put{$\sc\bullet$} [b] at -.4 2.45
\put{$\boxed{\hbox{Cosmic connections}}$} [lt] at .37 3.6
\plot .37 3 .37 3.6 /
\put{$\sc\bullet$} [b] at .36 2.95
\put{$\bullet$} at 0 0
\put{$\boxed{\hbox{\large Postulates}}$} [t] at 0 .8
\plot -1 .1 -1 .8 /
\plot 1 .1 1 .8 /
\put{$\triangle$} [b] at 1 -.1
\put{$\triangle$} [b] at -1 -.1
\normale
\plot .6 .35 .63 0 -.65 0 -.6 .35 /
\setplotsymbol (\spessinored)
\put{\textcolor{red}{$\bullet$}} at 0 3.7
\plot 0 0 0 .35 /
\plot 0 .85 0  3.7 /
\arrow <6pt> [.2,.6] from 0 1.7 to 0 2.5
\put{\firma} at 3 0
\endpicture
$$
\begin{figure} [H]
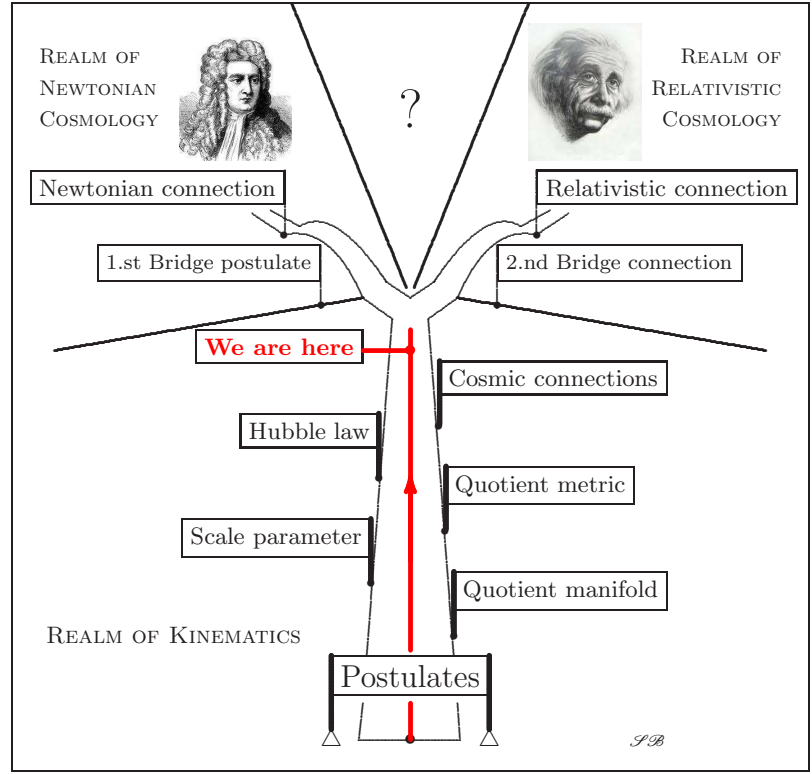

  \centering
\vskip -10mm
  \caption{The cosmic road.}
  \label{fig:bifurc}
\end{figure}

\begin{center}
\fbox{\begin{minipage}{.95\linewidth} \vspu
The assignment of a function $F(q^\Z)$ characterizing a cosmic connection must be the consequence of a postulate. In this regard we observe that a connection provides the basis of a dynamics, so that the postulate we are arguing will form a `bridge' between the Cosmic Kinematics and the Cosmic Dynamics.  We will consider two of these \index{bridge-postulates} {\bf bridge-postulates} that open the way to two different dynamics (Fig.\ \ref{fig:bifurc}). \vspd
\end{minipage}}
\end{center}
  
\section{The Newtonian cosmic connection} \label{s:1st}

\begin{center}
\fbox{\begin{minipage}{.95\linewidth} \vspu
{\bf $\bs 1^{st}$ Bridge$-$postulate}. The cosmic time $t$ is an affine parameter for the world-lines of the free particles. \vspd
\end{minipage}}
\end{center}
  
Remind that the world-line of a particle is (by definition) transversal to the spatial foliation (paragraph 2 of Section \ref{s:cosmmon}). Note that, due the first requirement of a cosmic connection (page \pageref{req:1}), the cosmic time is an affine parameter of the galactic geodesics. Thus, this bridge-postulate strictly concerns with free-particles which are not galaxies. Consequently, in the following the parametric functions $\gamma^a(q^\Z)$ are assumed to be not all constant.

\bt
The parameter $q^\Z=\kappa\,t$ is affine for all transversal geodesics of a cosmic connection if and only if $F(q^\Z)=0$.
\et

\bprf 
Write the first geodesic equation \eqref{e:geod-xi-4} for $\xi=q^\Z$ and put $\lambda=0$:
$$
F\,\wt g_{ab}\,\dfrac{d\gamma^a}{dq^\Z}\,\dfrac{d\gamma^b}{dq^\Z}
=0.\quad [*]
$$
$$
\wt g_{ab}\,\dfrac{d\gamma^a}{dq^\Z}\,\dfrac{d\gamma^b}{dq^\Z}=0
\iff
\dfrac{d\gamma^a}{dq^\Z}=0 \;\To\; \hbox{absurd}
$$
because of the transversality assumption. Then $[*] \To F=0$. \eprf

We conclude that there is a unique cosmic connection meeting the above postulate. We call it \index{Newtonian!cosmic connection} {\bf Newtonian cosmic connection} since a cosmic space-time equipped with this connection is a generalization of the Newtonian space-time of classical mechanics (Fig.\ \ref{fig:Newton}).

In the Newtonian space-time: 

1.\ The manifold $M$ is an affine four-dimensional space.

2.\ The spatial sections are Euclidean three-dimensional affine spaces.

3.\ The galactic world-lines are parallel straight lines and represents the motion of the so-called \index{fixed stars} {\bf fixed stars}. The congruence of these lines is an \index{inertial!reference frame} {\bf inertial reference frame}, as well as any other congruence of parallel lines transversal to the foliation $\mathcal S$.

4.\ The world-lines of the free-falling particles are transversal straight lines (law of inertia).

5.\ The cosmic time $t$ is the \index{absolute!time} {\bf absolute time}.

6.\ The expansion factor is constant and equal to 1, and the Hubble parameter is $h=0$; thus $H=0$. Consequently  if the coordinates $\wt q$ are Cartesian, then all symbols $\Gamma_{\alpha\beta}^\gamma$ vanishes. The cosmic connection is flat and coincides with the canonical connection of an affine space.

\begin{figure}[H]
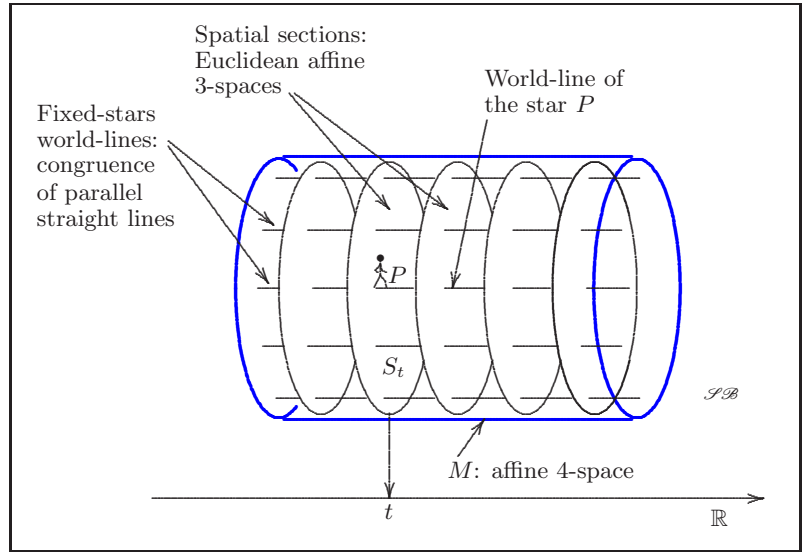

  \centering
$
\beginpicture
\setcoordinatesystem units <.7cm,.7cm>
\setplotarea x from -8.5 to 6.5, y from -5 to 5.4
\normalgraphs
\grid 1 1
\small
\setlinear
\setplotsymbol (\normaleblue)
\plot -3.3 2.5 3.3 2.5 /
\plot -3.3 -2.5 3.3 -2.5 /
\ellipticalarc axes ratio 1:3 360 degrees from 4.22 0 center at 3.4 0
\ellipticalarc axes ratio 1:3 115 degrees from -4.22 0 center at -3.4 0
\ellipticalarc axes ratio 1:3 -115 degrees from -4.22 0 center at -3.4 0
\setsolid\normale
\put{$M$: affine 4-space} [lt] at -.2 -3.3
\arrow <6pt> [.2,.6] from 0 -3.2  to .5 -2.5
\put{\small Fixed-stars} [lt] at -8 3.5
\put{\small world-lines:} [lt] at -8 3
\arrow <6pt> [.2,.6] from -5.5 2.8 to -3.5 1.2
\arrow <6pt> [.2,.6] from -5.5 2.7 to -3.6 .1
\put{\small congruence} [lt] at -8 2.4
\put{\small of parallel} [lt] at -8 2
\put{\small straight lines} [lt] at -8 1.5
\put{\small World-line of } [l] at .5 4
\put{\small the star $P$} [l] at .5 3.5
\arrow <6pt> [.2,.6] from .55 3.2 to -.1 0
\put{\small Spatial sections:} [lt] at -5 5
\put{\small Euclidean affine} [lt] at -5 4.5
\put{\small 3-spaces} [lt] at -5 4
\arrow <6pt> [.2,.6] from -3.2 3.7 to -1.3 1.5
\arrow <6pt> [.2,.6] from -3.1 3.7 to -.2 1.5
\startrotation by 0 -1 about 0 0
\put{\omino} at -.25 -1.55
\put{$P$} at -.25 -1.15
\ellipticalarc axes ratio 3:1 290 degrees from -1.4 -1.95 center at 0 -2.6
\ellipticalarc axes ratio 3:1 290 degrees from -1.4 -.65 center at 0 -1.3
\ellipticalarc axes ratio 3:1 290 degrees from -1.4 .65 center at 0 0
\ellipticalarc axes ratio 3:1 290 degrees from -1.4 1.95 center at 0 1.3
\ellipticalarc axes ratio 3:1 290 degrees from -1.4 3.25 center at 0 2.6
\ellipticalarc axes ratio 3:1 360 degrees from 2.4 2.6 center at 0 2.6
\setlinear
\plot -2.1 -3.45 -2.1 -3 /
\plot -2.1 -2.7  -2.1 -1.75 /
\plot -2.1 -1.4 -2.1 -.45 /
\plot -2.1 -.1 -2.1 .9 /
\plot -2.1 1.2 -2.1 2.15 /
\plot -2.1 2.5 -2.1 3.45 /
\plot 2.1 -3.45 2.1 -3 /
\plot 2.1 -2.7  2.1 -1.75 /
\plot 2.1 -1.4 2.1 -.45 /
\plot 2.1 -.1 2.1 .9 /
\plot 2.1 1.2 2.1 2.15 /
\plot 2.1 2.5 2.1 3.45 /
\plot 0 -3.8  0 -3.4 /
\plot 0 -2.75  0 -2.1 /
\plot 0 -1.55  0 -.85 /
\plot 0 -.25  0 .5 /
\plot 0 1.05  0 1.75 /
\plot 0 2.45  0 3.25 /
\plot -1.1 -3.7 -1.1 -3.3 /   
\plot -1.1 -2.85  -1.1 -2 / 
\plot -1.1 -1.55 -1.1 -.7 /
\plot -1.1 -.25 -1.1 .55 /
\plot -1.1 1.05 -1.1 1.9 /
\plot -1.1 2.35 -1.1 3.1 /
\plot 1.1 -3.7 1.1 -3.3 /   
\plot 1.1 -2.85 1.1 -2 / 
\plot 1.1 -1.55 1.1 -.7 /
\plot 1.1 -.25 1.1 .55 /
\plot 1.1 1.05 1.1 1.9 /
\plot 1.1 2.35 1.1 3.1 /
\arrow <6pt> [.2,.6] from 4 -5.8  to 4 5.8
\arrow <6pt> [.2,.6] from 2.38 -1.3  to 4 -1.3
\put{$t$} [l] at 4.25 -1.4
\put{$S_t$} [l] at 1.5 -1.45
\stoprotation
\put{$\mathbb R$} [lt] at 4.8 -4.2
\put{\firma} at 5 -2
\endpicture
$
 \caption{Newtonian space-time.}
  \label{fig:Newton}
\end{figure}

\section{The relativistic cosmic connection}\label{s:2nd} 

\begin{center}
\fbox{\begin{minipage}{.95\linewidth} \vspu
{\bf $\bs 2^{nd}$ Bridge$-$postulate}. There exist {\bf special particles} whose peculiar velocity \eqref{e:vpec} is a {\bf universal constant} $c$:
\be\label{e:spepar}
\bboxed{a(t)\,\dfrac{ds}{dt}=c=\rm constant}
\ee
\end{minipage}}
\end{center}

Since the peculiar velocity is in fact the velocity with respect to a local frame of reference, this postulate clearly falls within the relativistic vision. Thus, the connection we are going to define will be called {\bf relativistic cosmic connection}. \index{relativistic cosmic connection}

\bt
There is a unique cosmic connection compatible with the existence of special particles. The function $F(q^\Z)$ is given by
\be\label{e:defK}
\bboxed{F=\dfrac{\kappa^2}{c^2}\,A^2\,H}
\ee
\et

\begin{figure} [H]
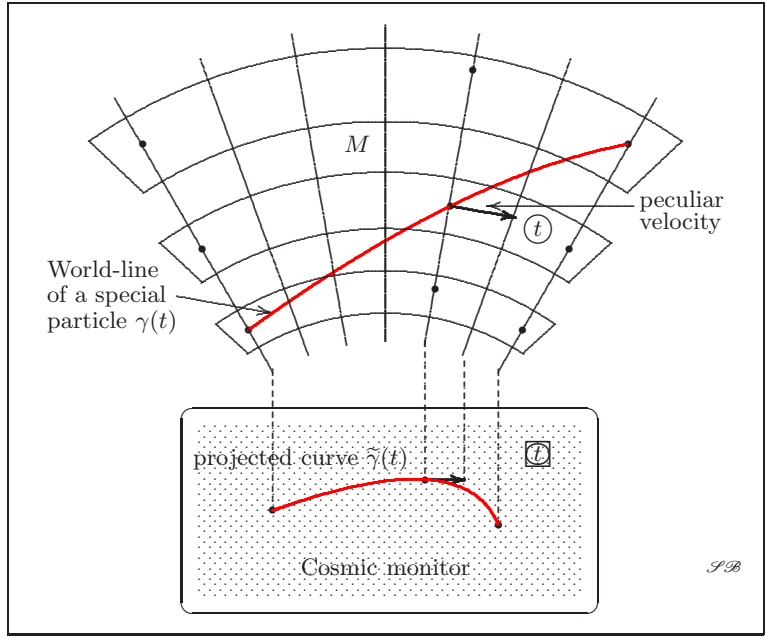
 
$$
\beginpicture
\setcoordinatesystem units <.75cm,.75cm>
\setplotarea x from -6.7 to 6.7, y from -4.2 to 7
\normalgraphs
\small
\grid 1 1
\setlinear
\plot 0 1 0 6.6 /
\startrotation by .984 -.173 about 0 -3 
\plot 0 1 0 6.6 /
\stoprotation
\put{$\sc\bullet$} at -4.3 4.5
\put{$\sc\bullet$} at -3.25 2.64
\put{$\sc\bullet$} at -2.43 1.2
\put{$\sc\bullet$} at 1.55 5.8
\put{$\sc\bullet$} at 1.15 3.4 
\put{$\sc\bullet$} at .88 1.93
\put{$\sc\bullet$} at 4.3 4.5
\put{$\sc\bullet$} at 3.25 2.64
\put{$\sc\bullet$} at 2.43 1.2
\put{World-line} [l] at -6 2.3
\put{of a special} [l] at -6 1.8
\put{particle $\gamma(t)$} [l] at -6 1.35
\arrow <6pt> [.2,.6] from -3.7 1.8 to -2 1.5
\put{projected curve $\wt\gamma(t)$} [rb] at .4 -1.3
\setplotsymbol (\normalered)
\setquadratic
\plot 
-2.43 1.2 
1.15 3.4
4.3 4.5 /
\setplotsymbol (\spessino)
\arrow <5pt> [.2,.6] from 1.15 3.4 to 2.3 3.2
\normale
\put{\large $\bigcirc$} at 2.7 3
\put{$t$} at 2.7 3
\put{peculiar} [l] at 4.5 3.5
\put{velocity} [l] at 4.5 3.1
\arrow <5pt> [.2,.6] from 4.4 3.4 to 1.8 3.4
\setlinear
\startrotation by .939 -.342 about 0 -3 
\plot 0 1 0 6.6 /
\stoprotation
\startrotation by .866 -.5 about 0 -3 
\plot 0 1 0 6.6 /
\stoprotation
\setlinear
\plot 0 1 0 6.6 /
\startrotation by .984 .173 about 0 -3 
\plot 0 1 0 6.6 /
\stoprotation
\startrotation by .939 .342 about 0 -3 
\plot 0 1 0 6.6 /
\stoprotation
\startrotation by .866 .5 about 0 -3 
\plot 0 1 0 6.6 /
\stoprotation
\circulararc 33 degrees from 0 1.5 center at 0 -3
\circulararc -33 degrees from 0 1.5 center at 0 -3
\circulararc 35 degrees from 0 2.25 center at 0 -3
\circulararc -35 degrees from 0 2.25 center at 0 -3
\setlinear
\plot -2.45 .8 -3 1.3 /
\plot 2.45 .8 3 1.3 /
\circulararc 33 degrees from 0 3 center at 0 -3
\circulararc -33 degrees from 0 3 center at 0 -3
\circulararc 35 degrees from 0 4 center at 0 -3
\circulararc -35 degrees from 0 4 center at 0 -3
\plot -3.25 2.05 -4 2.75 /
\plot 3.25 2.05 4 2.75 /
\circulararc 33 degrees from 0 4.9 center at 0 -3
\circulararc -33 degrees from 0 4.9 center at 0 -3
\circulararc 35 degrees from 0 6.2 center at 0 -3
\circulararc -35 degrees from 0 6.2 center at 0 -3
\plot -4.3 3.65 -5.25 4.55 /
\plot 4.3 3.65 5.25 4.55 /
\put{$M$} at -.5 4.5
\put{
\cornersize{.1}
\ovalbox{
\begin{minipage}{5.2cm}
\begin{minipage}{4.7cm}
\rule[-2.5cm]{0mm}{0cm}
\end{minipage}
\end{minipage}}
} at 0 -2
\put{
\setshadegrid span <2pt>
\setlinear
\vshade -3.5 -1.5 1.6 3.5 -1.5 1.6 /
} at 0 -2
\put{Cosmic monitor} at 0 -3
\put{$\sc\bullet$} [t] at -2 -1.92
\put{$\sc\bullet$} [t] at 2 -2.2
\put{$\sc\bullet$} [t] at .7 -1.4
\setplotsymbol (\spessino)
\arrow <5pt> [.2,.6] from .7 -1.45 to 1.4 -1.46

\normale
\setlinear
\setdashes <2pt>
\plot -2 .5 -2 -2 /
\plot .7 1 .7 -1.4 /
\plot 2 .5 2 -2.2 /
\plot 1.4 .65 1.4 -1.5 /
\setsolid
\setplotsymbol (\normalered)
\setquadratic
\plot -2 -2 
.7 -1.45 
2 -2.25 /
\normale
\put{\Large $\square$} at 2.7 -1
\put{$\bs\bigcirc$} at 2.7 -1
\put{$t$} at 2.7 -1
\put{\firma} at 6 -3
\endpicture
$$
\vskip -3mm
 \caption{Special particle.}
  \label{fig:spepar}
\end{figure}

\bprf
Due to \eqref{e:spepar}  the monitor speed \eqref{e:mon-speed} of a special particle is
\be\label{e:spepar1}
\bboxed{v(t)\Def\dfrac{ds}{dt}=\dfrac{c}{a(t)}}
\ee
Then in the parameter $q^\Z$ the monitor speed is expressed by the function
\be\label{e:spepar2}
V(q^\Z)\Def\dfrac{ds}{dq^\Z}=\dfrac{c}{\kappa\,A(q^\Z)}.
\ee
Note that this `speed' coincides with the function $V(q^\Z)$ defined in \eqref{e:defVq0} for which equation \eqref{e:VGH} holds,
$$
\dfrac{d\log V}{dq^\Z}-F\,V^2+2\,H=0.
$$
Due  to \eqref{e:spepar2} this equation is equivalent to
$$
\dfrac{d\log A^{-1}}{dq^\Z}+2\,H-F\,\dfrac{c^2}{\kappa^2}\,A^{-2}=0
$$
As $H=(\log A)'$, we have $H-F\,\dfrac{c^2}{\kappa^2}\,A^{-2}=0$ 
and we find equation \eqref{e:defK}. \eprf 

\br
The definition \eqref{e:defK} of $F$ shows that the cosmic connection  depends explicitly on $c$. (the functions $A$ and $H$ do not depend on $c$).  Then $c$ plays the role of {\bf universal constant}. \er

However, as an alternative to the second bridge-postulate,   one could consider  a third one:
\begin{center}
\fbox{\begin{minipage}{.95\linewidth} \vspu
{\bf $\bs 3^{rd}$ Bridge$-$postulate}. There exist two (or more) {\bf special particles} whose peculiar velocities \eqref{e:vpec} are universal constants:
\be\label{e:spepar-more}
\bboxed{a(t)\,\dfrac{ds_\uno}{dt}=c_\uno, \;\;
a(t)\,\dfrac{ds_\due}{dt}=c_\due, \; ...}
\ee
\end{minipage}}
\end{center}
Then we will have two (or more) superposed cosmic connections and two (or more) light-particles, with which we can build up something that we could call $\boxed{\hbox{\it multi-relativistic cosmic dynamics}}\;$ (science fiction?). 
\br
Remind that the constant $\kappa$ has been introduced at the beginning of our discussion for a dimensional consistency in the correlation between the cosmic time $t$ and the length-dimensional coordinate $q^\Z$: $q^\Z=\kappa\,t$. Its numerical value has been left arbitrary, but fixed. This constant has been present throughout our discussion, and it is also present in the definition \eqref{e:defK} of $F$. Then there is no loss of generality in considering 
$$
\bboxed{\kappa=c} 
$$
It follows from \eqref{e:spepar2} that for a special particle
\be\label{e:spepar2.}
\quad\bboxed{\dfrac{ds}{dq^\Z}=V=A^{-1}} \quad \bullet
\ee
\erx

The symbols of the relativistic cosmic connection are
\be\label{e:CoCo2nd}
\boxed{
\bc
\!\! \Gamma_{a\Z}^\Z=0
\ac
\!\! \Gamma_{\Z\Z}^c=0
\ac
\!\! \Gamma_{\Z\Z}^\Z=0
\ec
\bc
\!\! \Gamma_{a\Z}^b=H\,\delta_a^b
\ac
\!\! \Gamma^\Z_{ab}=A^2\,H\,\wt g_{ab}(\wt q)
\ac
\!\! \Gamma^c_{ab}=\wt\Gamma^c_{ab}
\ec}
\ee

\bt\label{t:spepar-geo}
The world-line of a special particle is a geodesic of the relativistic cosmic  connection.\footnote{
\kern 2pt In other words: {\it a special particle is a free particle of the relativistic cosmic connection}. }
\et

\vskip -5mm
\bprf
The world-line of a special particle is a curve transversal to the spatial sections. The characteristic equations \eqref{e:geod-rs} of the $\Gamma$-geodesics in the parameter $s$ read

\centerline{$
\dfrac{d\,}{ds}\dfrac{d\gamma^c}{ds}
+\Gamma_{ab}^c\,\dfrac{d\gamma^a}{ds}\,\dfrac{d\gamma^b}{ds}
=V^{-1}\left(\vpx V^2\,F-(\log V)'-2\,H\right)\dfrac{d\gamma^c}{ds}.
$}
Due to \eqref{e:spepar2.} $V=A^{-1}$, the coefficient at the right hand side vanishes: $V^2\,F-(\log V)'-2\,H=H+(\log A)'-2\,H=0$. It follows that $\dfrac{d^2\gamma^c}{ds^2}
+\Gamma_{ab}^c\,\dfrac{d\gamma^a}{ds}\,\dfrac{d\gamma^b}{ds}=0$ 
and we can apply Theorem \ref{t:cored}. \eprf

\begin{center}
\fbox{\begin{minipage}{.95\linewidth} \vspu
If we accept the existence of `special particles' then at the bifurcation point of Fig.\ \ref{fig:bifurc2} we turn right. On this way we are going towards a relativistic formulation of the isotropic cosmology. In fact at the first station we will find a surprise: as a consequence of our postulates the space-time admits in a canonical way a metric with Lorentzian signature. \vspd
\end{minipage}}
\end{center}
  
\begin{figure}[H]
  \centering
\kern -4mm $
\beginpicture
\setcoordinatesystem units <1cm,1.2cm> 
\setplotarea x from -5.05 to 5.05, y from -.5 to 7.6
\normalgraphs
\grid 1 1
\small
\setquadratic
\plot  
.6 4.2 
1.05 4.7
1.6 4.8 /
\setlinear
\plot .23 4 .6 4.2 /
\plot 1.6 4.8 2 5 /
\setquadratic 
\plot  
.3 4.35 
.9 4.85
1.4 4.88 /
\setlinear
\plot 0 4.2 .3 4.35 /
\plot 1.4 4.88 1.77 5.03 /
\setquadratic
\plot  
-.6 4.2 
-1.05 4.7
-1.6 4.8 /
\setlinear
\plot -.23 4 -.6 4.2 /
\plot -1.6 4.8 -2 5 /
\setquadratic 
\plot  
-.3 4.35 
-.9 4.85
-1.4 4.88 /
\setlinear
\plot 0 4.2 -.3 4.35 /
\plot -1.4 4.88 -1.77 5.03 /
\setlinear
\plot -.55 .8 -.23 4 /
\plot .55 .8 .23 4 /
\put{\beginpicture
\put{$\boxed{\hbox{\footnotesize 1.st Bridge postulate}}$} [rt] at -.5 2.1
\plot -.5 1.5 -.5 2.1 /
\endpicture} at -2.55 4.53
\put{$\sc\bullet$} at -1.13 4.13
\put{\beginpicture
\put{$\boxed{\hbox{Newtonian connection}}$} [rt] at -.5 2.1
\plot -.5 1.5 -.5 2.1 /
\endpicture} at -3.2 5.25
\put{$\sc\bullet$} at -1.6 4.8
\put{\includegraphics[width=1.5cm,keepaspectratio]{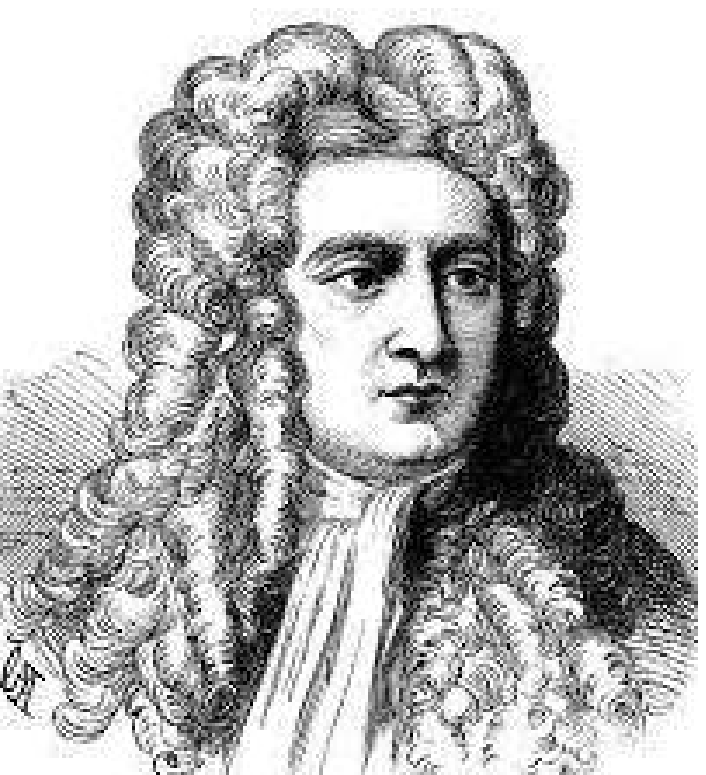}} [rb] at -1.5 5.5
\put{\footnotesize\scshape Realm of} [l] at -4.7 6.5
\put{\footnotesize\scshape Newtonian} [l] at -4.7 6.2
\put{\footnotesize\scshape Cosmology} [l] at -4.7 5.9
\put{\beginpicture
\put{$\boxed{\hbox{\footnotesize 2.nd Bridge postulate}}$} [lt] at .37 3.6
\plot .37 3 .37 3.6 /
\endpicture} at 2.65 4.53
\put{$\sc\bullet$} at 1.2 4.13
\put{\beginpicture
\put{$\boxed{\hbox{Relativistic connection}}$} [lt] at .37 3.6
\plot .37 3 .37 3.6 /
\endpicture} at 3.25 5.25
\put{$\sc\bullet$} at 1.6 4.8
\put{\includegraphics[width=1.5cm,keepaspectratio]{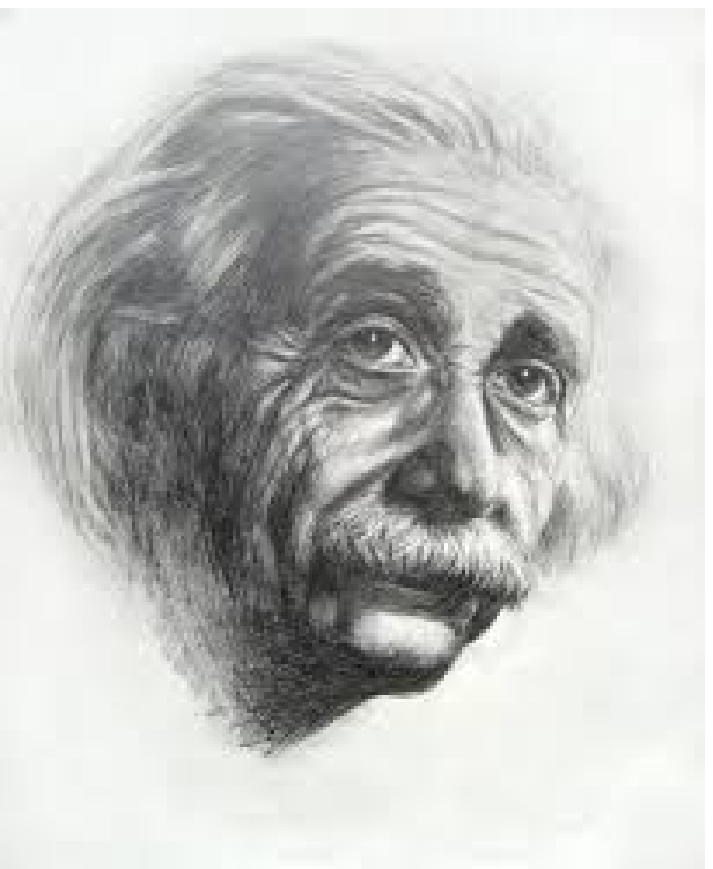}} [lb] at 1.5 5.5
\put{\footnotesize\scshape Realm of} [r] at 4.7 6.5
\put{\footnotesize\scshape Relativistic} [r] at 4.7 6.2
\put{\footnotesize\scshape Cosmology} [r] at 4.7 5.9
\put{\huge ?} at 0 6
\setplotsymbol (\spessino)
\plot -.6 4.2 -4.5 3.7  / 
\plot .6 4.2 4.5 3.7  / 
\plot .05 4.3 1.5 7 /
\plot -.05 4.3 -1.5 7 /
\put{$\boxed{\hbox{\textcolor{red}{We turn right}}}$} [rt] at -.6 3.9
\put{\textcolor{red}{$\bullet$}} at 0 3.7
\setplotsymbol (\spessinored)
\plot 0 3.7 0 4  /
\arrow <6pt> [.2,.6] from 0 4  to .5 4.37
\plot -.6 3.7 0 3.7 /
\put{\scshape Realm of Kinematics} at -3 1 
\setplotsymbol (\spessino)
\put{$\boxed{\hbox{Quotient manifold}}$} [lt] at .55 1.6
\plot .55 1 .55 1.6 /
\put{$\sc\bullet$} [b] at .55 .95
\put{$\boxed{\hbox{Scale parameter}}$} [rt] at -.5 2.1
\plot -.5 1.5 -.5 2.1 /
\put{$\sc\bullet$} [b] at -.5 1.45
\put{$\boxed{\hbox{Quotient metric}}$} [lt] at .45 2.6
\plot .45 2 .45 2.6 /
\put{$\sc\bullet$} [b] at .44 1.95
\put{$\boxed{\hbox{Hubble law}}$} [rt] at -.4 3.1
\plot -.4 2.5 -.4 3.1 /
\put{$\sc\bullet$} [b] at -.4 2.45
\put{$\boxed{\hbox{Cosmic connections}}$} [lt] at .37 3.6
\plot .37 3 .37 3.6 /
\put{$\sc\bullet$} [b] at .36 2.95
\put{$\bullet$} at 0 0
\put{$\boxed{\hbox{\large Postulates}}$} [t] at 0 .8
\plot -1 .1 -1 .8 /
\plot 1 .1 1 .8 /
\put{$\triangle$} [b] at 1 -.1
\put{$\triangle$} [b] at -1 -.1
\normale
\plot .6 .35 .63 0 -.65 0 -.6 .35 /
\setplotsymbol (\spessored)
\put{\textcolor{red}{$\bullet$}} at 0 3.7
\plot 0 0 0 .35 /
\plot 0 .85 0  3.7 /
\arrow <6pt> [.2,.6] from 0 1.7 to 0 2.5
\put{\firma} at 3 0
\endpicture
$
\vskip -2mm
 \caption{Towards the relativistic cosmology.}
  \label{fig:bifurc2}
\end{figure}

\section{The canonical cosmic metric}\label{s:CoMe}

\bt\label{t:CaCoMe}
The relativistic cosmic connection is the Levi-Civita connection of the Lorentzian metric\footnote{
\kern 2pt Note that the Newtonian connection, for which $F=0$, cannot admit a cosmic metric.}
\be\label{e:CaCoMe}
\bboxed{g_{\alpha\beta}\colon \!\!\!
\bc
\!\!g_{\Z\Z}=\alpha={\rm const.}
\ax
\!\!g_{a\Z}=0
\ax
\!\!g_{ab}=-\alpha\,A^2\,\wt g_{ab}
\ec \!\!\!\!\!\!}
\;\;
\bboxed{g_{\alpha\beta}\,dq^\alpha\,dq^\beta=\alpha\,(dq^\Z{}^2-A^2\,\wt g_{ab}\,dq^a\,dq^b)}
\ee
\et
\bprf
According to Theorem \ref{t:Sq0} the components of any metric have the form

$$
g_{\alpha\beta}:\,
\bc
g_{\Z\Z}={\alpha}(q^\Z),
\ax
g_{\Z a}=0,
\ax
g_{ab}={\beta}(q^\Z)\,\wt g_{ab}
\ec
\iff
g^{\alpha\beta}:\,\bc
g^{\Z\Z}={\alpha}^{-1},
\ax
g^{\Z a}=0,
\ax
g^{ab}={\beta}^{-1}\,\wt g^{ab}.
\ec
$$
Computation of the first-kind Christoffel $\Gamma_{\alpha\beta,\gamma}=\tfrac 12\,(\partial_\alpha g_{\beta\gamma}+\partial_\beta g_{\gamma\alpha}-\partial_\gamma g_{\alpha\beta})$:
$$
2\,\Gamma_{\Z\Z,\gamma}=\partial_\Z g_{\Z\gamma}+\partial_\Z g_{\gamma \Z}-\partial_\gamma g_{\Z\Z}=
\bc
\!\!
2\,\Gamma_{\Z\Z,\Z}=\partial_\Z g_{\Z\Z}+\partial_\Z g_{\Z\Z}-\partial_\Z g_{\Z\Z}={\alpha}'.
\ax
\!\!2\,\Gamma_{\Z\Z,a}=\partial_\Z g_{\Z a}+\partial_\Z g_{a\Z}-\partial_a g_{\Z\Z}=0.
\ec
$$
$$
2\,\Gamma_{\Z b,\gamma}=\partial_\Z g_{b\gamma}+\partial_b g_{\gamma\Z}-\partial_\gamma g_{\Z b}=
\bc
\!\!2\,\Gamma_{\Z b,\Z}=\partial_\Z g_{b\Z}+\partial_b g_{\Z\Z}-\partial_\Z g_{\Z b}=0.
\ax
\!\!2\,\Gamma_{\Z b,c}=\partial_\Z g_{bc}+\partial_b g_{c\Z}-\partial_c g_{\Z b}={\beta}'\,\wt g_{bc}.
\ec
$$
$$
\ba
2\,\Gamma_{ab,\gamma}=\partial_a g_{b\gamma}+\partial_b g_{\gamma a}-\partial_\gamma g_{ab}
=
\ac
\kern 2cm \bc
\!\!2\,\Gamma_{ab,\Z}=\partial_a g_{b\Z}+\partial_b g_{\Z a}-\partial_\Z g_{ab}=-{\beta}'\,\wt g_{ab}.
\ax
\!\!2\,\Gamma_{ab,c}=\partial_a g_{bc}+\partial_b g_{ca}-\partial_c g_{ab}
=2\,{\beta}\,\wt\Gamma_{ab,c}.
\ec
\ea
$$
Computation of the second-kind Christoffel $\Gamma_{\alpha\beta}^\gamma=g^{\gamma\delta}\,\Gamma_{\alpha\beta,\delta}$:
$$
\Gamma_{\Z\Z}^\gamma=g^{\gamma\delta}\,\Gamma_{\Z\Z,\delta}
=
\bc
\Gamma_{\Z\Z}^\Z=g^{\Z\delta}\,\Gamma_{\Z\Z,\delta}
=g^{\Z\Z}\,\Gamma_{\Z\Z,\Z}=\tfrac 12\,{\alpha}^{-1}\,{\alpha}'=\tfrac 12\,(\log {\alpha})'.
\ax
\Gamma_{\Z\Z}^c=g^{c\delta}\,\Gamma_{\Z\Z,\delta}=g^{cd}\,\Gamma_{\Z\Z,d}=0.
\ec
$$
$$
\Gamma_{a\Z}^\gamma=g^{\gamma\delta}\,\Gamma_{a\Z,\delta}
=\bc
\Gamma_{a\Z}^\Z=g^{\Z\delta}\,\Gamma_{a\Z,\delta}=g^{\Z\Z}\,\Gamma_{a\Z,\Z}=0.
\ax
\Gamma_{a\Z}^c=g^{c\delta}\,\Gamma_{a\Z,\delta}=g^{cd}\,\Gamma_{a\Z,d}
=\tfrac 12\, {\beta}^{-1}\,\wt g^{cd}\,{\beta}'\,\wt g_{ad}
\ax
\kern 6mm =\tfrac 12\, (\log {\beta})'\,\delta_a^c.
\ec
$$
$$
\Gamma_{ab}^\gamma=g^{\gamma\delta}\,\Gamma_{ab,\delta}
=
\bc
\Gamma_{ab}^\Z=g^{\Z\delta}\,\Gamma_{ab,\delta}=g^{\Z\Z}\,\Gamma_{ab,\Z}=-\tfrac 12\,{\alpha}^{-1}\,{\beta}'\,\wt g_{ab}.
\ax
=g^{c\delta}\,\Gamma_{ab,\delta}=g^{cd}\,\Gamma_{ab,d}
={\beta}^{-1}\,\wt g^{cd}\,{\beta}\,\wt\Gamma_{ab,d}
=\wt\Gamma_{ab}^c.
\ec
$$
$$
\hbox{Summary:}\;\;
\bc
\Gamma_{\Z\Z}^\Z=\tfrac 12\,(\log {\alpha})',
\ax
\Gamma_{\Z\Z}^c=0, 
\ax 
\Gamma^\Z_{a\Z}=0,
\ec
\quad
\bc
\Gamma_{a\Z}^c=\tfrac 12\,(\log {\beta})'\,\delta_a^c,
\ax
\Gamma^\Z_{ab}=-\tfrac 12\,{\alpha}^{-1}\,{\beta}'\,\wt g_{ab},
\ax
\Gamma_{ab}^c=\wt\Gamma_{ab}^c.
\ec
$$
These symbols coincide with those of the relativistic cosmic connection \eqref{e:CoCo2nd}
$$
\bc
\Gamma_{\Z\Z}^\Z=0,
\ax
\Gamma_{\Z\Z}^c=0, 
\ax 
\Gamma^\Z_{a\Z}=0,
\ec
\quad
\bc
\Gamma_{a\Z}^c=H\,\delta_a^c,
\ax
\Gamma^\Z_{ab}=K\,\wt g_{ab},
\ax
\Gamma_{ab}^c=\wt\Gamma_{ab}^c.
\ec
\quad
\bc
H=A^{-1}A'=(\log A)'
\ac
K=A\,A'=A^2\,H
\ec
$$
if and only if 
$$
\bc
{\alpha}={\rm constant}, 
\ax
\tfrac 12\,(\log {\beta})'=(\log A)',
\ax
-\tfrac 12\,{\alpha}^{-1}\,{\beta}'=A\,A'
\ec
\iff
\bc
{\alpha}={\rm constant},
\ax
A^{-2}\,\beta={\rm constant}=\gamma,
\ax
\beta'=-2\,\alpha\,A\,A'.
\ec
$$
$$
\iff
\bc
{\alpha}={\rm constant},
\ax
\beta=\gamma\, A^2,
\ax
\beta'=-2\,\alpha\,A\,A'.
\ec
\iff
\bc
{\alpha}={\rm constant},
\ax
\beta'=2\,\gamma\,A\,A',
\ax
\beta'=-2\,\alpha\,A\,A'.
\ec
$$
$$
\iff
\bc
{\alpha}={\rm constant},
\ax
\gamma=-\alpha,
\ax
\beta=-\alpha\,A^2.
\ec
\To \eqref{e:CaCoMe}  \eprfx
$$
Without loss of generality we can take $\alpha=-1$ and consider the metric 
\be\label{e:FRW}
\bboxed{\!g_{\alpha\beta}\colon \!\!\!
\bc
\!\!\!g_{\Z\Z}=-1
\ax
\!\!\!g_{\Z a}=0
\ax
\!\!\!g_{ab}=A^2(q^\Z)\,\wt g_{ab}(\wt q)
\ec\!\!\!\!\!\!}
\;\;
\bboxed{
g_{\alpha\beta}\,dq^\alpha\,dq^\beta=-\,dq^\Z{}^2+A^2\,\wt g_{ab}\,dq^a\,dq^b}
\ee
with signature $-+++$ as the {\bf canonical cosmic metric}. \index{canonical cosmic metric} The contravariant components are
\be\label{e:g-contra}
\bboxed{g^{\alpha\beta}: \bc
g^{\Z\Z}=-1
\ax
g^{\Z a}=0
\ax
g^{ab}=A^{-2}(q^\Z)\,\wt g^{ab}(\wt q)
\ec}
\ee

\bt\label{t:time}
The galactic world-lines are time-like geodesics of the canonical cosmic metric orthogonal to the spatial sections.
\et

\bprf
By virtue of the first requirement of a cosmic connection (p.\  \pageref{req:1}) the galactic world-lines are geodesics of the relativistic connection, thus of the cosmic metric. Since in co-moving coordinates $g_{\Z a}=0$, these world lines are orthogonal to the spatial sections, thus they are time-like. \eprf

\section{Light particles}

By the second bridge-postulate (Section \ref{s:2nd}) we have introduced the notion of special particle. 

\bt\label{t:light}
The world-lines of the special particles are light-like geodesics of the canonical cosmic metric.
\et

\bprf
For any world-line we have
$$
g_{\alpha\beta}\,\dfrac{d\gamma^\alpha}{dq^\Z}\,\,\dfrac{d\gamma^\alpha}{dq^\Z}
=
-\left(\dfrac{d\gamma^\Z}{dq^\Z}\right)^{\!2}
+g_{ab}\,\dfrac{d\gamma^a}{dq^\Z}\,\,\dfrac{d\gamma^b}{dq^\Z}
=
-\,1+A^2\,\wt g_{ab}\,\dfrac{d\gamma^a}{ds}\,\,\dfrac{d\gamma^b}{ds}\,
\left(\dfrac{ds}{dq^\Z}\right)^{\!2}.
$$
Since 
$$
\wt g_{ab}\,\dfrac{d\gamma^a}{ds}\,\,\dfrac{d\gamma^b}{ds}=1
$$
we get
\be\label{e:pp-1}
g_{\alpha\beta}\,\dfrac{d\gamma^\alpha}{dq^\Z}\,\,\dfrac{d\gamma^\alpha}{dq^\Z}=
A^2\,\left(\dfrac{ds}{dq^\Z}\right)^{\!2}-1.
\ee
For a photon equation \eqref{e:spepar2.} $\dfrac{ds}{dq^\Z}=A^{-1}$ holds. Then equation \eqref{e:pp-1} gives
$$
g_{\alpha\beta}\,\dfrac{d\gamma^\alpha}{dq^\Z}\,\,\dfrac{d\gamma^\alpha}{dq^\Z}=A^2\,A^{-2}-1=0. 
$$
This proves that the world-line of a special particle is a light-like curve. We already know that the world-line of a special particle is a geodesic (Theorem \ref{t:spepar-geo}, page \pageref{t:spepar-geo}). \eprf

Special particles are very strange particles: they are never at rest and have the same peculiar velocity in whatever local reference frame. This is in accordance with the theory of propagation of the electro-magnetic waves. Then by virtue of Theorem \ref{t:light} the concept of special particle can be identified with that of {\bf electro-magnetic signal} \index{electro-magnetic signal} or that of {\bf photon} \index{photon} in a broadest sense i.e., as a particle associated with light (visible or non-visible). 

\section{Sub-luminal particles}\label{s:mat-part}

\bd
A {\bf sub-luminal particle} is a particle ($2^{\rm nd}$ postulate of Kinematics, page \pageref{p:particle}) whose world-line $\gamma^\alpha(t)$ is a time-like curve admitting a parameter $\tau$, called {\bf proper time}, such that
\be\label{e:Vc2}
\bboxed{g_{\alpha\beta}\,\dfrac{d\gamma^\alpha}{d\tau}\,\dfrac{d\gamma^\beta}{d\tau}=-\,c^2}\;\;
\bboxed{\dfrac{d\tau}{dt}>0} \;{\vp}\footnote{
\kern 2pt The proper time is oriented towards the future, as the cosmic time $t$.}
\ee
\ed

\bt\label{t:dtau/dt}
Sub-luminal particles have peculiar velocity less than $c$ and along their world-lines 
$d\tau<dt$.\footnote{
\kern 2pt $d\tau<dt$: the proper time is runs slower than the cosmic time: twins paradox.} 
\et

\bprf \eqref{e:pp-1} $\iff$ 
$g_{\alpha\beta}\,\dfrac{d\gamma^\alpha}{d\tau}\,\,\dfrac{d\gamma^\alpha}{d\tau}\left(\dfrac{d\tau}{dq^\Z}\right)^{\!2}=A^2\,\left(\dfrac{ds}{dq^\Z}\right)^{\!2}-1.$

\eqref{e:Vc2} $\To$ $-c^2\left(\dfrac{d\tau}{dq^\Z}\right)^{\!2}=
A^2\,\left(\dfrac{ds}{dq^\Z}\right)^{\!2}-1$
$\To$ $1-\left(\dfrac{d\tau}{dt}\right)^{\!2}=
c^{-2}\,a^2\,\left(\dfrac{ds}{dt}\right)^{\!2}$.

\eqref{e:vpec}: $v_{\rm pec}\Def a(t)\dfrac{ds}{dt}$ $\To$
$1-\left(\dfrac{d\tau}{dt}\right)^{\!2}=
c^{-2}\,v_{\rm pec}^2$ $\To$
$$
\bc
\hbox{if $v_{\rm pec}\neq 0$}: 1-\left(\dfrac{d\tau}{dt}\right)^{\!2}>0 \;\;\hbox{i.e.} \;\; \left(\dfrac{d\tau}{dt}\right)^{\!2}<1.
\ac
\hbox{if $v_{\rm pec}=0$}: 1-\left(\dfrac{d\tau}{dt}\right)^{\!2}=0 \;\;\hbox{i.e.} \;\; \dfrac{d\tau}{dt}=1.
\ac
v_{\rm pec}^2=c^2\,\left[\vp\right.1-\left(\dfrac{d\tau}{dt}\right)^{\!2}\left.\vp\right]<c^2 \eprfx
\ec
$$
\br
Galaxies have zero peculiar velocity. Thus  galaxies are sub-luminal particles with $\tau=t$. \er

\section{Geodesics of the cosmic metric}

For later use we summarize here some fundamental equations concerning the geodesics of the relativistic cosmic metric. The symbols of the relativistic cosmic connection are given in \eqref{e:CoCo2nd}. The non-identically-vanishing symbols are
\be\label{e:geo-1}
\boxed{
\bc
\Gamma_{a\Z}^b=\dfrac{A'}A\,\delta_a^b=H\,\delta_a^b.
\ac
\Gamma^\Z_{ab}=A\,A'\,\wt g_{ab}=A^2\,H\,\wt g_{ab}
\ac
\Gamma^c_{ab}=\wt\Gamma^c_{ab}
\ec}
\ee
As a consequence, the geodesic equations \eqref{e:geod-xi-2} for a curve $\gamma^\alpha(\xi)$ with a generic parameter $\xi$ read
\be \label{e:geo-2}
\bc
\dfrac{d^2\gamma^\Z}{d\xi^2}
+A^2\,H\,\wt g_{ab}\,\dfrac{d\gamma^a}{d\xi}\,\dfrac{d\gamma^b}{d\xi}
=\lambda\,\dfrac{d\gamma^\Z}{d\xi},
\ac
\dfrac{d^2\gamma^c}{d\xi^2}
+\wt\Gamma_{ab}^c\,\dfrac{d\gamma^a}{d\xi}\,\dfrac{d\gamma^b}{d\xi}
+2\,H\,\dfrac{d\gamma^c}{d\xi}\,\dfrac{d\gamma^\Z}{d\xi}
=\lambda\,\dfrac{d\gamma^c}{d\xi}.
\ec
\ee
In the parameter $q^\Z$:
\be \label{e:geo-3}
\gamma^\Z=q^\Z\,\To\,\bc
A^2\,H\,\wt g_{ab}\,\dfrac{d\gamma^a}{dq^\Z}\,\dfrac{d\gamma^b}{dq^\Z}
=\lambda,
\ax
\dfrac{d^2\gamma^c}{dq^\Z{}^2}
+\wt\Gamma_{ab}^c\,\dfrac{d\gamma^a}{dq^\Z}\,\dfrac{d\gamma^b}{dq^\Z}
=\left(\lambda-2\,H\right)\dfrac{d\gamma^c}{dq^\Z}.
\ec
\ee
In the parameter $t$:
$$
\gamma^\Z=c\,t\, \To \, \bc
A^2\,H\,\wt g_{ab}\,\dfrac{d\gamma^a}{dt}\,\dfrac{d\gamma^b}{dt}
=\lambda\,c,
\ac
\dfrac{d^2\gamma^c}{dt^2}
+\wt\Gamma_{ab}^c\,\dfrac{d\gamma^a}{dt}\,\dfrac{d\gamma^b}{dt}
=\left(\lambda-2\,c\,H\right)\dfrac{d\gamma^c}{dt}.
\ec
$$
Since $A'=\dfrac{da}{dt}\,\dfrac{dt}{dq^\Z}=\dfrac{\dot a}{c}\;$, so that $H=A'/A=\dfrac{\dot a}{c\,a}=\dfrac hc$, these last equations become 

$
\To\;\bc
a^2\,h\,\wt g_{ab}\,\dfrac{d\gamma^a}{dt}\,\dfrac{d\gamma^b}{dt}
=\lambda\,c^2,
\ac
\dfrac{d^2\gamma^c}{dt^2}
+\wt\Gamma_{ab}^c\,\dfrac{d\gamma^a}{dt}\,\dfrac{d\gamma^b}{dt}
=\left(\lambda-2\,h\right)\dfrac{d\gamma^c}{dt}.
\ec
$

$
\To\;\bc
\lambda=\dfrac{a^2}{c^2}\,h\,\wt g_{ab}\,\dfrac{d\gamma^a}{dt}\,\dfrac{d\gamma^b}{dt},
\ac
\dfrac{d^2\gamma^c}{dt^2}
+\wt\Gamma_{ab}^c\,\dfrac{d\gamma^a}{dt}\,\dfrac{d\gamma^b}{dt}
=\left(\dfrac{a^2}{c^2}\,\wt g_{ab}\,\dfrac{d\gamma^a}{dt}\,\dfrac{d\gamma^b}{dt}-2\,h\right)h\,\dfrac{d\gamma^c}{dt}.
\ec
$

$\To$
\be\label{e:geo-4}
\bboxed{\bc
\gamma^\Z=c\,t,
\ax
\dfrac{d^2\gamma^c}{dt^2}
+\wt\Gamma_{ab}^c\,\dfrac{d\gamma^a}{dt}\,\dfrac{d\gamma^b}{dt}
=
\dfrac{h}{c^2}\left(a^2\,\wt g_{ab}\,\dfrac{d\gamma^a}{dt}\,\dfrac{d\gamma^b}{dt}-2\,c^2\right)\,\dfrac{d\gamma^c}{dt}
\ec}
\ee
These are the general geodesic equations in the cosmic time parameter $t$. 
Observe that for any world-line (transversal to the spatial sections)
\be \label{e:geo-5}
g_{\alpha\beta}\,\dfrac{d\gamma^\alpha}{dt}\,\,\dfrac{d\gamma^\alpha}{dt}
=
-\left(\dfrac{d\gamma^\Z}{dt}\right)^{\!2}
+a^2\,\wt g_{ab}\,\dfrac{d\gamma^a}{dt}\,\,\dfrac{d\gamma^b}{dt}
=
-c^2+a^2\,\wt g_{ab}\,\dfrac{d\gamma^a}{dt}\,\,\dfrac{d\gamma^b}{dt}.
\ee
For the world-line of a photon 
$$
g_{\alpha\beta}\,\dfrac{d\gamma^\alpha}{dt}\,\dfrac{d\gamma^\alpha}{dt}=0
\;\To\;
a^2\,\wt g_{ab}\,\dfrac{d\gamma^a}{dt}\,\,\dfrac{d\gamma^b}{dt}=c^2,
$$
and equations \eqref{e:geo-4} reduce to
\be\label{e:geo-6}
\bboxed{\bc
\gamma^\Z=c\,t,
\ax
\dfrac{d^2\gamma^c}{dt^2}
+\wt\Gamma_{ab}^c\,\dfrac{d\gamma^a}{dt}\,\dfrac{d\gamma^b}{dt}
=-h\,\dfrac{d\gamma^c}{dt}
\ec}
\ee
These are the light-like geodesic equations in the parameter $t$. Note that $t$ is \underbar{not} an affine parameter (because $h\neq 0$).

\section{Comments on the Weyl principle}

The standard texts of cosmology refer to the postulate of Weyl and the cosmological principle as basic statements on which to build up  models of the evolution of the universe. They are essentially formulated as follows:\footnote{
\kern 2pt See e.g.\ \cite{Narlikar}.}
\begin{center}
\fbox{\begin{minipage}{.95\linewidth} \vspu
{\bf Weyl's principle}: \it 
In cosmic space-time the world-lines of the galaxies form a bundle of non-intersecting \underbar{time-like} g\underline{eodesics} orthogonal to a series of space-like hyper-surfaces. \vspd \rm
\end{minipage}}
\end{center}
\begin{center}
\fbox{\begin{minipage}{.95\linewidth} \vspu
{\bf The cosmological principle}: \it On large scales the universe is spatially homogeneous and spatially isotropic. \vspd  \rm
\end{minipage}}
\end{center}

{\bf $\bs 1^{st}$ comment}: As mentioned in the Introduction, the Weyl postulate put cosmology in the framework of general relativity from the very beginning. In our approach the second part of this postulate is a theorem (Theorem \ref{t:time}) while the first part is contained in our third postulate. One might argue why not accept from the very beginning the Weyl postulate instead to spend so a long time starting from several postulates. The answer is that in this longer way we do not lose the knowledge of important facts which are not strictly pertinent to the theory of general relativity. For instance, the Hubble law  as well as several other results,\footnote{
\kern 2pt Concerning, for instance, the scale parameter and the quotient metric (Section \ref{s:qma}), the cosmic connections (this chapter) and the symmetric tensors (Chapter \ref{c:fst}).} 
are the subject of theorems valid regardless of any dynamical assumptions on the evolution of the galactic fluid.

{\bf $\bs 2^{nd}$ comment} about the cosmological principle: in our approach isotropy implies homogeneity (Theorem \ref{t:scalar}). In fact this follows at once from the fourth postulate concerning the existence of a metric on each spatial section. 
 
\begin{center}
\fbox{\begin{minipage}{.95\linewidth} \vspu
At the end of these first two chapters it is worth emphasizing that so far, as well as in the following discussion, we did not use any special coordinate system  like, for instance, one of those which are commonly used on manifolds with constant curvature. \vspd
\end{minipage}}
\end{center}


\chapter{Fundamental symmetric tensors}\label{c:fst}

\section{Symmetric tensors and conservation equations}

\bt\label{t:divT=0}
Let $\nabla_\alpha$ be the covariant derivative with respect to a general cosmic connection $\Gamma$ \eqref{e:tab-3}  and $T^{\alpha\beta}$  the components \eqref{e:Tq0} of an isotropic symmetric tensor. The four conservation equations \index{conservation equations} $\nabla_\alpha T^{\alpha\beta}=0$ are equivalent to the single equation
\be\label{e:divT=0}
\bboxed{\Phi'+3\,(H\,\Phi+F\,\Psi)=0}
\ee
\et

\bprf
It is sufficient to prove that
\be\label{e:divTq0}
\boxed{\;\ba
\nabla_\alpha T^{\alpha \Z}={\Phi}'+3\,(H\,{\Phi}+F\,{\Psi})
\ac
\nabla_\alpha T^{\alpha b}=0
\ea\;}
\ee
$\nabla_\alpha T^{\alpha \beta}=\partial_\alpha T^{\alpha\beta}+\Gamma^\alpha _{\alpha \gamma} T^{\gamma\beta} +\Gamma^\beta_{\alpha \gamma} T^{\alpha \gamma}\;\To$

$
\nabla_\alpha T^{\alpha \Z}=\partial_\alpha T^{\alpha \Z}
+\Gamma^\alpha _{\alpha \gamma} T^{\gamma \Z} 
+\Gamma^\Z_{\alpha \gamma} T^{\alpha \gamma}
=\partial_\Z T^{\Z\Z}
+\Gamma^\alpha _{\alpha\Z}\, T^{\Z\Z} 
+\Gamma^\Z_{\Z\Z} T^{\Z\Z}
+\Gamma^\Z_{ab} T^{ab}
$

$
={\Phi}'
+(\Gamma^\alpha _{\alpha\Z}+\Gamma^\Z_{\Z\Z})\,{\Phi}
+\Gamma^\Z_{ab}\,{\Psi}\,\wt g^{ab}
=\nabla_\alpha T^{\alpha b}=\partial_\alpha T^{\alpha b}+\Gamma^\alpha _{\alpha \gamma} T^{\gamma b} +\Gamma^b_{\alpha \gamma} T^{\alpha \gamma}
$

$
=\partial_a T^{ab}+\Gamma^\alpha _{\alpha a} T^{ab} 
+\Gamma^b_{a \gamma} T^{a\gamma}
+\Gamma^b_{\Z\gamma} T^{\Z\gamma}
=\partial_a T^{ab}+\Gamma^\alpha _{\alpha a} T^{ab} 
+\Gamma^b_{ac}\, T^{ac}
+\Gamma^b_{\Z\Z}\, T^{\Z\Z}
$

$
={\Psi}\,(\partial_a \wt g^{ab}+\Gamma^\alpha _{\alpha a}\,\wt g^{ab} 
+\Gamma^b_{ac}\,\wt g^{ac})
+\Gamma^b_{\Z\Z}\,{\Phi}.
$

Use \eqref{e:tab-3},
$$
\bc
\nabla_\alpha T^{\alpha \Z}={\Phi}'+3\,(H\,{\Phi}+F\,{\Psi}),
\ac
\nabla_\alpha T^{\alpha b}=
{\Psi}\,(\partial_a \wt g^{ab}+\wt\Gamma^c _{ca}\,\wt g^{ab} 
+\wt\Gamma^b_{ac}\,\wt g^{ac})={\Psi}\,\wt\nabla_a\wt g^{ab}=0.  \eprfx
\ec
$$

\section{Ricci tensor of a cosmic connection}

\bt
The  components of the Ricci tensor \index{Ricci!tensor of a cosmic connection} of a general cosmic connection \eqref{e:tab-3} are 
\be\label{e:Ricci-q0}
\bboxed{
\ba
R_{\Z\Z}=-3\,(H'+H^2)
\ax
R_{a\Z}=0
\ax
R_{ab}=(F'+H\,F+2\,\wt K)\,\wt g_{ab}
\ea
}
\ee
where $\wt K$ is the curvature constant of the quotient metric $\wt g$.
\et

\bprf 
The Ricci tensor components are defined by (see \eqref{e:Ricci-def})
$$
R_{\alpha\beta}=\partial_\mu\Gamma^\mu_{\alpha\beta}
-\partial_\beta\Gamma^\mu_{\alpha\mu}
+\Gamma^\mu_{\sigma\mu}\,\Gamma^\sigma_{\alpha\beta}
-\Gamma^\mu_{\sigma\beta}\,\Gamma^\sigma_{\alpha\mu}.
$$
1. Computation of $R_{\Z\Z}$:

$
R_{\Z\Z}=\partial_\mu\Gamma^\mu_{\Z\Z}-\partial_\Z\Gamma^\mu_{\Z\mu}
+\Gamma^\mu_{\sigma\mu}\,\Gamma^\sigma_{\Z\Z}
-\Gamma^\mu_{\sigma \Z}\,\Gamma^\sigma_{\Z\mu}
={}-\partial_\Z\Gamma^\mu_{\Z\mu}-\Gamma^\mu_{\sigma \Z}\,\Gamma^\sigma_{\Z\mu}
$

$
={}-\partial_\Z\Gamma^a_{\Z a}-\Gamma^a_{b \Z}\,\Gamma^b_{\Z a}
=-H'\,\delta_a^a-H\,\delta_b^a\,H\,\delta_a^b
=-3\,(H'+H^2).
$

2. Computation of $R_{ab}$:

$
R_{ab}=\partial_\mu\Gamma^\mu_{ab}
-\partial_b\Gamma^\mu_{a\mu}
+\Gamma^\mu_{\sigma\mu}\,\Gamma^\sigma_{ab}
-\Gamma^\mu_{\sigma b}\,\Gamma^\sigma_{a\mu}
$

$
=\partial_\Z\Gamma^\Z_{ab}+\partial_c\Gamma^c_{ab}
-\partial_b\Gamma^c_{ac}
+\Gamma^c_{\Z c}\,\Gamma^\Z_{ab}
+\Gamma^c_{d c}\,\Gamma^d_{ab}
-\Gamma^\Z_{c b}\,\Gamma^c_{a\Z}
-\Gamma^c_{\Z b}\,\Gamma^\Z_{ac}
-\Gamma^c_{db}\,\Gamma^d_{ac}.
$

Reordering:

$
{R_{ab}}=
\partial_c\wt\Gamma^c_{ab}
-\partial_b\wt\Gamma^c_{ac}
+\wt\Gamma^c_{d c}\,\wt\Gamma^d_{ab}
-\wt\Gamma^c_{db}\,\wt\Gamma^d_{ac}
+\partial_\Z\Gamma^\Z_{ab}
+\Gamma^c_{\Z c}\,\Gamma^\Z_{ab}
-\Gamma^\Z_{c b}\,\Gamma^c_{a\Z}
-\Gamma^c_{\Z b}\,\Gamma^\Z_{ac}
$

(the first four terms give the Ricci-tensor components $\wt R_{ab}$ of $\wt g$)

$
=\wt R_{ab}
+F'\,\wt g_{ab}+3\,H\,F\,\wt g_{ab}-F\,\wt g_{cb}\,H\,\delta_a^c
-H\,\delta_b^c\,H\,\wt g_{ac}
$

$
=\wt R_{ab}+F'\,\wt g_{ab}+3\,H\,F\,\wt g_{ab}-F\,\wt g_{ab}\,H
-H\,F\,\wt g_{ab}
$

$
=\wt R_{ab}+(F'+H\,F)\,\wt g_{ab}
\quad
(\hbox{apply \eqref{e:ccm2}, $\wt R_{ab}=2\,\wt K\,\wt g_{ab}$})
=(2\,\wt \Gamma+F'+H\,F)\,\wt g_{ab}. \eprfx
$

\bt
The covariant components of the Ricci tensors  of the cosmic connections  take the form 
\begin{gather}
\label{e:Ricci-first}
\bboxed{\hbox{\rm Newtonian}:
\bc
R_{\Z\Z}=-3\,A^{-1}\,A''
\ax
R_{ab}=2\,\wt K\,\wt g_{ab}
\ec}
\\
\label{e:Ricci-second}
\bboxed{
\hbox{\rm Relativistic}:
\bc
R_{\Z\Z}=-3\,A^{-1}\,A''
\ax
R_{ab}=\left(2\,A'^2+A\,A''+2\,\wt K\right)\,\wt g_{ab}
\ec}
\end{gather}
\et

\bprf
(i) For $\Gamma=0$ (Newtonian connection) equations \eqref{e:Ricci-q0} reduce to
$$
\bc
R_{\Z\Z}=-3\,(H'+H^2),
\ax
R_{ab}=2\,\wt K\,\wt g_{ab}.
\ec
$$
Since $H=(\log A)'=A^{-1}\,A'$, we find 
$$
H^2+H'=A^{-2}\,(A')^2-A^{-2}\,(A')^2+A^{-1}\,A''
=A^{-1}\,A'',
$$
and \eqref{e:Ricci-first} are proved. (ii) For $F=A^2\,H$ (relativistic connection)  

$
F'+H\,F=(A^2\,H)'+A^2\,H^2=2\,A\,A'\,H+A^2\,H'+A^2\,H^2
$

(use again $H=A^{-1}\,A'$)

$
=2\,(A')^2+A^2\,\left[A^{-1}\,A''-A^{-2}\,(A')^2\right]+(A')^2
=2\,(A')^2+A\,A''.$

Enter this result in equations \eqref{e:Ricci-q0} and \eqref{e:Ricci-second} are proved. \eprf

\section{Einstein tensor of the relativistic cosmic connection}

\vspace*{-4mm} For the relativistic cosmic connection we can compute the mixed and the contravariant components of the Ricci tensor by raising the indices of the components \eqref{e:Ricci-second} by means of the contravariant components \eqref{e:g-contra} of the metric. We find that
\begin{gather}
\label{e:Ricci-mixed}
\bboxed{
\bc
R_\Z^\Z=3\,A^{-1}\,A''
\\
R_a^b=A^{-2}\left(2\,(A')^2+A\,A''+2\,\wt K\right)\,\delta_a^b
\ec\;\,\,\,}
\\
\label{e:Ricci-contra}
\bboxed{
\bc
R^{\Z\Z}=-3\,A^{-1}\,A''
\\
R^{ab}=A^{-4}\left(2\,(A')^2+A\,A''+2\,\wt K\right)\,\wt g^{ab}
\ec}
\end{gather}
From \eqref{e:Ricci-mixed} we derive the Ricci scalar curvature
$$
R\Def R_\alpha^\alpha=
3\left[A^{-1}\,A''+A^{-2}\left(2\,(A')^2+A\,A''+2\,\wt K\right)\right] \;\To
$$
\vskip -5mm
\be\label{e:Ricci-scalar}
\boxed{R=6\,A^{-2}\left(A'{}^2+A\,A''+\wt K\right)}
\ee
As a consequence we can compute the contravariant\footnote{
As we will see, we are more interested in these components, rather than in the covariant or mixed ones.} 
 components of the Einstein tensor $G^{\alpha\beta}\Def R^{\alpha\beta}-\tfrac 12\,R\,g^{\alpha\beta}$: 
$$
\ba
G^{\Z\Z}=R^{\Z\Z}-\tfrac 12\,R\,g^{\Z\Z}
=
-3\,A^{-1}\,A''+3\,A^{-2}\left(A'{}^2+A\,A''+\wt K\right) 
\\
=
3\,A^{-2}\left(A'{}^2+\wt K\right).
\ea
$$

$
\ba
G^{ab}=R^{ab}-\tfrac 12\,R\,g^{ab}
=
A^{-4}\left(2\,A'{}^2+A\,A''+2\,\wt K\right)\,\wt g^{ab}
\ax
\qquad{}-3\,A^{-4}\left(A'{}^2+A\,A''+\wt K\right)\wt g^{ab}
\ax
=A^{-4}\left[2\,A'{}^2+A\,A''+2\,\wt K-3\left(A'{}^2+A\,A''+\wt K\right)\right]\wt g^{ab}
\ax
=-A^{-4}\left[A'{}^2+2\,A\,A''+\wt K\right]\wt g^{ab}. \;\To
\ea$
\be\label{e:G-Einstein}
\bboxed{
\bc
G^{\Z\Z}=3\,A^{-2}\left(A'{}^2+\wt K\right)
\\
G^{ab}
=-A^{-4}\left(A'{}^2+2\,A\,A''+\wt K\right)\,\wt g^{ab}
\ec}
\ee

\section{The intervention of the cosmic time}

So far we have used the coordinate $q^\Z$ as a parameter. However, in the perspective of applying in a dynamical context the formulas so far found, we should rewrite them using as a parameter the cosmic time $t$. For this purpose we observe that
\be\label{e:q0t-1}
\bc
q^\Z=c\,t\,\To\,\dfrac{d\,}{dq^\Z}=\dfrac 1c\,\dfrac{d\,}{dt}\,\To\,\dfrac{dt}{dq^\Z}=\dfrac 1c. 
\ac
A(q^\Z)=a(t)\,\To\,A'=\dfrac{\dot a}c,\quad A''=\dfrac{\ddot a}{c^2}.
\ac
H(q^\Z)=A^{-1}\,A'\,\To\,H(q^\Z)=\dfrac{\dot a}{c\,a}.
\ac
F(q^\Z)=A^2\,H=A\,A'\,\To\,\dfrac {a\,\dot a}c.
\ec
\quad
\bc
{\vpx}'=\dfrac{d\,}{dq^\Z}, 
\ac
{\vpx}^{\bs .}=\dfrac{d\,}{dt}.
\ec
\ee
The Ricci curvature \eqref{e:Ricci-scalar} and the contravariant components  \eqref{e:G-Einstein} of the Einstein tensor become
\begin{gather}
\label{e:Ricci-t}
\bboxed{R_{\rm icci}=6\,c^{-2}\,a^{-2}\left(\dot a^2+a\,\ddot a+c^2\,\wt K\right)}
\\
\label{e:Gt}
\boxed{\;\ba 
G^{\Z\Z}=\dfrac 3{c^2\,a^2}\left(\dot a^2+c^2\,\wt K\right)
\ac
G^{ab}=-\dfrac 1{c^2\,a^4}\left(2\,a\,\ddot a+\dot a^2+c^2\wt K\right)\,\wt g^{ab}
\ea\;}
\end{gather}

The  contravariant components \eqref{e:Tq0} of an isotropic symmetric two-tensor take the form
\be\label{e:Tt}
\bboxed{
\bc
T^{\Z\Z}={\phi}(t)=\hbox{\rm a function of $t$ only},
\ax
T^{\Z a}=0,
\ax
T^{ab}={\psi}(t)\,\wt g^{ab}(\wt q)=\hbox{\rm a function of $t$ times $\wt g^{ab}$},
\ec
}
\ee
We call $\phi(t)$ and $\psi(t)$ the {\bf characteristic functions} of the symmetric tensor $T^{\alpha\beta}$.
\begin{center}
\fbox{\begin{minipage}{.95\linewidth} 
\bt\label{t:divT=0-t}\vspu
For an isotropic symmetric tensor \eqref{e:Tt} the conservation equations $\nabla_\alpha T^{\alpha\beta}=0$  are equivalent to the single equation
\be\label{e:divT=0-1}
\bboxed{a\,\dot\phi+3\,\dot a\,\left(\phi+a^2\,\psi\right)=0}
\ee
In turn, this equation is equivalent to 
\be\label{e:divT=0-2}
\bboxed{(\phi\,a^3)^{\bs .}+3\,a^4\,\dot a\,\psi=0}
\ee
\et
\end{minipage}}
\end{center}

\bprf
Apply the rules \eqref{e:q0t-1} to equation \eqref{e:divT=0}:
$$
\Phi'+3\,(H\,\Phi+F\,\Psi)=0
\iff
\dfrac{\dot\phi}{c}+3\,\left(\dfrac {\dot a}{c\,a}\,\phi+\dfrac{a\,\dot a}{c}\,\psi\right)=0 \iff \eqref{e:divT=0-1}. \eprfx
$$


\chapter{Relativistic cosmic dynamics}

\section{The principles of the relativistic cosmic dynamics}

In the previous chapters we have constructed the geometrical background we need for passing to the foundation of the {\scshape cosmic dynamics}. We have seen that the evolution of the universe  can be described by a single function of the cosmic time, the scale parameter $a(t)$. Our goal is now to state physical laws governing the evolution of $a(t)$. 

\begin{center}
\fbox{\begin{minipage}{.95\linewidth} \vspu
{\bf $\bs 1^{st}$ Postulate}. We found the cosmic dynamics on the  principles of the cosmic kinematics (first chapter) and on the existence of photons ($2^{\rm nd}$ bridge-postulate). \vspd
\end{minipage}}
\end{center}
  The cosmic space-time is then equipped with the cosmic metric \eqref{e:FRW}
\be\label{e:CaCoMe-}
\bboxed{\!\!\bc
g_{\Z\Z}=-1
\ax
g_{a\Z}=0
\ax
g_{ab}=a^2(t)\,\wt g_{ab}(\wt q)
\ec\!\!\!}
\;\;
\bboxed{
g_{\alpha\beta}\,dq^\alpha\,dq^\beta=-c^2\,dt^2+a^2(t)\,\wt g_{ab}\,dq^a\,dq^b}
\ee
The contravariant components are 
\be\label{e:CaCoMeL+}
\bboxed{\bc
g^{\Z\Z}=-1
\ax
g^{a\Z}=0
\ax
g^{ab}=a^{-2}(t)\,\wt g^{ab}(\wt q)
\ec}
\ee

\begin{center}
\fbox{\begin{minipage}{.95\linewidth} \vspu
{\bf $\bs 2^{nd}$ Postulate}. The evolution of the scale parameter $a(t)$ is governed by the {\bf Einstein field equations} \index{Einstein!field equations}
\be\label{e:Ein-0}
\boxed{\vp\;R^{\alpha\beta}+\left(\Lambda-\tfrac 12\, R\right)\,g^{\alpha\beta}=\chi\,T^{\alpha\beta}\;}
\ee
\vskip -3mm
equivalent to
\vskip -5mm
\be\label{e:Ein-1}
\boxed{\vp\;G^{\alpha\beta}=\chi\,T^{\alpha\beta}-\Lambda\,g^{\alpha\beta}\;}
\ee
\vskip -2mm
where $T^{\alpha\beta}$ is the {\bf energy tensor}, \index{energy tensor} $\Lambda\geq 0$ is the {\bf cosmological constant} and the costant $\chi$ is given by
$$
\chi=\dfrac{8\,\pi\,G_N}{c^4},
$$ 
where $G_N$ is the {\bf Newtonian gravitational constant}. \vspd
\end{minipage}}
\end{center}
Dimensional analysis of the Einstein equations:
\begin{table} [H]
\centering
\begin{tabular}{|c|c|c|}
\hline
 $\vpt$ Object
& $\Dim$
& Note
\\
\hline\hline
$\vpt \Lambda$
& $L^{-2}$
& [{\tt a}]
\\
\hline
$\vpt \chi\,T^{\alpha\beta}$
& $L^{-2}$
&  [{\tt a}]
\\
\hline 
$\vpt T^{\alpha\beta}$
& $M\,L^{-1}\,T^{-2}$ (energy density)
&  [{\tt b}]
\\
\hline 
$\vpt \chi$
& $M^{-1}\,L^{-1}\,T^{2}$
&  [{\tt c}]
\\
\hline 
$\vpt G_N$
& $M\,L^3\,T^{-2}$
&  [{\tt d}]
\\
\hline 
\end{tabular}
\end{table}

[{\tt a}] According to our conventions (Section \ref{s:Dim}) the coordinates $q^\alpha$ are length-dimensional, thus the metric tensor components $g_{\alpha\beta}$, $g^{\alpha\beta}$ are dimensionless and $\Dim(R^{\alpha\beta})=\Dim(R)=L^{-2}$. Then from the Einstein equations \eqref{e:Ein-0} it follows that:
$\Dim(\Lambda)=\Dim(\chi\,T^{\alpha\beta})=L^{-2}$. 

\vskip 4pt
[{\tt b}] Equations \eqref{e:ETcomp} below shows that $\Dim(T^{\Z\Z})=\Dim(e)$ and $\Dim(T^{ab})=\Dim(p)$. From the entries of Table \ref{tab:Dim-2} (page \pageref{tab:Dim-2}) $\Dim(e)=\Dim(p)=M\,L^{-1}\,T^{-2}$.

\vskip 4pt
[{\tt c}] $\Dim(\chi)=\Dim(\chi\,T^{\alpha\beta})/\Dim(T^{\alpha\beta})=
L^{-2}/(M\,L^{-1}\,T^{-2})$.

\vskip 4pt
[{\tt d}] $\Dim(\chi)=L^{-4}\,T^4\cdot\Dim(G_N)$ $\To$ $\Dim(G_N)=\Dim(\chi)\,L^{4}\,T^{-4}$

$=M^{-1}\,L^{-1}\,T^{2}\,L^{4}\,T^{-4}=M^{-1}\,L^3\,T^{-2}$. 

\section{The energy-momentum tensor}

According the above postulates, the core of a cosmological model is the choice of an energy-momentum tensor $T^{\alpha\beta}$ for the galactic fluid.  As for any isotopic symmetric tensor, it must have the form \eqref{e:iT}:
\be\label{e:iT}
\bboxed{\bs T=\phi(t)\,\partial_\Z\otimes\partial_\Z+\psi(t)\,\wt g^{ab}\,\partial_a\otimes\partial_b} 
\bboxed{T^{\alpha\beta}\colon\bc
T^{\Z\Z}=\phi(t)
\ax
T^{a\Z}=0
\ax
T^{ab}=\psi(t)\,\wt g^{ab} 
\ec}
\ee
The two characteristic functions $\phi(t)$ and $\psi(t)$ must satisfy the conservation equation 
\be\label{e:claw1}
\boxed{\vp\;a\,\dot\phi+3\,\dot a\,\left(\phi+a^2\,\psi\right)=0\;}
\ee
which is equivalent to the four conservation equations $\nabla_\alpha T^{\alpha\beta}=0$ (Theorem \ref{t:divT=0-t}). An equivalent form of this equation is 
\be\label{e:claw2}
\boxed{\vp\;(\phi\,a^3)^{\dot{}}=-3\,a^4\,\dot a\,\psi\;}
\ee

\begin{center}
\fbox{\begin{minipage}{.95\linewidth} \vspu
Nothing can be changed in definition \eqref{e:iT}, in equation \eqref{e:claw1} and in equation \eqref{e:claw2} without violating our postulates. The only degrees of freedom we will have in the following will be the choice of the energy-momentum tensor  (Section \ref{s:ET}), which reduces to the choice of two characteristic functions (of time) and of an equation of state binding these functions (Section \ref{s:basic-bar}). \vspd
\end{minipage}}
\end{center}

\bt\label{t:ede}
Let $T^{\alpha\beta}$ be an isotropic energy-momentum tensor with characteristic functions $\phi(t)$ and $\psi(t)$. Then:

{\rm (i)} The ten Einstein equations are equivalent to the two differential equations 
\be\label{e:EFE0}
\bboxed{
\bc
\dfrac{\dot a^2}{c^2}=\tfrac 13\,a^2\,(\Lambda+\chi\,\phi)-\wt K
\ac
2\,\dfrac{\ddot a}{c^2}=a\left[\tfrac 23\,\Lambda-\chi\,(\psi\,a^2+\tfrac 13\phi)\right]
\ec
}
\ee
{\rm (ii)} Due to the conservation law \eqref{e:claw1} the second equation \eqref{e:EFE0} is a differential consequence of the first one.
\et

\bprf
(i) The contravariant components of the metric tensor and of the Einstein tensor are given in \eqref{e:Gt}.  Then the Einstein field equations
$$
\left\{\ba
G^{\Z\Z}=\chi\,T^{\Z\Z}-\Lambda\,g^{\Z\Z},
\ac
G^{ab}=\chi\,T^{ab}-\Lambda\,g^{ab}.
\ea\right.
$$
 are equivalent to
 
$\iff \left\{\ba
\dfrac 3{c^2\,a^2}\left(\dot a^2+c^2\,\wt K\right)=
\chi\,\phi+\Lambda
\ac
-\dfrac 1{c^2\,a^4}\left(2\,a\,\ddot a+\dot a^2+c^2\wt K\right)\,\wt g^{ab}=\chi\,\psi\,\wt g^{ab}-\Lambda \,a^{-2}\,\wt g^{ab}
\ea\right.
$

$\iff
\left\{\ba
 \dot a^2=\tfrac 13\,c^2\,a^2\,\Big(\chi\,\phi+\Lambda\Big)-c^2\,\wt K
\ac
\left(2\,a\,\ddot a+\dot a^2+c^2\wt K\right)\,\wt g^{ab}=c^2\,a^4\left(\Lambda\, a^{-2}-\chi\,\psi\right)\wt g^{ab} 
\ea\right.
$

$\iff
\left\{\ba
 \dot a^2=c^2\left[\tfrac 13\,a^2\,\left(\vpx\Lambda+\chi\,\phi\right)-\wt K\right]
\ac
2\,a\,\ddot a+\dot a^2+c^2\wt K=c^2\,a^2\,\Big(\Lambda-\chi\,\psi\,a^2\Big) 
\ea\right.
$

Substitute the first equation into the second one:

$
\iff
\left\{\ba
 \dot a^2=c^2\left[\tfrac 13\,a^2\,(\Lambda+\chi\,\phi)-\wt K\right]
\ac
2\,a\,\ddot a+c^2\left[\tfrac 13\,a^2\,(\Lambda+\chi\,\phi)-\wt K\right]+c^2\wt K=c^2\,a^2\,(\Lambda-\chi\,\psi\,a^2) 
\ea\right.
$

$\iff
\left\{\ba
 \dot a^2=c^2\left[\tfrac 13\,a^2\,(\Lambda+\chi\,\phi)-\wt K\right]
\ac
2\,a\,\ddot a+\tfrac 13\,c^2\,a^2\,(\Lambda+\chi\,\phi)=c^2\,a^2\,(\Lambda-\chi\,\psi\,a^2) 
\ea\right.
$

$\iff
\left\{\ba
 \dot a^2=c^2\left[\tfrac 13\,a^2\,(\Lambda+\chi\,\phi)-\wt K\right]
\ac
\ddot a=\tfrac 12\,c^2\,a\left[\tfrac 23\,\Lambda-\chi\,\Big(\psi\,a^2+\tfrac 13\phi\Big)\right].\quad \eprfx
\ea\right.
$

(ii) Differentiate the first equation \eqref{e:EFE0}:
\be\label{e:difffirst}
\dfrac{2\,\dot a\,\ddot a}{c^2}=\tfrac 23\,(\chi\,\phi+\Lambda)\,a\,\dot a+\tfrac 13\,\chi\,a^2\,\dot\phi.
\ee
Substitute into this equation the conservation law \eqref{e:claw1} in the form $a\,\dot\phi=-3\,\dot a\,\left(\phi+a^2\,\psi\right)$. Then

$
\eqref{e:difffirst} \;\To\;
\dfrac{2\,\dot a\,\ddot a}{c^2}=\tfrac 23\,(\chi\,\phi+\Lambda)\,a\,\dot a-\chi\,\left(\phi+a^2\,\psi\right)\,a\,\dot a
$

$
\To\;
\dfrac{2\,\ddot a}{c^2\,a}=\tfrac 23\,(\chi\,\phi+\Lambda)-\chi\,\left(\phi+a^2\,\psi\right)
$

$
\To\;
\dfrac{2\,\ddot a}{c^2\,a}=\tfrac 23\,\Lambda-\chi\,(\tfrac 13\,\phi+a^2\,\psi).
\;\iff\;\hbox{second equation \eqref{e:EFE0}}. \eprfx 
$

\section{The energy-momentum tensor of the galactic fluid}\label{s:ET}

\bt\label{t:ET}
Let $\mathcal V(t)$ be the volume of any arbitrary co-moving portion of the galactic fluid, $\epsilon(t)$ and $p(t)$ the {\bf  energy density} and the {\bf pressure} in that portion. Then the energy conservation law
\be\label{e:e-claw-1}
\bboxed{\dfrac{d\,}{dt}\left(\epsilon\,\mathcal V\right)=-p\,\dfrac{d\mathcal V}{dt}}
\ee
holds if and only if the energy-momentum tensor of the galactic fluid is that of a {\bf perfect fluid}\footnote{
\kern 2pt See \cite{Weinberg} p.\ 127, \cite{MTW} p.\ 132, \cite{Lichnerowicz-1955} p.\ 14, \cite{Lichnerowicz-1967} p.\ 23. Due to the different conventions, there are changes of sign.} 
\be\label{e:ET}
\boxed{\vp\;T^{\alpha\beta}=\left(e+p\right)\,U^\alpha\,U^\beta+p\,g^{\alpha\beta}\;}
\ee
where $U^\alpha$ is the unitary four-velocity of the galactic fluid  
$$
U^\alpha\Def c^{-1}\,\dfrac{d\gamma^\alpha}{dt}:\, 
\bc
U^\Z=1,
\ax
U^a=0,
\ec
\quad g_{\alpha\beta}\,U^\alpha\,U^\beta=-1.
$$ 
\et
Note that
\be\label{e:ETcomp}
\bc
T^{\Z\Z}=\epsilon+p-p=\epsilon,
\ax
T^{a\Z}=0,
\ax
T^{ab}=p\,g^{ab}.
\ec
\ee

\blm
The conservation law \eqref{e:e-claw-1} is equivalent to the equation
\be\label{e:claw-1}
\bboxed{a\,\dot \epsilon+3\,(\epsilon+p)\,\dot a=0}
\ee
\elm

\bprf
Due to \eqref{e:Va3}, $\dfrac{\mathcal V(t)}{a^3(t)}={\rm const.}=\dfrac{\mathcal V(t_*)}{a^3(t_*)}.$
$$
\mathcal V=\dfrac{\mathcal V(t_*)}{a^3(t_*)}\,a^3(t), \quad
\dot{\mathcal V}=3\,\dfrac{\mathcal V(t_*)}{a^3(t_*)}\,a^2(t)\,\dot a(t)=3\,\dfrac{\dot a}a\,\mathcal V.
$$
$$
\eqref{e:claw-1} \iff \dot \epsilon\,\mathcal V+\epsilon\,\dot{\mathcal V}=-p\,\dot{\mathcal V} \iff \dot \epsilon\,\mathcal V+(\epsilon+p)\,\dot{\mathcal V}=0
$$
$$
\iff \dot \epsilon+3\,(\epsilon+p)\,h=0. \eprfx
$$
{\bf Poof of Theorem \ref{t:ET}}. 
Equation \eqref{e:claw-1} is in perfect agreement with equation \eqref{e:claw1} $a\,\dot\phi+3\,\dot a\,\left(\phi+a^2\,\psi\right)=0$ with 
\be\label{e:subst}
\bboxed{\phi=\epsilon(t), \quad \psi=a^{-2}\,p}
\ee
Then from \eqref{e:iT} we get \eqref{e:ETcomp}. \eprf

\br
Since $U^\beta\,U_\beta=-1$, the four-velocity $U^\alpha$ is an eigenvector of $T^{\alpha\beta}$ with eigenvalue $-\epsilon$:
$$
T^{\alpha\beta}\,U_\beta=-\epsilon\,U^\alpha.
$$
Moreover, any vector $X^\alpha$ orthogonal to $U^\alpha$ is an eigenvector of $T^{\alpha\beta}$ with eigenvalue $p$:
$$
T^{\alpha\beta}\,X_\beta=p\,X^\alpha. \quad\bullet
$$
\erx

\br\label{r:dyn}
With the substitution \eqref{e:subst} $\phi=\epsilon(t), \quad \psi=a^{-2}\,p$ the dynamical equations \eqref{e:EFE0} read respectively
\begin{gather}
\bboxed{\dfrac{\dot a^2}{c^2}=\tfrac 13\,a^2\,(\Lambda+\chi\,\epsilon)-\wt K}
\label{e:EFE-*}
\\
\bboxed{\dfrac{\ddot a}{c^2}=\tfrac 12\,a\left[\vpx\tfrac 23\,\Lambda-\chi\,(p+\tfrac 13\,\epsilon)\right]}
\label{e:EFE-**}
\end{gather}
Equation \eqref{e:EFE-**} is a differential consequence of  \eqref{e:EFE-*} and of the conservation law \eqref{e:claw-1} (Theorem \ref{t:ede}, item (iii)).\footnote{
\kern 2pt Equations \eqref{e:EFE-*} and \eqref{e:EFE-**} are called {\bf Friedman equations}, see Section \ref{s:Fr}.} 
 \er

\br\label{r:forces}
The {\bf acceleration equation} \eqref{e:EFE-**} highlights a salient feature of the relativistic cosmic dynamics: it can be interpreted as a Newtonian dynamical equation of a point subjected to three forces competing with each other, namely
$$
\bc
F_\Lambda(a)\Def\tfrac 13\,c^2\,\Lambda\,a,
\ax
F_p(a)\Def -\tfrac 12\,c^2\,\chi\,p\,a,
\ax
F_\epsilon(a)\Def -\tfrac 16\,c^2\,\chi\,\epsilon\,a.
\ec 
$$
Since $\Lambda$ and $\chi$ are positive constants, $F_\Lambda$ acts as a centrifugal force (with center $a=0$), as well as $F_p$ and $F_\epsilon$ if $p(t)$ and $\epsilon(t)$ are negative. On the contrary, $F_p$ and $F_e$ are attractive towards $a=0$ when $p(t)$ and $\epsilon(t)$ are positive.
\er

\section{Comments on the Friedman equations}\label{s:Fr}

The equations of Friedman are definitely the most cited equations  in the texts of cosmology, where they appear written in various different forms. Actually, by {\bf Friedman equations} one should understand the dynamical equations  appearing 
in the original work {\it \"Uber die Kr\"ummung des Raumes} by A.\ Friedman (1922) which are written exactly as follows: 
\be\label{e:Fr-1}
\bboxed{\bc
(4)\quad \dfrac{{R'}^2}{R^2}+\dfrac{2\,R\,R''}{R^2}+\dfrac{c^2}{R^2}-\lambda=0
\ac
(5) \quad \dfrac{3\,{R'}^2}{R^2}+\dfrac{3\,c^2}{R^2}-\lambda=\varkappa\,c^2\varrho
\ec}
\quad 
\bc
R'=\dfrac{dR}{dx_4},
\ac
R''=\dfrac{d^2R}{dx_4^2}.
\ec
\ee
where $\varrho$ is declared to be the density of mass and $\varkappa$  {\it eine Konstante}.  The coordinate $x_4$ is time-dimensional and the signature of the metric is $(---+)$. These equations come from the Einstein field equations\footnote{
\kern 2pt The comparison with our Einstein equations \eqref{e:Ein-0} 
$R^{\alpha\beta}+\left(\Lambda-\tfrac 12\, R\right)\,g^{\alpha\beta}=\chi\,T^{\alpha\beta}$ shows a difference of sign in the right side. This is due to the different signature of the metric. 
} 
$$
{\rm(A)} \quad R_{ik}-\tfrac 12\,g_{ik}\,\bar R+\lambda\,g_{ik}=-\varkappa\,T_{ik}, \quad 
\bc
\bar R=g^{ik}\,R_{ik}
\ax
T_{ik}=0, \quad i,k\neq 4
\ax
T_{44}=c^2\,\varrho\,g_{44}
\ec
$$
for $i=k=1,2,3\;$ and for $i=k=4$, respectively. Looking at the energy tensor components we observe that (i) {\it Friedman takes into account the cosmological constant}, (ii) the kinetic pressure $p$ is not present, so {\it Friedman deals with a dust galactic fluid}. Furthermore, as proved below, (iii) {\it Friedman deals with a positive spatial curvature}.

In our theory we have seen that the Einstein equations determined by the energy-momentum tensor \eqref{e:ET} reduce to the differential equations \eqref{e:EFE-*} and \eqref{e:EFE-**}, namely
\be\label{e:Fr-2}
\bc
\dfrac{\dot a^2}{c^2}=\tfrac 13\,a^2\,(\Lambda+\chi\,e)-\wt K,
\ac
\dfrac{\ddot a}{c^2}=\tfrac 12\,a\left[\vpx\tfrac 23\,\Lambda-\chi\,(p+\tfrac 13\,e)\right].
\ec
\ee
Let us compare these equations with the Friedman equations \eqref{e:Fr-1}. To do this we rewrite them as 
\be\label{e:Fr-3}
\bc
[4]\quad 2\,R\,R''+{R'}^2+c^2-\lambda\,R^2=0,
\ax
[5] \quad {R'}^2+c^2-\tfrac 13\,\left(\lambda+\varkappa\,c^2\varrho\right)R^2=0,
\ec
\ee
Subtract side by side $[4]-[5]$:
$$
2\,R\,R''-\lambda\,R^2+\tfrac 13\,\left(\lambda+\varkappa\,c^2\varrho\right)R^2=0.
$$
Since $R\neq 0$, we get $2\,R''-\tfrac 23\,\lambda\,R+\tfrac 13\,\varkappa\,c^2\varrho\,R=0$, i.e.
\be\label{e:Fr-4}
R''=\tfrac 16\,R\,\left(2\,\lambda-\varkappa\,c^2\varrho\right).
\ee
If in \eqref{e:Fr-4} we put
\be\label{e:Fr-5}
\bc
dx_4=dt
\ax
R=a
\ec
\quad
\bc
R'=\dot a
\ax
R''=\ddot a
\ec
\ee
then  we get the equation $\ddot a=\tfrac 16\,a\,\left(2\,\lambda-\varkappa\,c^2\varrho\right)$ which coincides with the second equation \eqref{e:Fr-2} with $p=0$,
$$
\dfrac{\ddot a}{c^2}=\tfrac 16\,a\left(2\,\Lambda-\chi\,\epsilon\right)
$$
provided that $2\,\lambda-\varkappa\,c^2\varrho=c^2\left(2\,\Lambda-\chi\,\epsilon\right)$ i.e.
\be\label{e:Fr-6}
\lambda=c^2\,\Lambda,\quad
\varkappa\,\varrho=\chi\,\epsilon.
\ee
In turn, due to the substitutions \eqref{e:Fr-5}, the second equation [5] in \eqref{e:Fr-3} reads
$$
\dot a^2=\tfrac 13\,\left(\lambda+\varkappa\,c^2\varrho\right)a^2-c^2.
$$
Due to \eqref{e:Fr-6},
$$
\dot a^2=\tfrac 13\,c^2\,a^2\left(\Lambda+\chi\,\epsilon\right)-c^2.
$$
This equation coincides with our first  equation \eqref{e:Fr-2}  
$$
\dfrac{\dot a^2}{c^2}=\tfrac 13\,a^2\,(\Lambda+\chi\,\epsilon)-\wt K
$$
provided that $\wt K=1$. This proves item (iii) above.


\chapter{Barotropic dynamics of a single-component universe}\label{c:bar}

\section{Preamble}\label{s:basic-bar}

The dynamics of the scale factor $a(t)$ is so far governed by the two first-order differential equations \eqref{e:EFE-*} and \eqref{e:claw-1}. These two equations involve three unknown functions $a(t)$, $\epsilon(t)$ and $p(t)$.  We need a further equation. This equation should express the physical characteristics of the galactic fluid and should be dictated by physical arguments. The galactic fluid may have different components (typically mass, radiation, etc.). Assuming that the energy densities $\epsilon_i$ of these components are additive, we can write the total energy density $\epsilon$ as the sum  
$$
\epsilon=\textstyle \sum_i \epsilon_i.
$$
Beside the densities we have consider the pressures $p_i$ of each component.  Then we must assume that all of these variables are bound each other by means of certain {\bf equations of state}. In the case in which there is no interaction between the components, these equations are separated between them, that is to say of the type
$$
p_i=f_i(\epsilon_i).
$$
In the simplest case, we can consider linear equations
$$
p_i=w_i\,\epsilon_i
$$
where $w_i$ are dimensionless constants called {\bf barotropic parameters}.
In this chapter we confine our analysis to a single-component universe with the equation of state
\be\label{e:linbar}
\bboxed{p=w\,\epsilon}
\ee
 As a consequence, the {\bf barotropic dynamics} will involve two functions  in the cosmic time $t$, $a(t)>0$ and $\epsilon(t)$, together with a constant parameter $w$. This dynamics is  governed by the differential equations
\begin{gather} 
\label{e:fluid-e}
\bboxed{a\,\dot \epsilon+3\,(w+1)\,\epsilon\,\dot a=0}
\\
\label{e:vel-e}
\bboxed{\dfrac{\dot a^2}{c^2}=\tfrac 13\,a^2\,(\Lambda+\chi\,\epsilon)-\wt K}
\\
\label{e:acc-e}
\bboxed{\dfrac{\ddot a}{c^2}=\tfrac 12\,a\left[\vpx\tfrac 23\,\Lambda-\chi\,(w+\tfrac 13)\,\epsilon\right]}
\end{gather}
which are respectively called {\bf fluid equation}, {\bf velocity equation} and {\bf acceleration equation}. Equations  \eqref{e:fluid-e} and \eqref{e:acc-e} come from equations \eqref{e:claw-1} and \eqref{e:EFE-**} with the substitution $p=w\,\epsilon$. Equation \eqref{e:vel-e} is nothing but equation \eqref{e:EFE-*}. The acceleration equation \eqref{e:acc-e} is a differential consequence of the fluid and the velocity equations (Remark \ref{r:dyn}). The velocity equation \eqref{e:vel-e} can be written in the form
\be\label{e:vel-e*}
\bboxed{h^2=\tfrac 13\,c^2\,(\Lambda+\chi\,\epsilon)-\dfrac{c^2\,\wt K}{a^2}}
\ee

Our purpose is to find and classify all possible solutions $a(t)$ of these dynamical equations. We call them {\bf profiles of the universe}. Such a classification should depend on the value of the parameter $w$ and on the value  of the spatial curvature $\wt K$, specially on its sign. Moreover, two profiles differing by a translation along the $t$-axis have to be considered equivalent.

\section{Basic theorems}

In the analysis of a barotropic dynamics it turns out to be convenient to replace the parameter $w$ with the new parameter\footnote{
\kern 2pt 
Table of conversion: 
\begin{center}
\begin{tabular}{|c|c|c|c|c|c|c|c|c|}
  \hline
 $\vpt u=w+1 $ 
& $-\tfrac 13$
& $0$
& $\tfrac 13$ 
& $\tfrac 23$
& $1$
& $\tfrac 43$
& $\tfrac 53$
\\
\hline
 $\vpt w$
& $-\tfrac 43$
& $-1$ 
& $-\tfrac 23$ 
& $-\tfrac 13$
& $0$
& $\tfrac 13$
& $\tfrac 23$
\\
  \hline 
\end{tabular}
\end{center}
The parameter $u$ is used by other authors, and it is denoted also by $\gamma$ or $\Gamma$.}  
\be\label{e:u-def}
\bboxed{u\Def w+1}
\ee
and introduce the new constants
\be\label{e:lambda,mu}
\bboxed{\ba
\lambda\Def \tfrac 13\,c^2\Lambda>0
\ax
\mu_\sharp\Def \tfrac 13\, \chi\,c^2\,\epsilon_\sharp>0
\ea}\;\;
\bboxed{\Dim(\lambda)=\Dim(\mu_\sharp)=T^{-2}}
\ee
$\epsilon_\sharp$ being the value of $\epsilon(t)$ at the normalization time $t_\sharp$. 


\bt\label{t:bWe} 
The evolution $a(t,t_\sharp)$ of the scale factor is governed by the single  first-order differential equation
\be\label{e:bWe}
\bboxed{\dot a^2=\lambda\;a^2+\mu_\sharp\;a^{2-3u}-c^2\,\wt K}
\ee
which can also be written as
\be\label{e:X1}
\bboxed{h^2=\dfrac{\dot a^2}{a^2}=\lambda+\dfrac{\mu_\sharp}{a^{3u}}-c^2\,\dfrac{\wt K}{a^2}}
\ee
\et
\bprf
$\eqref{e:fluid-e} \iff \dfrac{\dot\epsilon}\epsilon+3u\,\dfrac{\dot a}a=0 
\;\iff \; d\log \epsilon+3u\,d\log a=0$ $\iff$
\be\label{e:etat}
\bboxed{\epsilon(t)\,a^{3u}(t,t_\sharp)=\hbox{\rm constant in $t$}}
\ee
As $a(t_\sharp,t_\sharp)=1$, equation \eqref{e:etat} is equivalent to
\be\label{e:etat-t0}
\bboxed{\epsilon(t)\,a^{3u}(t,t_\sharp)=\epsilon_\sharp}
\ee
Substituting the expression $\epsilon(t)=\epsilon_\sharp\,a^{-3u}(t,t_\sharp)$ coming from this last equation into equation \eqref{e:vel-e*} we get \eqref{e:bWe} and \eqref{e:lambda,mu}. \eprf

\br\label{r:inv-1}
Remind that $\wt K=K(t_\sharp)$ and $\epsilon_\sharp=\epsilon(t_\sharp)$. The dynamical equation \eqref{e:X1} must be invariant under any change $t_\sharp\mapsto t_\flat$ of the normalization time of the scale parameter $a(t,t_\sharp)$, as explained in Remark \ref{r:test}, page \pageref{r:test}. We know that the definition of the Hubble parameter $h$ is invariant. Since also the constant $\lambda$ is invariant, the invariant condition of \eqref{e:X1} reduces to equation 
\be\label{e:inv-1}
\tfrac 13\,\chi\, \dfrac{\epsilon(t_\sharp)}{a^{3u}(t,t_\sharp)}-\dfrac{K(t_\sharp)}{a^2(t,t_\sharp)}
=\tfrac 13\,\chi\, \dfrac{\epsilon(t_\flat)}{a^{3u}(t,t_\flat)}-\dfrac{K(t_\flat)}{a^2(t,t_\flat)},
\ee
to be satisfied for all $t,t_\sharp,t_\flat$. For $t=t_\flat$ we get 
$$
\tfrac 13\,\chi\,\dfrac{\epsilon(t_\sharp)}{a^{3u}(t_\flat,t_\sharp)}-\dfrac{K(t_\sharp)}{a^2(t_\flat,t_\sharp)}=\tfrac 13\,\chi\,\epsilon(t_\flat)-K(t_\flat).
$$
Due to \eqref{e:Kt*Kt0}, $K(t_\flat)=\dfrac{K(t_\sharp)}{a^2(t_\flat,t_\sharp)}$, so that this last equation reduces to 
$$
\epsilon(t_\sharp)=\epsilon(t_\flat)\,a^{3u}(t_\flat,t_\sharp).
$$
This equation holds for all values of $(t_\flat,t_\sharp)$,  and by putting $t_\flat=t$ we get equation \eqref{e:etat-t0} that, as we have seen above,  is a consequence of the fluid equation \eqref{e:fluid-e}. This proves that {\it the dynamical equation \eqref{e:X1} satisfies the required invariance condition}. \er

\br
By virtue of  Theorem \eqref{t:muu} (page \pageref{t:muu}) the following  equations are equivalent:
\begin{gather}
\label{e:inv-3}
\epsilon(t)\,\mathcal V^u(U,t)={\rm const.} \;\; \forall\; \hbox{$U=$ co-moving domain)},
\ac
\label{e:inv-2bis}
\epsilon(t)\,a^{3u}(t,t_\sharp)=\hbox{constant in $t$} 
\ac
\label{e:inv-4}
a\,\dot\epsilon+3\,u\,\epsilon\,\dot a=0\quad (\hbox{fluid equation}),
\\
\label{e:inv-5}
h=-\tfrac 1{3u}\,\dfrac{\dot\epsilon}\epsilon. 
\end{gather}
This shows that the energy density $\epsilon(t)$ is a conserved density of order equal to the barotropic parameter $u$. \er

\bt \label{t:negative}
If the spatial curvature is negative then any physical length has a permanent superluminal expansion or contraction.
\et

Note that this theorem holds whatever $u$.

\bprf
Let us  go back to Section \ref{s:Hl}, equation \eqref{e:elldot}, $\dot\ell(t)=\dot a(t,t_\sharp)\,\wtell$, where $\wtell$ is the length of a curve (not necessarily a geodesic) on the quotient manifold and $\ell(t)$ is the corresponding length on the spatial section $S_t$. Due to equation \eqref{e:bWe} we have
\be\label{e:dot-ell}
\dot\ell^2(t)=\dot a^2(t,t_\sharp)\,\wtell^2=\left[\vp\lambda\;a^2+\mu_\sharp\;a^{-(3w+1)}-c^2\,\wt K\right]\wtell^2.
\ee
If $\wt K<0$  then $\dot\ell^2=\left[\vp\lambda\;a^2+\mu_\sharp\;a^{-(3\,w+1)}+c^2\,|\wt K|\right]\,\wtell^2$. This shows that $\dot\ell^2>c^2$. \eprf

\begin{center}
\fbox{\begin{minipage}{.95\linewidth} 
\br\label{r:negative}\vspu
The superluminal condition $\ell >c^2$ of any physical length during all the whole duration of the universe has no physical sense. Thus, our theory leads to consider \underbar{inadmissible} the barotropic models with \underbar{ne}g\underbar{ative} spatial curvature. Later on (Section \ref{s:noK>0}) it will be shown that also the models with p\underbar{ositive} spatial curvature are inadmissible. \vspd 
\er
\end{minipage}}
\end{center}
 
\section{The profiles of the barotropic flat models}

\begin{center}
\fbox{\begin{minipage}{.95\linewidth} \vspu
In the vastness of the cosmological mathematical models, the flat barotropic models have the rare property that the evolution in time of the scale factor admits an analytical expression in terms of elementary functions (exponentials, or hyperbolic functions), whatever the value of the parameter $u$. \vspd
\end{minipage}}
\end{center}
   
We will denote by $a(u;t,t_\sharp)$ the scale factor of a barotropic universe with barotropic parameter $u=w+1$ and reference time $t_\sharp$. 

\bt\label{t:aut-0}
The profiles $a(u;t,t_\sharp)$ of a barotropic flat model admit the following two equivalent representations
\begin{gather} 
\label{e:aut-e} 
\bboxed{a(u;t,t_\sharp)=\left[\tfrac 14\,\dfrac\chi\Lambda\,\epsilon_\sharp\,\dfrac{(e^{u\beta t}-1)^2}{e^{u\beta t}}\right]\!{\vp}^{\tfrac 1{3u}}
\quad 
\bc 
\!\hbox{\rm exponential}
\ax
\!\hbox{\rm form}
\ec
} 
\\
\label{e:aut-c} 
\bboxed{a(u;t,t_\sharp)=\left[\tfrac 12\,\dfrac\chi\Lambda\,\epsilon_\sharp\,\big(\cosh(u\beta t)-1\big)\right]\!{\vp}^{\tfrac 1{3u}}
\quad 
\bc
\!\hbox{\rm hyperbolic}
\ax
\!\hbox{\rm form}
\ec
}
\end{gather}
where
\be\label{e:def-beta}
\bboxed{\vp\beta\Def\sqrt{3\,\Lambda}\,c}\;\;
\bboxed{\vphantom{\Def}\Dim(\beta)=T^{-1}}
\ee
and $\epsilon_\sharp$ is the value of the energy density $\epsilon(t)$ at the refrence time $t_\sharp$.
\et

\bprf 
With $K_\sharp=0$ equation \eqref{e:X1} reads
\be\label{e:X1.}
\dfrac{\dot a^2}{a^2}=\lambda+\dfrac{\mu_\sharp}{a^{3u}}
\ee
and is equivalent to
\be\label{e:We-1.}
\dfrac{da}{a\,\sqrt{\left(1+b\,a^{-3u}\right)}}=\sqrt\lambda\,dt, \quad b\Def\dfrac{\mu_\sharp}\lambda.
\ee
The left-hand side is integrable in terms of elementary functions:
$$
\dint\dfrac{da}{a\,\sqrt{\left(1+b\;a^{-3u}\right)}} 
=\tfrac 1{3u}\,\log\dfrac{\sqrt{a^{3u}+b}+\sqrt{a^{3u}}}{\sqrt{a^{3u}+b}-\sqrt{a^{3u}}}+\rm constant.
$$
Thus from \eqref{e:We-1.} we get
$$
\log\dfrac{\sqrt{a^{3u}+b}+\sqrt{a^{3u}}}{\sqrt{a^{3u}+b}-\sqrt{a^{3u}}}=3u\,\sqrt\lambda\,(t-t_*)
$$
with an arbitrary $t_*$. However, there is no loss of generality in assuming $t_*=0$. In this case $a(0)=0$.\footnote{
\kern 2pt The physical meaning of a scale factor is invariant under  translations along the $t$-axis.} 
 By setting $\beta=3\,\sqrt\lambda=\sqrt{3\,\Lambda}\,c$ we can write 
$$
\dfrac{\sqrt{a^{3u}+b}+\sqrt{a^{3u}}}{\sqrt{a^{3u}+b}-\sqrt{a^{3u}}}=e^{u\,\beta\,t}.
$$
In order to solve this equation with respect to $a^{3u}$ we put
\be\label{e:def-Xt}
\bboxed{X\Def e^{u\beta t}}
\ee
and
$$
A\Def\sqrt{a^{3u}+b}, \quad B\Def\sqrt{a^{3u}}.
$$
Note that $A^2-B^2=b$. Then we have the following sequence of implications:

$\dfrac{A+B}{A-B}=X$ $\To$ $A+B=X\,(A-B)$ $\To$ $b=X\,(A-B)^2$ $[\dag]$

$\To$ $b=X\,(A^2+B^2-2\,A\,B)$ $\To\,b=X\,\left(2\,A^2+b-2\,A\,B\right)$

$\To$ $\tfrac 12\,\dfrac{X-1}X\,b=B\,(A-B)$. Due to $[\dag]$, $\;A-B=\sqrt{\dfrac bX}$. Then
 
$\To$ $\tfrac 12\,\dfrac{X-1}X\,b=B\,\sqrt{\dfrac bX}$ 
$\To$ $\tfrac 12\,\dfrac{X-1}{\sqrt X}\,\sqrt b=B$
$\To$ $B^2=\tfrac 14\,\dfrac{(X-1)^2}{X}\,b$

As $B^2=a^{3u}$, $X=e^{u\,\beta\,t}$ and $b=\dfrac{\mu_\sharp}\lambda$, we finally get
$$
a^{3u}=\tfrac 14\,\dfrac{\mu_\sharp}\lambda\dfrac{(e^{u\beta t}-1)^2}{e^{u\beta t}}.
$$
Due to the definitions \eqref{e:lambda,mu} we have
\be\label{e:mu/lambda} 
\dfrac{\mu_\sharp}\lambda=\dfrac\chi\Lambda\,\epsilon_\sharp,
\ee
$\epsilon_\sharp$ being the value of $\epsilon(t)$ at the normalization time $t_\sharp$, and \eqref{e:aut-e} is proved. The profile \eqref{e:aut-c} follows from \eqref{e:aut-e} by observing that
$$
2\,(\cosh(z)-1)=e^z+e^{-z}-2
=e^{-z}\left[e^{2z}+1-2\,e^z\right]
=\dfrac{(e^z-1)^2}{e^z}. \eprfx
$$

\br
The profiles \eqref{e:aut-e} and \eqref{e:aut-e} show the convenience of introducing the {\bf dimensionless time} 
\be\label{e:def-x}
\bboxed{x\Def \beta\, t}
\ee
so that they assume the form
\begin{gather} 
\label{e:aux-e} 
\bboxed{a(u;x,x_\sharp)=\left[\tfrac 14\,\dfrac\chi\Lambda\,\epsilon_\sharp\,\dfrac{(e^{ux}-1)^2}{e^{ux}}\right]\!{\vp}^{\tfrac 1{3u}}
\quad 
\bc 
\!\hbox{\rm exponential}
\ax
\!\hbox{\rm form}
\ec} 
\\
\label{e:aux-c}
\bboxed{a(u;x,x_\sharp)=\left[\tfrac 12\,\dfrac\chi\Lambda\,\epsilon_\sharp\,\big(\cosh(ux)-1\big)\right]\!{\vp}^{\tfrac 1{3u}}
\quad 
\bc 
\!\hbox{\rm hyperbolic}
\ax
\!\hbox{\rm form}
\ec} 
\end{gather}
These profiles will be plotted later on (Section \ref{s:aux0}) since we need more information about the magnitude of the constants that are involved. \er

\section{The profiles of the Hubble parameter}

\bt
The profiles $h(u;x)$ of the Hubble parameter of a barotropic flat model admit the following two equivalent representations
\begin{gather}
\label{e:hux}
\bboxed{h(u;x)\Def\dfrac 1a\,\dfrac{da}{dx}=\tfrac 13\,\dfrac{e^{ux}+1}{e^{ux}-1}}
\\
\label{e:hut}
\bboxed{h(u;t)\Def\dfrac 1a\,\dfrac{da}{dt}=\tfrac 13\,\beta\,\dfrac{e^{u\beta t}+1}{e^{u\beta t}-1}}
\end{gather}
\et

\bprf
By setting, as above, $X\Def e^{ux}$ and observing that $X'=u\,x$ from 
\eqref{e:aux-e} we get 
$$
3\,u\,\dfrac{d\log a}{dX}=2\,\dfrac{d\log(X-1)}{dX}-\dfrac{d\log X}{dX}
=\dfrac 2{X-1}-\dfrac 1X=\dfrac {X+1}{X\,(X-1)} \To
$$
\be\label{e:dadX}
\bboxed{\dfrac{da}{dX}=\dfrac a{3u}\,\dfrac {X+1}{X\,(X-1)}}
\ee
$\To$ $\dfrac{da}{dx}=\dfrac{da}{dX}\,u\,X=\dfrac a{3u}\,\dfrac {X+1}{X\,(X-1)}\,u\,X$ $\To$
\be\label{e:dadx}
\quad\bboxed{\dfrac{da}{dx}=\tfrac 13\,a\,\dfrac {X+1}{X-1}=\tfrac 13\,a\,\dfrac {e^{ux}+1}{e^{ux}-1}}
\ee
$\To$ \eqref{e:hux}. As $da/dt=\beta\,da/dx$ we get \eqref{e:hut}. \eprf
\begin{figure} [H]
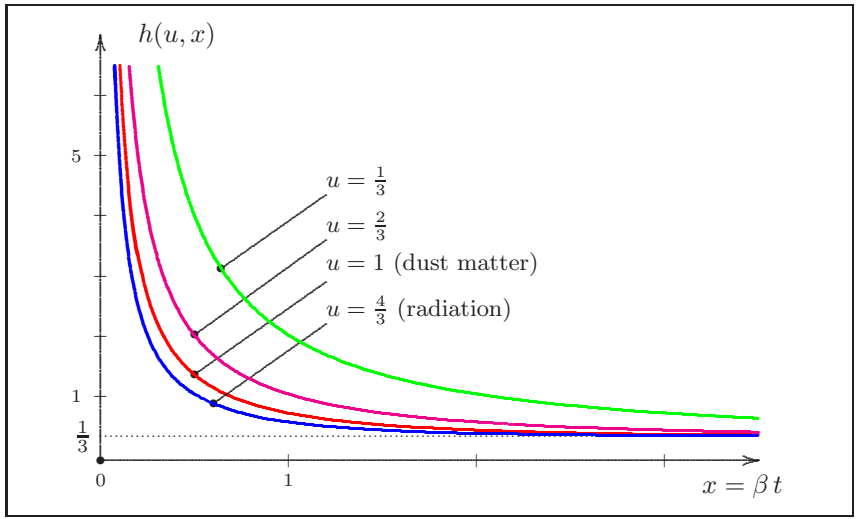

\begin{center}
$
\beginpicture
\setcoordinatesystem units <2.5cm,.8cm>
\setplotarea x from 0 to 3.5, y from 0 to 6.5
\normalgraphs
\putrectangle corners at -.5 7.5 and 4 -1
\arrow <6pt> [.2,.6] from 0 -.07 to 3.5 -.07
\arrow <6pt> [.2,.6] from 0 0 to 0 7
\put{$\sc\bullet$} at 0 -.07
\put{$x=\beta\,t$} [lt] at 3.2 -.3
\put{$h(u,x)$} [l] at .2 7
\setlinear
\setdots <2pt>
\plot 0 .333 3.5 .333 /
\setsolid
\put{$\tfrac 13$} [r] at -.05 .333
\put{$\sc 0$} [t] at 0 -.3
\put{$\sc -$} at 0 1
\put{$\sc -$} at 0 2
\put{$\sc -$} at 0 3
\put{$\sc -$} at 0 4
\put{$\sc -$} at 0 5
\put{$\sc -$} at 0 6
\put{$\sc 1$} [r] at -.1 1
\put{$\sc 5$} [r] at -.1 5
\put{$\scc |$} at 1 -.07
\put{$\scc |$} at 2 -.07
\put{$\scc |$} at 3 -.07
\put{$\sc 1$} [t] at 1 -.3
\inboundscheckon
\put{$\sc\bullet$} at .5 1.361
\setlinear \plot .5 1.361 1.2 2.9 /
\put{\small $u=1$ (dust matter)} [lb] at 1.2 3
\setquadratic
\setplotsymbol(\normalered)
\plot
0.10	6.6722
0.15	4.4528
0.20	3.3444
0.25	2.6805
0.30	2.2389
0.35	1.9242
0.40	1.6888
0.45	1.5064
0.50	1.3610
0.55	1.2425
0.60	1.1442
0.65	1.0615
0.70	0.9910
0.75	0.9302
0.80	0.8773
0.85	0.8310
0.90	0.7901
0.95	0.7538
1.0	0.7213
1.2	0.6207
1.4	0.5515
1.6	0.5020
1.8	0.4654
2.0	0.4377
2.2	0.4164
2.4	0.3998
2.6	0.3868
2.8	0.3765
3.0	0.3683
3.2	0.3617
3.4	0.3564
3.6	0.3521
3.8	0.3486
4.0	0.3458
4.2	0.3435
4.4	0.3416
4.6	0.3401
4.8	0.3389
5.0	0.3379
5.2	0.3370
5.4	0.3364
5.6	0.3358
5.8	0.3354
/
\normale
\put{$\sc\bullet$} at .6 .877
\setlinear \plot .6 .877 1.2 2.2 /
\put{\small $u=\tfrac 43$ (radiation)} [lb] at 1.2 2.2
\setplotsymbol(\normaleblue)
\setquadratic
\plot 
0.04	12.503
0.06	8.338
0.08	6.256
0.10	5.007
0.12	4.176
0.14	3.582
0.16	3.137
0.18	2.791
0.20	2.515
0.22	2.289
0.24	2.101
0.26	1.942
0.28	1.806
0.30	1.689
0.32	1.586
0.34	1.496
0.36	1.415
0.38	1.344
0.40	1.279
0.42	1.221
0.44	1.169
0.46	1.121
0.48	1.077
0.50	1.037
0.6	0.877
0.8	0.683
1.0	0.572
1.2	0.502
1.4	0.455
1.6	0.423
1.8	0.400
2.0	0.383
2.2	0.371
2.4	0.362
2.6	0.355
2.8	0.350
3.0	0.346
3.2	0.343
3.4	0.341
3.6	0.339
3.8	0.338
4.0	0.337
4.2	0.336
4.4	0.335
4.6	0.335
4.8	0.334
5.0	0.334
/
\normale
\put{$\sc\bullet$} at .5 2.018
\setlinear \plot .5 2.018 1.2 3.6 /
\put{\small $u=\tfrac 23$} [lb] at 1.2 3.6
\setquadratic
\setplotsymbol(\normalemag)
\plot 
0.10	10.004
0.12	8.338
0.14	7.148
0.16	6.256
0.18	5.562
0.20	5.007
0.22	4.554
0.24	4.176
0.26	3.856
0.28	3.582
0.30	3.344
0.32	3.137
0.34	2.954
0.36	2.791
0.38	2.646
0.40	2.515
0.42	2.396
0.44	2.289
0.46	2.191
0.48	2.101
0.50	2.018
0.52	1.942
0.54	1.872
0.56	1.806
0.58	1.746
0.60	1.689
0.62	1.636
0.64	1.586
0.66	1.540
0.68	1.496
0.70	1.454
0.72	1.415
0.74	1.379
0.76	1.344
0.78	1.311
0.80	1.279
0.82	1.250
0.84	1.221
0.86	1.194
0.88	1.169
0.90	1.144
0.92	1.121
0.94	1.098
0.96	1.077
0.98	1.056
1.00	1.037
1.02	1.018
1.04	1.000
1.06	0.982
1.08	0.966
1.10	0.949
1.12	0.934
1.14	0.919
1.16	0.905
1.18	0.891
1.20	0.877
1.22	0.864
1.24	0.852
1.26	0.840
1.28	0.828
1.30	0.817
1.32	0.806
1.34	0.795
1.36	0.785
1.38	0.775
1.40	0.765
1.42	0.756
1.44	0.747
1.46	0.738
1.48	0.730
1.50	0.721
1.52	0.713
1.54	0.705
1.56	0.698
1.58	0.690
1.60	0.683
1.62	0.676
1.64	0.669
1.66	0.663
1.68	0.656
1.70	0.650
1.72	0.644
1.74	0.638
1.76	0.632
1.78	0.626
1.80	0.621
1.82	0.615
1.84	0.610
1.86	0.605
1.88	0.600
1.90	0.595
1.92	0.590
1.94	0.585
1.96	0.581
1.98	0.576
2.0	0.572
2.2	0.533
2.4	0.502
2.6	0.476
2.8	0.455
3.0	0.438
3.2	0.423
3.4	0.410
3.6	0.400
3.8	0.391
4.0	0.383
4.2	0.376
4.4	0.371
4.6	0.366
4.8	0.362
5.0	0.358
/
\normale

\put{$\sc\bullet$} at .64 3.12
\setlinear \plot .64 3.12 1.2 4.35 /
\put{\small $u=\tfrac 13$} [lb] at 1.2 4.35
\setplotsymbol(\normalegreen)
\plot 
0.22	9.095
0.24	8.338
0.26	7.697
0.28	7.148
0.30	6.672
0.32	6.256
0.34	5.889
0.36	5.562
0.38	5.270
0.40	5.007
0.42	4.770
0.44	4.554
0.46	4.356
0.48	4.176
0.50	4.009
0.52	3.856
0.54	3.714
0.56	3.582
0.58	3.459
0.60	3.344
0.62	3.237
0.64	3.137
0.66	3.043
0.68	2.954
0.70	2.870
0.72	2.791
0.74	2.716
0.76	2.646
0.78	2.579
0.80	2.515
0.82	2.454
0.84	2.396
0.86	2.341
0.88	2.289
0.90	2.239
0.92	2.191
0.94	2.145
0.96	2.101
0.98	2.059
1.00	2.018
1.02	1.980
1.04	1.942
1.06	1.906
1.08	1.872
1.10	1.839
1.12	1.806
1.14	1.775
1.16	1.746
1.18	1.717
1.20	1.689
1.22	1.662
1.24	1.636
1.26	1.611
1.28	1.586
1.30	1.562
1.32	1.540
1.34	1.517
1.36	1.496
1.38	1.475
1.40	1.454
1.42	1.435
1.44	1.415
1.46	1.397
1.48	1.379
1.50	1.361
1.52	1.344
1.54	1.327
1.56	1.311
1.58	1.295
1.60	1.279
1.62	1.264
1.64	1.250
1.66	1.235
1.68	1.221
1.70	1.208
1.72	1.194
1.74	1.181
1.76	1.169
1.78	1.156
1.80	1.144
1.82	1.132
1.84	1.121
1.86	1.109
1.88	1.098
1.90	1.088
1.92	1.077
1.94	1.067
1.96	1.056
1.98	1.047
2.0	1.037
2.2	0.949
2.4	0.877
2.6	0.817
2.8	0.765
3.0	0.721
3.2	0.683
3.4	0.650
3.6	0.621
3.8	0.595
4.0	0.572
4.2	0.552
4.4	0.533
4.6	0.517
4.8	0.502
5.0	0.489
/
\inboundscheckoff
\endpicture
$
\end{center}
\vskip -5mm
\caption{Graphs of $h(u;x)$.}
 \label{fig:hux}
\end{figure}

\br
The evolution of the Hubble parameter does not depend on the   normalization time $x_\sharp$, in accordance with Theorem \ref{t:Hnotsharp}, page \pageref{t:Hnotsharp}. \er

\section{Cosmological data}

The dimensionless time $x=\beta\,t$  has no practical significance as long as we do not know the value of the constant $\beta$. The estimates of $\beta$ and other constants introduced in this theory are listed in Table \ref{tab:Data-supp} and are inferred from Table \ref{tab:Data} containing basic cosmological data taken from \cite{CODATA}.

\begin{table}[H]
\centering
\begin{tabular}{|l|c|c|c|}
\hline
$\vpt$ name & symbol & estimate & note
\\
\hline \hline
$\vpt$ tropical year (2011)
& 
& $31.5569522\cdot 10^6\,s$
&
\\
\hline
$\vpt$ speed of light
& $c$
& $299.792458\cdot 10^3\,km\,s^{-1}$ 
&
\\
\hline
$\vpt$ age of the universe
& $t_\Z$
& $\simeq 13.81\pm 0.05\;Gyr$ 
&
\\
\hline
$\ba\hbox{Hubble parameter} \vpt
\\[-6pt] \hbox{today} \vpt \ea$
& $H_\Z$
& $\simeq 0.6882972691\cdot 10^{-10}\,yr^{-1}$ 
&
\\
\hline
$\vpt$ gravitational constant
& $G_N$
& $\simeq 6.67408\cdot 10^{-11}\,m^3\,kg^{-1}s^{-2}$
&
\\
\hline
$\vpt$ dark energy density
& $\Omega_\Lambda$
& $0.685^{+0.017}_{-0.016}$
&
\\
\hline
\end{tabular}
 \caption{Basic cosmological data.}\label{tab:Data}
\end{table}
\begin{table}[H]
\centering
\begin{tabular}{|c|c|l|c|}
\hline
$\vpt$ constant & dimension & estimate & note
\\
\hline\hline
\rule[-7pt]{0mm}{20pt}
$\Lambda$
& $L^{-2}$
& $\simeq 1.087769524444\cdot 10^{-52}\,m^{-2}$
& [1]
\\
\hline
$\rule[-5mm]{0mm}{12mm}\beta$
& $T^{-1}$
& $\simeq\bc 
5.41563983302\cdot 10^{-18}\,s^{-1}
\ax
0.170901087343\,Gyr^{-1}
\\[-3pt]
\ec$
&[2]
\\
\hline
\rule[-5mm]{0mm}{12mm} $\dfrac 1\beta$ & T & 
$\simeq
\bc 
0.18465038865819 \cdot 10^{18}\,s
\ax
5.8513378442864\,Gyr
\\[-3pt]
\ec$
&
\\
\hline
$\rule[-7pt]{0mm}{20pt} \chi$ 
&  $\!\!\!L^{-1}M^{-1}T^2\!\!\!$
& $\simeq 2.07657899185574 \cdot 10^{-43}\,m^{-1}\,kg^{-1}\,s^2$
& [3]
\\
\hline
$\rule[-5mm]{0mm}{12mm} \dfrac{\Lambda}{\chi}$ 
& $L^{-1}M\,T^{-2}$
& $\simeq 5.238276649771\cdot 10^{-10}\,m^{-1}\,kg\,s^{-2} $
&
\\
\hline
\end{tabular}
 \caption{Supplementary data.}\label{tab:Data-supp}
\end{table}

\newpage
{\bf Notes}.

$[1]$ {\bf Estimate of} $\,\Lambda\Def \dfrac{3H_\Z^2}{c^2}\,\Omega_\Lambda$:

$H_\Z\simeq 0.6882972691\cdot 10^{-10}\,yr^{-1}$

$H_\Z^2\simeq 0.47375313065051\cdot 10^{-20}\,yr^{-2}$

$3\,H_\Z^2\simeq 1.42125939195\cdot 10^{-20}\,yr^{-2}$

$3\,H_\Z^2/c \simeq 0.004740811031176774967434  \cdot 10^{-20}\,yr^{-2}   \cdot 10^{-6}\,m^{-1}\,s$

$3\,H_\Z^2/c^2 \simeq 1.581364342119899149509\cdot 10^{-5} \cdot 10^{-20}\,yr^{-2}   \cdot 10^{-12}\,m^{-2}\,s^2$ [*]

$\Lambda=3\,H_\Z^2/c^2\cdot\Omega_\Lambda \simeq 1.581364342119899149509\cdot 10^{-37} \,yr^{-2}\,m^{-2}\,s^2
\cdot 0.685$

$\Lambda\simeq 1.08324574352130917414\cdot 10^{-37} \,yr^{-2}\,m^{-2}\,s^2$

$yr/s=3.15569522\cdot 10^7$, \quad $yr^2/s^2=9.9584123215308484\cdot 10^{14}$

$\boxed{\Lambda\simeq 0.1087769524444422833994\cdot 10^{-51}\,m^{-2}}$ \eprf

$[2]$ {\bf Estimate of} $\beta\Def\sqrt{3\,\Lambda}\,c$:

$3\,\Lambda\simeq 3.263308573333268501982 \cdot 10^{-52}\,m^{-2}$

$\sqrt{3\,\Lambda}\simeq 1.806463000820462002629 \cdot 10^{-26}\,m^{-1}$

$c=299.792458\cdot 10^6\,m\,s^{-1}$

$\beta=\sqrt{3\,\Lambda}\,c \simeq 541.5639833020223204638 \cdot 10^{-26}\,m^{-1}\cdot 10^6\,m\,s^{-1}$

$\simeq 5.415639833020223204638 \cdot 10^{-18}\,s^{-1}$.

$1\,yr = 31.5569522\cdot 10^6\,s$.

$\beta\simeq 1.709010873430351653021\cdot 10^{-10}\,yr^{-1}$. \eprf

$[3]$ {\bf Estimate of} $\chi\Def \dfrac{8\,\pi\,G_N}{c^4}$:

$c^{-4}\simeq 1.237990147236120239125\cdot  10^{-34} \,m^{-4}\,s^4$

$\pi\simeq 3.1415926535897932384626433$

$8\,\pi\,c^{-4}\simeq 31.11408601418833454066 \cdot 10^{-34}\,m^{-4}\,s^4$

$G_N=6.67408(31)\cdot 10^{-11}\,m^3\,kg^{-1}s^{-2}$ 

$\chi\simeq 207.657899185574 \cdot 10^{-34}\,m^{-4}\,s^4\cdot 10^{-11}\,m^3\,kg^{-1}s^{-2}$ 

$\boxed{\chi\simeq 2.07657899185574 \cdot 10^{-43}\,m^{-1}\,s^2\,kg^{-1}}$  \eprf 

\section{The age of the universe}

\bt\label{t:D}
If $H_\Z$ is the present-day value of the Hubble parameter, then the age of the universe is
\be\label{e:H0t0}
\boxed{\bboxed{\vpud x_\Z=\beta\,t_\Z=\dfrac 1u\,\log\dfrac{3\,H_\Z+\beta}{3\,H_\Z-\beta}}}
\ee
\et

\bprf 
Equation \eqref{e:hux} is solvable with respect to $X=e^{ux}$:

$h(u,x)=\tfrac 13\,\dfrac{X+1}{X-1}$  $\To$ $3\,(X-1)\,h(u,x)=X+1$ 

$\To$ $[3\,h(u,x)-1]\,X=3\,h(u,x)+1$ $\To$ $X=\dfrac{3\,h(u,x)+1}{3\,h(u,x)-1}$. 

Because of \eqref{e:hut}, $h(u,x)=\beta^{-1}\,h(u,t)$ $\To$   
$X=\dfrac{3\,h(u,t)+\beta}{3\,h(u,t)-\beta}$.

As $X=e^{ux}$ $\To$ 
\be\label{e:xh}
\bboxed{x=\beta\,t=\dfrac 1u\,\log\dfrac{3\,h(u,t)+\beta}{3\,h(u,t)-\beta}}
\ee
The profile \eqref{e:aut-e} satisfies the initial condition $a(0)=0$ with $t_\alpha=0$. Then the beginning of the universe corresponds to $t=x=0$, so that equation \eqref{e:xh} applied to the present epoch  provides the age of the universe. \eprf

According to the formula \eqref{e:H0t0}, for computing the age of the universe we only need the values of $H_\Z$ and $\beta=\sqrt{3,\Lambda}\;c$: 
$$\left.\ba
H_\Z\simeq 0.6882972691\cdot 10^{-10}\,yr^{-1}
\ax
\beta\simeq 1.70901087343\cdot 10^{-10}\,yr^{-1}
\ea\right\}\To
\ba 
u\,x_\Z\simeq \log 10.6043969244822
\ax
\simeq 2.361268719306985270849
\ea\,\To
$$
\begin{gather}
\label{e:x0}
\bboxed{u\,x_\Z\simeq 2.3612687193}
\\
\label{e:t0}
\bboxed{u\,t_\Z=\dfrac {u\,x_\Z}{\beta}\,\simeq 13.81658101781\cdot 10^{9} \, yr}
\end{gather}

\begin{center}
\fbox{\begin{minipage}{.95\linewidth} 
\br \vspu
For $u=1$ this estimate is very close to that supplied by the astronomers \cite{CODATA} (2015) $t_\Z\simeq 13.81\pm 0.05\;Gyr$. This means that the primordial phase of radiation dominance has an irrelevant influence on the evaluation of the present-day age of the universe. \vspd\er 
\end{minipage}}
\end{center}
  
\section{Inadmissibility of the positive curvature}\label{s:noK>0}

The evaluation of equation \eqref{e:vel-e*} at the present day $t_\Z$ gives
$$
\dfrac{3\,H_\Z^2}{c^2}=\Lambda+\chi\,\epsilon_\Z-\dfrac{3\,\wt K}{a^2(t_\Z,t_\sharp)}.
$$
This equation is invariant under the choice of the reference time (Remark \ref{r:inv-1}). Then for $t_\sharp=t_\Z$ we get
$$
\dfrac{3\,H_\Z^2}{c^2}-\Lambda-\chi\,\epsilon_\Z=-\,3\,\wt K.
$$
If we assume $\wt K>0$ then the following inequalities hold:
$$
\dfrac{3\,H_\Z^2}{c^2}-\Lambda-\chi\,\epsilon_\Z<0
\iff
\dfrac{3\,H_\Z^2}{c^2}-\Lambda-\chi\,\epsilon_\Z<0
\iff
\dfrac{3\,H_\Z^2}{c^2}-\Lambda<\chi\,\epsilon_\Z\iff
$$
\be\label{e:e0>e*}
\bboxed{\epsilon_\Z>\dfrac 1\chi\left(\dfrac{3\,H_\Z^2}{c^2}-\Lambda\right)\Def \epsilon_*}
\ee
Go back to [*]:  

$3\,H_\Z^2/c^2 \simeq 1.581364342119\cdot 10^{-37} \cdot \,m^{-2}\,s^2\,yr^{-2}$, \quad $1\,yr = 31.5569522\cdot 10^6\,s$ 

$\To$ $3\,H_\Z^2/c^2 \simeq \dfrac{1.581364342119}
{(31.5569522)^2} \cdot 10^{-43}\,m^{-2}
\simeq 1.58796833\cdot 10^{-46}\,m^{-2}$.

Due to the estimate of $\Lambda$ and $\chi$,

$\boxed{\vpt\Lambda\simeq 0.10877695\cdot 10^{-51}\,m^{-2}}$ \quad
$\boxed{\vpt\chi\simeq 2.07657899 \cdot 10^{-43}\,m^{-1}\,s^2\,kg^{-1}}$

we get

$\epsilon_*\simeq
\dfrac 1{2.07657899} \cdot 10^{43}
\left(1.58796833\cdot 10^{-46}
-0.10877695\cdot 10^{-51}\right)\,m^{-1}\,kg\,s^{-2}$

$\simeq
\dfrac 1{2.07657899} \cdot 10^{43}\cdot 10^{-51}
\left(1.58796833\cdot 10^{5}
-0.10877695\right)\,m^{-1}\,kg\,s^{-2}$

$\simeq
\dfrac {1.58796833}{2.07657899} \cdot 10^{43}\cdot 10^{-51}
\cdot 10^{5}
\;\To$
$$
\bboxed{\epsilon_*\simeq 0.764704\cdot 10^{-3}\,m^{-1}\,kg\,s^{-2}}
$$
This energy density corresponds to a mass density $\rho_*=\epsilon_*\,c^{-2}$. Since $c=299.792458\cdot 10^6\,m\,s^{-1}$ and 
$0.764704/(299.792458)^2\simeq 8.50847\cdot 10^{-6}$, we get

$\rho_*\simeq \dfrac{0.764704\cdot 10^{-3}\,m^{-1}\,kg\,s^{-2}}{(299.792458)^2\cdot 10^{12}\,m^2\,s^{-2}}
\simeq 8.50847\cdot 10^{-6}\,\dfrac{ 10^{-3}\,kg}{10^{12}\,m^3}
$
$$
\bboxed{\rho_*\simeq 8.50847\cdot 10^{-21}\,\dfrac{kg}{m^3}}
$$
A rough present-day estimate of the mass density is 
$$
\rho_\Z\simeq (9.2\pm 1.8)\cdot 10^{-27}\,\dfrac{kg}{m^3}\simeq  10^{-26}\,\dfrac{kg}{m^3}.
$$
Then we see that the inequality \eqref{e:e0>e*} $\epsilon_\Z>\epsilon_*$, which is equivalent to $\rho_\Z>\rho_*$, is far to be satisfied even if the estimate of $\rho_\Z$ is not accurate. This proves that the assumption $\wt K>0$ is in strong contrast with the astronomical observations.

\section{The \emph{`exact'} profiles of the flat barotropic universes}\label{s:aux0}

\begin{center}
\fbox{\begin{minipage}{.95\linewidth} \vspu
If we consider sufficiently reliable the estimate of the age of the universe  found above, then we should consider equally reliable the choice of the present-day time $t_\Z$  as  reference time for the scale factor. In doing so we get a `sufficiently reliable' (or {\it `exact'$\,$}) numerical evaluation of the universe profile,  for any value of the state parameter $u$. \vspd
\end{minipage}}
\end{center}
 
\bt \label{t:aux-0}
With the present-day reference time $x_\Z$ the profiles of the universe are
\begin{gather}
\label{e:aux-c-0}
\bboxed{a(u;x,x_\Z)=\left[\vp\ff c_\Z\,\big(\cosh(ux)-1\big)\right]\!{\vp}^{\tfrac 1{3u}}=\left[\vp\tfrac 12\,\ff c_\Z\dfrac{(e^{ux}-1)^2}{e^{ux}}\right]\!{\vp}^{\tfrac 1{3u}}\rule[-4mm]{0mm}{9mm}}
\\
\label{e:def-c0}
\bboxed{\ff c_\Z\Def\dfrac 1{\cosh(ux_\Z)-1}\simeq0.2299194811} 
\end{gather}
\et

\bprf With $x_\sharp=x_\Z$ the profiles \eqref{e:aux-c} read
$$
a(u;x,x_\Z)=\left[\tfrac 12\,\dfrac\chi\Lambda\,\epsilon_\Z\,\big(\cosh(ux)-1\big)\right]\!{\vp}^{\tfrac 1{3u}}.
$$
By imposing the normalization condition $a(u;x_\Z,x_\Z)=1$ we get
\be\label{e:eps-0}
\tfrac 12\,\dfrac\chi\Lambda\,\epsilon_\Z\,\big(\cosh(ux_\Z)-1\big)=1
\ee
i.e.\ 
\be\label{e:c0}
\tfrac 12\,\dfrac\chi\Lambda\,\epsilon_\Z=\ff c_\Z
\ee
with $\ff c_\Z$ defined as in \eqref{e:def-c0}. \eprf

The graphs of $a(u;x,x_\Z)$ are plotted in Fig.\ \ref{fig:au-ux} with respect to the variable $ux$  for some relevant values of $u$. Whatever $u$, they all pass through the point $(x_\Z,1)$, as expected. In Fig.\ \ref{fig:aux-x} the profiles are plotted with respect to the variable $x$. In both representation we observe (i) a different way of approaching the origin $x=0$ and (ii) the presence of inflection points $x_{ip}$ for certain values of $u$.

\begin{figure} [H]
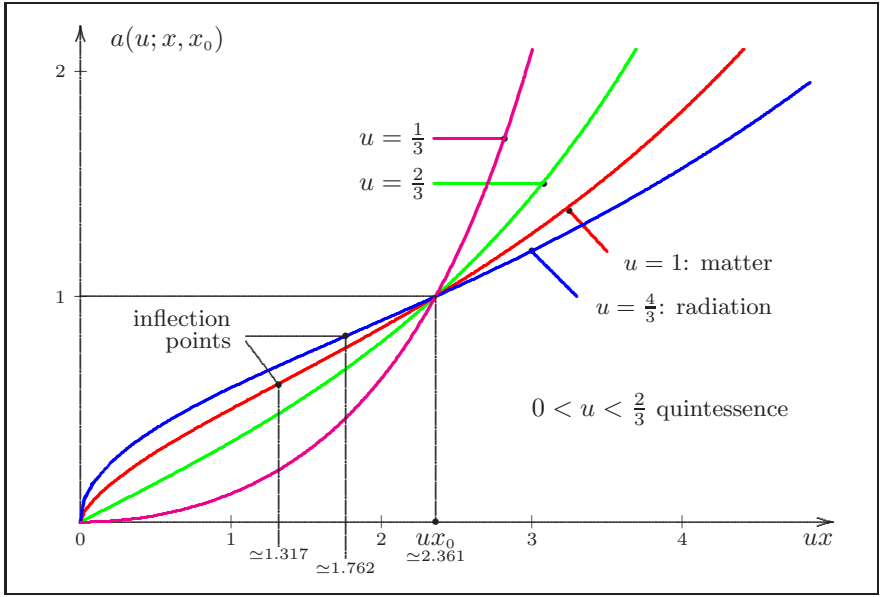

\begin{center}
$
\beginpicture
\setcoordinatesystem units <2cm,3cm>
\setplotarea x from 0 to 4.85, y from 0 to 2.1
\normalgraphs
\putrectangle corners at -.5 2.3 and 5.3 -.32
\arrow <6pt> [.2,.6] from 0 0  to 5 0
\arrow <6pt> [.2,.6] from 0 0  to 0 2.2
\put{$ux$} [lt] at 4.8 -.05
\put{$a(u;x,x_\Z)$} [lt] at .2 2.2
\put{$\scc |$} at 1 0
\put{$\scc |$} at 2 0
\put{$\scc |$} at 3 0
\put{$\scc |$} at 4 0
\put{$\sc 0$} [t] at 0 -.05 
\put{$\sc 1$} [t] at 1 -.05
\put{$\sc 2$} [t] at 2 -.05
\put{$\sc 3$} [t] at 3 -.05
\put{$\sc 4$} [t] at 4 -.05
\put{$\sc 1$} [r] at -.1 1
\put{$\sc -$}  at 0 1
\put{$\sc 2$} [r] at -.1 2
\put{$\sc -$}  at 0 2
\inboundscheckon
\setplotsymbol (\normalegreen)
\put{\ssmall $u=\tfrac 23$} [r] at  2.3 1.5
\setlinear
\plot 2.35 1.5 3.08 1.5 /
\put{$\scc\bullet$} at 3.08 1.5
\setquadratic
\plot
0.0	0.0000
0.2	0.0679
0.4	0.1366
0.6	0.2065
0.8	0.2786
1.0	0.3534
1.2	0.4318
1.4	0.5145
1.6	0.6023
1.8	0.6962
2.0	0.7971
2.2	0.9059
2.4	1.0238
2.6	1.1519
2.8	1.2915
3.0	1.4441
3.2	1.6112
3.4	1.7943
3.6	1.9955
3.8	2.2165
4.0	2.4598
4.2	2.7277
4.4	3.0229
4.6	3.3484
4.8	3.7073
/
\setplotsymbol (\normalered)
\put{\small $u=1$: matter} [r] at 4.6 1.15
\setlinear
\plot 3.25 1.38 3.5 1.2 /
\put{$\scc\bullet$} at 3.25 1.38
\setquadratic
\plot 
0.0	0.0000 
0.02	0.0358
0.04	0.0569
0.06	0.0745
0.08	0.0903
0.10	0.1048
0.12	0.1184
0.14	0.1312
0.16	0.1434
0.18	0.1552
0.2	0.1665
0.4	0.2652
0.6	0.3494
0.8	0.4265
1.0	0.4999
1.2	0.5713
1.4	0.6421
1.6	0.7132
1.8	0.7855
2.0	0.8597
2.2	0.9362
2.4	1.0158
2.6	1.0989
2.8	1.1860
3.0	1.2776
3.2	1.3743
3.4	1.4766
3.6	1.5850
3.8	1.7000
4.0	1.8222
4.2	1.9522
4.4	2.0907
4.6	2.2381
/
\setplotsymbol (\normaleblue)
\put{\small $u=\tfrac 43$: radiation} [r] at 4.6 .95
\setlinear
\plot  3.3 1 3 1.2 /
\put{$\scc\bullet$} at 3 1.2
\setquadratic
\plot 
0.0	0.0000
0.02	0.0824
0.04	0.1165
0.06	0.1427
0.08	0.1647
0.1	0.1842
0.2	0.2606
0.3	0.3196
0.4	0.3695
0.5	0.4139
0.6	0.4545
0.7	0.4922
0.8	0.5278
0.9	0.5618
1.0	0.5945
1.2	0.6571
1.4	0.7173
1.6	0.7761
1.8	0.8344
2.0	0.8928
2.2	0.9518
2.4	1.0118
2.6	1.0733
2.8	1.1365
3.0	1.2017
3.2	1.2693
3.4	1.3395
3.6	1.4126
3.8	1.4888
4.0	1.5684
4.2	1.6516
4.4	1.7387
4.6	1.8299
4.8	1.9254
5.0	2.0257
/
\setplotsymbol (\normalemag)
\put{\ssmall $u=\tfrac 13$} [r] at  2.3 1.7
\setlinear
\plot 2.35 1.7 2.8 1.7 /
\put{$\scc\bullet$} at 2.82 1.7
\setquadratic
\plot 
0.0	0.0000
0.2	0.0046
0.4	0.0186
0.6	0.0427
0.8	0.0776
1.0	0.1249
1.2	0.1864
1.4	0.2647
1.6	0.3628
1.8	0.4847
2.0	0.6353
2.2	0.8206
2.4	1.0481
2.6	1.3268
2.8	1.6681
3.0	2.0855
3.2	2.5959
3.4	3.2196
3.6	3.9818
3.8	4.9131
4.0	6.0507
4.2	7.4405
4.4	9.1380
4.6	11.2116
4.8	13.7443
/
\inboundscheckoff
\normale
\setlinear
\plot 2.361 0 2.361 1 0 1 /
\put{$\scc\bullet$}  at 2.361 0
\put{$ux_\Z$} [t]  at 2.361 -.05
\put{\small $\sc\simeq 2.361$} [t]  at 2.361 -.13
\plot 1.317 -.11 1.317 .61 /
\put{$\scc\bullet$}  at 1.317 .61
\put{$\scc\bullet$}  at 1.762 .825
\plot 1.762 -.16 1.762 .825 /
\put{\small inflection} [rb] at 1 .87
\put{\small points} [rb] at 1 .75
\plot 1.1 .825 1.762 .825 /
\plot 1.1 .8 1.317 .61 /
\put{\small $\sc\simeq 1.317$} [t]  at 1.317 -.12
\put{\small $\sc\simeq 1.762$} [t]  at 1.762 -.18
\put{$0<u<\tfrac 23$ \small quintessence} [l] at 3 .5
\endpicture
$
\end{center}
\vskip -5mm
\caption{Graphs of $a(u;x,x_\Z)$ in the variable $ux$.}
 \label{fig:au-ux}
\end{figure}

\bt
{\rm (i)} The profiles approach the beginning of the universe $x=0$ in different ways:
\be\label{e:lim0}
\left\{\begin{array}{lll}
u>\tfrac 23 & \To & \lim_{x\to 0}\dfrac{da}{dx}= +\infty.
\ac
u=\tfrac 23 & \To & \lim_{x\to 0}\dfrac{da}{dx}=\tfrac 13\sqrt{2\,\ff c_\Z}.
\ac
u<\tfrac 23 & \To & \lim_{x\to 0}\dfrac{da}{dx}= 0.
\ea\right.
\ee
{\rm (ii)} For $u>\tfrac 23$ there is an inflection point at the time
\be\label{e:xip}
\bboxed{x_{ip}(u)=\dfrac 1u\,\log\left[\vp 3\,u-1+\sqrt{\vpx(3\,u-1)^2-1}\,\right]=\dfrac 1u\,\arccosh(3\,u-1)}
\ee
\et

\br
This theorem indicates the value $u=\tfrac 23$ (corresponding to  $w=-\tfrac 13$) as a {\bf threshold  parameter}: for any small variation of this value the profile $a(u,x)$ changes radically. Due to this sort of `instability' we should consider {\bf inadmissible} the case $u=\tfrac 23$. \er

\begin{figure} [H]
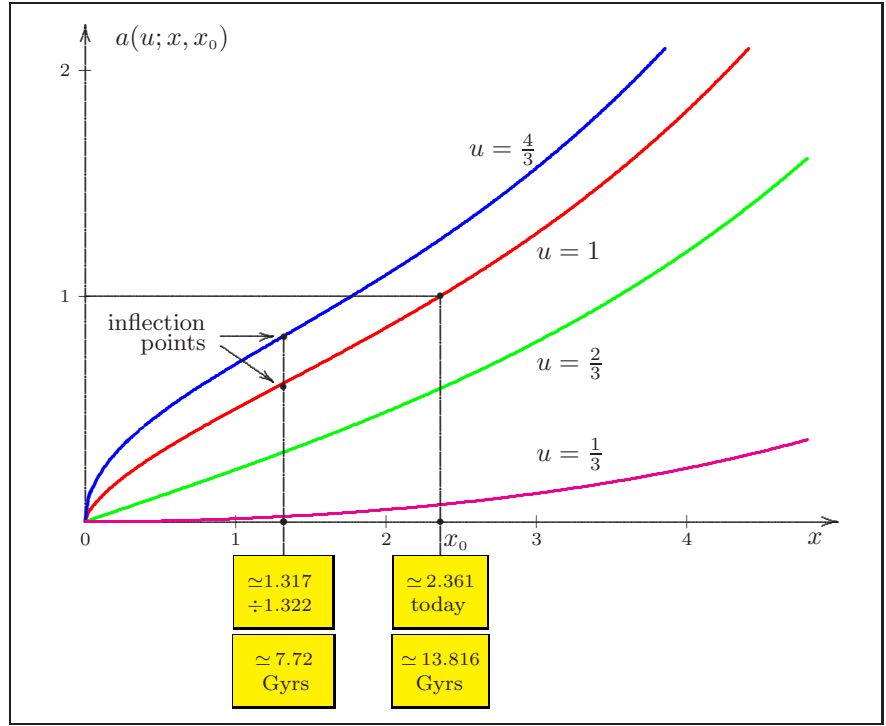

\begin{center}
$
\beginpicture
\setcoordinatesystem units <2cm,3cm>
\setplotarea x from 0 to 4.85, y from -.2 to 2.1
\normalgraphs
\putrectangle corners at -.5 2.3 and 5.3 -.9
\arrow <6pt> [.2,.6] from 0 0  to 5 0
\arrow <6pt> [.2,.6] from 0 0  to 0 2.2
\put{$x$} [lt] at 4.8 -.05
\put{$a(u;x,x_\Z)$} [lt] at .2 2.2
\put{$\scc |$} at 1 0
\put{$\scc |$} at 2 0
\put{$\scc |$} at 3 0
\put{$\scc |$} at 4 0
\put{$\sc 0$} [t] at 0 -.05 
\put{$\sc 1$} [t] at 1 -.05
\put{$\sc 2$} [t] at 2 -.05
\put{$\sc 3$} [t] at 3 -.05
\put{$\sc 4$} [t] at 4 -.05
\put{$\sc 1$} [r] at -.1 1
\put{$\sc -$}  at 0 1
\put{$\sc 2$} [r] at -.1 2
\put{$\sc -$}  at 0 2
\inboundscheckon
\setquadratic
\setplotsymbol (\normalegreen)
\put{$u=\tfrac 23$} [l] at 3 .7
\plot
0.0	0.0000
0.2	0.0452
0.4	0.0907
0.6	0.1365
0.8	0.1830
1.0	0.2302
1.2	0.2785
1.4	0.3281
1.6	0.3791
1.8	0.4317
2.0	0.4863
2.2	0.5431
2.4	0.6022
2.6	0.6641
2.8	0.7289
3.0	0.7969
3.2	0.8685
3.4	0.9439
3.6	1.0236
3.8	1.1078
4.0	1.1969
4.2	1.2913
4.4	1.3915
4.6	1.4979
4.8	1.6109
/
\setplotsymbol (\normalered)
\put{$u=1$} [l] at 3 1.2
\plot 
0.00	0.0000
0.02	0.0358
0.04	0.0569
0.06	0.0745
0.08	0.0903
0.10	0.1048
0.12	0.1183
0.14	0.1312
0.16	0.1434
0.18	0.1552
0.20	0.1665
0.22	0.1774
0.24	0.1881
0.26	0.1984
0.28	0.2086
0.30	0.2184
0.32	0.2281
0.34	0.2376
0.36	0.2470
0.38	0.2561
0.40	0.2651
0.42	0.2740
0.44	0.2828
0.46	0.2915
0.48	0.3000
0.50	0.3084
0.52	0.3168
0.54	0.3251
0.56	0.3332
0.58	0.3413
0.6	0.3494
0.8	0.4265
1.0	0.4998
1.2	0.5712
1.4	0.6420
1.6	0.7131
1.8	0.7854
2.0	0.8596
2.2	0.9361
2.4	1.0157
2.6	1.0987
2.8	1.1858
3.0	1.2775
3.2	1.3742
3.4	1.4765
3.6	1.5848
3.8	1.6998
4.0	1.8220
4.2	1.9520
4.4	2.0904
4.6	2.2379
4.8	2.3951
5.0	2.5628
/
\setplotsymbol (\normaleblue)
\put{$u=\tfrac 43$} [r] at 3 1.65
\plot 
0.00	0.0000
0.02	0.0951
0.04	0.1345
0.06	0.1647
0.08	0.1902
0.10	0.2127
0.12	0.2330
0.14	0.2518
0.16	0.2692
0.18	0.2856
0.20	0.3011
0.22	0.3159
0.24	0.3301
0.26	0.3437
0.28	0.3568
0.30	0.3695
0.4	0.4278
0.6	0.5278
0.8	0.6157
1.0	0.6974
1.2	0.7760
1.4	0.8537
1.6	0.9319
1.8	1.0117
2.0	1.0940
2.2	1.1796
2.4	1.2692
2.6	1.3634
2.8	1.4629
3.0	1.5683
3.2	1.6800
3.4	1.7988
3.6	1.9253
3.8	2.0600
4.0	2.2037
4.2	2.3569
4.4	2.5205
4.6	2.6951
4.8	2.8817
/
\setplotsymbol (\normalemag)
\put{$u=\tfrac 13$} [l] at 3 .3
\plot 
0.0	0.0000
0.2	0.0005
0.4	0.0020
0.6	0.0046
0.8	0.0082
1.0	0.0129
1.2	0.0186
1.4	0.0255
1.6	0.0335
1.8	0.0426
2.0	0.0530
2.2	0.0646
2.4	0.0776
2.6	0.0919
2.8	0.1076
3.0	0.1249
3.2	0.1437
3.4	0.1642
3.6	0.1864
3.8	0.2105
4.0	0.2365
4.2	0.2646
4.4	0.2949
4.6	0.3276
4.8	0.3627
/
\inboundscheckoff
\normale
\setlinear
\plot 0 1 2.361 1 2.361 -.15 /
\put{$\scc\bullet$} at 2.361 1
\put{$\scc \bullet$} at 2.361 0
\put{$x_\Z$} [lt]  at 2.38 -.06
\put{\fcolorbox{black}{yellow}{$\!\!\ba
\sc\simeq \,2.361
\\[-1mm]
\,\hbox{\footnotesize today}
\ea\!\!$}} [t] at 2.361 -.15
\put{\fcolorbox{black}{yellow}{$\!\!\!\ba
\sc\simeq \,13.816
\\[-1mm]
\;\;\hbox{\footnotesize Gyrs}
\ea\!\!\!$}} [t] at 2.361 -.5
\put{$\scc\bullet$} at 1.317 0
\put{$\scc\bullet$} at 1.317 .6
\put{$\scc\bullet$} at 1.322 0
\put{$\scc\bullet$} at 1.322 .82
\plot 1.32 -.15 1.32 .82 /
\put{\fcolorbox{black}{yellow}{$\!\!\ba
\sc\simeq 1.317 
\\[-1mm]
\sc \div 1.322
\ea$}} [t] at 1.32 -.15
\put{\small inflection} [rb] at .8 .85
\put{\small points} [rb] at .8 .74
\arrow <5pt> [.2,.6] from .9 .82  to 1.25 .82
\arrow <5pt> [.2,.6] from .9 .78 to 1.25 .62
\put{\fcolorbox{black}{yellow}{$\ba
\sc\simeq \,7.72
\\[-1mm]
\;\hbox{\footnotesize Gyrs}
\ea$}} [t] at 1.32 -.5

\endpicture
$
\end{center}
\vskip -5mm
\caption{Graphs of $a(u;x,x_\Z)$ in the variable $x$.}
 \label{fig:aux-x}
\end{figure}

\bprf
(i) \eqref{e:hux} and \eqref{e:aux-c-0} $\To$ 
$$
\ba
\dfrac{da}{dx}=\tfrac 13\,a\,\dfrac {e^{ux}+1}{e^{ux}-1}
=\tfrac 13\,\left(\tfrac 12\,\ff c_\Z\right)^{\frac 1{3u}}\left[\,\dfrac{(e^{ux}-1)^2}{e^{ux}}\right]\!{\vp}^{\tfrac 1{3u}}\,\dfrac {e^{ux}+1}{e^{ux}-1} 
\ac
=\tfrac 13\,\left(\tfrac 12\,\ff c_\Z\right)^{\frac 1{3u}}\,e{\vpx}^{-\tfrac x3}\,
\big(e^{ux}+1\big)\,\big(e^{ux}-1\big){\vpx}^{\tfrac{2}{3u}-1}. 
\ea
$$
$\ba 
\Lim_{x\to 0}\dfrac{da}{dx}
=\tfrac 23\left(\tfrac 12\,\ff c_\Z\right)^{\frac 1{3u}}
\big(e^{ux}-1\big){\vpx}^{\tfrac{2}{3u}-1}={}
\ac
\kern 15mm =
\left\{\begin{array}{llcl}
0 & \iff & \tfrac{2}{3u}-1>0
&  \iff \tfrac 23>u,
\ax
\tfrac 13\sqrt{2\,\ff c_\Z} &  \iff  & \tfrac{2}{3u}-1=0
&  \iff \tfrac 23=u,
\ax
+\infty & \iff & \tfrac{2}{3u}-1<0
&  \iff \tfrac 23<u.
\ea\right.
\ea
$

Note that $\tfrac 13\sqrt{2\,\ff c_\Z}\simeq 0.2260378$.

(ii)  \eqref{e:dadx} $\To$
$\dfrac{d^2a}{dx^2}=\tfrac 13 \left[\dfrac{da}{dx}\,\dfrac {X+1}{X-1}+a\,\dfrac{d\;}{dx}\dfrac {X+1}{X-1}
\right]$

$=\tfrac 13\,a \left[\tfrac 13\,\dfrac {(X+1)^2}{(X-1)^2}
+\dfrac {u\,X\,(X-1)-(X+1)\,u\,X}{(X-1)^2}\right]
=\tfrac 13\,a\, \dfrac{\tfrac 13\,(X+1)^2-2\,u\,X}{(X-1)^2}.$

$\dfrac{d^2a}{dx^2}=0 \iff \tfrac 13\,(X+1)^2-2\,u\,X=0
\iff X^2+2\,(1-3u)\,X+1=0$

$\iff X=3u-1\pm \sqrt{(3u-1)^2-1}$. With the $-$ sign $X<1$, rejected. $X=e^{ux}$ $\To$ \eqref{e:xip}. \eprf

\br
The inflection point marks the transition from decelerated to accelerated expansion.
\begin{table}[H]
\centering
\begin{tabular}{|l|c|c|c|c|}
\hline
Component
& $\vpt w$ 
& $\vpt u$
& $x_{ip}(u)$
& $t_{ip}(u)=\beta^{-1}\,x_{ip}(u)$
\\
\hline \hline
Matter
& $\vpt 0 $ 
& $\vpt 1 $ 
& 1.316958
& $7.70601\,Gyr$
\\
\hline
Radiation 
& $\vpt \tfrac 13 $
& $\vpt \tfrac 43 $ 
& 1.322067
& $7.73591\,Gyr$
\\
\hline
\end{tabular}
 \caption{Estimate of the inflection time.}\label{tab:ip}
\end{table}
The two times $x_{ip}(1)$ and $x_{ip}(\tfrac 43)$ are very close.  The phase of accelerated expansion starts $\simeq 6.08\div 6.11$ billion years ago. Note that $x_{ip}$ does not depend on the choice of the normalization time. \er
\section{The profiles of the energy density}

\bt
In the barotropic flat models the profiles $e(u;x)$ of the energy density do not depend on the reference time $t_\sharp$ and admit the following two equivalent representations
\be
\label{e:eu-x}
\bboxed{\epsilon(u;x)=\dfrac{\Lambda}{\chi}\,\dfrac{4\,e^{ux}}{(1-e^{ux})^2}}\;\;
\bboxed{\epsilon(u;x)=\dfrac\Lambda\chi\,\dfrac 2{\cosh(ux)-1}}
\ee
\et

\bprf
$
\bc
\eqref{e:aux-e}\;\To\;a^{3u}(u;x,x_\sharp)=\tfrac 14\,\epsilon_\sharp\,\dfrac\chi\Lambda\,\dfrac{(e^{ux}-1)^2}{e^{ux}}.
\ac
\eqref{e:aux-c}\;\To\;a^{3u}(u;x,x_\sharp)=\tfrac 12\,\epsilon_\sharp\,\dfrac\chi\Lambda\,\big(\cosh(ux)-1\big).
\ec
$

\eqref{e:etat-t0}  $\To\; \epsilon(u;x,x_\sharp)=\dfrac{\epsilon_\sharp}{a^{3u}(u;x,x_\sharp)}$ $\To$ \eqref{e:eu-x}. \eprf

\br
The evolution of the energy density does not depend   on the choice of the normalization time but only on the  parameter $u$ and on the ratio $\Lambda/\chi$. As a consequence, since we know a `reliable' numerical value of $\Lambda/\chi$ (see Table \ref{tab:Data-supp})
\be\label{e:lambda/chi} 
\dfrac{\Lambda}{\chi}\simeq 5.238276649771\cdot 10^{-10}\,m^{-1}\,kg\,s^{-2} 
\ee
then we can get a `reliable' numerical estimate of the evolution of the energy density for any value of the parameter $u$ (Fig.\ \ref{fig:ed}). The formula to be used for plotting 
$\epsilon(u;x)$ is
$$
\epsilon(u;x)=5.238276649771*2\,\dfrac 1{\cosh(u*x)-1}. \quad \bullet
$$
\erx

\begin{figure} [H]
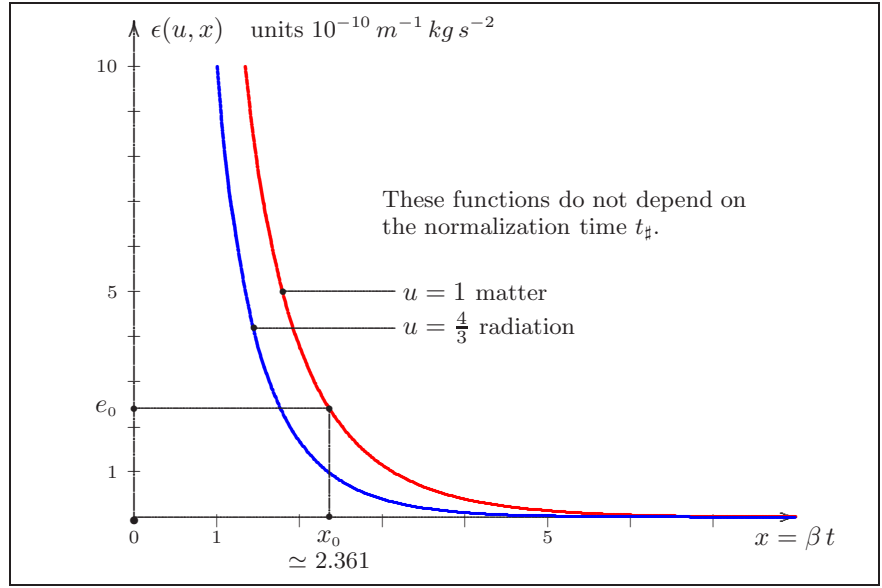

\centering
$
\beginpicture
\setcoordinatesystem units <1.1cm,.6cm>
\setplotarea x from -1 to 8, y from 0 to 10
\normalgraphs
\putrectangle corners at -1.5 11.4 and 9 -1.5
\arrow <6pt> [.2,.6] from 0 0 to 8 0
\arrow <6pt> [.2,.6] from 0 0 to 0 11
\put{$\sc\bullet$} at 0 -.07
\put{$x=\beta\,t$} [lt] at 7.5 -.2
\put{$\epsilon(u,x)$\quad \small units $10^{-10}\,m^{-1}\,kg\,s^{-2} $} [lb] at .2 10.5
\put{$\sc 0$} [t] at 0 -.3
\put{$\sc -$} at 0 1
\put{$\sc -$} at 0 2
\put{$\sc -$} at 0 3
\put{$\sc -$} at 0 4
\put{$\sc -$} at 0 5
\put{$\sc -$} at 0 6
\put{$\sc -$} at 0 7
\put{$\sc -$} at 0 8
\put{$\sc -$} at 0 9
\put{$\sc -$} at 0 10
\put{$\sc 1$} [r] at -.2 1
\put{$\sc 5$} [r] at -.2 5
\put{$\sc 10$} [r] at -.2 10
\put{$\scc |$} at 1 -.07
\put{$\scc |$} at 2 -.07
\put{$\scc |$} at 3 -.07
\put{$\scc |$} at 4 -.07
\put{$\scc |$} at 5 -.07
\put{$\scc |$} at 6 -.07
\put{$\scc |$} at 7 -.07
\put{$\sc 1$} [t] at 1 -.3
\put{$\sc 5$} [t] at 5 -.3
\inboundscheckon
\setquadratic
\setplotsymbol (\normalered) 
\plot
1.2	12.924
1.4	9.103
1.6	6.641
1.8	4.971
2.0	3.793
2.2	2.936
2.4	2.299
2.6	1.816
2.8	1.444
3.0	1.155
3.2	0.928
3.4	0.748
3.6	0.605
3.8	0.490
4.0	0.398
4.2	0.324
4.4	0.264
4.6	0.215
4.8	0.175
5.0	0.143
5.2	0.117
5.4	0.095
5.6	0.078
5.8	0.064
6.0	0.052
6.2	0.043
6.4	0.035
6.6	0.029
6.8	0.023
7.0	0.019
7.2	0.016
7.4	0.013
7.6	0.010
7.8	0.009
8.0	0.007
8.2	0.006
8.4	0.005
8.6	0.004
8.8	0.003
/
\setplotsymbol (\normaleblue) 
\plot
0.8	16.765
1.0	10.185
1.2	6.641
1.4	4.534
1.6	3.193
1.8	2.299
2.0	1.681
2.2	1.244
2.4	0.928
2.6	0.697
2.8	0.526
3.0	0.398
3.2	0.302
3.4	0.230
3.6	0.175
3.8	0.134
4.0	0.102
4.2	0.078
4.4	0.060
4.6	0.046
4.8	0.035
5.0	0.027
5.2	0.020
5.4	0.016
5.6	0.012
5.8	0.009
6.0	0.007
6.2	0.005
6.4	0.004
6.6	0.003
6.8	0.002
7.0	0.002
7.2	0.001
7.4	0.001
7.6	0.001
7.8	0.001
8.0	0.000
8.2	0.000
8.4	0.000
8.6	0.000
8.8	0.000
9.0	0.000
9.2	0.000
9.4	0.000
9.6	0.000
9.8	0.000
10.0	0.000
/
\inboundscheckoff
\normale
\put{$\scc\bullet$} at 2.361 0
\put{$x_\Z$} [t] at 2.361 -.3
\put{\small $\simeq$ 2.361} [t] at 2.361 -.8
\setlinear 
\plot 2.361 0 2.361 2.408 0 2.408 /
\put{$\scc\bullet$} at 2.361 2.408
\put{$\scc\bullet$} at 0 2.408
\put{$e_\Z$} [r] at -.2 2.408
\put{\small These functions do not depend on} [l] at 3 7
\put{\small the normalization time $t_\sharp$.} [l] at 3 6.4
\put{$u=1$ \small matter} [l] at 3.25 5 
\put{$\scc\bullet$} at 1.8 5
\plot 1.8 5 3.15 5 /
\put{$u=\tfrac 43$ \small radiation} [l] at 3.25 4.2 
\put{$\scc\bullet$} at 1.45 4.2
\plot 1.45 4.2 3.15 4.2 /
\endpicture
$
\vskip -2mm
\caption{Graphs of $\epsilon(u,x)$.}
  \label{fig:ed}
\end{figure}

\br
The present-day value $\epsilon_\Z$ of the energy density does not depend on the barotropic parameter $u$. Due to \eqref{e:c0}, \eqref{e:def-c0} and \eqref{e:lambda/chi} we have
$$
\epsilon_\Z=2\,\ff c_\Z\,\dfrac\Lambda\chi\simeq
2*0.2299194811*5.238276649771\cdot 10^{-10}\,m^{-1}\,kg\,s^{-2}, 
$$
\be\label{e:e0}
\bboxed{\epsilon_\Z\simeq 2.408763697\cdot 10^{-10}\,m^{-1}\,kg\,s^{-2}}\quad\bullet
\ee
\erx

\section{The vanishing of the cosmological constant}

With $\lambda=0$ the dynamical equation \eqref{e:X1.} reads
$$
\dfrac{\dot a^2}{a^2}=\dfrac{\mu_\sharp}{a^{3u}}
$$
and is equivalent to
$$
\dfrac{da}{\sqrt{a^{2-3u}}}=\sqrt{\mu_\sharp}\,dt.
$$
Let us consider the case $u=1$ (dust matter):
$$
\sqrt a\,da=\sqrt{\mu_\sharp}\,dt \;\To\;
\tfrac 23\,a^{\frac 32}=\sqrt{\mu_\sharp}\,t \;\To\;
a^{\frac 32}=\tfrac 32\,\sqrt{\mu_\sharp}\,t \;\To\;
$$
$$
\bboxed{a(1;t,t_\sharp)=\sqrt[3]{\tfrac 94\,\mu_\sharp}\;t^{\frac 23}}
$$

\begin{figure} [H]
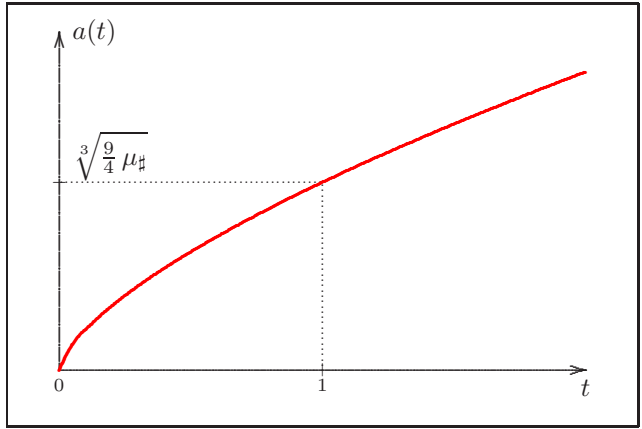

\centering
$
\beginpicture
\setcoordinatesystem units <3.5cm,2.5cm>
\setplotarea x from -.2 to 2.2, y from -.3 to 1.95
\normalgraphs
\grid 1 1
\arrow <6pt> [.2,.6] from 0 0 to 2 0
\arrow <6pt> [.2,.6] from 0 0 to 0 1.8
\put{$a(t)$} [l] at .05 1.8
\put{$\sc 0$} [t] at 0 -.05
\put{$\sc -$} at 0 1
\put{$\sqrt[3]{\tfrac 94\,\mu_\sharp}$} [lb] at .05 1.01
\put{$\scc |$} at 1 0
\put{$\sc 1$} [t] at 1 -.05
\put{$t$} [t] at 2 -.05
\setlinear
\setdots<2pt>
\plot 1 0 1 1 0 1 /
\setsolid
\inboundscheckon
\setquadratic
\setplotsymbol (\normalered) 
\plot
0.00	0.000
0.05	0.136
0.10	0.215
0.15	0.282
0.20	0.342
0.25	0.397
0.30	0.448
0.35	0.497
0.40	0.543
0.45	0.587
0.50	0.630
0.55	0.671
0.60	0.711
0.65	0.750
0.70	0.788
0.75	0.825
0.80	0.862
0.85	0.897
0.90	0.932
0.95	0.966
1.00	1.000
1.05	1.033
1.10	1.066
1.15	1.098
1.20	1.129
1.25	1.160
1.30	1.191
1.35	1.221
1.40	1.251
1.45	1.281
1.50	1.310
1.55	1.339
1.60	1.368
1.65	1.396
1.70	1.424
1.75	1.452
1.80	1.480
1.85	1.507
1.90	1.534
1.95	1.561
2.00	1.587
/
\inboundscheckoff
\endpicture
$
\vskip -2mm
\caption{\small Trend of $a(t)$ for $u=1$ and $\Lambda=0$.}
  \label{fig:l=0}
\end{figure}
The Hubble factor evolution is
$$
h=\dfrac{\dot a}{a}=\tfrac 23\,t^{-1}.
$$
The evaluation of this formula for $t=t_\Z$ (today) gives the age of the universe:
$$
t_\Z=\tfrac 23\,H_\Z^{-1}=\tfrac 23\,t^H_\Z\;  \quad\hbox{($t^H_\Z\Def H_\Z^{-1}$ is the {\bf Hubble time}) }
$$
Since $H_\Z\simeq 0.6882972691\cdot 10^{-10}\,yr^{-1}$ we get 
$$
\bboxed{t_\Z\simeq 9.685737494475 \cdot 10^{9}\,yr}
$$
which is far from the current estimate of the age of the universe. This is one of the many reasons that make it unacceptable to consider $\Lambda=0$.

\section{Superluminal recession speed and the Hubble radius}

Let $\ell_\sharp$ be the distance of two galaxies at the reference  time $t_\sharp$.  This distance evolves with time according to the law \eqref{e:ell-t} 
\be\label{e:sle-1.}
\ell(u;t,t_\sharp)=a(u;t,t_\sharp)\,\ell_\sharp,
\ee
with an expansion speed $\dot\ell(u;t,t_\sharp)=\dot a(u;t,t_\sharp)\,\ell_\sharp$ 
that may becomes greater than the light speed $c$. 

\bt\label{t:sle-1}
The recession speed $\dot\ell(u;t,t_\sharp)$ is superluminal  
$$
\dot\ell(u;t,t_\sharp)\geq c
$$
in the time interval defined by the inequality
\be\label{e:sle-2.}
\bboxed{(X+1)^{3u}\,(X-1)^{2-3u}\geq\ff C\,X \vphantom{\Def}}
\;\;\bboxed{X\Def e^{ux}} \;\;\bboxed{x\Def \beta t}
\ee
where the dimensionless constant $\ff C$ is defined by 
\be\label{e:sle-3.}
\ff C\Def4\cdot 3^{\tfrac{3u}{2}}\,\dfrac{\Lambda^{1-\tfrac{3u}{2}}}{\chi\,\epsilon_\sharp\,\ell_\sharp^{3u}}.
\ee
This constant does not depend on the choice of the reference time $t_\sharp$.
\et

\bprf
(i) $\;\dot\ell(u;t,t_\sharp)=\beta\,\ell'(u;x,x_\sharp)$ and 
$\ell'=a'\,\ell_\sharp$ $\To$
$$
\dot\ell\geq c \iff \ell'\geq\dfrac c\beta=\dfrac 1{\sqrt{3\Lambda}} \iff
a'\,\ell_\sharp\geq \dfrac 1{\sqrt{3\Lambda}}
\iff
$$
\be\label{e:sle-4.}
a'\geq \dfrac 1{\sqrt{3\Lambda}\;\ell_\sharp}.
\ee
(ii) \eqref{e:dadx}: $a'=\dfrac{da}{dx}=\tfrac 13\,a\,\dfrac {X+1}{X-1}$.  
\eqref{e:aux-e}: $a(u;x,x_\sharp)
=\left[\tfrac 14\,\dfrac\chi\Lambda\,\epsilon_\sharp\,\dfrac{(X-1)^2}{X}\right]\!{\vp}^{\tfrac 1{3u}}$ 
 
$\To$ $a'=\tfrac 13\,\left[\tfrac 14\,\dfrac\chi\Lambda\,\epsilon_\sharp\,\dfrac{(X-1)^2}{X}\right]\!{\vp}^{\tfrac 1{3u}}\,\dfrac {X+1}{X-1}$ 

$\To$ $[a']^{3u}=[\tfrac 13]^{3u}\,\tfrac 14\,\dfrac\chi\Lambda\,\epsilon_\sharp\,\dfrac{(X-1)^2}{X}\,\dfrac {(X+1)^{3u}}{(X-1)^{3u}}$.

(iii) \eqref{e:sle-4.} $\iff [a']^{3u}\geq \dfrac 1{[\sqrt{3\Lambda}]^{3u}\;\ell_\sharp^{3u}}$

$\iff\; [\tfrac 13]^{3u}\,\tfrac 14\,\dfrac\chi\Lambda\,\epsilon_\sharp\,\dfrac{(X-1)^2}{X}\,\dfrac {(X+1)^{3u}}{(X-1)^{3u}}\geq \dfrac 1{[\sqrt{3\Lambda}]^{3u}\;\ell_\sharp^{3u}}$

$\iff\; \dfrac{(X-1)^{2-3u}\,(X+1)^{3u}}{X}\geq 
\dfrac{4\cdot3^{3u}\;\Lambda}{\chi\,[\sqrt{3\Lambda}]^{3u}\;\epsilon_\sharp\,\ell_\sharp^{3u}}
=\dfrac{4\cdot 3^{\tfrac{3u}2}}{\chi\,\Lambda^{\tfrac{3u-2}2}\;\epsilon_\sharp\,\ell_\sharp^{3u}}$  $\To$
$$
\dot\ell\geq c \iff 
(X-1)^{2-3u}\,(X+1)^{3u}\geq \ff C\,X,
$$
where $\ff C$ is defined as in \eqref{e:sle-3.}. (iv) By virtue of \eqref{e:etat-t0} $\;\epsilon(t)\,a^{3u}(t,t_\sharp)\,\ell_\sharp^{3u}=\epsilon_\sharp\,\ell_\sharp^{3u}= \hbox{constant in $t$}$, i.e.
$$
\epsilon(t)\,\ell^{3u}(t)=\epsilon_\sharp\,\ell_\sharp^{3u}, \quad \forall \;t.
$$
This shows that the product $\epsilon_\sharp\,\ell_\sharp^{3u}$ 
does not depend on the choice of the reference time $t_\sharp$. Hence, also $\ff C$ is independent. \eprf

\bt
The constant $\ff C$ has the form
\begin{gather}
\label{e:sle-4*}
\bboxed{\ff C=\left(\dfrac{\ff L_\sharp}{\ell_\sharp}\right)\!\!{\vpt}^{3u}\rule[0mm]{0mm}{6mm}} 
\\
\label{e:sle-L}
\bboxed{\ff L_\sharp\Def
\sqrt{\dfrac 3\Lambda}\,
\left(\dfrac{(1-e^{ux_\sharp})^2}{e^{ux_\sharp}}\right)^{\tfrac 1{3u}}
=
\sqrt{\dfrac 3\Lambda}
\,\left[\vp 2\,\big(\cosh(ux_\sharp)-1\big)\right]^{\tfrac 1{3u}}}
\end{gather}
\et
\bprf \eqref{e:sle-3.}: $\ff C\Def4\cdot 3^{\tfrac{3u}{2}}\,\dfrac{\Lambda^{1-\tfrac{3u}{2}}}{\chi\,\epsilon_\sharp\,\ell_\sharp^{3u}}$.

\eqref{e:eu-x}: $\epsilon(u;x)=\dfrac{\Lambda}{\chi}\,\dfrac{4\,e^{ux}}{(1-e^{ux})^2}=\dfrac\Lambda\chi\,\dfrac 2{\cosh(ux)-1}$

$\ff C=4\cdot 3^{\tfrac{3u}{2}}\,\dfrac{\Lambda^{1-\tfrac{3u}{2}}}{\chi\,\ell_\sharp^{3u}}
\cdot\dfrac{\chi}{\Lambda}\,\dfrac{(1-e^{ux_\sharp})^2}{4\,e^{ux_\sharp}}
=3^{\tfrac{3u}{2}}\,\dfrac{\Lambda^{-\tfrac{3u}{2}}}{\ell_\sharp^{3u}}
\cdot\dfrac{(1-e^{ux_\sharp})^2}{e^{ux_\sharp}}
$

$
\ff C=\left(\dfrac 3\Lambda\right)\!\!{\vp}^{\tfrac{3u}2}\dfrac{(1-e^{ux_\sharp})^2}{e^{ux_\sharp}}\cdot\dfrac 1{\ell_\sharp^{3u}}
=\left[\left(\dfrac 3\Lambda\right)\!\!{\vp}^{\tfrac 12}
\left(\dfrac{(1-e^{ux_\sharp})^2}{e^{ux_\sharp}}\right)^{\tfrac 1{3u}}\cdot\dfrac 1{\ell_\sharp^{3u}}\right]^{3u} \To \eqref{e:sle-L}.
$

The alternative expression of $\ff L_\sharp$ follows from $2\,(\cosh(z)-1)=\dfrac{(e^z-1)^2}{e^z}$. \eprf

We can continue the analysis of the superluminal recession speed only with the specification of the barotropic parameter $u$. We will consider the case $u=1$: dust-matter universe.

\bt
For $u=1$ the superluminal expansion condition \eqref{e:sle-2.} is equivalent to
\be\label{e:sle-5.}
\bboxed{
\ba
\dot\ell(1;t,t_\sharp)\geq c 
\\
\iff 
f(\ff C;X)\Def X^3+(3-\ff C)\,X^2+(3+\ff C)\,X+1\geq 0
\ea}
\ee
with the constant $\ff C$ given by
\be\label{e:sle-C}
\bboxed{\ff C=\left(\dfrac{\ff L_\sharp}{\ell_\sharp}\right)\!\!{\vpt}^3\rule[-4mm]{0mm}{10.5mm}} 
\quad\bboxed{\ff L_\sharp=
\sqrt{\dfrac 3\Lambda}\,\sqrt[3]{\dfrac{(X_\sharp-1)^2}{X_\sharp}}\rule[-4mm]{0mm}{10.5mm}}
\quad\bboxed{\ba
X\Def e^x
\\
X_\sharp\Def e^{x_\sharp}
\ea
\rule[0mm]{0mm}{7mm}}
\ee
\et
\bprf
\eqref{e:sle-2.} $\To$ $(X+1)^3\,(X-1)^{-1}\geq\ff C\,X$
 $\iff$ $(X+1)^3\geq\ff C\,X\,(X-1)$

$\iff X^3+3\,X^2+3\,X+1\geq \ff C\,X^2-\ff C\,X$

$\iff X^3+3\,X^2-\ff C\,X^2+\ff C\,X+3\,X+1\geq 0$ $\iff$ \eqref{e:sle-5.}. \eprf

According to this theorem the analysis of the occurrence of the superluminal phenomenon is reduced to the analysis of the roots of the cubic polynomial $f(\ff C;X)$, whose coefficients depends on the constant $\ff C$. The graphs of $f(\ff C,X)$ are plotted in Fig.\ \ref{fig:sle-1} for various values of $\ff C$.  Some relevant facts must be highlighted:

\begin{figure} [H]
\centering
$
\beginpicture
\setcoordinatesystem units <.88cm,.3cm>
\setplotarea x from -.5 to 12.5, y from -7.5 to 18
\normalgraphs
\small
\putrectangle corners at -.7 20 and 12.5 -10.5
\arrow <6pt> [.2,.6] from -.5 0  to 12 0
\arrow <6pt> [.2,.6] from 0 -5  to 0 19
\put{$X=e^{\beta t}$} [lb] at 11.1 .5
\put{$f(\ff C;X)$} [lt] at .2 19
\put{$\scc |$} at 1 0
\put{$\scc |$} at 2 0
\put{$\scc |$} at 3 0
\put{$\scc |$} at 4 0
\put{$\scc |$} at 5 0
\put{$\scc |$} at 6 0
\put{$\scc |$} at 7 0
\put{$\scc |$} at 8 0
\put{$\scc |$} at 9 0
\put{$\scc |$} at 10 0
\put{$\sc 0$} [lt] at .05 -.25 
\put{$\sc 1$} [t] at 1.1 -.5
\put{$\sc 2$} [t] at 2 -.5
\put{$\sc 3$} [t] at 3 -.5
\put{$\sc 4$} [t] at 4 -.5
\put{$\sc 5$} [t] at 5 -.5
\put{$\sc 6$} [t] at 6 -.5
\put{$\sc 7$} [t] at 7 -.5
\put{$\sc 9$} [t] at 9 -.5
\put{$\sc 10$} [t] at 10 -.5
\put{$\sc 1$} [r] at -.1 1
\put{$-$}  at 0 1
\put{$\sc 4$} [r] at -.1 4
\put{$\sc -$}  at 0 4
\put{$\sc 8$} [r] at -.1 8
\put{$\sc -$}  at 0 8
\put{$\sc 12$} [r] at -.1 12
\put{$\sc -$}  at 0 12
\put{$\sc 16$} [r] at -.1 16
\put{$\sc -$}  at 0 16
\setlinear
\setshadegrid span <2pt>
\vshade 0 -4 18 1 -4 18 /
\put{$t<0$} at .5 -2
\plot 0 8 1 8 /
\plot 1 -4 1 18 /
\inboundscheckon
\setquadratic
\normale
\put{\ssmall $\ff C=$ 6} [r] at 2 16
\plot  
-0.4	-3.144
-0.2	-0.928
0.0	1.000
0.2	2.688
0.4	4.184
0.6	5.536
0.8	6.792
1.0	8.000
1.2	9.208
1.4	10.464
1.6	11.816
1.8	13.312
2.0	15.000
2.2	16.928
2.4	19.144
2.6	21.696
2.8	24.632
3.0	28.000
3.2	31.848
3.4	36.224
3.6	41.176
3.8	46.752
4.0	53.000
4.2	59.968
4.4	67.704
4.6	76.256
4.8	85.672
5.0	96.000
5.2	107.288
5.4	119.584
5.6	132.936
5.8	147.392
6.0	163.000
/
\normale
\put{\ssmall 10} [r] at 4.3 8
\plot 
-0.4	-5.384
-0.2	-1.888
0.0	1.000
0.2	3.328
0.4	5.144
0.6	6.496
0.8	7.432
1.0	8.000
1.2	8.248
1.4	8.224
1.6	7.976
1.8	7.552
2.0	7.000
2.2	6.368
2.4	5.704
2.6	5.056
2.8	4.472
3.0	4.000
3.2	3.688
3.4	3.584
3.6	3.736
3.8	4.192
4.0	5.000
4.2	6.208
4.4	7.864
4.6	10.016
4.8	12.712
5.0	16.000
5.2	19.928
5.4	24.544
5.6	29.896
5.8	36.032
6.0	43.000
6.2	50.848
6.4	59.624
/
\normale
\put{\ssmall 8} [r] at 2.5 13.5
\plot 
-0.4	-4.264
-0.2	-1.408
0.0	1.000
0.2	3.008
0.4	4.664
0.6	6.016
0.8	7.112
1.0	8.000
1.2	8.728
1.4	9.344
1.6	9.896
1.8	10.432
2.0	11.000
2.2	11.648
2.4	12.424
2.6	13.376
2.8	14.552
3.0	16.000
3.2	17.768
3.4	19.904
3.6	22.456
3.8	25.472
4.0	29.000
4.2	33.088
4.4	37.784
4.6	43.136
4.8	49.192
5.0	56.000
5.2	63.608
5.4	72.064
5.6	81.416
5.8	91.712
6.0	103.000
6.2	115.328
6.4	128.744
/
\normale
\put{\ssmall 9} [r] at 3.05 10.8
\plot 
-0.4	-4.824
-0.2	-1.648
0.0	1.000
0.2	3.168
0.4	4.904
0.6	6.256
0.8	7.272
1.0	8.000
1.2	8.488
1.4	8.784
1.6	8.936
1.8	8.992
2.0	9.000
2.2	9.008
2.4	9.064
2.6	9.216
2.8	9.512
3.0	10.000
3.2	10.728
3.4	11.744
3.6	13.096
3.8	14.832
4.0	17.000
4.2	19.648
4.4	22.824
4.6	26.576
4.8	30.952
5.0	36.000
5.2	41.768
5.4	48.304
5.6	55.656
5.8	63.872
6.0	73.000
/
\put{$\sc\bullet$} at 3.7324 0
\put{$X_\Delta$} [t] at 3.7324 -.7
\put{\small 3.\ssmall 7324} [t] at 3.72 -1.8
\setplotsymbol (\normalered)
\setquadratic
\plot 
-0.4	-5.602
-0.2	-1.982
0.0	1.000
0.2	3.390
0.4	5.238
0.6	6.590
0.8	7.494
1.0	8.000
1.2	8.154
1.4	8.006
1.6	7.602
1.8	6.990
2.0	6.220
2.2	5.338
2.4	4.394
2.6	3.434
2.8	2.506
3.0	1.660
3.2	0.942
3.4	0.402
3.6	0.086
3.8	0.042
4.0	0.320
4.2	0.966
4.4	2.030
4.6	3.558
4.8	5.598
5.0	8.200
5.2	11.410
5.4	15.278
5.6	19.850
5.8	25.174
6.0	31.300
/
\normale
\put{\ssmall 11} [l] at 5.7 5.5
\plot
-0.4	-5.9440
0.0	1.0000
0.4	5.3840
0.8	7.5920
1.2	8.0080
1.6	7.0160
2.0	5.0000
2.4	2.3440
2.8	-0.5680
3.2	-3.3520
3.6	-5.6240
4.0	-7.0000
4.4	-7.0960
4.8	-5.5280
5.2	-1.9120
5.6	4.1360
6.0	13.0000
6.4	25.0640
6.8	40.7120
/
\put{\ssmall 13} [l] at 8.1 4
\plot
-0.4	-7.0640
0.0	1.0000
0.4	5.8640
0.8	7.9120
1.2	7.5280
1.6	5.0960
2.0	1.0000
2.4	-4.3760
2.8	-10.6480
3.2	-17.4320
3.6	-24.3440
4.0	-31.0000
4.4	-37.0160
4.8	-42.0080
5.2	-45.5920
5.6	-47.3840
6.0	-47.0000
6.4	-44.0560
6.8	-38.1680
7.2	-28.9520
7.6	-16.0240
8.0	1.0000
8.4	22.5040
/
\put{$\sc\bullet$} at 2.72 0
\put{$X_\uno$} [rt] at 2.83 -.6
\put{$\sc\bullet$} at 5.35 0
\put{$X_\due$} [lt] at 5.3 -.6
\setplotsymbol (\normaleblue)
\plot 
0.0	1.000
0.2	4.183
0.4	6.426
0.6	7.778
0.8	8.287
1.0	8.000
1.2	6.966
1.4	5.232
1.6	2.847
1.8	-0.142
2.0	-3.686
2.2	-7.738
2.4	-12.248
2.6	-17.171
2.8	-22.457
/
\plot
10.0	-49.870
10.2	-34.859
10.4	-18.388
10.6	-0.408
10.8	19.129
11.0	40.270
11.2	63.064
/
\inboundscheckoff
\put{$\sc\bullet$} at 10.6 0
\put{$X_\Z=e^{x_\Z}$} [lt] at 10.7 -.7
\put{\small 10.\ssmall 6043} [lt] at 10.7 -1.85
\put{\small today} [lt] at 10.7 -2.9
\put{$f(\ff C_\Delta;X)$} [lt] at 6.4 14
\normale
\setlinear
\plot 6.3 13.2 5.25 12.5 /
\put{$\scc\bullet$} at 5.25 12.5
\setlinear
\put{\small $\ff C_\Delta \sc =$ \ssmall 10.\ssmall 3923} [l] at 6.3 12
\put{$f(\ff C_H;X)$} [rb] at 10.1 7
\put{$\scc\bullet$} at 10.65 6.5
\plot 10.2 7 10.65 6.5 /
\put{\ssmall 15.343} [l] at 10.8 4.4
\put{$\ff C_H=$} [lb] at 10.8 5

\put{$\sc\bullet$} at 1.791 0
\arrow <5pt> [.2,.6] from 1 -5.5  to 1.77 -.3
\put{first root $X_\bigstar$} [t] at 1 -5.7
\put{of $f(\ff C_H;X)$} [t] at 1 -7
\put{$\simeq$ 1.791299} [t] at 1 -8.5

\put{second root $X_0$} [t] at 9.2 -5.7
\put{of $f(\ff C_H;X)$} [t] at 9.2 -7
\arrow <5pt> [.2,.6] from 9.2 -5.5  to 10.5 -.3

\endpicture
$
\vskip -1mm
\caption{Graphs of $f(\ff C;X)$.}
 \label{fig:sle-1}
\end{figure}

\begin{enumerate}
  \item As shown by \eqref{e:sle-C} $\ff C$ is the cube of the ratio of two lengths, $\ff L_\sharp$ and $\ell_\sharp$. $\ff L_\sharp$ has the property of being computable regardless of the given value of galactic distance $\ell_\sharp$ (see item 7 below).  

\item  The analysis makes sense only in the interval $X=e^x\geq 1$, corresponding to $t\geq 0$, since $t=0$ is the date of birth of the universe.
  
 \item Whatever $\ff C$, $f(\ff C,0)=1$ and $f(\ff C,1)=8$: all graphs pass through the points (0,1) and (1,8). Furthermore, $f(\ff C,X)$ has a real negative root close to $X=0$ and the other real roots (if any) are located in the unbounded interval $X>1$.

  \item There exists a discriminant value 
$$
\boxed{\ff C_\Delta\simeq 10.3923}
$$ 
for which $f(\ff C_\Delta,X)$ is tangent to the $X$-axis at a point $X_\Delta\simeq 3.7324$. The point $X_\Delta$ corresponds to the value\footnote{
\kern 2pt $X_\Delta=e^{\beta t_\Delta}$ $\To$ $\beta\,t_\Delta=\log X_\Delta\simeq 1.317051$ $\To$ $t_\Delta=\beta^{-1}\,\log X_\Delta\simeq 7.706513$.}
$$
\boxed{t_\Delta\simeq 7.706513\;Gyr}
$$
of the cosmic time. 

\item For $\ff C<\ff C_\Delta$, there are no real roots $X>1$ of $f(\ff C;X)$  and we have a permanent superluminal recession speed. For $\ff C>\ff C_\Delta$ the polynomial $f(\ff C;X)$ has two (positive) simple roots $X_1<X_2$. The recession speed is subluminal in the interval $(X_1,X_2)$ delimited by these roots and containing $X_\Delta$. 
   
\item There exists a special value $\ff C_H>\ff C_\Delta$ given by\footnote{
\kern 2pt \bprf
 $f(\ff C,X_\Z)\Def X_\Z^3+(3-\ff C)\,X_\Z^2+(3+\ff C)\,X_\Z+1$ 
with $X\Def e^{x_\Z}$, $x_\Z\Def \beta t_\Z$.
\vskip 2pt
$
\ba 
f(\ff C,X_\Z)=0 \iff X_\Z^3+(3-\ff C)\,X_\Z^2+(3+\ff C)\,X_\Z+1=0
\ac
\iff X_\Z^3+3\,X_\Z^2-\ff C\,X_\Z^2+3\,X_\Z+\ff C\,X_\Z+1=0
\ac
\iff X_\Z^3+3\,X_\Z^2+3\,X_\Z+1=\ff C\,X_\Z\,(X_\Z-1) 
\iff (X_\Z+1)^3=\ff C\,X_\Z\,(X_\Z-1)
\ax
\iff \ff C_H=\dfrac{(X_\Z+1)^3}{X_\Z\,(X_\Z-1)}. \quad \ff C_H\simeq \dfrac{10.60439692+1)^3}{(10.60439692\cdot(10.60439692-1)}
\ax
{}\simeq 15.34304824.
\ea
$

The estimate \eqref{e:Xstar} of the first root $X_1<X_\Delta$ is a matter of numerical analysis. 

It follows that 
$$
t_\star=\beta^{-1}\log X_\star
\simeq \dfrac{0.582941054}{0.170901087343}\;Gyr\simeq 3.41098505\;Gyr. \eprfx
$$
}
\be\label{e:ell-3}
\bboxed{\ff C_H=\dfrac{(e^{x_\Z}+1)^3}{e^{x_\Z}\,(e^{x_\Z}-1)}\simeq 15.34304824}
\ee
such that the polynomial $f(\ff C_H,X)$ has a root  
\be
X_0\simeq 10.604396924>X_\Delta
\ee  
corresponding to the present-day time 
\be
t_\Z\simeq 13.816581\;Gyr \quad (x_\Z \simeq 2.3612687193) 
\ee
and a root 
\be\label{e:Xstar}
X_\star\simeq 1.791299<X_\Delta
\ee
corresponding to the cosmic time 
\be\label{e:tstar}
t_\star\simeq  3.41098505\;Gyr \quad (x_\star \simeq 0.582941054).
\ee

\item Since the definition of $\ff C$ does not depend on the choice of the reference time $t_\sharp$ (Theorem \ref{t:sle-1}) we can write the definition \eqref{e:sle-C} for $\ff C=\ff C_H$ by taking the reference times $t_\star$ and $t_\Z$ of item 6 above:
\be\label{e:sle-CH}
\bc
\ff C_H=\left(\dfrac{\ff L(t_\star)}{\ell(t_\star)}\right)\!\!{\vpt}^3,
\quad\ff L(t_\star)=\sqrt{\dfrac 3\Lambda}\,\sqrt[3]{\dfrac{(X_\star-1)^2}{X_\star}}.
\ac
\ff C_H=\left(\dfrac{\ff L(t_\Z)}{\ell(t_\Z)}\right)\!\!{\vpt}^3,
\quad\ff L(t_\Z)=\sqrt{\dfrac 3\Lambda}\,\sqrt[3]{\dfrac{(X_\Z-1)^2}{X_\Z}}.
\ec
\ee
Since $\ff L(t_\star)$ and $\ff L(t_\Z)$ are computable (see below), and $\ff C_H$ is known (item 4), from \eqref{e:sle-CH} we can derive the lengths $\ell(t_\star)$ and $\ell(t_\Z)$:
\be\label{e:sle-ellH}
\bc
\ell(t_\star)=\dfrac{\ff L(t_\star)}{\sqrt[3]{\vphantom\sum\ff C_H}}, 
\\[5mm]
\ell(t_\Z)=\dfrac{\ff L(t_\Z)}{\sqrt[3]{\vphantom\sum \ff C_H}},
\ec
\quad \ff C_H=\dfrac{(e^{x_\Z}+1)^3}{e^{x_\Z}\,(e^{x_\Z}-1)}\simeq 15.34304824.
\ee
\end{enumerate}

\begin{center}
\fbox{\begin{minipage}{.95\linewidth} \vspu
For the way in which it has been defined, $\ell(t_\Z)$ is the present-day distance of two galaxies crossing the boundary beyond which the recession velocity exceeds the speed of light.  This boundary is called {\bf Hubble radius} (of the {\bf Hubble sphere}).  In turn, $\ell(t_\star)$ is the distance at the time $t_*$ at the early universe when the recession speed of these two galaxies crossed this boundary in the opposite sense: from superluminal to subluminal recession speed. \vspd
\end{minipage}}
\end{center}
 
\br {\bf Computation of $\ell(t_\star)$ and $\ell(t_\Z)$}.

1. $\Lambda\simeq 1.087769524444422834\cdot 10^{-52}\,m^{-2}$ (table \ref{tab:Data-supp}) $\To$

$\To$ $\dfrac 3\Lambda\simeq 2.7579371664528356 \cdot 10^{52}\,m^2$ 

$\To$ $\sqrt{\dfrac 3\Lambda} \simeq 1.66070381659489\cdot 10^{26}\,m =1.66070381659489\cdot 10^{23}\,km$.

Conversion to light-years: $10^{23}\,km\simeq 10.570234105227\,Gly$ $\To$ 

$\bboxed{\sqrt{\dfrac 3\Lambda} \simeq 17.554028120851\,Gly}$ 

2. $\ff C_H\simeq 15.34304824$ $\To$ $\bboxed{\sqrt[3]{\vphantom\sum \ff C_H}\simeq 2.4848712052}$

3. $X_\star \simeq 1.791299$ $\To$ $\dfrac{(X_\star-1)^2}{X_\star}\simeq 
0.349553$ 
$\To$ $\sqrt[3]{\dfrac{(X_\star-1)^2}{X_\star}}\simeq 0.7044297$

$\To$ $\ff L(t_\star)\simeq 17.55402812\cdot 0.7044297\;Gly$ $\To$
 $\bboxed{\ff L(t_\star)\simeq 12.36557\;Gly}$

4. $\ell(t_\star)=\dfrac{\ff L(t_\star)}{\sqrt[3]{\vphantom\sum\ff C_H}} \simeq\dfrac{12.36557}{2.4848712052}\;Gly$ $\To$
$\bboxed{\ell(t_\star) \simeq 4.976342\;Gly}$

5. $X_0\!\simeq\! 10.604396924 \To \dfrac{(X_\Z-1)^2}{X_\Z}\!\simeq\! 
8.69869743 \!\To\!\sqrt[3]{\dfrac{(X_\Z-1)^2}{X_\Z}}\!\simeq 2.056607463$
$\To$ $\ff L(t_\Z)\simeq 17.55402812\cdot 2.056607463 \;Gly$ $\To$

 $\bboxed{\ff L(t_\Z)\simeq 36.10174523\;Gly}$

6. $\ell(t_\Z)=\dfrac{\ff L(t_\Z)}{\sqrt[3]{\vphantom\sum\ff C_H}} \simeq\dfrac{36.10174523}{2.4848712052}\;Gly$ $\To$
$\bboxed{\ell(t_\Z) \simeq 14.52861828\;Gly}$ \er

We  obtain the evolution of the Hubble radius in the cosmic time by suppressing the subscript 0 in \eqref{e:sle-CH} and \eqref{e:sle-ellH}:

$\left.\ba 
\ff L(t)=\sqrt{\dfrac 3\Lambda}\,\sqrt[3]{\dfrac{(e^x-1)^2}{e^x}}
\ac 
\ff C_H=\dfrac{(e^x+1)^3}{e^x\,(e^x-1)}
\ea\right\}
$ 
$\To\,\ell_H(t)=\dfrac{\ff L(t)}{\sqrt[3]{\vphantom\sum \ff C_H}}
=\sqrt{\dfrac 3\Lambda}\,\dfrac
{\sqrt[3]{(e^x-1)^3}}
{\sqrt[3]{(e^x+1)^3}}\,\To$
\be\label{e:lHt}
\bboxed{\ell_H(t)
=\sqrt{\dfrac 3\Lambda}\,\dfrac
{e^{\beta t}-1}{e^{\beta t}+1}}
\ee
\begin{figure} [H]
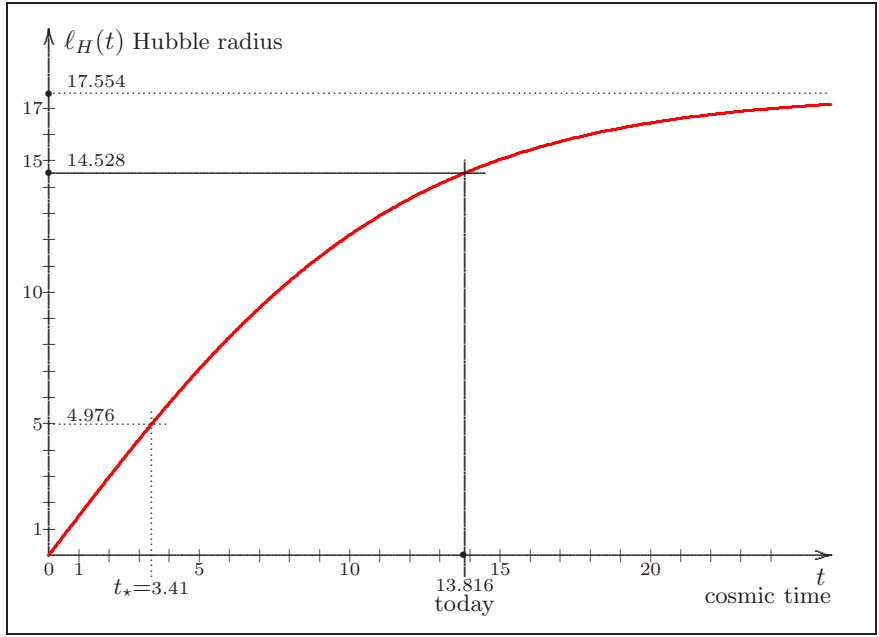

\centering
$
\beginpicture
\setcoordinatesystem units <.4cm,.35cm>
\setplotarea x from 0 to 26, y from -2 to 21
\normalgraphs
\putrectangle corners at -1.4 21 and 27.5 -3
\arrow <6pt> [.2,.6] from 0 0 to 26 0
\arrow <6pt> [.2,.6] from 0 0 to 0 20
\put{$\sc 0$} [t] at 0 -.3
\put{$\sc -$} at 0 1
\put{$\sc -$} at 0 2
\put{$\sc -$} at 0 3
\put{$\sc -$} at 0 4
\put{$\sc -$} at 0 5
\put{$\sc -$} at 0 6
\put{$\sc -$} at 0 7
\put{$\sc -$} at 0 8
\put{$\sc -$} at 0 9
\put{$\sc -$} at 0 10
\put{$\sc -$} at 0 11
\put{$\sc -$} at 0 12
\put{$\sc -$} at 0 13
\put{$\sc -$} at 0 14
\put{$\sc -$} at 0 15
\put{$\sc -$} at 0 16
\put{$\sc -$} at 0 17
\put{$\sc 1$} [r] at -.2 1
\put{$\sc 5$} [r] at -.2 5
\put{$\sc 10$} [r] at -.2 10
\put{$\sc 15$} [r] at -.2 15
\put{$\sc 17$} [r] at -.2 17
\put{$\scc |$} at 1 0
\put{$\scc |$} at 2 0
\put{$\scc |$} at 3 0
\put{$\scc |$} at 4 0
\put{$\scc |$} at 5 0
\put{$\scc |$} at 6 0
\put{$\scc |$} at 7 0
\put{$\scc |$} at 8 0
\put{$\scc |$} at 9 0
\put{$\scc |$} at 10 0
\put{$\scc |$} at 11 0
\put{$\scc |$} at 12 0
\put{$\scc |$} at 13 0
\put{$\scc |$} at 14 0
\put{$\scc |$} at 15 0
\put{$\scc |$} at 16 0
\put{$\scc |$} at 17 0
\put{$\scc |$} at 18 0
\put{$\scc |$} at 19 0
\put{$\scc |$} at 20 0
\put{$\scc |$} at 21 0
\put{$\scc |$} at 22 0
\put{$\scc |$} at 23 0
\put{$\scc |$} at 24 0
\put{$\sc 1$} [t] at 1 -.3
\put{$\sc 5$} [t] at 5 -.3
\put{$\sc 10$} [t] at 10 -.3
\put{$\sc 15$} [t] at 15 -.3
\put{$\sc 20$} [t] at 20 -.3
\inboundscheckon
\setquadratic
\setplotsymbol (\normalered) 
\plot
0	0.000
1	1.496
2	2.971
3	4.404
4	5.777
5	7.075
6	8.286
7	9.404
8	10.425
9	11.347
10	12.172
11	12.906
12	13.553
13	14.120
14	14.614
15	15.043
16	15.413
17	15.732
18	16.006
19	16.240
20	16.440
21	16.610
22	16.755
23	16.878
24	16.983
25	17.071
26	17.146
27	17.210
28	17.263
29	17.309
30	17.347
/
\inboundscheckoff
\normale
\setlinear
\plot 0 14.528  14.5 14.528 /
\plot  13.816 -.8  13.816 15 /
\put{$\scc\bullet$} at 13.78 0
\put{$\scc\bullet$} at 0 14.528
\put{\ssmall 13.816} [t] at 13.816 -.9
\put{\small today} [t] at 13.816 -1.5
\put{\ssmall 14.528} [lb] at .6 14.8
\put{$\scc\bullet$} at 0 17.554
\put{\ssmall 17.554} [lb] at .6 17.8
\setdots <2pt>
\plot 0 17.554 26 17.554 /

\plot 3.41 -.8 3.41 5.5 /
\plot 0 4.976 4 4.976 /
\put{\small $t_\star\!\!=$\ssmall 3.41} [t] at 3.4 -.9
\put{\ssmall 4.976} [lb] at .6 5.1
\put{$\ell_H(t)\;$\small Hubble radius} [l] at .5 19.5
\put{$t$} [lt] at 25.5 -.5
\put{\small cosmic time} [rt] at 26 -1.2
\endpicture
$
\vskip -1mm
\caption{Graph of the Hubble radius $\ell_H(t)$.}
  \label{fig:Hr}
\end{figure}


\chapter{Transmission of photons}

A human lives in a celestial body $O$ and observes in its sky another celestial body $B$ by means of optical or electro-magnetic devices. This requires that $B$ is capable of emitting electro-magnetic signals in form of `photons' in a broad sense i.e., in form of particles whose world-lines in space-time are light-like geodesics. We also assume that the world-lines of $B$ and $O$ are time-like geodesics belonging to the galactic fluid.\footnote{
\kern 2pt In other words, we assume that $B$ and $O$ have to be considered as particles of the galactic fluid.} 
 The observer asks a number of questions:

 How far is $B$? On what date the photons that I am receiving now  have been issued by $B$? When $B$ has appeared in my sky? As long as it will be visible? 

The focus of this chapter is to give an answer to these (and related) questions.

\section{Emission-reception of photons}\label{s:erp}
A photon is emitted by $B$ at the time $t_e$ and received by $O$ at the time $t_r$. Its world-line is depicted in Fig.\ \ref{fig:erp-0}. At any intermediate time $t$ (in the figure two of them are marked: $t_\uno<t_\due$) a length $\ell(t_e,t)$ is defined: it is the distance traveled by the photon till the time $t$ measured on the spatial section $S_t$.\footnote{
\kern 2pt It is a synchronous distance at the time $t$, as defined in  Section \ref{s:Hl}.} 
 
If we look at the photon progression through the cosmic monitor, i.e.\ on the quotient manifold (Fig.\ \ref{fig:erp-00}) then we observe that it moves along a geodesic joining $B$ to $O$ with a speed given by equation \eqref{e:spepar1}:
\be\label{e:erp-1.}
v(t)=\dfrac{ds}{dt}=\dfrac{c}{a(t,t_\sharp)}.
\ee

On the other hand, the synchronous traveled distance in space-time 
(depicted in Fig. \ref{fig:erp-0}) is given by
\be\label{e:erp-2.}
\ell(t_e,t)= a(t,t_\sharp)\,\ell(t_e,t;t_\sharp),
\ee

\begin{figure} [H]
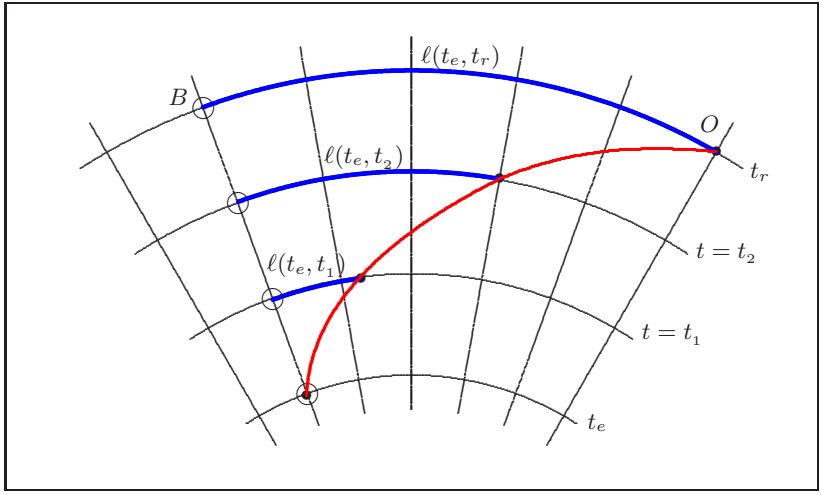

  \centering
$
\beginpicture
\setcoordinatesystem units <.9cm,.9cm>
\setplotarea x from -6 to 6, y from -.2 to 7
\normalgraphs
\small
\grid 1 1
\setlinear
\plot 0 1 0 6.5 /
\startrotation by .984 -.173 about 0 -3 
\plot 0 1 0 6.5 /
\stoprotation
\startrotation by .939 -.342 about 0 -3 
\plot 0 1 0 6.5 /
\stoprotation
\startrotation by .866 -.5 about 0 -3 
\plot 0 1 0 6.5 /
\stoprotation
\setlinear
\plot 0 1 0 6.5 /
\startrotation by .984 .173 about 0 -3 
\plot 0 1 0 6.5 /
\stoprotation
\startrotation by .939 .342 about 0 -3 
\plot 0 1 0 6.5 /
\put{\small $\bigcirc$} at 0 1.5
\put{\small $\bigcirc$} at 0 3
\put{\small $\bigcirc$} at 0 4.5
\put{\small $\bigcirc$} at 0 6
\stoprotation
\startrotation by .866 .5 about 0 -3 
\plot 0 1 0 6.5 /
\stoprotation
\circulararc 33 degrees from 0 1.5 center at 0 -3
\circulararc -33 degrees from 0 1.5 center at 0 -3
\circulararc 33 degrees from 0 3 center at 0 -3
\circulararc -33 degrees from 0 3 center at 0 -3
\circulararc 33 degrees from 0 4.5 center at 0 -3
\circulararc -33 degrees from 0 4.5 center at 0 -3
\circulararc 33 degrees from 0 6 center at 0 -3
\circulararc -33 degrees from 0 6 center at 0 -3
\put{$B$} [br] at -3.3 5.5
\put{$O$} [br] at 4.55 5.1
\put{$\bullet$} at -1.55 1.2
\put{$\bullet$} at -.75 2.93
\put{$\bullet$} at 1.3 4.4
\put{$\bullet$} at 4.5 4.8
\setplotsymbol (\spessinoblue)
\circulararc 13 degrees from -.75 2.93 center at 0 -3
\circulararc 30 degrees from 1.3 4.4 center at 0 -3
\circulararc 50 degrees from 4.5 4.8 center at 0 -3
\put{$t_e$} [l] at 2.6 .8
\put{$t=t_\uno$} [l] at 3.4 2.1
\put{$t=t_\due$} [l] at 4.2 3.3
\put{$t_r$} [l] at 5 4.5
\put{$\ell(t_e,t_\uno)$} [b] at -1.55 2.95
\put{$\ell(t_e,t_\due)$} [b] at -.7 4.55
\put{$\ell(t_e,t_r)$} [b] at .73 6.03
\setplotsymbol (\normalered)
\setquadratic
\plot
 -1.55 1.2
-.8 2.9
1.3 4.4
2.8 4.8
4.5 4.8 
/
\normale
\endpicture
$
\vskip -1mm
  \caption{Photon world-line from $B$ to $O$.}
  \label{fig:erp-0}
\end{figure}
\vskip -3mm
\begin{figure} [H]
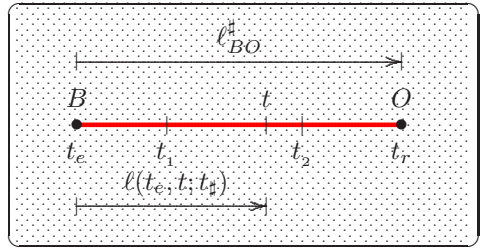

  \centering
$
\beginpicture
\setcoordinatesystem units <1.2cm,1.2cm>
\setplotarea x from -3.3 to 3.4, y from -1.3 to 1.3
\normalgraphs
\put{
\cornersize{.1}
\ovalbox{
\begin{minipage}{5.9cm}
\begin{minipage}{5cm}
\rule[-3cm]{5mm}{0cm}
\end{minipage}
\end{minipage}}
} at 0 0
\setshadegrid span <2pt>
\setlinear
\vshade -2.5 -1.4 1.4 2.6 -1.4 1.4 /
\setplotsymbol (\spessinored)
\setlinear
\plot -1.8 0 1.8 0 /
\put{$\bullet$} at -1.8 0
\put{$\bullet$} at 1.8 0
\put{$B$} [b] at -1.8 .2
\put{$O$} [b] at 1.8 .2
\put{$t_e$} [t] at -1.8 -.2
\put{$t_r$} [t] at 1.8 -.2
\put{$\sc |$} at -.8 0
\put{$\sc |$} at .7 0
\put{$t_\uno$} [t] at -.8 -.2
\put{$t_\due$} [t] at .7 -.2
\normale
\arrow <6pt> [.2,.6] from -1.8 -.9 to .3 -.9
\put{$\sc |$} at .3 0
\put{$\sc |$} at -1.8 -.9
\put{$\sc |$} at .3 -.9 
\put{$t$} [b] at .3 .2
\put{$\ell(t_e,t;t_\sharp)$} [b] at -.7 -.8
\arrow <6pt> [.2,.6] from -1.8 .7 to 1.8 .7
\put{$\sc |$} at -1.8 .7
\put{$\sc |$} at 1.8 .7
\put{$\ell^\sharp_{BO}$} [b] at 0 .8
\endpicture
$
\vskip -1mm
  \caption{Photon trajectory observed on the cosmic monitor.}
  \label{fig:erp-00}
\end{figure} 

\noindent where $\ell(t_e,t;t_\sharp)$ is the traveled distance on the cosmic monitor, whose metric is that of the spatial section at the reference time $t_\sharp$.\footnote{
\kern 2pt The symbol $t_\sharp$ has been inserted in this notation to emphasize the dependence on the reference time of the scale factor.} 
 Then at the time $t>t_e$ the traveled distance on the monitor is given by
\be\label{e:erp-1}
\ell(t_e,t;t_\sharp)=c\dint_{t_e}^t\dfrac {dt'}{a(t',t_\sharp)}.
\ee

Consequently, the photon will reach $O$ at the time $t_r$ if and only if
\be\label{e:erp-1*}
\bboxed{\ell(t_e,t_r;t_\sharp)\Def c\dint_{t_e}^{t_r}\dfrac {dt'}{a(t',t_\sharp)}}=
\bboxed{\ell^\sharp_{BO}\Def
\hbox{distance from $B$ to $O$}}
\ee
provided that the integral exists as a finite number. In this case we say that $B$ is {\bf visible to  $O$ at the time $t_r$}. This property is independent from the choice of $t_\sharp$ and  has a significant geometrical interpretation illustrated in Fig.\ \ref{fig:erp-2}: the shaded area delimited by the graph of $c/a(t,t_\sharp)$ upon the emission-reception interval  $[t_e,t_r]$ represents the integral \eqref{e:erp-1*} hence the co-moving distance $\ell^\sharp_{BO}$. Note that, from the dimensional view-point, this `area' is in fact a \underbar{time} times a \underbar{velocit}y quantity, i.e.\ a length-dimensional quantity.

\begin{figure} [H]
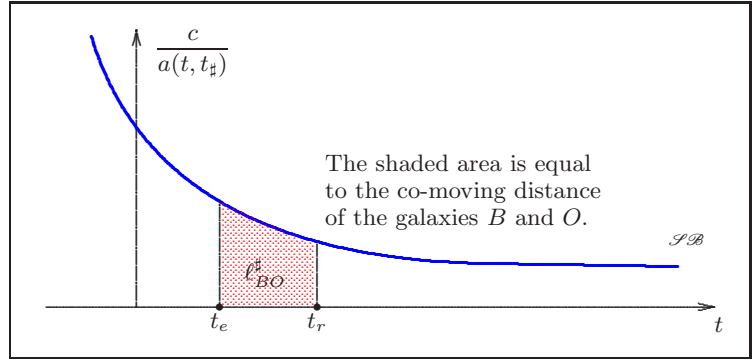
 
  \centering
$
\beginpicture
\setcoordinatesystem units <2.4cm,.45cm>
\setplotarea x from 0 to 4.5, y from -1.8 to 9
\normalgraphs
\small
\putrectangle corners at .3 9 and 4.4 -1.5
\arrow <6pt> [.2,.5] from .5 0 to 4.2 0
\arrow <6pt> [.2,.5] from 1 0 to 1 8.2
\put{$t$} [lt] at 4.2 -.3
\put{$\dfrac c{a(t,t_\sharp)}$} [l] at 1.1 7.5
\setplotsymbol(\normaleblue)
\setquadratic
\plot
.75 8
1.5 3
3 1.28
3.4 1.25
4 1.2
/
\normale
\setlinear
\plot 2 0 2 1.9 /
\put{$\sc\bullet$} at 2 0
\put{$t_r$} [t] at 2 -.2
\plot 1.46 0 1.46 3.15 /
\put{$\sc\bullet$} at 1.46 0
\put{$t_e$} [t] at 1.46 -.2
\textcolor{red}{
\setshadegrid span <1pt>
\vshade 1.46 0 3.12 2 0 1.8  /}
\put{$\ell^\sharp_{BO}$} [lb] at 1.55 .6
\put{The shaded area is equal} [l] at 2 4.2
\put{to the co-moving distance} [l] at 2 3.4
\put{of the galaxies $B$ and $O$.} [l] at 2 2.6
\put{\firma} at 4 2
\endpicture
$
\vskip -2mm 
  \caption{\small Geometrical interpretation of equation \eqref{e:erp-1*}.}
  \label{fig:erp-2}
\end{figure}
\vskip -3mm
\begin{figure} [H]
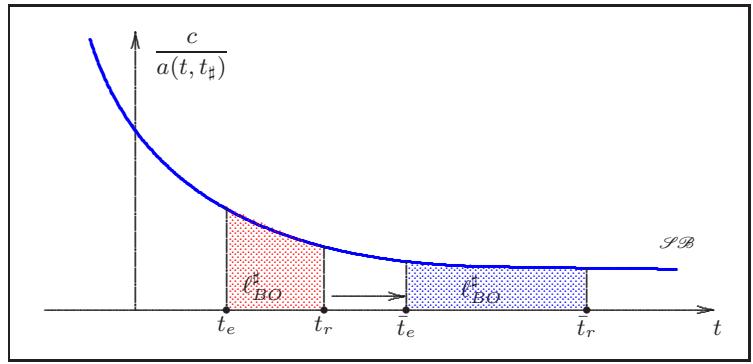
 
  \centering
$
\beginpicture
\setcoordinatesystem units <2.4cm,.45cm>
\setplotarea x from 0 to 4.5, y from -1.7 to 9
\normalgraphs
\small
\putrectangle corners at .3 9 and 4.4 -1.5
\arrow <6pt> [.2,.5] from .5 0 to 4.2 0
\arrow <6pt> [.2,.5] from 1 0 to 1 8.2
\put{$t$} [lt] at 4.2 -.3
\put{$\dfrac c{a(t,t_\sharp)}$} [l] at 1.1 7.5
\setplotsymbol(\normaleblue)
\setquadratic
\plot
.75 8
1.5 3
3 1.28
3.4 1.25
4 1.2
/
\normale
\setlinear
\plot 2.5 0 2.5 1.4 /
\put{$\sc\bullet$} at 2.5 0
\put{$\bar t_e$} [t] at 2.5 -.2
\plot 3.5 0 3.5 1.23 /
\put{$\sc\bullet$} at 3.5 0
\put{$\bar t_r$} [t] at 3.5 -.2
\put{$\ell^\sharp_{BO}$} [lb] at 2.8 .2
\put{\firma} at 4 2
\textcolor{blue}{
\setshadegrid span <1pt>
\vshade 2.5 0 1.4 3.5 0 1.23 /}
\plot 2 0 2 1.9 /
\put{$\sc\bullet$} at 2 0
\put{$t_r$} [t] at 2 -.2
\plot 1.46 0 1.46 3.05 /
\put{$\sc\bullet$} at 1.46 0
\put{$t_e$} [t] at 1.46 -.2
\textcolor{red}{
\setshadegrid span <1pt>
\vshade 1.46 0 3.05 2 0 1.8  /}
\put{$\ell^\sharp_{BO}$} [lb] at 1.5 .3
\arrow <6pt> [.2,.5] from 2 .4 to 2.4 .4 
\put{the shaded area.} [l] at 2 -3
\endpicture
$
\vskip -10mm 
  \caption{\small Shift of the emission-reception intervals preserving}
\label{fig:erp-3}
\end{figure}

If a second photon is emitted at $\bar t_e>t_e$ then we have 
a different reception time $\bar t_r>t_r$. The two emission-reception intervals have (in general) different magnitudes: $\bar t_r-\bar t_e\neq t_r-t_e$. However, the two shaded areas over these intervals remain unchanged since both of them are equal to $\ell^\sharp_{BO}$, according to equation \eqref{e:erp-1} (Fig.\ \ref{fig:erp-3}). In other words, the shaded area behaves as a {\it planar incompressible fluid} constrained to stay under the graph of $c/a(t,t_\sharp)$ and upon the emission-reception interval $[t_e,t_r]$.

\section{Event horizon}

The topic of the previous section is based on the existence of a finite reception time $t_r$. However, one could object that the photon might not have enough time to reach $O$ before the end of the universe (at the time $t_\omega$). In fact, this happens when
\be\label{e:erp-1**}
\bboxed{\ell(t_e,t_\omega;t_\sharp)\Def c\dint_{t_e}^{t_\omega}\dfrac {dt'}{a(t',t_\sharp)}<
\ell^\sharp_{BO}}
\ee
as illustrated in Fig.\ \ref{fig:erp-000}:
\begin{figure} [H]
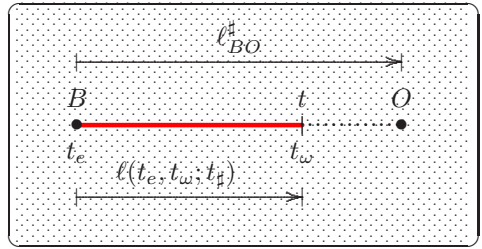

  \centering
$
\beginpicture
\setcoordinatesystem units <1.2cm,1.2cm>
\setplotarea x from -3.3 to 3.4, y from -1.3 to 1.3
\normalgraphs
\put{
\cornersize{.1}
\ovalbox{
\begin{minipage}{5.9cm}
\begin{minipage}{5cm}
\rule[-3cm]{5mm}{0cm}
\end{minipage}
\end{minipage}}
} at 0 0
\setshadegrid span <2pt>
\setlinear
\vshade -2.5 -1.4 1.4 2.6 -1.4 1.4 /
\setplotsymbol (\spessinored)
\setlinear
\plot -1.8 0 .7 0 /
\setplotsymbol (\spessino)
\setdots<3pt>
\plot .7 0 1.8 0 /
\setsolid
\put{$\bullet$} at -1.8 0
\put{$\bullet$} at 1.8 0
\put{$B$} [b] at -1.8 .2
\put{$O$} [b] at 1.8 .2
\put{$t_e$} [t] at -1.8 -.2
\put{$\sc |$} at .7 0
\put{$t_\omega$} [t] at .7 -.2
\normale
\arrow <6pt> [.2,.6] from -1.8 -.8 to .7 -.8
\put{$\sc |$} at .7 0
\put{$\sc |$} at -1.8 -.8
\put{$\sc |$} at .7 -.8 
\put{$t$} [b] at .7 .2
\put{$\ell(t_e,t_\omega;t_\sharp)$} [b] at -.7 -.7
\arrow <6pt> [.2,.6] from -1.8 .7 to 1.8 .7
\put{$\sc |$} at -1.8 .7
\put{$\sc |$} at 1.8 .7
\put{$\ell^\sharp_{BO}$} [b] at 0 .8
\endpicture
$
\vskip -1mm
  \caption{Graphic representation of \eqref{e:erp-1**}.}
  \label{fig:erp-000}
\end{figure} 

\vskip -8pt
Alternatively, one can consider the limit case where the reception time coincides with the finish time of the universe. In this case, denoting by $t_\star$ the emission time, we have
\be\label{e:erp-1***}
\bboxed{\ell(t_\star,t_\omega;t_\sharp)\Def c\dint_{t_\star}^{t_\omega}\dfrac {dt'}{a(t',t_\sharp)}=
\ell^\sharp_{BO}}
\ee
\vskip -8pt
\begin{figure} [H]
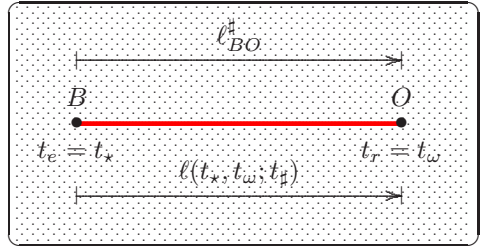

  \centering
$
\beginpicture
\setcoordinatesystem units <1.2cm,1.2cm>
\setplotarea x from -3.3 to 3.4, y from -1.3 to 1.3
\normalgraphs
\put{
\cornersize{.1}
\ovalbox{
\begin{minipage}{5.9cm}
\begin{minipage}{5cm}
\rule[-3cm]{5mm}{0cm}
\end{minipage}
\end{minipage}}
} at 0 0
\setshadegrid span <2pt>
\setlinear
\vshade -2.5 -1.4 1.4 2.6 -1.4 1.4 /
\setplotsymbol (\spessinored)
\setlinear
\plot -1.8 0 1.8 0 /
\put{$\bullet$} at -1.8 0
\put{$\bullet$} at 1.8 0
\put{$B$} [b] at -1.8 .2
\put{$O$} [b] at 1.8 .2
\put{$t_e=t_\star$} [t] at -1.8 -.2
\put{$t_r=t_\omega$} [t] at 1.8 -.2
\normale
\arrow <6pt> [.2,.6] from -1.8 -.8 to 1.8 -.8
\put{$\sc |$} at -1.8 -.8
\put{$\sc |$} at 1.8 -.8
\put{$\ell(t_\star,t_\omega;t_\sharp)$} [b] at 0 -.7
\arrow <6pt> [.2,.6] from -1.8 .7 to 1.8 .7
\put{$\sc |$} at -1.8 .7
\put{$\sc |$} at 1.8 .7
\put{$\ell^\sharp_{BO}$} [b] at 0 .8
\endpicture
$
\vskip -1mm
  \caption{Graphic representation of \eqref{e:erp-1***}.}
  \label{fig:erp-0000}
\end{figure} 

\vskip -8pt
If $\bar t_e>t_\star$ is a second emission time, then $\ell(\bar t_e,t_\omega;t_\sharp) < \ell(t_\star,t_\omega;t_\sharp)$ i.e.
$$
\ell(\bar t_e,t_\omega;t_\sharp) <\ell^\sharp_{BO}
$$
and we fall in the previous case: {\it any photon emitted after $t_\star$ never reaches $O$}. So, {\it $t_\star$ is the boundary after which no event occurring on $B$ will be observed from $O$}: it is the {\bf event boundary} of $B$.

This general argument finds a concrete and fruitful application  in the case of a dust-matter flat universe. Indeed, in the case we have an estimate of the age of the universe $t_\Z$, so that we can take $t_\Z$ as reference time. Hence, by virtue of \eqref{e:aux-c-0} and \eqref{e:def-c0}, the boundary $t_\star$ for which \eqref{e:erp-1***} holds,
\be\label{e:eh-1***}
\ell(t_\star,+\infty;t_\Z)\Def c\dint_{t_\star}^{+\infty}\dfrac {dt'}{a(t',t_\Z)}=
\ell^\Z_{BO}
\ee
is implicitly defined by equation
\be\label{e:eh-1*}
\dfrac c{\sqrt[3]{\ff c_\Z}}\!\dint_{t_\star}^{+\infty}\!\!\dfrac {dt'}{\sqrt[3]{\cosh(\beta\, t')-1}}=\ell^\Z_{BO},
\ee
where $\ff c_\Z\simeq 0.2299194811$ (dimensionless) and $\beta \simeq 0.170901087343\;Gyr^{-1}$.

For $t_\star=t_\Z$ (today) we find what is called the (present-day) {\bf radius of the event horizon}:
\be\label{e:e-horizon}
\bboxed{R_{eh}(t_\Z)\Def \ell(t_\Z,+\infty;t_\Z)\simeq 16.702920561\;Gly}
\ee
{\it The {\bf event horizon} encloses the set of bodies $B$ from which the photons emitted by $t_\Z$ onwards  will never be received by an observer $O$ in the 
future}.\footnote{
\kern 2pt
 If the time $t$ is expressed in Gyr units and we want  $\ell_{eh}(t_\Z)$ expressed in light-years units, then we have to put $c=1$ in \eqref{e:ph-3.} 
(the speed of light is equal to one light-year per year). Since 
$$
\dfrac 1{\sqrt[3]{\ff c_\Z}}\simeq \dfrac {1}{\sqrt[3]{0.2299194811}}\;Gly
\simeq 1.6323304282\;Gly,
$$
$\beta \simeq 0.170901087343\;Gyr^{-1}$, and $t_\Z \simeq 13.81658101781\;Gyr$, the formula to be used for the estimation of the radius is
$$
1.6323304282*\dint_{13.81658101781}^{\hbox{\ssmall very large}}\!\!
\dfrac{dx}{\sqrt[3]{\cosh( 0.170901087343*x)-1}}.
$$
Our estimate is in agreement with that provided in \cite{Lineweaver}: $\boxed{\simeq 16,4\;Glyr}$.
}

\section{Particle horizon}

According to the physics of the early universe there is a date $t_\star>0$ 
 in which the photons began to spread freely in the universe, as a consequence of a phenomenon called {\it recombination}, whose current estimate is $t_\star\simeq 378,000 \;yr$.  Thus, according to \eqref{e:erp-1}, the distance traveled by a photon from $t_\star$ to a time $t$ is given by
\be\label{e:ph-1.}
\ell(t_\star,t;t_\sharp)= c\!\dint_{t_\star}^{t}\!\!\dfrac {dz}{a(z,t_\sharp)}.
\ee
If at the time $t$ the co-moving distance $\ell^\sharp_{BO}$ is greater than $\ell(t_\star,t;t_\sharp)$, 
\be\label{e:ph-2.}
\ell(t_\star,t;t_\sharp)<\ell^\sharp_{BO},
\ee
then the body $B$ is {\it not yet visible} to the observer $O$. In this regard it must be noted that, due to \eqref{e:at0t*}, 
$$
\ell(t_\star,t;t_\sharp)=c\!\dint_{t_\star}^{t}\!\!\dfrac {dz}{a(z,t_\sharp)}
=\dfrac c{a(t_*,t_\sharp)}\!\dint_{t_\star}^{t}\!\!\dfrac {dz}{a(z,t_\flat)}=a(t_\sharp,t_\flat)\, \ell(t_\star,t_\Z;t_\flat)
$$
and that, due to \eqref{e:qma-3}, $\ell^\sharp_{BO}=a(t_\sharp,t_\flat)\,\ell^\flat_{BO}$. Hence, in changing the reference time, the inequality \eqref{e:ph-2.} remains invariant. More precisely: both members are multiplied by the same factor $a(t_\sharp,t_\flat)$.

In the case of a dust-matter flat model, for which we have an estimate of the age of the universe $t_\Z$, we can take $t_\Z$ as reference time. Hence, by virtue of \eqref{e:aux-c-0} and \eqref{e:def-c0}, from \eqref{e:ph-1.} we get  
\be\label{e:ph-3.}
\ell(t_\star,t;t_\Z)=c\!\dint_{t_\star}^{t}\!\!\dfrac {dt'}{a(t',t_\Z)}
=\dfrac c{\sqrt[3]{\ff c_\Z}}\!\dint_{t_\star}^{t}\!\!\dfrac {dt'}{\sqrt[3]{\cosh(\beta\, t')-1}}
\ee
where $\ff c_\Z\simeq 0.2299194811$. For $t=t_\Z$ we find\footnote{
\kern 2pt
For this evaluation we follow the same procedure as for $R_{eh}(t_\Z)$, footnote of page \pageref{e:e-horizon}. Since $t_\star\simeq 0.000378\;Gyr$
the formula to be used for this computation is 
$$
1.63233*\dint_{0.000378}^{ 13.816581}\!\!
\dfrac{dx}{\sqrt[3]{\cosh( 0.1709*x)-1}}.
$$
}
\be\label{e:p-horizon}
\bboxed{R_{ph}(t_\Z)\Def\ell(t_\star,t_\Z;t_\Z)\simeq 45.627196784\;Gly}
\ee
The meaning of this length is the following: {\it a radiating body $B$ is currently not visible by an observer $O$  if 
\be\label{e:ph-4.}
R_{ph}(t_\Z)<\ell^\Z_{BO},
\ee
where $\ell^\Z_{BO}$ is the present-day proper distance of $B$ and $O$}. $R_{ph}(t_\Z)$ is called the (present-day) {\bf radius of the visible universe} or {\bf radius of the particle horizon}.\footnote{
\kern 2pt See for instance  \cite{Mukhanov} (Section 2.2) and \cite{Ryden}.}  
The result \eqref{e:p-horizon} is in good agreement with the current estimate of $\boxed{\simeq\; 46}$ billion light years.

\section{Red-shift}

\begin{figure} [H]
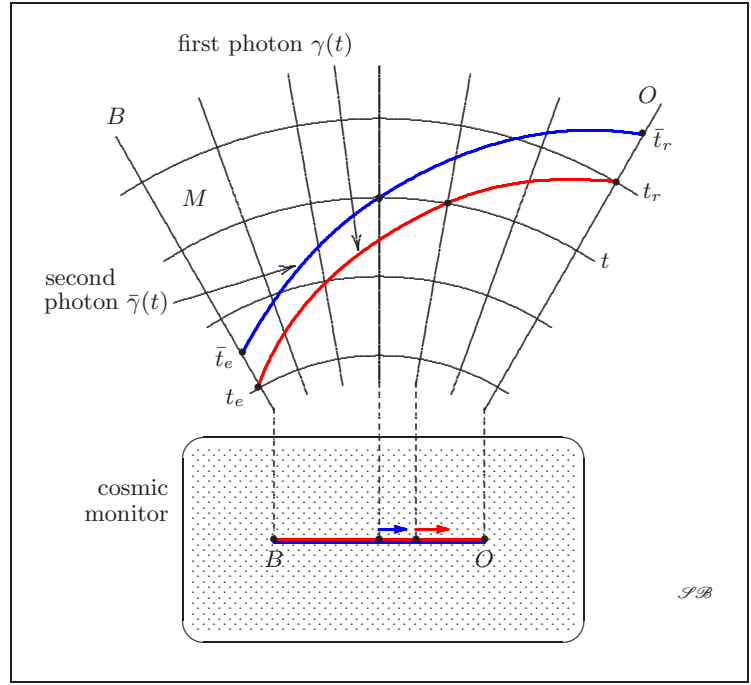
 
  \centering
$
\beginpicture
\setcoordinatesystem units <.7cm,.7cm>
\setplotarea x from -7 to 7, y from -4.7 to 8.2
\normalgraphs
\small
\grid 1 1
\setlinear
\plot 0 1 0 7 /
\startrotation by .984 -.173 about 0 -3 
\plot 0 1 0 7 /
\stoprotation
\startrotation by .939 -.342 about 0 -3 
\plot 0 1 0 7 /
\stoprotation
\startrotation by .866 -.5 about 0 -3 
\plot 0 1 0 7.7 /
\stoprotation
\setlinear
\plot 0 1 0 7 /
\startrotation by .984 .173 about 0 -3 
\plot 0 1 0 7 /
\stoprotation
\startrotation by .939 .342 about 0 -3 
\plot 0 1 0 7 /
\stoprotation
\startrotation by .866 .5 about 0 -3 
\plot 0 1 0 7 /
\stoprotation
\circulararc 33 degrees from 0 1.5 center at 0 -3
\circulararc -33 degrees from 0 1.5 center at 0 -3
\circulararc 33 degrees from 0 3 center at 0 -3
\circulararc -33 degrees from 0 3 center at 0 -3
\circulararc 33 degrees from 0 4.5 center at 0 -3
\circulararc -33 degrees from 0 4.5 center at 0 -3
\circulararc 33 degrees from 0 6 center at 0 -3
\circulararc -33 degrees from 0 6 center at 0 -3
\put{$t_e$} [r] at -2.55 .7
\put{$\bar t_e$} [r] at -2.75 1.45
\put{$t$} [l] at 4.2 3.3
\put{$t_r$} [l] at 5 4.6
\put{$\bar t_r$} [l] at 5.2 5.6
\setplotsymbol (\normalered)
\setquadratic
\plot
-2.3 .9
-1 2.9
1.3 4.4
2.8 4.8
4.5 4.8 
/
\setplotsymbol (\normaleblue)
\setquadratic
\plot
-2.6 1.55
-1.4 3.3
0 4.5
2.5 5.6
5 5.7
/
\normale
\put{$\sc\bullet$} at 0 4.5
\put{$\sc\bullet$} at -2.3 .9 
\put{$\sc\bullet$} at 1.3 4.4
\put{$\sc\bullet$} at 4.5 4.8
\put{$\sc\bullet$} at -2.6 1.55 
\put{$\sc\bullet$} at 0 4.5
\put{$\sc\bullet$} at 5 5.72
\put{$M$} at -3.5 4.5
\put{$B$} [b] at -5 5.9
\put{$O$} [b] at 5.1 6.3
\put{first photon $\gamma(t)$} [r] at -.5 7.4
\arrow <6pt> [.2,.6] from -.85 7 to -.4 3.5
\put{second} [r] at -5 2.99
\put{photon $\bar\gamma(t)$} [r] at -4 2.5
\arrow <6pt> [.2,.6] from -3.9 2.5 to -1.6 3.2
\setlinear
\setdashes <2pt>
\plot -2 .5 -2 -2 /
\plot 2 .5 2 -2 /
\plot .7 1 .7 -2 /
\plot 0 1 0 -2 /
\setsolid
\put{
\cornersize{.2}
\ovalbox{
\begin{minipage}{5cm}
\begin{minipage}{4.5cm}
\rule[-2.5cm]{0mm}{0cm}
\end{minipage}
\end{minipage}}
} at 0 -2
\put{
\setshadegrid span <2pt>
\setlinear
\vshade -3.7 -1.8 1.8 3.7 -1.8 1.8 /
} at 0 -2
\put{cosmic} [r] at -4 -1
\put{monitor} [r] at -4 -1.5
\setplotsymbol (\normalered)
\setlinear
\plot -2 -2  2 -2 /
\setplotsymbol (\normaleblue)
\setlinear
\plot -2 -2.05  2 -2.05 /
\put{$\sc\bullet$} [t] at -2 -1.92
\put{$B$} [t] at -2 -2.2
\put{$\sc\bullet$} [t] at 2 -1.92
\put{$O$} [t] at 2 -2.2
\put{$\sc\bullet$} [t] at .7 -1.92
\put{$\sc\bullet$} [t] at 0 -1.92
\setplotsymbol (\normaleblue)
\arrow <4pt> [.2,.6] from 0 -1.8 to .5 -1.8
\setplotsymbol (\normalered)
\arrow <4pt> [.2,.6] from .7 -1.8 to 1.3 -1.8
\normale
\put{\firma} at 6 -3
\endpicture
$
  \caption{\small Two photons observed on the cosmic monitor.}
  \label{fig:rsh-1}
\end{figure}
\vskip -3mm

Let us go back to the end of Section \ref{s:erp}, p.\ \pageref{fig:erp-3}. Assume that the emission time $\bar t_e$ of the second photon is very close to $t_e$ (Fig.\ \ref{fig:rsh-1}). From Fig.\ \ref{fig:rsh-2} we infer that the areas over the intervals $I_{er}=[t_e,t_r]$ and $\bar I_{er}=[\bar t_e,\bar t_r]$ are both equal to the co-moving  distance $\ell^\sharp_{BO}$. Since the central blank area is a common part of these two  areas, these two shaded areas are equal. If the magnitude of the intervals $I_e=[t_e,\bar t_e]$ and $I_r=[t_r,\bar t_r]$ are `extremely smaller' than the intervals $I_{er}$ and $\bar I_{er}$, 
 then the shaded areas can be considered equal to {\it width $\times$ height} of the rectangles where they are contained. So, we can write with `great precision'\footnote{
\kern 2pt This argument is taken, with minor modifications, from \cite{Liddle}, pp.126-127.} 

$$
\dfrac{I_e}{a(t_e,t_\sharp)}=\dfrac{I_r}{a(t_r,t_\sharp)} 
$$
i.e.,
\be\label{e:red1}
\boxed{\vp\;\dfrac{a(t_r,t_\sharp)}{a(t_e,t_\sharp)}=\dfrac{I_r}{I_e}\;}
\ee

\begin{figure} [H]
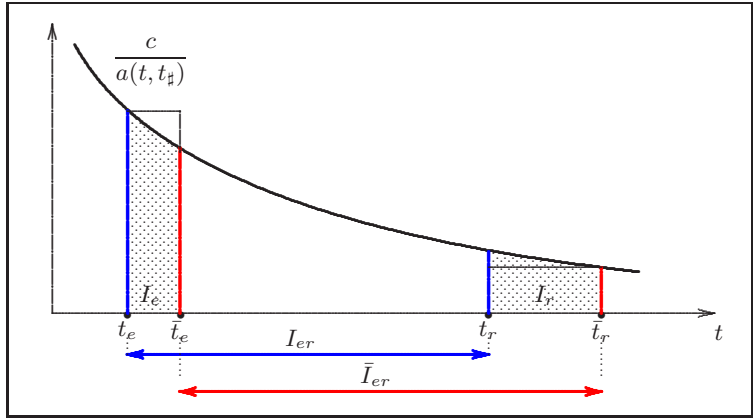
 
  \centering
$
\beginpicture
\setcoordinatesystem units <1cm,.55cm>
\setplotarea x from 0 to 9, y from -2.5 to 8
\normalgraphs
\small
\putrectangle corners at -.4 7.5 and 9.5 -2.5
\arrow <6pt> [.2,.5] from .2 0  to 9 0
\arrow <6pt> [.2,.5] from .2 0  to .2 7
\put{$t$} [lt] at 9 -.3
\put{$\dfrac c{a(t,t_\sharp)}$} [lb] at 1 5.5
\setplotsymbol(\spessino)
\setquadratic
\plot
.5 6.5
3 3
8 1 /
\normale
\put{$t_e$} [t] at 1.2 -.2
\put{$\sc\bullet$} [t] at 1.2 .05
\put{$\bar t_e$} [t] at 1.9 -.2
\put{$\sc\bullet$} [t] at 1.91 .05
\put{$t_r$} [t] at 6 -.2
\put{$\sc\bullet$} [t] at 6 .05
\put{$\bar t_r$} [t] at 7.5 -.2
\put{$\sc\bullet$} [t] at 7.5 .05
\put{$I_e$} [b] at 1.5 .2
\put{$I_r$} [b] at 6.75 .2
\setlinear
\setplotsymbol (\normaleblue)
\plot 1.2 0 1.2 4.9 /
\plot 6 0 6 1.5 /
\arrow <6pt> [.2,.5] from 1.2 -1  to 6 -1
\arrow <6pt> [.2,.5] from 6 -1  to 1.2 -1
\put{$I_{er}$} [b] at 3.5 -.85
\setplotsymbol (\normalered)
\plot 1.9 0 1.9 4 /
\plot 7.5 0 7.5 1.1 /
\arrow <6pt> [.2,.5] from 1.9 -1.9  to 7.5 -1.9
\arrow <6pt> [.2,.5] from 7.5 -1.9  to 1.9 -1.9
\put{$\bar I_{er}$} [b] at 4.5 -1.75
\normale
\setdots<2pt>
\plot 1.2 -.6 1.2 -1 /
\plot 6 -.6 6 -1 /
\plot 1.9 -.6 1.9 -1.6 /
\plot 7.5 -.6 7.5 -1.6 /
\setsolid
\setshadegrid span <1.5pt>
\vshade 1.2 0 4.9 1.9 0 3.9 /
\plot 1.2 4.9 1.9 4.9 1.9 3.9 /
\vshade 6 0 1.5 7.5 0 1 /
\plot 6 1.1 7.5 1.1 /
\endpicture
$
\vskip -1mm  
  \caption{\small The two shaded areas are equal.}
  \label{fig:rsh-2}
\end{figure}

This argument can be correctly applied when the two events of emission of the two photons  correspond to successive crests of a monochromatic light-wave emitted by $B$ with wavelength 
$$
\lambda_e=c\,I_e.
$$
In this case the two reception-events correspond to successive crests of the same light-wave received by $O$ with wavelength 
$$
\lambda_r=c\,I_r.
$$
Then \eqref{e:red1} is translated into equation 
\be\label{e:red2}
\bboxed{\dfrac{a(t_r,t_\sharp)}{a(t_e,t_\sharp)}=\dfrac{\lambda_r}{\lambda_e}}
\ee
This formula describes the well-known {\bf spectral shift phenomenon}: 
$$
\left\{
\begin{array}{lccc}
a(t_r,t_\sharp)>a(t_e,t_\sharp) &\iff& \lambda_r>\lambda_e &\iff 
\ax
a(t_r,t_\sharp)<a(t_e,t_\sharp) &\iff& \lambda_r<\lambda_e &\iff 
\ea
\right.
$$
$$
\left\{
\begin{array}{ll}
\iff& \hbox{shift of the original wavelength towards the red}
\ax
\iff& \hbox{shift of the original wavelength towards the blue}
\ea
\right.
$$

\br 
The term `shift' sounds like `translational displacement', and this may cause a misunderstanding. In fact, if we write \eqref{e:red2} in the form
$$
\lambda_r=\dfrac{a(t_r,t_\sharp)}{a(t_e,t_\sharp)}\,\lambda_e
$$
then we observe that the spectrum of a galaxy is multiplied by $a(t_r,t_\sharp)/a(t_e,t_\sharp)$ and not translated as a whole. \er

\section{Red-shift versus the emission-time}

If we introduce the so-called  {\bf red-shift parameter}
\be\label{e:red3}
\bboxed{z\Def\dfrac{\lambda_r-\lambda_e}{\lambda_e}=\dfrac{\lambda_r}{\lambda_e}-1}
\ee
then equation \eqref{e:red2} can be written as
\be\label{e:red4}
\bboxed{\dfrac{a(t_r,t_\sharp)}{a(t_e,t_\sharp)}=1+z}
\ee
This formula can be used for determining the time of emission $t_e$ of a photon from a galaxy $B$ knowing the red-shift $z$ observed from another galaxy $O$. Indeed, since the reception time $t_r$ is equal to the today time $t_\Z$, then by taking the reference time $t_\sharp$ of the profile $a(t,t_\sharp)$ equal to $t_\Z$ equation \eqref{e:red4} reduces to
\be\label{e:red4-0}
\bboxed{\dfrac 1{a(t_e,t_\Z)}=1+z}
\ee
because $a(t_\Z,t_\Z)=1$. As a consequence, if we know the analytic expression of the profile  $a(t,t_\Z)$ then from \eqref{e:red4-0} we can extract the emission time $t_e$ as a function of $z$.

We apply this result to the barotropic flat model with $u=1$ (pure matter). 
\bt
If we know the present-day red-shift $z$ of a galaxy $B$ observed from a galaxy $O$, then the dimensionless emission time $x_e$ is given by 
\begin{gather}
\label{e:xe-1}
\bboxed{x_e=\arccosh(y)=\log\left(y+\sqrt{y^2-1}\right)
}
\\
\label{e:xe-2}
\bboxed{y\Def\dfrac 1{\ff c_\Z\,(1+z)^3}+1
=\dfrac 1{\ff c_\Z}\left(\dfrac{\lambda_e}{\lambda_r}\right)\!\!{\vpt}^3
+1, \rule[-10pt]{0mm}{25pt}
\quad \ff c_\Z\simeq 0.2299194811}
\end{gather} 
\et 
\bprf From \eqref{e:aux-c-0} we derive $\;\;a^3(1;x,x_\Z)=\ff c_\Z\,\big(\cosh(x)-1\big)$ 

$\To$ 
$\ff c_\Z^{-1}\,a^3(1;x,x_\Z)=\cosh(x)-1$ 
$\To$ 
$\cosh(x)=\ff c_\Z^{-1}\,a^3(1;x,x_\Z)+1$

$\To$ 
$x=\arccosh[\ff c_\Z^{-1}\,a^3(1;x,x_\Z)+1]$ 
$\To$ $x_e=\arccosh[\ff c_\Z^{-1}\,a^3(1;x_e,x_\Z)+1]$.

By virtue of \eqref{e:red4-0} we obtain the emission dimensionless time $x_e$. \eprf

The emission time $t_e$ in Gyr is given by $t_e=x_e/\beta$ with 
$\beta^{-1}\simeq 5.8513378\,Gyr$ (table \ref{tab:Data-supp}, page \pageref{tab:Data-supp}). Thus, the formula to be used for computing the emission cosmic time $t_e$ is 
\be\label{e:tez}
\bboxed{t_e(z)=5.8513378 * \arccosh\left(\vp \dfrac 1{0.22991948 *(1+z)^3}+1    \right)}
\ee

\begin{figure} [H]
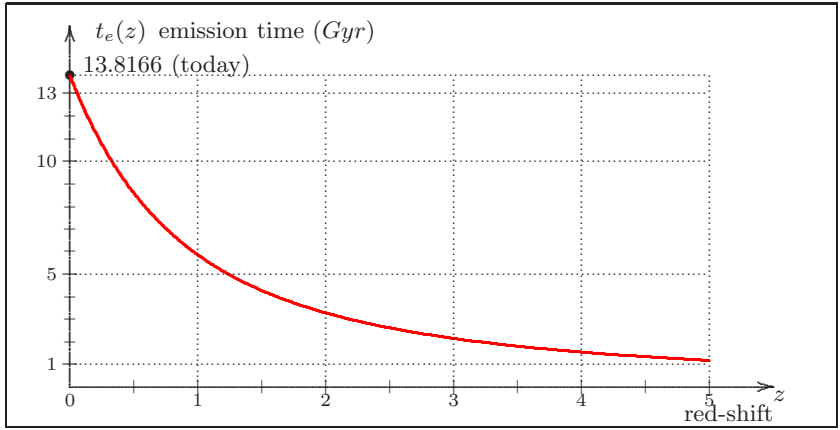

\centering
$
\beginpicture
\setcoordinatesystem units <1.7cm,.3cm>
\setplotarea x from -.5 to 6, y from -1.8 to 17
\normalgraphs
\small
\grid 1 1
\arrow <6pt> [.2,.6] from 0 0 to 5.5 0
\arrow <6pt> [.2,.6] from 0 0 to 0 16
\put{\small $t_e(z)\,$  emission time ($Gyr$)} [lb] at .2 15.2
\put{$z$} [lt] at 5.5 -.2
\put{\small red-shift} [rt] at 5.5 -.8
\put{$\bullet$} at 0.0	13.8166
\put{\small 13.8166  (today)} [l] at .1 14.2
\put{\ssmall 0} [t] at 0 -.3
\put{$\sc |$} at 1 0
\put{\ssmall 1} [t] at 1 -.3
\put{$\sc |$} at 2 0
\put{\ssmall 2} [t] at 2 -.3
\put{$\sc |$} at 3 0
\put{\ssmall 3} [t] at 3 -.3
\put{$\sc |$} at 4 0
\put{\ssmall 4} [t] at 4 -.3
\put{$\sc |$} at 5 0
\put{\ssmall 5} [t] at 5 -.3
\put{$-$} at 0 1
\put{\ssmall 1} [r] at -.1 1
\put{$-$} at 0 5
\put{\ssmall 5} [r] at -.1 5
\put{$-$} at 0 10
\put{\ssmall 10} [r] at -.1 10
\put{$-$} at 0 13
\put{\ssmall 13} [r] at -.1 13
\put{$\sc -$} at 0 2
\put{$\sc -$} at 0 3
\put{$\sc -$} at 0 4
\put{$\sc -$} at 0 6
\put{$\sc -$} at 0 7
\put{$\sc -$} at 0 8
\put{$\sc -$} at 0 9
\put{$\sc -$} at 0 11
\put{$\sc -$} at 0 12
\put{$\scc |$} at .5 0
\put{$\scc |$} at 1.5 0
\put{$\scc |$} at 2.5 0
\put{$\scc |$} at 3.5 0
\put{$\scc |$} at 4.5 0
\setquadratic
\setplotsymbol (\normalered) 
\plot
0.0	13.8166
0.1	12.4646
0.2	11.2925
0.3	10.2722
0.4	9.3808
0.5	8.5990
0.6	7.9107
0.7	7.3027
0.8	6.7633
0.9	6.2832
1.0	5.8543
1.1	5.4696
1.2	5.1234
1.3	4.8109
1.4	4.5278
1.5	4.2705
1.6	4.0360
1.7	3.8216
1.8	3.6251
1.9	3.4445
2.0	3.2782
2.1	3.1246
2.2	2.9824
2.3	2.8505
2.4	2.7280
2.5	2.6138
2.6	2.5073
2.7	2.4078
2.8	2.3146
2.9	2.2272
3.0	2.1452
3.1	2.0680
3.2	1.9953
3.3	1.9267
3.4	1.8620
3.5	1.8007
3.6	1.7428
3.7	1.6878
3.8	1.6357
3.9	1.5862
4.0	1.5391
4.1	1.4943
4.2	1.4517
4.3	1.4110
4.4	1.3721
4.5	1.3350
4.6	1.2996
4.7	1.2657
4.8	1.2332
4.9	1.2021
5.0	1.1723
/
\normale
\setlinear
\setdots <2pt>
\plot 0 13.81 5 13.81 5 0 /
\plot 0 13 5 13 /
\plot 0 10 5 10 /
\plot 0 5 5 5 /
\plot 0 1 5 1 /
\plot 1 0 1 13.81 /
\plot 2 0 2 13.81 /
\plot 3 0 3 13.81 /
\plot 4 0 4 13.81 /
\endpicture
$
\vskip -1mm
\caption{Red-shift $z$ versus the emission time $t_e$.}
  \label{fig:z-te}
\end{figure}

\begin{minipage}[c]{1 \textwidth} 
\begin{table}[H] 
\centering
\begin{tabular}{|c|c|}
\hline
$\vpt$ red-shift & emission time 
\\
\rule[-2mm]{0mm}{3mm}$z$  & $t_e\,(Glyr)$  
\\
\hline\hline
\rule[-1mm]{-1.5mm}{5mm} 0.0 & 13.8166 
\\
0.0005 & ...
\\
0.001 & ...
\\
0.002 & ...
\\
0.003 & ...
\\
0.004 & ...
\\
0.005 & ...
\\
0.006 & ...
\\
0.1 &	12.4646
\\
0.2 &	11.2925
\\
0.3 &	10.2722
\\
0.4 &	9.3808
\\
0.5 &	8.5990
\\
0.6 &	7.9107
\\
0.7 &	7.3027
\\
0.8 &	6.7633
\\
\rule[-2mm]{-1.5mm}{3mm} 0.9 &	6.2832
\\
\hline
\end{tabular}
\quad
\begin{tabular}{|c|c|}
\hline
$\vpt$ red-shift & emission time 
\\
\rule[-2mm]{0mm}{3mm}$z$  & $t_e\,(Glyr)$  
\\
\hline\hline
\rule[-1mm]{-1.5mm}{5mm} 
1.0 &	5.8543
\\
1.1 & 5.4696
\\
1.2 & 5.1234
\\
1.3 &	4.8109
\\
1.4 &	4.5278
\\
1.5 &	4.2705
\\
1.6 &	4.0360
\\
1.7 &	3.8216
\\
1.8 &	3.6251
\\
1.9 &	3.4445
\\
2.0 &	3.2782
\\
2.1 &	3.1246
\\
2.2 &	2.9824
\\
2.3 &	2.8505
\\
2.4 &	2.7280
\\
2.5 &	2.6138
\\
\rule[-2mm]{-1.5mm}{3mm} 2.6 &	2.5073
\\
\hline
\end{tabular}
\caption{Red-shift $z$ versus the emission time $t_e$.}\label{tab:z-te}
\end{table}
\end{minipage}


\begin{thebibliography}{90}


\bibitem{Besse} 
Besse A.L., {\it Einstein Manifolds}, Springer, 1987.



\bibitem{CODATA}
CODATA, {\it Recommended Values of the Fundamental Physical 
Constants}, 2014 (July 2015).


\bibitem{Davis-Lineweaver} Davis T.M., Lineweaver C.H.,  {\it Expanding confusion: common misconceptions of cosmological horizons and the superluminal expansion of the universe}, Publications of the Astronomical Society of Australia, 2004, 21, 97-109.

\bibitem{Eisenhart}
Eisenhart L.P., {\it Riemannian Geometry}, Princeton Univ.\ Press, 1997.

\bibitem{Friedman}
Friedman  A., {\it \"Uber die Kr\"ummung des Raumes}, Zeitschrift f\"ur Physik, 10 (1), 1922, 377-386.





\bibitem{Lichnerowicz-1955}
Lichnerowicz A., {\it Théories relativistes de la gravitation  et de l'électromagnetisme}, Masson \& C.ie, 1955.

\bibitem{Lichnerowicz-1967}
Lichnerowicz A., {\it Relativistic Hydrodynamics and Magnetohydrodynamics}, Benjamin, Inc, 1967.

\bibitem{Liddle}
Liddle A., {\it An Introduction to Modern Cosmology}, Wiley 1998.

\bibitem{Lineweaver}
Lineweaver C.H., Egan C.A., {\it Dark energy and the entropy of the observable universe}, Invisible Universe Conference Proceedings 2010, AIP Conference Proceedings 1241, p.645, Springer.

\bibitem{MTW}
Misner W., Thorne K.S.,  Wheeler J.A., {\it Gravitation}, W.H.\ Freeman \& Company, N.Y., 1973.


\bibitem{Mukhanov}
Mukhanov V., {\it Physical foundations of cosmology}, Cambridge Univ.\ Press, 2005.

\bibitem{Narlikar}
Narlikar J.L., {\it An Introduction to Cosmology}, Cambridge Univ.\ Press, 2002.



\bibitem{Perlmutter}
Perlmutter S., Supernovae, Dark Energy, and the Accelerating Universe. {\it Physics Today}, April 2003.

\bibitem{Riess}
Riess A.G., My path to the accelerating universe. {\it The Nobel Lecture}, December 8, 2011.

\bibitem{Ryden}
Ryden B., {\it Introduction to Cosmology}, Addison Wesley, 2003.




\bibitem{Weinberg}
Weinberg S., {\it Gravitation and Cosmology},  John Wiley \& Sons, 1972.

\bibitem{Wolf}
Wolf J.A., {\it Spaces of constant curvature},  Publish or Perish, Inc., 1984.
\end{thebibliography}
\end{document}